\documentclass[a4paper,twoside]{ociamthesis}
\title{Precision QCD at the LHC: from the structure of the proton to all-order resummations}
\author{Luca Rottoli}
\college{New College}

\degree{Doctor of Philosophy}
\degreedate{Trinity 2018}

	\newcommand{\rd}{\textsuperscript{rd}}
	
\usepackage{xspace}	
\newcommand{\amper}{\protect\textsc{\&}\xspace}
\usepackage{slashed}
\usepackage{cancel}

\newcommand{\PRLsep}{\noindent\makebox[\linewidth]{\resizebox{0.3333\linewidth}{1pt}{$\bullet$}}\bigskip}

\usepackage[OT1]{fontenc}

\usepackage[upint,libertine]{newtxmath}
\usepackage{libertine}
\usepackage{inconsolata}
\usepackage[scr=rsfso]{mathalfa}
\usepackage{bm}

\renewcommand*{\chapnumfont}{%
  \usefont{OML}{fxl}{b}{n}\fontsize{100}{130}\selectfont
  \color{chaptergrey}%
}

\usepackage{lettrine}

\usepackage{titlesec}
\titleformat*{\section}{\large\scshape}
\titleformat*{\subsection}{\itshape}
\titleformat*{\subsubsection}{\itshape}

\usepackage{tocloft}
\addto\captionsenglish{

}

\usepackage[labelfont=it]{caption}

\newcommand{\Vsc}{V}

\newcommand{\Ndat}{N_{\text{dat}}}
\newcommand{\LambdaQ}{\Lambda_{\text{QCD}}}
\newcommand{\MSbar}{\overline{\text{MS}}}
\newcommand{\CSS}{\text{CSS}}
\newcommand{\mH}{m_H}
\newcommand{\mt}{m_t}
\newcommand{\muF}{\mu_{\text{F}}}
\newcommand{\muR}{\mu_{\text{R}}}
\newcommand{\muH}{\mu_{\text{H}}}
\newcommand{\muS}{\mu_{\text{S}}}

\def\({\left(}
\def\){\right)}
\def\[{\left[}
\def\]{\right]}

\DeclareRobustCommand{\ensuremathrm}[1]{\ensuremath{\text{#1}}\xspace}
\DeclareRobustCommand{\rd}{\ensuremathrm{d}} 

\newcommand{\eps}{\epsilon}

\newcommand{\pth}{p_{t}^{\text{ H}}}

\newcommand{\ptgo}{p_{t}^{\gamma_1}}
\newcommand{\ptgt}{p_{t}^{\gamma_2}}
\newcommand{\ptgg}{p_{t}^{\gamma\gamma}}

\newcommand{\etag}{\eta^{\gamma_i}}
\newcommand{\ptg}{\eta^{\gamma_i}}
\newcommand{\Sud}{\mathcal{S}}

\def\bea#1\eea{\begin{align}#1\end{align}}
\def\bean#1\eean{\begin{align*}#1\end{align*}}
\def\beq{\begin{equation}}  
\def\eeq{\end{equation}}

\newcommand{\dZ}{\rd\mathcal{Z}[\{{R}', k_i\}]}

\newcommand{\M}{\mathcal M}

\newcommand{\as}{\alpha_s}
\newcolumntype{C}[1]{>{\centering\let\newline\\\arraybackslash\hspace{0pt}}m{#1}}

\allowdisplaybreaks

\addto{\captionsenglish}{%
  
}
\usepackage{url}
\usepackage{cite}
\usepackage{blindtext}
\usepackage{booktabs}
\usepackage[normalem]{ulem}

\begin{document}


\setlength{\textbaselineskip}{22pt plus2pt}

\setlength{\frontmatterbaselineskip}{17pt plus1pt minus1pt}

\setlength{\baselineskip}{\textbaselineskip}


\setcounter{secnumdepth}{2}
\setcounter{tocdepth}{2}


\begin{romanpages}

\maketitle


\begin{abstract}
	\vspace*{2cm}

\begin{center}
	\huge Abstract
\end{center}

\vspace{1cm}

Experiments at the LHC are collecting a wealth of data with an unprecedented level of precision.
As a consequence, the theoretical error is now starting to lag behind the experimental one, and a ceaseless effort is required to reduce the theory uncertainty to match the precision of the data. 
At hadron colliders, QCD predictions are obtained by convoluting perturbative parton-level results with non-perturbative parton distribution functions (PDFs), whose precise determination is crucial to reach percent accuracy in the theoretical predictions.

At the parton level, cross sections are obtained through a perturbative expansion in the strong coupling parameter. 
In some cases, large terms appear at all orders, spoiling the convergence of the series. 
The perturbative description is rescued by performing an all-order resummation of the series, thereby making theory calculations accurate in regions where a fixed-order treatment is not sufficient.

In this thesis I first present two global PDF sets where fixed-order calculations are supplemented by threshold and high-energy resummation, respectively.
In the first case, it is found that including resummation into PDFs can compensate for the enhancement in the partonic cross sections, with implications for high-mass resonance searches.
In the second case, resummation quantitatively improves the description of the HERA structure functions, thus providing evidence of the onset of a new dynamical regime of QCD in the HERA data.

I then focus on Higgs production in gluon fusion.
The effect of threshold resummation on the total cross section is found to significantly improve the convergence of the perturbative series and to provide a robust method for estimating missing higher order uncertainty. 
Finally, I present predictions for the Higgs transverse-momentum spectrum both in the inclusive case and within fiducial cuts, exploiting a novel approach where transverse-momentum resummation is performed in direct space.

\end{abstract}

\begin{statement}
	\vspace*{0.5cm}

\begin{center}
	\huge Declaration
\end{center}
	
\vspace{1cm}

This thesis is a record of the work performed during my DPhil at the University of Oxford and contains research originally published as follows.

\vspace{0.5cm}

\noindent The discussion in section~\ref{sec:factheavyquarks} contains results presented in

\begin{itemize}


	\item 
  	R.~D.~Ball, M.~Bonvini and L.~Rottoli,
  	\textit{Charm in Deep-Inelastic Scattering},
  JHEP { 1511} (2015) 122
  [arXiv:1510.02491 [hep-ph]].
  
\end{itemize}

\noindent Part of section~\ref{sec:threshold} includes material discussed in  
 
\begin{itemize}
	\item M.~Bonvini and L.~Rottoli,
  \textit{Three loop soft function for $N^3LL^\prime$ gluon fusion Higgs production in soft-collinear effective theory},
  Phys.\ Rev.\ D {91} (2015) no.5,  051301
  [arXiv:1412.3791 [hep-ph]].
\end{itemize}

\noindent Section~\ref{sec:radish} contains results originally appeared in
 
\begin{itemize}
	\item 
  W.~Bizon, P.~F.~Monni, E.~Re, L.~Rottoli and P.~Torrielli,
  \textit{Momentum-space resummation for transverse observables and the Higgs p$_{\perp}$ at $N^{3}LL+NNLO$},
  JHEP {1802} (2018) 108
  [arXiv:1705.09127 [hep-ph]].
\end{itemize}

\noindent Chapter~\ref{ch-respdfs} is based upon worked published in

\begin{itemize}
\item   M.~Bonvini, S. Marzani, J. Rojo, L. Rottoli, M. Ubiali {\it et al.},
  \textit{Parton distributions with threshold resummation},
  JHEP {1509} (2015) 191
  [arXiv:1507.01006 [hep-ph]].

\item
  R.~D.~Ball, V.~Bertone, M.~Bonvini, S.~Marzani, J.~Rojo and L.~Rottoli,
  \textit{Parton distributions with small-x resummation: evidence for BFKL dynamics in HERA data,}
  Eur.\ Phys.\ J.\ C {78} (2018) no.4,  321
  [arXiv:1710.05935 [hep-ph]].

\end{itemize}

\noindent Chapter~\ref{ch-respheno} is based upon results presented in

\begin{itemize}
\item
  M.~Bonvini, S.~Marzani, C.~Muselli and L.~Rottoli,
 \textit{On the Higgs cross section at $N^{3}LO+N^{3}LL$ and its uncertainty,}
  JHEP {1608} (2016) 105
  [arXiv:1603.08000 [hep-ph]].

\item  W.~Bizon, X. Chen, A. Gehrmann-De Ridder, T. Gehrmann, N. Glover, A. Huss, P. F. Monni, E. Re, L. Rottoli and P. Torielli,
  \textit{Fiducial distributions in Higgs and Drell-Yan production at $N^3LL+NNLO$},
  arXiv:1805.05916 [hep-ph].

\end{itemize}

\noindent Unless explicitly stated in the caption, I have produced all the figures that appear in this work, which will often closely follow those in the corresponding publications.

%
%

%
%

\cleardoublepage

\end{statement}

\begin{dedication}
\textit{Alle mie nonne e ai miei nonni}
\end{dedication}

\begin{acknowledgements}
 	\vspace*{-1.5cm}

\begin{flushright}
	\huge Acknowledgements
\end{flushright}

\vspace{1cm}

First and foremost, I am extremely grateful to my supervisor Juan Rojo.
I learnt a lot from him, and I thank him for his constant support, for being a guide in the academic world, and for always encouraging me to pursue a variety of research interests during my DPhil.
Without him, this thesis could not exist. 
I am also deeply indebted to Stefano Forte, who first introduced me to the research world and who always has the time, and the patience, to answer to my questions, no matter how trivial, and to teach me something new. 
I would like to thank a lot also Giovanni Ridolfi, who has been an inspiring figure and a kind friend, whose advice is always precious.

Among my collaborators, a very special role has been played by Marco Bonvini, Pier Monni, and Simone Marzani.
Marco has been way more than a collaborator: he has been a teacher, a mentor and a friend: I feel very lucky to have met him already at the time of my Master's thesis.
Almost all I know about threshold and small-$x$ resummation I owe to his clear and concise explanations.
Moreover, my plots and my code would not look the same without his caustic, punctual, yet always sincere comments --- though sometimes I find his drive for (or perhaps I should say delusion of) perfection just a tad too excessive.
Pier always manages to make many intricacies of resummation as simple as it can be, but not simple(r).  
I always discover a lot about the technical aspects of QCD whenever we speak, and at the same time I realize how much I still need to learn. 
I have learnt a lot about QCD also from Simone in all the conversations we had --- mostly over Skype --- in the past few years.
Besides, I thank Simone because he is always serene and even-tempered; his disposition has been an example in the (occasionally) opinionated and arrogant academic world.

I am deeply grateful to all the members of the NNPDF collaboration for what they taught me (still, one should probably ask elsewhere for a trouble-proof conference call system).
I would like to thank especially the people with whom I had the luck to work in Oxford in these four years for their constant help and for having answered hundreds of questions: Nathan Hartland, who helped me to grasp the basics of the NNPDF language (and tried to improve my English as well); Valerio Bertone, despite he makes fun of the way I pronunce {\it schiscetta}; Emanuele Nocera, {\it vir bonus dicendi peritus}, whose pungent irony and sense of humor I will always envy. 
I would also like to extend my gratitude to Richard Ball and Jos\'e Ignacio Latorre for many clarifying discussions, and to Stefano Carrazza and Zahari Kassabov for their constant support and assistance.
Very very special thanks go to Emma Slade, who shared with me the thrill of being a NNPDF DPhil student in Oxford, for listening to my rants, for all the whimsical British words she taught me, and for kindly helping me proofread this thesis.

Throughout my DPhil I have been fortunate to work with many other exceptional people such as Alba Cervera-Lierta, Rhorry Gauld, Ulrich Haisch, Claudio Muselli, V. Ravindran, Emanuele Re, Subir Sarkar, Jim Talbert, and Paolo Torrielli.
I would like to extend my sincere thanks to all of them for the many inspiring discussions we had and for the very productive collaborations in the past years. 
I also wish to acknowledge Fady Bishara, Fabrizio Caola, and Giulia Zanderighi for many interesting discussion and their advice.
Andrea Banfi and Lucian Harland-Lang have kindly agreed to be my examiners and provided me with very valuable advice to improve this thesis and to clarify various discussions: they both deserve huge thanks.
I also wish to thank all the people I had the opportunity to meet all over the world during my DPhil, either at conferences or at summer and winter schools.
Many thanks especially go to Matteo Tuveri for sharing beers, food, thoughts, and his love for music, guitar and Sardinia with me.

My daily life in the department would not have been the same without all the wonderful people I met, who allowed me to spend my working hours in such a great environment. 
Thanks a lot to all the fellow students, postdocs and staff members who made amazing these four years at the Rudolf Peierls Centre for Theoretical Physics.
Thanks to Michelle Bosher and Natalie Wells for being always helpful, friendly and cheerful and for creating the cake club, albeit quite at the very end of my time in Oxford.
I will miss Michelle's laugh echoing around the Beecroft building.
I could not present many results collected in this thesis without Jonathan Patterson, who helped me run code on the Hydra cluster.
Thanks to Joe Conlon for taking the co-supervisor role after Juan moved to Amsterdam and for being always available for a chat or advice. 

I wish to thank William Astill and Wojciech `the wizard' Bizon, my loyal office mates in the past four years. 
In particular, I am extremely grateful to Wojtek, for all I learnt in the disparate discussions we had which spanned from basic physics to the many little details of QCD, for being always a calm and reassuring presence whenever I was complaining about something (i.e. basically every day), for sharing the delight and burden of tutoring C6, and for being such a good friend.

%
Thanks to Francesco Giuli (aka Giulia) who knows a lot about everything, has been everywhere, and knows everyone. 
I will always remember the blissful time we spent together sharing the same office before he decided he would rather be \sout{a caveman} an experimentalist.
It has been great fun watching Ruggero, the G.O.A.T., invent tennis at Wimbledon, or cheering for The Rocket at the World Snooker Championship whilst our programs were running. 
I suppose I should also thank Alessandro Geraldini (aka Rocco), my --- barely decent --- housemate.
Despite our \textit{more uxorio} cohabitation, I regret to inform that the reports of our bromance are in fact greatly exaggerated.
He comes from Roma Nord, and that is all I have to say about him.
The word \textit{svro} best captures our acquaintance. 

I wish to thank also all the other friends I met in Oxford, who made my life in this city an unforgettable experience: Karl for being such a great housemate, though he lives in the East Coast time zone and tried to kill me and Ale with his home-brewed beer; Jutta, Gaia, Lorenzo, Ferdinando, Michael, Niccol\`o, Maria, Peter, for all the laughs we shared, the dinners we had, the pizza we ate and all the time we spent together; Manon and her cute French lilt; \.Idil, Patty, Jan, Virginia, Caro, Anabel, Meindert, Markus, and all the friends from New College and from the other Oxford colleges. 
And of course, I am also grateful to all my water polo team mates, with whom I spent so many evenings, for all the fun I had in the past years trying to save their shots.  

I must also thank all my old friends from Como: it has always been fun to come back even for a few days and laugh with them as when we were young. 
Thanks to Ricky, Fonta, Yas, Teo, Michele, and Marianna.
Thanks to Alberto for always being himself; Mezzule, especially because she came visiting against all the odds; and Chiara with whom I shared all the doubts about doing a PhD in Physics.
I would also like to thank all my fellow physics friends from Via Celoria: no matter if we are scattered around the word, it is always nice to bump into each other once in a while.

The last --- and the most obvious --- thanks go to those who, despite the distance, were always the closest, and who have always put up with me and with all my idiosyncrasies. 
I am very grateful to my family and relatives; these few words cannot really convey my gratitude to my parents and my sister for all the things they have done and for their unconditional support. 
There are not enough words to thank Franci, who is always \textit{here}, who understands me as no one else does, whose love makes all possible. 

\vspace{0.5cm}

\PRLsep

\vspace{0.2cm}

My DPhil has received funding from the European Research Council (ERC), under the European Union's Horizon 2020 research and innovation programme `Parton Distributions in the Higgs Boson Era', No.~335260.

\cleardoublepage

\end{acknowledgements}

\dominitoc 

\flushbottom


\setlength{\cftbeforetoctitleskip}{-3em}
\tableofcontents




\end{romanpages}



\flushbottom
\begin{savequote}[8cm]
\textlatin{The aim of science is not to open the door to infinite wisdom, but to set some limit on infinite error.}

  \qauthor{--- Bertold Brecht, \textit{Life of Galileo}}
\end{savequote}

\chapter*{Introduction}
\addcontentsline{toc}{chapter}{Introduction}
\markboth{Introduction}{Introduction}



\lettrine[lines=2]{W}{ith} the announcement of the discovery of the Higgs boson on 4 July 2012, the long-sought completion of the Standard Model (SM) of particle physics was finally achieved.
The SM took a long time to build: more than a century separates the discovery of the Higgs boson from that of the electron by J.J. Thomson, back in 1897.
This major achievement required the joint effort of thousands of experimental and theoretical physicists and many years spent in planning and building a machine powerful enough to finally detect the only missing particle in the SM puzzle: the Large Hadron Collider (LHC), the world's largest particle accelerator. 
In this circular accelerator, protons are currently collided at a centre-of-mass energy of 13 TeV.
The particles produced in the collisions are measured by four main experiments: ATLAS (A Toroidal Lhc ApparatuS), CMS (Compact Muon Solenoid), ALICE (A Large Ion Collider Experiment), and LHCb (LHC beauty).


Six years after the historical announcement of the Higgs boson discovery, the physics programme at the LHC delved into a new phase.
Indeed, though the SM successfully explains most of the known phenomena in elementary particle physics, there is strong evidence that it is not complete, as it fails to explain a number of observed phenomena, such as the baryon asymmetry in the universe, the existence of dark matter, neutrino masses, et cetera.
Explaining these phenomena demands new physics Beyond the Standard Model (BSM). 

The increasing luminosity delivered by the LHC --- which exceeds $100$ fb$^{-1}$ at 13~TeV in June 2018 --- brings both discovery reach and precision. 
On the one hand, this makes it possible to push forward the energy frontier in the hope of finding conclusive experimental signatures of BSM physics.
On the other hand, the large data sample allows the LHC to test several properties of the SM with increasing precision.
Since data collected so far suggest that new physics will not likely manifest as a direct signal, hints of new physics may indirectly appear as small deviations from the expected SM behaviour.
As a consequence, accurate theory calculations for collider processes are crucial to interpret the precise experimental data and to discern whether experimental measurements differ from what the SM predicts.
To match the precision of the data, theory uncertainties should be reduced to the few percent level.

In hadron colliders such as the LHC, the dominant theory uncertainties are those related to strong interaction physics, described by Quantum Chromo-Dynamics (QCD).
QCD is characterized by a coupling \mccorrect{parameter} which vanishes at large scales.
In this limit, cross sections can be computed using perturbative methods; yet, at the LHC the initial state consists of protons, which cannot be treated perturbatively.
General cross sections at the LHC are thus a combination of short- and long-distance behaviour, and as such they are not computable directly using perturbative QCD.
However, factorization theorems allow one to separate the short-distance from the long-distance behaviour.
The latter is encoded into universal parton distribution functions (PDFs): these are non-perturbative objects, which have to be determined from data and describe the distribution of quarks and gluons (collectively called partons) in the proton.
The quest for precision at the LHC thus requires the improvement of both the accuracy of perturbative calculations and our understanding of the initial (and final) state in hadron collisions.
Moreover, to obtain consistent predictions it is crucial that the advancement in perturbative calculations go hand in hand with PDFs determined with equal (or higher) theoretical accuracy.

To reduce the theory uncertainties to the few percent level, it is often necessary to have theoretical predictions at or beyond next-to-next-to-leading order (NNLO) accuracy, which is becoming today's state of the art for fixed-order computations.
In principle, to reach the desired accuracy in perturbative computations it should be sufficient to compute a finite number of terms in the perturbative expansion.
%
In practice, however, the computation of higher-order terms in the series becomes rapidly very challenging. 
Currently, only two hadron-collider processes are known at N$^3$LO, and both of them in some approximation. 
Moreover, there are cases in which large contributions --- for instance, logarithms of different energy scales on which the cross section depends --- appear at all orders.
In this case, the convergence of the perturbative series is spoiled.
The series must be reorganized and the enhanced terms must be summed to all  orders, an operation known as resummation.

Since most of the observables which are studied at the LHC generically depend on more than one energy scale, all order calculations often become necessary to describe the kinematic regimes in which the logarithms of ratios of these scales become large. 
Logarithmic enhancements of higher-order perturbative contributions require, in general, different resummation techniques.
For instance, such enhancements take place when the invariant mass of the final state approaches the kinematic threshold: this region is known as the threshold region, and the resummation of logarithms from this region is known as threshold resummation. 
Another class of logarithmic enhancement appears when the centre-of-mass energy of the partonic collision is much higher than the hard scale of the process.
The logarithms being resummed are known as high-energy logarithms.

Both kinematic regions are of particular interest in the context of parton distribution fits.
PDFs depend on a dimensionful scale $Q$, the hard scale of the process, and a dimensionless scale $x$, which represents the proton momentum fraction carried by the parton.
The extraction of precise PDFs thus depends not only on the precision of the available data, but also on a reliable theoretical description of the physical observables in the whole $(x,Q^2)$ range probed by the experiments.
Threshold and high-energy logarithms spoil the convergence of the perturbative series in the large-$x$ and small-$x$ region, respectively (for this reason, they are also called large-$x$ and small-$x$ logarithms). 
As a consequence, a fixed-order perturbative description of these regions may not be reliable enough to describe all the processes included in PDF fits.

In this thesis we will argue that the inclusion of resummation effects in PDF fits is beneficial for two main reasons.
Firstly, it allows one to perform consistent calculations in kinematic regions where the logarithmic enhancement is taken into account both in the hard cross sections and in the PDFs.
Secondly, the inclusion of resummation effects is useful to improve the quality of the fits in regions where a fixed-order description does not accurately reproduce the experimental data. 

Supplementing theoretical calculations with resummation becomes mandatory when one considers increasingly differential observables.
For differential distributions, the number of energy scales increases and so does the number of the kinematic regions which may develop logarithmic enhancements. 
Among differential observables, a primary example is the transverse-momentum distribution of a system with high-invariant mass $M$, produced with extra QCD radiation.
It is well known that the fixed-order description of the transverse-momentum distribution fails at small values of the transverse momentum $p_t$, due to the appearance of large logarithms of $p_t/M$ in the partonic cross sections.
In this thesis, we will introduce a novel approach to perform the resummation of this class of logarithms, focusing on the case of the Higgs $p_t$ spectrum.


Besides being crucial to describe the Higgs transverse-momentum spectrum in the kinematic region marred by large logarithms, resummed calculations are also advantageous for the total Higgs cross section in gluon fusion.
Indeed, even if Higgs production at the LHC happens far from threshold, it turns out that the threshold-enhanced terms, though somewhat moderate in size, represent a substantial part of the higher-order corrections to the cross section.   
Their resummation thus significantly improves the convergence of the perturbative series; moreover, it can provide a rather conservative --- yet robust --- way to assess the uncertainty from missing higher orders.

This thesis is divided into three main parts. 
In the first part, we review some aspects of perturbative QCD and all-order resummation techniques.
The second part treats the effect of resummation on parton distribution functions.
In the last part we focus on Higgs production in gluon fusion and we discuss the effect of threshold and of transverse-momentum resummation for the total cross section and the differential spectrum, respectively.

In particular, in chapter~\ref{ch-qcd} we introduce some basic ingredients of QCD.
We focus on the collinear factorization framework and in particular on the role of heavy quarks, especially the charm quark. 
Part of the material discussed in this chapter was first presented in ref.~\cite{Ball:2015dpa} in collaboration with Richard Ball and Marco Bonvini.

In chapter~\ref{ch-res} we present a review of various aspects of resummation in perturbative QCD. 
We first discuss threshold and transverse-momentum resummation, introducing a number of concepts which will be used for phenomenological studies in the second and third part of the thesis.
We then analyse some key features of high-energy resummation, with the aim of introducing the ingredients necessary to construct PDF sets with high-energy resummation.
Among other things, this chapter also discusses ideas and concepts appeared in various forms in refs.~\cite{Bonvini:2014tea,Bizon:2017rah} in collaboration with Wojciech Bizon, Marco Bonvini, Pier Monni, Emanuele Re, and Paolo Torielli.

In chapter~\ref{ch-respdfs} we present two PDF fits which include threshold and high-energy resummation, respectively, building upon the ingredients introduced in chapter~\ref{ch-res}.
The fits were first presented in ref.~\cite{Bonvini:2015ira}, in collaboration with Richard Ball, Valerio Bertone, Marco Bonvini, Stefano Carrazza, Nathan Hartland, Simone Marzani, Maria Ubiali, and Juan Rojo, and in ref.~\cite{Ball:2017otu}, in collaboration with Richard Ball, Valerio Bertone, Marco Bonvini, Simone Marzani, and Juan Rojo . 

In chapter~\ref{ch-respheno} we present phenomenological results for Higgs production in gluon fusion.
We first discuss the case of the total cross section focusing on the impact of threshold resummation and its curative properties on the perturbative series.
Most of this work was done in collaboration with Marco Bonvini, Simone Marzani and Claudio Muselli and previously presented in ref.~\cite{Bonvini:2016frm}. 
We then focus on the impact of transverse-momentum resummation on the Higgs spectrum.
This latter work was presented in ref.~\cite{Bizon:2018foh}, using the formalism discussed in ref.~\cite{Bizon:2017rah}, in collaboration with Wojciech Bizon, Pier Monni, Emanuele Re, Paolo Torielli, and with the authors of the \texttt{NNLOJET} code.

We finally summarize the work presented in this thesis and provide some concluding remarks in chapter~\ref{ch-summ}. 

%
%


\mtcaddchapter
\begin{savequote}[8cm]
\textlatin{The purest and most thoughtful minds are those which love colour the most.}

  \qauthor{---  John Ruskin, \textit{The Stones of Venice} }
\end{savequote}

\chapter{\label{ch-qcd}QCD \amper factorization} 


\lettrine[lines=2]{Q}{uantum} Chromo-Dynamics (QCD) is the modern theory of strong interactions. 
It describes the interactions of quarks and gluons, collectively called partons, and the processes by which they bind to form hadrons. 
The accurate understanding of QCD over a vast range of scales --- which span from the proton mass ($\sim$$1$ GeV) to the centre-of-mass energy of the hard-scattering processes (few TeV) --- is an essential requirement for precision physics at hadron colliders.

One of the key property of QCD is asymptotic freedom: the effective coupling of QCD goes to zero at short distances.
In this limit, partons can be considered quasi-free particles and perturbation theory can be exploited to calculate partonic cross sections by expanding in powers of the strong coupling. 
Conversely, at long distances the strong coupling grows and the perturbative picture breaks down. 

High-energy scattering cross sections are in general a combination of short- and long-distance behaviours, and are therefore not directly computable in perturbative QCD.
The predictive power of QCD is restored thanks to the concept of collinear factorization, which allows one to separate the short-distance, perturbative, process-dependent physics from the long-distance, non-perturbative physics, encoded in universal objects: the Parton Distribution Functions (PDFs). 

In this chapter we present a brief overview of QCD and we introduce some notation we shall use in this thesis. 
A comprehensive description of QCD and its properties is beyond the purposes of this chapter; we refer the interested reader to many excellent text books, e.g.~\cite{Altarelli:1981ax,Ellis:1991qj,Muta:1998vi,Dissertori:2003pj,Collins:2011zzd,Wells:2017uxd}.

\section{Basics of QCD}

In the early 1960s, the possibility to describe the strong interaction in terms of
a conventional quantum field theory was questioned by most physicists.
Despite the success of QED, the renormalization procedure was considered 
suspect, and searches for alternative approaches to renormalizable 
quantum field theories began.
One of the most promising candidates was the Veneziano model~\cite{Veneziano:1968yb}, which
however failed to explain the parton-like behaviour of the strong interaction
seen in certain kinematic limits. 
Between 1970 and 1974 it became clear that a renormalizable
Yang-Mills gauge theory~\cite{Yang:1954ek} could describe the experimental features observed, and QCD finally emerged as a successful theory of the 
strong interaction\footnote{A recollection of the discoveries of the 
early '70s and their contribution to the rehabilitation of
quantum field theories are presented for instance in~\cite{tHooft:1998qmr}.}.

\subsection{The QCD Lagrangian \amper its simmetries}

QCD is an unbroken non-abelian gauge theory based on the gauge group $SU(3)_{\textrm {colour}}$. 
Its matter fields, the quarks, come in six different types, called flavours.
The quarks are the basic constituents of the hadrons, whilst the binding is provided
by the eight massless gauge bosons, the gluons.

The rich dynamical content of QCD results from its deceptively simple Lagrangian,
which at the classical level reads
\begin{equation}
\label{eq:QCDlagr}
	\mathcal L_{\textrm{YM}} = -\frac{1}{4} \sum_{a=1}^8 F^{a\mu\nu} F^a_{\mu\nu} + 
	\sum_{j=1}^{n_f} \bar q_{j} (i \slashed D -m_j ) q_j,
\end{equation}
\mccorrect{where $q_j$ are the fermionic fields in the fundamental representation of $SU(3)_{\textrm {colour}}$, with masses $m_j$ and $n_f$ different flavours.}
The covariant derivative and the field strength are defined in terms of the 
gluonic field $A_\mu$ as
\begin{align}
	 D_\mu &= \partial_\mu + i g_s A_\mu^a t^a \\
	 F_{\mu \nu} &= (\partial_\mu A^a_\nu - \partial_\nu A^a_\mu -g_s f^{abc} A_\mu^b A_\nu^c ) t^a,
\end{align}
where $g_s$ is the gauge coupling. 
The matrices $t^a$ are the generators of the $SU(3)_{\textrm {colour}}$ Lie algebra in the fundamental representation
and satisfy
\begin{align}
	&[t^a,t^b]=i f^{abc}t^c, & {\text{ tr}} [t^a, t^b] = T_R \delta^{ab},
\end{align}
where $f^{abc}$ are the real structure constants of the algebra, $a,b,c=1,\ldots 8$, and it is customary to choose $T_R=1/2$. 
The invariants $C_F$ and $C_A$ are defined by
\begin{align}
	& \sum_a t^a_{ik} t^a_{kj} = C_F  \delta_{ij}, & \sum_{ab} f^{abc} f^{abd} = C_A \delta^{cd} . 
\end{align}

\begin{mccorrection}
The quantization of the classical Lagrangian equation requires a gauge fixing term
$\mathcal L_{\textrm{gauge-fixing}}$ and a ghost term $\mathcal L_{\textrm{ghost}}$. 
The reader can find details about the quantization procedure in any QFT textbook; 
here we simply recall that in the common covariant gauge one has
\begin{align}
	&\mathcal L_{\textrm{gauge-fixing}}= -\frac{1}{2\xi}(\partial^\mu A_\mu)^2,
	&\mathcal L_{\textrm{ghost}}= \partial^\mu {\bar \eta}^a (\partial_\mu \delta^{ab} + g f^{abc} A^c_\mu )\eta^b,
\end{align}
where $\xi$ is an arbitrary gauge parameter and the ghost fields $\eta$ are complex
scalar fields which obey Fermi-Dirac statistics. 
Another important example is the set of axial gauges $n^\mu A_\mu^a =0$, where $n$ is an arbitrary four-vector and the gauge-fixing and the ghost terms have the form (see e.g.~\cite{Leibbrandt:1987qv})
\begin{equation}
	\mathcal L_{\textrm{gauge-fixing}}= -\frac{1}{2\xi}(n^\mu A_\mu)^2, \qquad
	\mathcal L_{\textrm{ghosts}}= \bar \eta^a n^\mu  (\partial_\mu \delta^{ab} + g f^{abc} A^c_\mu ) \eta^b.	
\end{equation}
In the common light-cone gauge, $n^2=0$, $\xi =0$.
This class of gauges has the important property that ghosts are decoupled and can thus be ignored.
Though the gauge-fixing and the ghost terms are not invariant under gauge transformations, the full, gauge-fixed action fulfils the Becchi-Rouet-Stora-Tyutin (BRST) symmetry~\cite{Becchi:1975nq,Tyutin:1975qk}.
This ensures that all the new terms which appear at the quantum level are constrained by BRST invariance, which guarantees the renormalizability of the theory (see for instance~\cite{Weinberg:1996kr}).
\end{mccorrection}

Along with the local $SU(3)_{\textrm {colour}}$ gauge symmetry, the QCD Lagrangian is characterized by additional global symmetries.
The conservation of the baryon number corresponds to the exact, accidental\footnote{In renormalizable theories such as QCD, a symmetry is accidental if it is broken by non-renormalizable terms.} $U(1)$ symmetry. 
Given that the down and up quarks are almost degenerate in mass, the QCD Lagrangian has also an approximate, but accurate, $SU(2)$ symmetry: isospin. 
Less accurate is the $SU(3)$ symmetry, which would be exact if the masses of the three lightest flavours were equal; nevertheless, this symmetry is the basis of the Gell-Mann quark model~\cite{GellMann:1964nj}, proposed well before the birth of QCD.

In the limit of $n_f$ massless quarks, the two chiral components of the quark fields are independent and the QCD Lagrangian has the global chiral symmetry
\begin{equation}
	SU(n_f)_V \otimes U(1)_V \otimes SU(n_f)_A \otimes U(1)_A,
\end{equation}
which we have written using vector (V) and axial (A) combinations. 
If we consider $n_f =2$, we can identify $U(1)_V$ with the portion of baryon number carried by down and up quarks, and $SU(2)_V$ with isospin. 
As we do not observe any approximate symmetry which corresponds to $U(2)_A= SU(2)_A \otimes U(1)_A$, the axial symmetry must be broken. 
The symmetry is in fact spontaneously broken and one can identify the pions with the would-be massless Goldstone bosons associated with the breaking of $U(2)_L \otimes U(2)_R$ to $SU(2)_V  \otimes U(1)_V \otimes U(1)_A$. 

What is the physical realization of $U(1)_A$? 
The symmetry cannot be realized with the Wigner-Weyl mechanism, otherwise all strongly-interacting particles would have a partner with opposite parity.
However, it cannot be spontaneously broken as Steven Weinberg first observed in his renowned article {\it The $U(1)$ problem}~\cite{Weinberg:1975ui}. 
The solution to the long-standing issue of the $U(1)_A$ symmetry relies on the highly non-trivial QCD vacuum topology (instantons); at the quantum level, the axial symmetry is explicitly broken by the Adler-Bell-Jackiw anomaly~\cite{Adler:1969gk,Bell:1969ts}, which induces in the QCD Lagrangian a term proportional to $\varepsilon_{\mu\nu\rho\sigma} F^{\rho\sigma} F^{\mu \nu}$, whose strength is determined by a parameter $\theta$.
This extra term violates CP invariance in the strong sector. 
Whereas one would a priori expect a quantity of $\mathcal O(1)$, the experimental bound on the value of $\theta$ is particularly strong and $\theta \lesssim 10^{-10}$.
The lack of explanation for such a small value of $\theta$ is known as the `$\theta$-problem' or `Strong-CP problem'; among the possible solutions, a particularly elegant proposal by Peccei and Quinn involves a dynamical mechanism to explain the value of $\theta$ through the introduction of a (pseudo-)scalar field, the axion~\cite{Peccei:1977hh}. 

\subsection{The running of the strong coupling parameter}

By inspecting the QCD Lagrangian eq.~\eqref{eq:QCDlagr} we immediately note that the strength of the interaction is ruled by a single parameter, the gauge coupling $g_s$. 
It is customary to introduce the strong coupling parameter
\begin{align}
 \as = \frac{g_s^2}{4 \pi};
\end{align}
the quantization of the QCD Lagragian then allows one to derive the Feynman rules and to obtain a formal perturbative expansion in terms of $\as$. 
However, as infinities are encountered in the calculation of loop diagrams, it is necessary to regularize and to renormalize the theory in order to compute predictions for physical observables.

Whilst several regularization procedures are possible, the most common choice is dimensional regularization (DR), which preserves gauge and Lorentz invariance. 
In DR, integrals are made finite in the ultraviolet (UV) region by formulating the theory in $D=4-2\varepsilon$ dimensions and the divergences are isolated in poles in the parameter $\varepsilon$. 
Unless otherwise stated, we shall use DR in this thesis. 
In a renormalizable theory such as QCD the dependence on the cutoff $\varepsilon$ can be reabsorbed into a redefinition of the gauge coupling, particle masses, and wave function normalizations. 

Whereas one would expect QCD to be scale invariant in the limit of massless quarks, scale invariance symmetry is (anomalously) broken at the quantum level and the renormalized lagrangian parameters acquire a logarithmic dependence on a renormalization scale $\mu$, ruled by Renormalization Group Equations (RGEs).
In particular, the running coupling obeys the RGE
\begin{equation}
	\mu^2 \frac{\partial \as( \mu^2)}{\partial \mu^2} = \beta \left(\as ( \mu^2)\right),
\end{equation}
where $ \beta \left(\as( \mu^2) \right)$ is known as the QCD $\beta$-function and can be expressed as a power series in $\as ( \mu^2)$
\begin{equation}
	\beta\left(\as ( \mu^2)\right) = - \as^2 ( \mu^2)  \left(\beta_0+ \beta_1 \as ( \mu^2) + \beta_2 \as^2 ( \mu^2)+ \ldots\right).
\end{equation}
Provided that $\as ( \mu^2)$ is small, the coefficients of the $\beta$-function are computable in perturbation theory; currently, the $\beta$-function is known up to five loops~\cite{Herzog:2017ohr}. 

The RGE for the running coupling can be solved iteratively. 
At the lowest (leading logarithmic) order the solution is
\begin{equation}\label{eq:as1loop}
	\as(\mu^2)  = \frac{\as(\mu_0^2) }{1 + \beta_0 \as(\mu_0^2)\ln \frac{\mu^2}{\mu_0^2}},
\end{equation}
where $\mu_0$ is an initial scale at which the coupling is known or measured. 
The slope of the coupling depends on the sign of the $\beta$-function; in particular, the strength of the coupling decreases with $\mu^2$ if $\beta_0$ is positive, as it is the case in QCD\footnote{Since $\beta_0=(11 C_A-2n_f)/(12 \pi)$ this is true as long as the number of active flavours is less than 17.}. 
This property is called \emph{asymptotic freedom}. 
It has been proven~\cite{Coleman:1973sx} that in four spacetime dimensions no renormalizable field theory without non-Abelian gauge fields can be asymptotically free.

In the limit $\mu^2 \rightarrow 0$, however, the strong coupling parameter grows and one cannot rely on perturbative methods.
In QCD this behaviour is often parametrized by trading the scale $\mu_0$ with a scale  $\LambdaQ$ where the strong coupling diverges (Landau pole):
\begin{align}
		\frac{1}{\as(\LambdaQ^2)}  = 0.
\end{align}
It is clear that the value of $\LambdaQ$ is determined by the definition of $\as$.
Therefore, it depends on the scale $\mu_0$, on the renormalization scheme, and on the order of the $\beta$-function; in particular, through the $\beta$-function it depends also on the number of active flavours $n_f$. 

The scale $\LambdaQ$ marks the boundary between the perturbative and non-perturbative regimes. 
At this scale, which is typically of the order of a few hundred MeV, QCD becomes strongly coupled. 
This is consistent with the property of \emph{confinement}: the impossibility of separating colour charges. 
Whilst lattice simulations provide valuable insight on our understanding of QCD at low energies, a complete explanation of confinement is still missing. 

\subsection{Infrared \amper collinear singularities in QCD}\label{sec: IRC safety}

As we have seen, although one would naively expect massless QCD to be scale invariant, departures from scaling appear. 
However, these deviations are computable, scale logarithmically and are asymptotically suppressed. 
Asymptotic freedom allows us to use perturbative methods at sufficiently high energies and to compute scattering amplitudes in massless QCD as an expansion in the parameter $\as$. 
In this regime, quarks and gluons behave as free particles, and perturbative QCD offers an excellent approximation to the exact theoretical description. 

In massive QCD one needs to introduce additional mass corrections proportional to the masses of the quarks, which are free parameters of the theory. 
As the RG evolution for quark masses has the same asymptotic behaviour of the running coupling RGE, namely
\begin{equation}
	\lim_{\mu \rightarrow \infty} m_q (\mu^2) = 0, 
\end{equation}
quark masses can be neglected as long as the energy of the process is much larger than the mass of the quarks. 
Therefore, in many processes at hadron colliders one typically takes into account only the masses of the top and bottom quarks.

As a consequence, at the parton level we can use the asymptotic predictions of massless QCD for the class of processes where all relevant energies $E_i$ are sufficiently large compared to $\Lambda_{\rm QCD}$.
In massive QCD there would be additional mass corrections suppressed by powers of $m_q/E_i$ (as long as the processes are finite in the limit $m_q\rightarrow 0$). 
To guarantee that the predictions are meaningful one must also ensure that the processes being computed are well-defined in perturbation theory.
This translates into the requirement of Infrared and Collinear (IRC) safety: processes must be free from infrared and collinear divergences.

\begin{figure}[t]
\begin{center}
\includegraphics{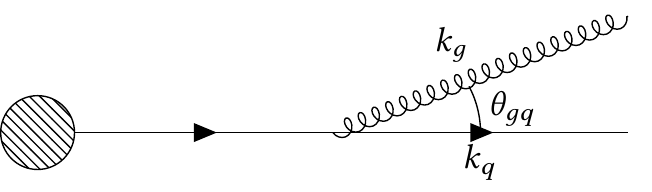}
 \caption{Splitting of a virtual quark into a gluon and a quark.}
  \label{fig:qsplitting}	
\end{center}
 \end{figure}

These divergences arise naturally both in virtual and in real diagrams. 
In virtual diagrams, infrared singularities appear when the loop momentum in loop integrals probes the IR region. 
In real diagrams, singularities appear when integrating over the phase space of the emitted particles.
For instance, let us consider the splitting of a virtual quark line into a quark and a gluon line, as shown in fig.~\ref{fig:qsplitting}.
In the case of massless quarks, the internal propagator is
\begin{equation}
	\frac{1}{(k_g+k_q)^2} = \frac{1}{2 E_g E_q} \frac{1}{1-\cos \theta_{gq}} \overset{\theta_{gq} \ll 1}{\sim} \frac{1}{E_q} \frac{1}{E_g \theta^2_{gq}},
\end{equation}
which diverges if the energy of the massless gluon goes to zero (soft singularity), or if the angle between the emitted gluon and the quark goes to zero (collinear singularity), or both.

More precisely, one can show that in the soft limit the squared matrix element and the phase space for the gluon emission off a $q \bar q$ system factorize (see e.g.~\cite{Salam:2010zt})
\begin{equation}
	|\mathcal M_{q\bar q g}|^2 d \Phi_{q \bar q g} \approx |\mathcal M_{q\bar q}|^2 d \Phi_{q \bar q} d \omega.
\end{equation}
The probability of soft gluon emission $d\omega$ in the soft and collinear limit is
\begin{equation}
	d \omega = \frac{\as C_F}{\pi} \frac{dE}{E} \frac{d \theta^2}{\theta^2} \frac{d \phi}{2 \pi },
\end{equation}
where $\theta$ and $\phi$ are the polar and the azimuthal angle of the gluon with respect to the quark, respectively, and $E$ is the energy of the gluon.
This result shows that each singularity gives rise to a single logarithmic divergence, leading to a \emph{double} logarithmic divergence in the soft and collinear limit.
Though discussed in this simple context, the soft and collinear singularities are an essential feature of QCD which appear whenever soft gluons are radiated, irrespective of the process. 
Let us note that the mass of the quark $m_q$ acts as a regulator for collinear singularities; in this case, the propagator reads
\begin{equation}
	\frac{1}{(k_g+k_q)^2-m_q^2} = \frac{1}{2 E_g E_q} \frac{1}{1-\beta_q \cos \theta_{gq}},
\end{equation}
where $\beta_q = \sqrt{1-m_q^2/E_q^2}$. 
For this reason, sometimes collinear singularities are also called mass singularities.

There exist two important theorems about the cancellation of collinear and soft singularities, which are based on the quantum-mechanical concept of summing over all possible configurations which lead to the same configuration in the final state. 
The first, known as the Bloch-Nordsieck theorem~\cite{Bloch:1937pw}, was formalized in the QED framework and states that IR singularities cancel when one adds up all the final states which appear as indistinguishable given a certain detector resolution. 
The theorem was later generalized to QCD by Kinoshita, Lee, and Nauenberg~\cite{Kinoshita:1962ur,Lee:1964is}. 
The Kinoshita-Lee-Nauenberg (KLN) theorem states that divergences related to mass singularities cancel when the summation over all the degenerate states (i.e. which have the same mass) is carried out.
That is, in the case of a one-particle final state with mass $m$, one should also add all final states that in the limit $m\rightarrow 0$ have the same mass and in particular gluons and massless quark pairs. 

Whereas the KLN theorem ensures that for completely inclusive observables \mccorrect{(for instance, a total cross section, where there is no need to identify all the particles in the final state and their properties)} IR singularities in QCD cancel, in real life we would like to compute also more exclusive observables to obtain additional information about the physics of the collisions. 
Therefore, one needs to define precisely which properties observables should have to allow for their computation in perturbation theory. 
The first definition of IRC safety was proposed by Sterman and Weinberg in a seminal paper~\cite{Sterman:1977wj}.
In words, it states that the quantity that is computed or measured should be insensitive to the radiation of soft and collinear gluons, i.e. its value should not change when adding one (or many) soft gluons, or replacing a particle with two (or many) collinear gluons.
This is yet another statement that well-defined observables cannot resolve long-distance physics. 

\begin{figure}[t]
\begin{center}
\includegraphics{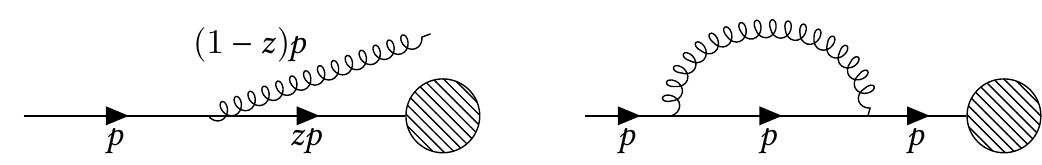}
 \caption{Kinematic differences between real and virtual contributions to initial state splitting are irrelevant in the soft limit ($z \rightarrow 1$) but relevant in the collinear limit $z \neq 1$.}
  \label{fig:IRinitial}	
\end{center}
 \end{figure}

Observables to which perturbative QCD  may apply must fulfill the IRC safety criterion.
For such observables, only mass singularities associated to the initial state particles remain. 
Indeed, in the case of initial-state IR divergences, collinear singularities do not cancel because of the different hard-scattering kinematics of the real and the virtual contributions (see fig.~\ref{fig:IRinitial}).
\mccorrect{Indeed, the momentum which flows in the hard process is different for real and virtual terms ($zp$ and $p$, respectively), and therefore the difference between the real and the virtual contributions becomes irrelevant only in the strictly soft limit  ($z \rightarrow 1$).}
As we shall see in section~\ref{sec:factorization}, these singularities are absorbed into universal, non-perturbative functions.

\subsection{Jets \amper hadronization}

\begin{mccorrection}
So far we have discussed QCD at the partonic level, focusing on the fields which appear in the Lagrangian.
In real life, however, free quarks and gluons do not exist; in particular, the final state consists of bunches of collimated hadrons: these objects are called \emph{jets}. 
Jets are defined through jet algorithms, which are sets of rules which allow one to cluster the particles which appear in the final state.
Some general properties for their definitions were first collected in the `Snowmass accord' in 1990~\cite{Huth:1990mi}.

During the years, many jet algorithms have been proposed. 
They usually depend on a set of parameters which allow one to determine whether two particles belong to the same jet or not and on a recombination scheme, which associates a 4-momentum to the combination of two particles.
Jet algorithms can be broadly divided into two categories: cone algorithms and sequential recombination jet algorithms; for an exhaustive review, see~\cite{Salam:2009jx}.
Good jet definitions can be applied both to experimental measurements as well as to the output of a parton level calculation; the physical observables are then constructed using these new objects.

To match experimental measurements and parton-level predictions it is further necessary to describe \emph{hadronization}: the transition between what can be calculated theoretically, in terms of partons, into an experimentally measurable quantity, in terms of hadrons.
Currently, a complete theoretical description of hadronization is still missing, though various phenomenological models are available. 
The general approach to hadronization is based on the concept of \emph{duality}: partons and hadrons are considered as equivalent sets of complete states.
This is quite safe to assume in the case of totally inclusive distributions (global duality) and it is supposed to also hold in the case of more exclusive computations (local duality).
Within this assumption, the observables are defined at the parton level and considered equivalent to what is observed in a real detector. 
Corrections to this picture are generally `higher twist', that is suppressed by powers of $\LambdaQ$ over some characteristic scale of the process, and therefore get smaller at higher energies. 
However, at present collider energies some non-perturbative hadronization model is needed to obtain precise predictions to be compared with experimental data.
\end{mccorrection}

\section{Factorization in hadronic collisions}\label{sec:factorization}

A short introduction to QCD cannot be complete without discussing in some depth how to compute theoretical predictions in the presence of initial state hadrons, which is paramount at hadron colliders such as the LHC.
This section will be thus dedicated to the cardinal concept of collinear factorization.

%

\subsection{Deep inelastic scattering}\label{sec:DISkin}

The simplest process which involves a hadron in the initial state is inclusive Deep Inelastic Scattering (DIS): the scattering between a lepton and a proton in which an electroweak boson with large virtuality is exchanged. 
The DIS process historically played, and still plays, a very important role in strong-interaction physics and in our understanding of the structure of the nucleon.
Furthermore, it provides a rich theoretical and experimental laboratory to quantitatively study and test QCD. 

\begin{figure}[t]
\begin{center}
\includegraphics{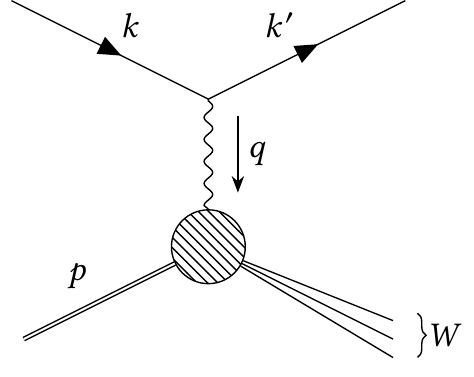}
 \caption{Deep inelastic scattering kinematics.}
  \label{fig:DISkin}	
\end{center}
 \end{figure}

We parametrize the scattering as in fig.~\ref{fig:DISkin} .
The (space-like) lepton transverse momentum is $q = k - k'$, with $Q^2 \equiv - q^2 > 0 $.
The centre-of-mass energy squared $s$ is $(p+k)^2$ and we define the invariant mass squared of the hadronic final state as $W^2=(p+q)^2$.
In the deep-inelastic regime $Q^2 \gg m^2_{\text{hadron}}$, $W^2 \gg m^2_{\text{hadron}} $ and we can neglect hadron and lepton masses.
Deep inelastic scattering allows one to probe the structure of the nucleon up to short distances $\lambda \sim1/Q \ll 1/m_{\text{hadron}} $.

It is customary to introduce the two dimensionless variables
\begin{align}\label{eq:xandy}
	&x = \frac{Q^2}{2 p \cdot q}, 
	&y = \frac{2 p \cdot q}{s}.
\end{align}
The variable $x$ is known as Bjorken variable and from kinematics it can take values between 0 and 1, where 1 is the elastic limit. 
The inelasticity $y$ represents the fraction of energy transferred by the scattered lepton \mccorrect{in the target rest frame} and is also kinematically allowed to take values between 0 and 1. 
At fixed $s$, the differential cross section is a function of $x$ and $y$, or $x$ and $Q^2$. 

The differential cross section, which is proportional to the matrix element squared, can be factorized into the product of two contributions, called respectively leptonic ($L_{\mu \nu}$) and hadronic ($W_{\mu \nu}$) tensors:
\begin{align}
	d \sigma \sim L_{\mu \nu} W^{\mu \nu}.
\end{align}
Whereas the leptonic tensor can be easily obtained from the lowest-order electroweak vertex (plus its radiative corrections), the complicated strong interactions are wholly contained in the hadronic tensor. 
Its general form is constrained by Lorentz invariance and current conservation.
In the case of neutral current (NC) unpolarized scattering, and considering only photon exchange (which dominates as long as $Q^2\ll m_Z^2$) the most general hadronic tensor can be written as
\begin{align}
	W^{\mu \nu} (p,q) &= \left( -g^{\mu \nu} - \frac{q^\mu q^\nu}{q^2} \right) F_1 (x, Q^2) + \frac{1}{p \cdot q} \left( p^\mu - \frac{p\cdot q}{q^2} q^\mu\right) \left( p^\nu - \frac{p\cdot q}{q^2} q^\nu\right) F_2  (x,Q^2),
\end{align}
where $F_{1,2}$ are two scalar functions, known as structure functions; in the case of charge currents (CC), or by allowing also a $Z$ boson to be exchanged in the neutral-current case, a parity-violating structure function $F_3$ is needed to parametrize the hadronic tensor, see e.g.~\cite{Patrignani:2016xqp}.

The results are often expressed as a function of $F_2$ and $F_L \equiv F_2 - 2x F_1$, which is related to the longitudinal component of the polarized virtual photon;
by computing explicitly the leptonic tensor one can show that
\begin{equation}\label{eq:dissigma}
	\frac{d \sigma}{d Q^2 d x} = \frac{2 \pi \alpha^2}{x Q^4} \left\{ [1+(1-y)^2] F_2 (x, Q^2) - y^2 F_L (x,Q^2 \right\},
\end{equation}
where $\alpha$ is the fine structure constant. 
In eq.~\eqref{eq:dissigma}, the prefactor and the $y$-dependent parts encode the dependence on the electroweak dynamics and on the kinematics, while the strong-interaction dynamics is embodied in the structure functions.

\subsection{The parton model}

As the first DIS data were collected, structure functions were observed to obey an approximate scaling law, known as Bjorken scaling~\cite{Bjorken:1968dy}: in the limit $Q^2 \rightarrow \infty$, the structure functions depend only on $x$ (as we shall see, this scaling is logarithmically broken in QCD)
\begin{align}
	F_i (x,Q^2) \xrightarrow{Q^2 \rightarrow \infty} 	F_i (x).
\end{align}
This scaling suggests that the virtual boson scatters off pointlike particles, as otherwise a dependence on the ratio $Q/Q_0$ would appear, where $1/Q_0$ is some intrinsic length scale characterizing the size of the constituents. 

Feynman therefore proposed to calculate the hadronic tensor by calculating the incoherent scattering between the boson and point-like constituents, which he called \emph{partons}, carrying a fraction $\xi$ of the momentum of the proton~\cite{Feynman:1969ej,Feynman:1969wa,Feynman:1973xc}.
In this simple model, the structure functions are obtained as
\begin{align}\label{eq:partonstruct}
	& F_2 (x) =  \sum_a e_a^2 x f_a(x), 	& F_L (x) = 0,
\end{align}
where the sum runs over the parton species with charges $e_a$, and $f_a(x)$ are parton densities or parton distribution functions (PDFs), which represent the probability to find a parton $a$ within the proton.
The second result in eq.~\eqref{eq:partonstruct}, known as Callan-Gross relation~\cite{Callan:1969uq}, is a consequence of the fact that the cross section for scattering of a longitudinal boson off on-shell massless spin-$\frac{1}{2}$ particles vanishes (in the QCD-improved parton model $F_L$ is only non-zero at leading order in perturbation theory).

The observation of Bjorken scaling in early DIS experiments and the success of the naive parton model were instrumental in establishing quarks --- which were introduced a few years earlier by Gell-Mann and Zwieg to explain the complicated features of the hadron spectrum~\cite{GellMann:1964nj,Zweig:1964jf,Zweig:1981pd} --- as the fundamental constituents of nucleons. 
The parton model predicts that parton densities are directly measureable in DIS experiments.
By varying the lepton probe, different combinations of quark densities with different electroweak coupling enter in the calculation of the structure function.
In particular, experiments showed that in the scaling region quarks carry only about $50\%$ of the momentum of the proton
\begin{equation}
	\sum_q \int_0^1 d x\ x [q(x)+\bar q (x)] \simeq 0.5,
\end{equation} 
and that therefore half of the momentum of the proton is carried by electroweak-neutral partons, the gluons. 

Nowadays, we understand the parton model as an approximate consequence of the leading-order perturbative treatment of QCD.
As we shall see in the next section, the parton model is not consistent within a perturbative approach due to the appearance of collinear singularities.

\subsection{Collinear singularities in processes with hadrons in the initial state}  

The parton model recipe can be summarized by the following schematic formula for a generic structure function $F$, which is valid at leading twist \mccorrect{(that is,  further corrections are suppressed by powers of $\Lambda_{\text{QCD}}/Q$ and are therefore intrinsically non-perturbative)}:
\begin{equation}\label{eq:convolutionparton}
	F (x, Q^2) = \sum_q \int_x^1 \frac{dz}{z} q (z) \hat \sigma_q \left(\frac{x}{z}, Q^2\right), 
\end{equation}
where the partonic cross section $\hat \sigma_q \left(z,Q^2\right) $ is the cross section for the scattering of a parton (a quark, in particular) $q$ off a virtual photon.

\begin{figure}[t]
\begin{center}
\includegraphics[width=\textwidth]{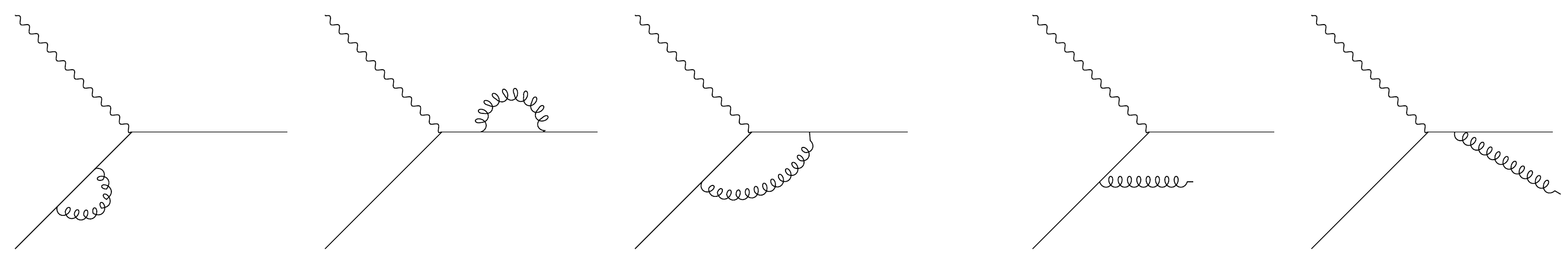}
 \caption{Virtual (left) and real (right) corrections to the process  $\gamma^* q  \rightarrow q $ entering at order $\as$.}
  \label{fig:gq_as}	
\end{center}
 \end{figure}

At the lowest order, the cross section for the process $\gamma^* q  \rightarrow q $ is proportional to $e_q^2 \delta(z-1) $.
At the next order in $\as$, we have to consider two different classes of QCD corrections to the process $\gamma^* q  \rightarrow q $: virtual contributions involving gluon loops and diagrams with real gluon emissions (see fig.~\ref{fig:gq_as}).
An explicit calculation shows that the order $\as$ contribution to the cross section has the following structure:
\begin{equation}\label{eq:DISalpha}
	\hat \sigma^{(1)}_q (x, Q^2) = e_q^2 \as \left[P(x) \ln \frac{Q^2}{Q_0^2} + R(x) \right],
\end{equation}
where $Q_0$ is a scale we introduced to regularize the logarithmic divergence,  $R(x)$ is a calculable function and 
\begin{align}\label{eq:splitting}
	P(x) =  \frac{C_F}{2 \pi} \left[ \frac{1+x^2}{(1-x)_+} + \frac{3}{2} \delta (1-x) \right]
\end{align}
is known as splitting function.
The subscript `$+$' in eq.~\eqref{eq:splitting}, known as the plus prescription, is defined by
\begin{align}
	\int_0^1 dz \, [g(z)]_+ f(z) = \int_0^1 dz\,  g(z)[(f(z) - f(1)]  .
\end{align}
The result of equation eq.~\eqref{eq:DISalpha} shows that beyond leading order the structure function acquires a logarithmic dependence on $Q$, thus violating Bjorken scaling.

\begin{figure}[t]
\begin{center}
\includegraphics[width=0.25\textwidth]{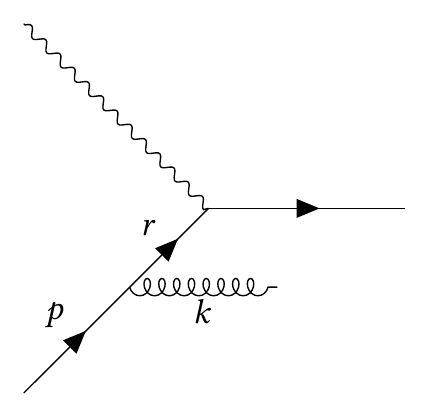}
 \caption{One of the real diagrams contributing to the process  $\gamma^* q  \rightarrow q $ at order $\as$.}
  \label{fig:gq_as_real}	
\end{center}
 \end{figure}

\begin{mccorrection}
To understand the origin of the divergence in eq.~\eqref{eq:DISalpha}, let us focus on the diagram in which the gluon is emitted off the incoming quark leg (fig.~\ref{fig:gq_as_real}).
Whilst in a generic gauge this is not the only diagram which contributes, in axial gauge it is the sole among the real diagrams to give rise to the log-proportional term for $0 < x < 1$.
Let us write the momentum of the outgoing gluon using the Sudakov parameterization 
\begin{align}
	& k^\mu= (1-z) p^\mu - \kappa_t^\mu + y \eta^\mu,
\end{align}
such that $ p^2=\eta^2 = 0$, $p \cdot \eta \neq 0$, and $\kappa_t \cdot p = \kappa_t \cdot \eta = 0$. 
Since $k^2=0$, $y = |\kappa_t|^2/[2 p \cdot \eta (1-z)]$ and the (negative) virtuality of the intermediate quark is equal to 
\begin{equation}
	r^2 = (p-k)^2 = - |\kappa_t|^2 -2 y z p \cdot \eta = - \frac{|\kappa_t|^2}{1-z} \equiv - \frac{k_t^2}{1-z},
\end{equation}
which is therefore related to its transverse momentum squared $k_t^2$.  
As a result, the differential cross section behaves as $1/k_t^2$: a factor of  $1/k_t^4$ comes from the propagator squared and a factor of $k_t^2$ appears in the numerator due to helicity conservation along the quark line (in the collinear limit, the emitted gluon and the initial and final quarks are all aligned; due to the vectorial nature of the strong interaction, the quark helicity cannot flip and the gluon should carry zero helicity, whilst real gluons have $\pm 1$ helicity, see e.g.~\cite{Altarelli:1981ax}) and thus for the real cross section we obtain
\begin{equation}
	\hat \sigma_{q, \text{real}}^{(1)} \sim \int^{Q^2} \frac{d k_t^2}{k_t^2} \as(k_t^2) \sim  \as \ln \frac{Q^2}{Q_0^2} ,
\end{equation}
where $Q_0$ is an IR regularization scale.
This collinear singularity is associated with the mass singularity connected to the initial quark line, which does not cancel (see the discussion at the end of sect.~\ref{sec: IRC safety}).
\end{mccorrection}

Therefore, it seems that by considering QCD corrections to the parton model we have to face two problems. 
Firstly, we acquired a sensitivity to the IR cutoff $Q_0 \sim \LambdaQ$ --- and hence to low-energy physics --- that we wanted to avoid in first place.
Secondly, due to parton radiation over a large $k_t$-gap perturbative convergence seems to be spoiled; at all orders in the strong coupling the terms $\left(\as(Q^2) \ln \frac{Q^2}{Q_0^2}\right)^n$ are $\mathcal O(1)$ and therefore must be resummed.

\subsection{Factorization of collinear singularities}

The key to solving the problem of initial-state collinear singularities is to realize that the small-$k_t$ limit corresponds to a sensitivity to long-range strong interactions, which cannot be calculated in perturbative QCD. 
In particular, we can consider the parton densities which appear in the naive parton model of eq.~\eqref{eq:convolutionparton} as unmeasurable, \emph{bare} parton densities. 
In analogy with renormalization, one can thus redefine them to absorb the collinear singularities into physical parton densities. 
This way, the long-distance behaviour is factorized inside non-perturbative PDFs, whereas the properly defined partonic cross sections are free of collinear singularities and can be computed order by order in perturbation theory.

The ability to factorize the collinear singularities into universal objects is a fundamental property of QCD.
Whilst we will not discuss factorization theorems in this thesis, the interested reader may for instance refer to~\cite{Collins:1989gx,Collins:2011zzd}.
Here we will briefly show how to achieve factorization at order $\alpha_s$ in the DIS case. 

\begin{mccorrection}
The structure function at order $\as$ is given by the convolution of the partonic cross sections with the parton densities.
Using the notation of the previous section, we write
\begin{equation}
		F (x, Q^2) =  e_q^2 \left[ f (x) + \as  \int_0^1 \frac{dz}{z} \left(  P (z)  \ln \frac{Q^2}{Q_0^2} + R (z) \right) f \left( \frac{x}{z} \right) + \mathcal O(\as^2)  \right],
\end{equation}
where for simplicity we consider the case in which there is only one parton species with coupling $e_q$. 
We realize that we can absorb the divergence by replacing the bare parton density $f(x) $ with a physical parton density measured at the scale $\muF$
\begin{equation}
	f(x, \muF^2) = f(x) + \as  \int_x^1 \frac{dz}{z}  f\left( \frac{x}{z} \right) \Gamma \left(z,\frac{\muF^2}{Q_0^2} \right) + \mathcal O(\as^2),
\end{equation}
where $\Gamma$ is a counter term whose divergent part is uniquely specified by requiring the removal of the collinear divergence. 
We thus see that the divergence can be subtracted by setting
\begin{equation}\label{eq:factsubr}
	\Gamma \left(z,\frac{\muF^2}{Q_0^2} \right) = P(z)\ln  \frac{\muF^2}{Q_0^2} + \Gamma_{\text{regular}} (z),
\end{equation}
where $\Gamma_{\text{regular}} (z)$ collects regular terms.
This way, any emission which is characterized by $k_t \lesssim \muF$ is absorbed --- factorized -- into the parton density, which thus becomes a function of the \emph{factorization scale} $\muF$.
As a result, the structure function becomes
\begin{equation}\label{eq:structfunc1}
F(x,Q^2) = e_q^2 \left[ f(z,\muF^2) 	+\as  \int_x^1 \frac{dz}{z} f\left(\frac{x}{z},\muF^2\right) \left( P(z) \ln \frac{Q^2}{\mu_F^2} + \tilde R(z)  \right)  + \mathcal O(\as^2) \right],
\end{equation}
where the form of $\tilde R$ depends on the specific choice of the regular counterterms in eq.~\eqref{eq:factsubr}, known as \emph{factorization scheme}.
It is customary to collect the finite contribution which remains after factorization by defining a \emph{coefficient function} $C\left(z, \frac{Q^2}{\mu_F^2}\right)$ such that the structure function reads
\begin{equation}\label{eq:structfunccoeff}
F(x,Q^2) = e_q^2 \left[ f(z,\muF^2) 	+\as  \int_x^1 \frac{dz}{z} f  \left(\frac{x}{z},\muF^2\right) C^{(1)}\left( z, \frac{Q^2}{\muF^2}\right)  + \mathcal O(\as^2) \right]. 
\end{equation}
The most common choice for the factorization scheme is the Modified Minimal Subtraction ($\MSbar$) scheme.
In this scheme, DR is used to regularize the divergence and an additional, universal contribution $ \ln 4 \pi - \gamma_E$ is subtracted along with the divergent piece.
\end{mccorrection}

Whereas we have discussed factorization in a simple case and up to order $\as$, the formul\ae\ which we found hold more generally.
In the DIS case, structure functions can be calculated as
\begin{equation}\label{eq:DISfact}
	F(x,Q^2) = \sum_{i} \int_x^1 \frac{dz}{z} C_i \left(\frac{x}{z}, \frac{Q^2}{\muF^2},\as \right)\, f_i (z, \muF^2) + \mathcal O((\LambdaQ/Q)^p),
	\end{equation} 
which differs from the naive parton-model formula in two respects: firstly, the sum now runs over all the parton species (both quarks and gluons) to allow for contributions which may enter at higher orders; secondly, it explicitly shows that Bjorken scaling is now broken by logarithms of the hard scale $Q$. 
The term $\mathcal O((\LambdaQ/Q)^p)$ generically collects non-perturbative contributions, such as hadronization effects, multiparton interactions, higher-twists, et cetera.
The factorized expression in eq.~\eqref{eq:DISfact} is often expressed in a more compact form as
\begin{equation}\label{eq:DISfactConv}
	F(x,Q^2) = \sum_{i} C_i \left(x, \frac{Q^2}{\muF^2},\as \right) \otimes f_i (x, \muF^2)+ \mathcal O((\LambdaQ/Q)^p),
	\end{equation} 
where we introduced the convolution symbol $\otimes$.

Coefficient functions depend on the factorization (and on the renormalization) scheme, as well as on the structure function under consideration. 
However, once properly defined, they can be computed order by order in perturbation theory
\begin{equation}
	C_{i}(x, \as) = C^{(0)}_i(x) + \as C^{(1)}_i(x)  + \as^2 C^{(2)}_i(x)  + \ldots \, .
\end{equation}
The first contribution is known as leading order (LO) contribution, the second as next-to-leading order (NLO) contribution, and so on.
The dependence on the factorization and renormalization scales $\muF$ and $\muR$ of the coefficient functions can be calculated by requiring that physical cross sections are independent of $\muF$ and $\muR$ order by order in $\alpha_s$.
Theoretical predictions will however display a (higher-order) dependence on the scales, whose choice is somewhat arbitrary.
To ensure a reliable estimate, the choice of $\muR$ and $\muF$ is performed such that the logarithms of  $\frac{Q}{\muF}$ and $\frac{Q}{\muR}$ which appear in the coefficient functions are of order one.
By varying the factorization and renormalization scales around $Q$ one could obtain an estimate of the uncertainty of the predictions due to missing higher orders.

Although we have discussed factorization in the case of one hadron in the initial state, the same logic applies to (inclusive enough) hard-scattering processes in hadron-hadron collisions,
\begin{equation}\label{eq:hadroncoll}
	h_1(p_1) + h_2(p_2) \rightarrow H (Q, \ldots) + X.
\end{equation}
In eq.~\eqref{eq:hadroncoll} $p_1$ and $p_2$ are the momenta of the incoming hadrons $h_1$ and $h_2$, $H$ represents the hard particle(s) produced (Higgs or vector bosons, heavy quarks, hard jets...), $Q$ is the relevant hard scale of the process, the ellipsis indicate any other relevant scales or kinematic variables, and $X$ denotes \mccorrect{other particles appearing in the final state}. 
In this case, the cross section can be computed as
\begin{align}\label{eq:hadronfact1}
	\sigma (p_1, p_2, Q, \ldots ) =& \sum_{a,b} \int_{\tau}^1 dx_1 dx_2\, f_{a/h_1} (x_1, \muF^2) f_{b/h_2} (x_2, \muF^2) \hat \sigma_{ab} \left(x_1 p_1 , x_2 p_2 , Q, \ldots, \muF^2\right) \nonumber \\
	&+ \mathcal O((\LambdaQ/Q)^p),
\end{align}
where $\tau = Q^2/s$ and $s=(p_1+p_2)^2$ is the centre-of-mass energy squared of the collision. 
In eq.~\eqref{eq:hadronfact1} we have indicated with $f_{a/h_1}$ and $f_{b/h_2}$ the parton densities relative to the two colliding hadrons and $a,b = q, \bar q, g$.
It is usually convenient to recast the cross section for a generic hadroproduction process as
\begin{equation}\label{eq:hadronfact2}
	\sigma (p_1, p_2, Q, \ldots ) = \sum_{a,b} \sigma^0_{ab}  \int_{\tau}^1 \frac{dz}{z} \mathcal L_{ab} (z, \muF^2)\, C_{ab} \left(\frac{\tau}{z} , Q, \ldots, \muF^2\right)+ \mathcal O((\LambdaQ/Q)^p) ,
\end{equation}
where we introduced the {\it parton luminosity}
\begin{equation}\label{eq:partonlumi}
	\mathcal L_{ab} (z, \muF^2)  \equiv \int_{z}^1\frac{dw}{w}  f_{a/h_1} \left(\frac{z}{w}, \muF^2\right ) f_{b/h_2} (w, \muF^2)  ,
\end{equation}
and we extracted a prefactor $\sigma^0_{ab}$ such that at LO the coefficient functions are a Dirac delta for the partons which couple to the final state at LO, or zero otherwise.

Eqs.~\eqref{eq:DISfact} and \eqref{eq:hadronfact1} represent the two master formul\ae\ needed to compute theoretical predictions in processes with hadrons in the initial state.
From our discussion, it is evident that parton distribution functions are an essential ingredient to compute hard-scattering processes.
Though they cannot yet be computed from first principles\footnote{PDFs can be calculated using lattice QCD, but their accuracy is not yet competitive with that of PDFs extracted from a fit to experimental data, see e.g.~\cite{Lin:2017snn} for recent developments.}, cross section measurements allow their extraction once the partonic cross sections have been perturbatively computed.
Once determined, PDFs can in turn be used to predict cross sections for other processes. 

\subsection{DGLAP evolution}
 
The value of the parton densities $f(x,\muF^2)$ at a fixed scale $\muF$ and their dependence on the momentum fraction $x$ are not calculable in perturbation theory (we refer the reader to appendix~\ref{app:PDF} for a brief review on PDF determination).
However, the scale dependence of PDFs is computable in perturbative QCD.
A rigorous proof of the PDF scale dependence can be obtained by using the operator product expansion (OPE) formalism; here we will present a simplified derivation based on the factorization scale independence of the physical cross sections.

Let us start by considering a generic structure function in DIS. 
For simplicity, let us assume that there is only one species of quarks in the target, and thus the structure function can be written as in eqs.~(\ref{eq:structfunc1}-\ref{eq:structfunccoeff}).
Since the structure functions must be independent of the factorization scale,
\begin{equation}
	\muF^2 \frac{\partial}{\partial \muF^2} F (x, Q^2) = 0,
\end{equation}
we obtain the following renormalization group equation (RGE) for the parton density $q(x,\mu^2)$:
\begin{equation}\label{eq:simpleDGLAP}
	\mu^2 \frac{\partial}{\partial \mu^2} q(x,\mu^2) =\as(\mu^2)   \int_x^1 \frac{dz}{z} P\left( \frac{x}{z} \right) q(z,\mu^2)  + \mathcal O \left(\as^2(\mu^2)\right),
\end{equation}
where we dropped the subscript $\rm F$. 
Note that the coupling is evaluated at $\mu^2$, thus by solving eq.~\eqref{eq:simpleDGLAP} all the leading-log terms $\as^n (\mu^2) \ln^n \mu^2$ are resummed.
In the case of DIS collisions, this result can be proven at all orders in perturbation theory using a rigorous treatment based on the renormalization group equation and OPE~\cite{Georgi:1951sr,Gross:1974cs}.

In this simplified derivation we have considered only the possibility that quarks radiate gluons.
However, one has to take into account also the presence of gluons in the target hadron, as the quark struck by the electroweak boson can be produced by gluon splitting.
Moreover, one must include quarks of different flavours.
The evolution equation~\eqref{eq:simpleDGLAP} thus takes the general form
\begin{equation}\label{eq:AP}
		\mu^2 \frac{\partial}{\partial \mu^2} f_i(x,\mu^2) =\int_x^1 \frac{dz}{z} P_{ij}\left(\frac{x}{z},  \as(\mu^2) \right) f_j(z,\mu^2), 
\end{equation}
which is the renowned DGLAP equation~\cite{Gribov:1972ri,Lipatov:1974qm,Altarelli:1977zs,Dokshitzer:1977sg}.
In eq.~\eqref{eq:AP}, $P_{ij} (z, \as)$ is a $(2n_f +1) \times (2n_f +1)$ matrix of Altarelli-Parisi splitting functions, which can be expanded in powers of the strong coupling as
\begin{equation}
	P_{ij} (z, \as) = \as \left[ P^{(0)}_{ij} (z) +  \as P^{(1)}_{ij} (z) +  \as^2 P^{(2)}_{ij} (z) + \mathcal O(\as^3) \right].
\end{equation}
The coefficients $P^{(k)}_{ij}$ can be calculated in perturbation theory and are currently known at NNLO accuracy~\cite{Moch:2004pa,Vogt:2004mw}. 
%
%
Due to charge conjugation and $SU(n_f)$ flavour symmetry, the following relations hold:
\begin{equation}
	P_{q_i q_j} = P_{\bar q_i \bar q_j}, \qquad P_{q_i\bar q_j} = P_{\bar q_i q_j}, \qquad P_{q_i g} = P_{\bar q_i g} \equiv P_{qg}/(2 n_f),  \qquad P_{g q_i} = P_{g \bar q_i} \equiv P_{gq}.
\end{equation}

The DGLAP equations can be substantially simplified by moving to a basis in which the majority of equations decouple.
The combinations of PDFs which decouple are known as \emph{non-singlet}; in particular, since QCD is flavour blind the gluon contribution to the valence distributions\footnote{With a slight abuse of notation, we use the symbol $j$ to identify the PDF $f_j$. }
\begin{equation}
	V_i = q_i - \bar q_i,
\end{equation}
and to the differences between quark sea distributions $q^+ = q + \bar q$
\begin{align}
	T_3 = u^+ - d^+, \qquad \qquad 
	T_5 = u^+ + d^+ - 2 s^+, \qquad \qquad
	T_{15} = u^+ + d^+ + s^+ -3 c^+, \nonumber \\
	T_{24} = u^+ + d^+ + s^+ +c^+ - 4b^+, \qquad \qquad
	T_{35} = u^+ + d^+ + s^+ + c^+ + b^+ - 5 t^+,
\end{align}
cancel, thereby largely diagonalizing the splitting matrix.
The evolution of the $V_i$ and $T_i$ combinations are governed by the splitting functions $P^-$ and $P^+$ respectively, which at LO read
\begin{equation}
	P^{+ (0) } (z) = 	P^{- (0) } (z) = \frac{C_F}{2 \pi} \left(\frac{1+z^2}{1-z} \right)_+.
\end{equation}

The only combination which couples to the gluon is the so-called \emph{singlet} quark PDF $\Sigma = \sum_i q^+_i  =\sum_i (q_i + \bar q_i) $.
The evolution of the gluon and singlet PDF is thus controlled by the matrix
\begin{equation}
	\left( \begin{array}{cc}
		P_{gg} & P_{gq} \\
		P_{qg} & P_{qq}
	\end{array} 
	\right).
\end{equation}
At LO, $P^{(0)}_{qq} (z) $ is given by eq.~\eqref{eq:splitting} and 
\begin{align}\label{eq:APsplittingLO}
	P_{gg}^{(0)} (z) &= \frac{C_A}{\pi} \left[ \frac{z}{(1-z)_+} + \frac{1-z}{z} + z (1-z) \right] + \frac{11 C_A-2 n_f}{12 \pi} \delta(1-z), \nonumber \\
	P_{gq}^{(0)} (z) &= \frac{C_F}{2\pi} \left[\frac{1+(1-z)^2}{z} \right], \nonumber \\
	P_{qg}^{(0)} (z) &= \frac{n_f}{2 \pi} \left[ z^2 + (1-z)^2 \right].
\end{align}

It is worth noticing that the convolution structure appearing in eq.~\eqref{eq:AP}, and which we have already encountered previously, has the form of a Mellin convolution
\begin{equation}
(f \otimes g) (x) \equiv \int_x^1 \frac{dz}{z} f(z) g\left( \frac{x}{z} \right),	
\end{equation}
which is symmetric under the change of variable $z \rightarrow x/z$. 
The convolution is diagonal under a Mellin transform
\begin{equation}
	\mathcal { M } [f] (N) \equiv {\bm f} (N) = \int_0^1 dz \, z^{N-1} f(z),
\end{equation}
such that in Mellin space (also called $N$-space) one has
\begin{equation}
	\mathcal { M } [f \otimes g] (N) = {\bm f} (N) {\bm g} (N).
\end{equation}
By taking the Mellin moments of eq.~\eqref{eq:AP} therefore
\begin{equation}\label{eq:APN-space}
	\mu^2 \frac{\partial}{\partial \mu^2} {\bm f} _j (N, \mu^2) = \gamma_{jk} (N-1,\as(\mu^2)) {\bm f}_k (N, \mu^2) ,
\end{equation}
where $\gamma_{ij}$ are known as anomalous dimensions.
We have used the convention (customarily used in small-$x$ physics)
\begin{equation}
	\gamma_{ij} (N, \as)  = \int_0^1 dz \, z^N P_{ij} (z, \as) .
\end{equation}

The universality of the splitting functions --- which appears naturally in the RGE approach --- can be understood by giving a physical interpretation to $P_{ij}$~\cite{Altarelli:1977zs}.
By recasting eq.~\eqref{eq:simpleDGLAP} as
\begin{equation}
	q (x, \mu^2) + d q (x, \mu^2) = \int_0^1 dy \int_0^1 dz \delta(zy - x) q (y, \mu^2) \left[\delta (z - 1) + \alpha_s(\mu^2)  P^{(0)}_{qq}(z) d \ln \frac{\mu^2}{\mu_0}\right],
\end{equation}
%
where $\mu_0$ is a reference scale, we can interpret the quantity 
\begin{equation}
	\delta (z - 1) + \alpha_s(\mu^2) P^{(0)}_{qq}(z) d \ln \frac{\mu^2}{\mu_0^2} = \mathcal P_{qq} (z)
\end{equation}
as the probability density of finding a quark inside a quark, with a fraction $z$ of the longitudinal momentum of the parent quark and with transverse momentum smaller than $\mu$.
We can thus interpret the splitting function as the order $\as$ expression for the variation per unit $\ln \frac{\mu^2}{\mu_0^2}$ of the probability density.
Note that since there are $\delta(1-z)$ terms also at order $\as$ --- see eq.~\eqref{eq:splitting} --- the probability for a quark to remain a quark with the same energy is lowered by the interaction. 

Interpreting splitting functions as probabilities allows one to readily compute the splitting functions from the QCD vertices $q \rightarrow q g$, $g \rightarrow q \bar q$, $g \rightarrow gg$. 
Typically, one first computes the regular part for $z < 1$ and fixes the $\delta (1-z)$ terms by imposing charge and momentum conservation.
The probabilistic interpretation implies that splitting functions are positive definite for $z < 1$.

The behaviour of the splitting functions is further constrained by very general arguments, known as sum rules. 
Sum rules are based on the observation that there are some quantities that cannot be modified by DGLAP evolution, as the QCD Lagrangian separately conserves fermion number, flavour and momentum.
For instance, though the distribution of the momentum among quarks and gluons may vary, the total momentum shared by partons should be equal to the total momentum of the proton.
Similarly, the total baryonic number --- computed as the number of quarks minus the number of antiquarks --- must be conserved.
Therefore we have
\begin{align}
	&\int_0^1 dz \, z [\Sigma (z,\mu^2) + g (z,\mu^2) ]  = 1,\\
	&\int_0^1 dz\, [q(z,\mu^2) - \bar q (z,\mu^2) ] = v_q ,
\end{align}
where $v_q = 2, 1, 0 \ldots$ for $q =u$, $d$, $s$, $\ldots$ quarks. 
By considering the Mellin moments of these equations, it is possible to recast these sum rules into constraints on the splitting functions:
\begin{align}
	\int_0^1 dz\, P_{qq}  (z) = 0, \qquad \int_0^1 dz\, z [ P_{qq}  (z) + P_{gq} (z) ] = 0, \qquad \int_0^1 dz\, z [ P_{qg}  (z) + P_{gg} (z) ] &= 0.
\end{align}
These relations constrain both the qualitative behaviour and the solutions of the evolution equations and also simplify the derivation of the splitting function as they are now related to each other. 


\section{Factorization \amper heavy quarks}\label{sec:factheavyquarks}

In the collinear factorization framework, quarks are generally classified as `light' or `heavy', though such a distinction has a certain degree of arbitrariness.
A convenient definition is based on the relative magnitude of the quark mass with respect to $\LambdaQ$: quarks whose mass $m \lesssim \LambdaQ $ are light, quarks whose mass $m \gg \LambdaQ $ are heavy. 
According to such a definition, up, down, and strange are light, whereas bottom and top are heavy.
The charm quark is usually classified as heavy, though its mass is relatively close to the scale $Q_0$ where the non-perturbative behaviour sets in.

Hitherto, we have worked in massless QCD and we have tacitly assumed that the quarks were massless. 
This approximation becomes however unjustified when the hard scale of the process $Q$ is comparable the quark masses.
In particular, the hard scale of the process $Q$ can be bigger, comparable or smaller than the mass of one or more `heavy' quarks. 
In this case, the use of a mass-independent scheme such as $\MSbar$ is not justified; moreover, precision predictions demand the inclusion of mass corrections when $Q\sim m$.
As a consequence, one must resort to so-called variable flavour number schemes (VFNSs), which combine computations performed with a different number of active flavours to obtain a valid theoretical description both close to and above the quark mass.

In this section we will start by discussing firstly how such calculations are performed, considering for simplicity the case of three and four active flavours. 
We will then discuss and compare different VFNSs which have been proposed for the calculation of DIS structure functions.
Finally, we will analyse the special role played by the charm quark. 

\subsection{Massive \amper massless schemes}

The computation of physical quantities, such as cross sections or structure functions, for a scattering process characterized by a hard scale $Q$ can be performed in different factorization or renormalization schemes.
Though the results must be the same to all orders, in general they differ order by order in perturbation theory by higher-order corrections and display a dependence on the renormalization and factorization scales $\muR$ and $\muF$, typically chosen to be $\sim Q$.
The presence of a heavy quark of mass $m_h$ introduces a new scale.
Therefore, the problem can be split into three main regions as follows:
\begin{itemize}
	\item $Q \ll m_h$: the heavy quark should decouple lest unphysical effects appear at scales much smaller than the heavy quark mass.
	\item $Q \sim m_h$ (threshold region): full dependence on the heavy quark mass is required for precision calculation.
	\item $Q \gg m_h$: the partonic cross sections contain large collinear logarithms of $m_h/Q$, which spoil the accuracy of any fixed-order computation.
\end{itemize}

The description of these three regions requires suitable renormalization and factorization schemes to ensure manifest decoupling at low scales and the resummation of collinear logarithms at scales much higher than the quark mass, as well as a prescription to include power-suppressed $m_h/Q$ terms relevant for precision phenomenology at $Q \sim m_h $. 
We will now briefly summarize the necessary ingredients to construct a scheme (or rather, a sequence of subschemes) to overcome the complications which arise when heavy quarks are involved.
For definitiveness, we consider here the case of three massless quarks and one single massive quark.

Let us first notice that from the point of view of renormalization the quark mass does not play an important role; whilst the quark mass acts as an IR regulator, massless-quark loops would still be UV-divergent even if the quark were massive.
However, a mass-independent renormalization scheme such as $\MSbar$ produces spurious heavy-quark effects at $Q \ll m$ as all flavours participate in $\as$ evolution at any scale. 
For this reason, it is appropriate to use the $\MSbar$ scheme only for quarks whose mass is lighter than $\muR$, whereas it is necessary to use another scheme to ensure the decoupling of heavier quarks. 

Such a scheme has been proposed by Collins, Wilczek and Zee (CWZ) in the context of neutral-current processes~\cite{Collins:1978wz} and was later straightforwardly applied to QCD~\cite{Collins:1998rz}. 
The CWZ (or decoupling) scheme is composed by a sequence of subschemes in which the $n_l$ quarks lighter than the renormalization scale $\muR$ are renormalized with $\MSbar$ terms, whereas UV divergences associated to quarks heavier than the renormalization scale are subtracted at zero external momentum.
Since the logarithms of $\muR/m_h$ are power suppressed as $\muR^2/m_h^2$, the CWZ scheme ensures that at scales $\muR \ll m_h$ the graphs containing heavy flavours decouple, whilst the same logarithms are resummed when $\muR \gg m_h$.
This way, at scales  $\muR < m_h$ the running of the strong coupling is identical to the $\MSbar$ running, whith $n_f=n_l$.  

The CWZ scheme is effectively a VFNS, as the number of `light' (active, or partonic) and `heavy' (inactive, or non-partonic) flavours depends on the scale and changes at the matching scale $\mu_h\sim m_h$.  
At the matching scale, the strong couplings in the schemes with three and four active flavours are related by the matching condition
\begin{equation}
	\as^{[3]}(\mu_h^2) = \as^{[4]} (\mu_h^2) + \sum_{j=2}^\infty c_j\, \left( \as^{[4]} (\mu_h^2) \right)^j,  
\end{equation}
whose coefficients $c_j$ can be computed order by order in perturbation theory.
%
This relation can be used to relate the coefficients of the perturbative expansion of a generic observable $F( \as)$ in the schemes with three and four active flavours
\begin{equation}
	F( \as) = \sum_k \left(\as^{[3]}\right)^k  F^{[3](k)}  = \sum_k \left(\as^{[4]}\right)^k  F^{[4](k)} 
\end{equation}
by expanding and matching the two expressions order by order once both are written in terms of $\as^{[4]}(\mu_h^2) $ (or, equivalently, $\as^{[3]} (\mu_h^2)$). 

Since the presence of the mass regularizes the collinear divergences, radiative corrections which involve massive quarks are IR finite. 
As a consequence, it is possible to factorize only the collinear logarithms associated to massless quarks while retaining the massive collinear logarithms in the partonic cross sections.
This choice, combined with the use of the decoupling scheme for the heavy quarks, leads to the so-called 3 flavour scheme (3FS) or `massive scheme'. 
In this scheme, only three quarks participate in DGLAP evolution. 
By denoting with $\Gamma_{ij}^{[3]}(Q^2)$ the solution of the DGLAP equation with 3 active flavours, the PDFs thus evolve according to
\begin{equation}
	f_i^{[3]}(Q^2) = \sum_{j, q, \bar q} \Gamma_{ij}^{[3]} (Q^2, Q^2_0) \otimes f_j^{[3]}(Q_0^2).
\end{equation}

A generic structure function in this scheme reads
\begin{equation}\label{eq:3fssf}
	F^{[3]} (Q^2, m_h^2) = \sum_{i=g,q,\bar q, h, \bar h} C_i^{[3]} \left(\frac{m_h^2}{Q^2}, \as^{[3]} (Q^2)  \right) \otimes f_i^{[3]}(Q^2). 
\end{equation}
The coefficient functions are evaluated at fixed order and thus contain the collinear logarithms due to gluon and light quark splittings.
The effects of the heavy quark mass is retained both in tree-level diagrams and in loop diagrams (e.g. virtual heavy quark loop in gluon propagators, or a gluon splitting into a heavy quark pair).
Note that the sum in eq.~\eqref{eq:3fssf} runs also over $i=h, \bar h$: though only three PDFs participate in DGLAP equation in the 3FS, we allow for an initial-state contribution from an \emph{intrinsic} heavy quark PDF which could be generated by nonperturbative effects at the scale $Q_0$ below the scale of perturbative heavy quark production.  
The heavy quarks appear only through virtual loops and decouple at scales below the production threshold when they are generated perturbatively.

Whereas the 3FS is appropriate for scales $Q \gtrsim m_h$, it becomes inaccurate at scales $Q \gg m_h$, since the collinear logarithms become increasingly large and spoil the perturbative behaviour of the 3FS result.
As a consequence, when the hard scale is much higher than the quark mass it is more suitable to factorize the massive logarithms into the definition of the PDFs and to resum them with DGLAP evolution, leading to a 4 flavour (or `massless') scheme.
The PDFs thus evolve according to
\begin{equation}
	f_i^{[4]}(Q^2) = \sum_{j=g, q, \bar q, h, \bar h} \Gamma_{ij}^{[4]} (Q^2, Q^2_0) \otimes f_j^{[4]}(Q_0^2).
\end{equation}
where $ \Gamma_{ij}^{[4]} $ is the solution of the DGLAP equation with four active flavours, which thus resums also the logarithms generated by heavy quark splittings. 
At scales $Q \gg m_h$, the structure functions are readily obtained using standard $\MSbar$ counterms for all the active quarks.
They can be written as 
\begin{equation}\label{eq:4fssf1}
	F^{[4]} (Q^2, m_h^2) = 	F^{[4]} (Q^2, 0) + \mathcal O \left(\frac{m_h^2}{Q^2} \right) =  \sum_{i=g,q,\bar q, h, \bar h} C_i^{[4]} \left(0, \as^{[4]} (Q^2)  \right) \otimes f_i^{[4]}(Q^2) + \mathcal O \left(\frac{m_h^2}{Q^2} \right) . 
\end{equation}
where $C_i^{[4]} \left(0, \as^{[4]} (Q^2)  \right) $ are massless coefficient functions, and are now finite as the logarithmic divergence has been subtracted.  

The PDFs in the 3FS and in the 4FS are related by the matching conditions 
\begin{equation}\label{eq:pdfmatching}
	f_i^{[4]} (\mu_h^2) = \sum_{j=g, q, \bar q, h, \bar h}  K_{ij} \left( \frac{m_h^2}{Q^2} \right) \otimes f_j^{[3]} (\mu_h^2).
\end{equation}
By inserting this expression into eq.~\eqref{eq:4fssf1} we recover \eqref{eq:3fssf} up to power-suppressed contributions provided
\begin{equation}\label{eq:masslesslimitcf}
 \sum_{i=g, q, \bar q, h, \bar h} C_i^{[4]} (0)  \otimes   K_{ij} \left( \frac{m_h^2}{Q^2} \right) = C_j^{[3,0]} \left( \frac{m_h^2}{Q^2} \right),
\end{equation}
where $C_i^{[3,0]} $ is the 3FS massive coefficient function with all power-suppressed terms set to zero. 
The matching condition can therefore be computed simply by calculating DIS structure functions in the 3FS and in the 4FS and expanding the result order by order in (the same) $\alpha_s$.

\subsection{Combining fixed order \amper resummation}

The use of eq.~\eqref{eq:4fssf1} above threshold, combined with the 3FS results below threshold, leads to the so-called Zero-Mass (ZM) scheme. 
Eq.~\eqref{eq:4fssf1} is appropriate only at scales much higher than the heavy quark mass, but as it neglects all power-suppressed corrections it fails to reproduce the fixed-order calculations at the threshold.
At that scale mass corrections are important and should be restored, leading to a structure function
\begin{equation}\label{eq:4fssf2}
	F^{[4]} (Q^2, m_h^2)  =  \sum_{i=g,q,\bar q, h, \bar h} \tilde C_i^{[4]} \left(\frac{m_h^2}{Q^2}, \as^{[4]} (Q^2)  \right) \otimes f_i^{[4]}(Q^2)
\end{equation}
which depends on coefficient functions $\tilde C_i^{[4]} \left(m_h^2/Q^2, \as^{[4]} (Q^2)  \right)$, which now include mass effects. 
There exist several prescriptions to reinstate the mass corrections, which lead to different formulations of VFNSs\footnote{To distinguish them from the ZM scheme, these schemes are usually called General-Mass (GM) schemes.}. 
The freedom to incorporate the mass corrections in the 4FS result essentially derives from the assumption that the heavy-quark PDF is perturbatively generated and vanishes at threshold.
This is equivalent to imposing a boundary condition on the perturbative evolution, as terms involving the heavy quarks in eq.~\eqref{eq:pdfmatching} can be dropped.
As a consequence, various incarnations of VFNSs may differ by subleading terms which are small but nevertheless present at any order in perturbation theory.
As we will see in the next section, this ambiguity is not present if the sum in eq.~\eqref{eq:pdfmatching} runs also over the heavy quark.

Let us now briefly summarize the main VFNS currently in use. 
For a review and a comparison among these schemes, see e.g. sect. 22 of ref.~\cite{Binoth:2010nha}.

\begin{itemize}
	\item {\scshape ACOT}. The ACOT scheme~\cite{Aivazis:1993pi,Aivazis:1993kh,Collins:1998rz,Tung:2001mv} is based on a factorization theorem proven to all orders by Collins~\cite{Collins:1998rz} which can be considered as an extension of the massless $\MSbar$ scheme to the massive case. 
	The basic idea of this scheme is to retain full mass dependence in the Wilson coefficient; the coefficient functions are obtained by using standard massless collinear counterterms for massless partons, whilst massive collinear counterterms are used for heavy quarks.  
	%
	
	\item {\scshape S-ACOT}. Under the assumption that the heavy-quark PDF is generated perturbatively, the ACOT scheme can be simplified by moving power corrections between $C_{h,\bar h}^{[4]}$ and  $C_i^{[4]}$ ($i=g,q,\bar q$)  without spoiling the formal accuracy in eq.~\eqref{eq:4fssf2}. 
	This freedom is used to obtain a simplified version of the ACOT scheme in which the heavy quark Wilson coefficients are computed in the massless limit, $\bar C_{h, \bar h}^{[4]} \left(m_h^2/Q^2  \right) = C_{h, \bar h}^{[4]} \left(0 \right) $~\cite{Kramer:2000hn} . 
	By rescaling the heavy-quark Bjorken $x$ to take into account the massive heavy-quark kinematics one obtains a variant of the scheme dubbed S-ACOT-$\chi$~\cite{Tung:2001mv,Guzzi:2011ew}.	
	
	\item {\scshape FONLL}. The original FONLL scheme~\cite{Forte:2010ta,Cacciari:1998it} is based on the observation that to obtain a description of physical quantities valid both at threshold and at high energy it is sufficient to add the 3FS and the 4FS results and to subtract double counting. 
	This way, the FONLL scheme does not require novel factorization schemes as it is based on quantities calculated in well-defined schemes.  
	A phenomenological damping factor is then introduced to (artificially) suppress formally subleading contributions close to threshold.	
	The scheme has been extended to include contributions from initial-state massive quarks in ref.~\cite{Ball:2015tna,Ball:2015dpa}. 
	If such contributions are included, the FONLL scheme is formally equivalent to ACOT to all orders in perturbation theory, whereas the original formulation is equivalent to S-ACOT~\cite{Ball:2015dpa}. 
	\item {\scshape TR}. The TR scheme~\cite{Thorne:1997ga} puts the emphasis on the threshold behaviour.  
	The ambiguity in the definition of the massive Wilson coefficient is removed by requiring that physical observables are continuous in the matching region.
	To this end, a formally higher-order, $Q$-independent contribution is added to the massless result above threshold.
	As a consequence, the result differs both from S-ACOT and ACOT by higher-order terms.	
	The scheme has been explicitly extended at NNLO in ref.~\cite{Thorne:2006qt} (TR$^\prime$ scheme).
	
	\item {\scshape BPT}. Recently, an alternative scheme has been introduced in ref.~\cite{Bonvini:2015pxa} in the context of $b \bar b H$ production using an effective field theory (EFT) approach.
	EFTs are a standard tool to describe processes characterized by separate scales and are therefore particularly convenient to describe processes with heavy quarks in the initial state. 
	The scheme is formally equivalent to S-ACOT and to the original formulation of FONLL. 
		
\end{itemize}

We observe that another source of difference among the VFNSs is related to the perturbative counting. 
For instance, whilst ACOT-like schemes perform the perturbative counting in a standard way by counting the explicit powers of $\as$ in the coefficient functions, in the TR schemes the lowest non-vanishing orders in the 3FS and in the 4FS appear at the same time (i.e. the $\mathcal O(\as^0) $ quark-initiated contribution in the 4FS is combined with the $\mathcal O(\as^1) $ gluon-initiated contribution in the 3FS). 
Similarly, the FONLL scheme exists in several incarnations, dubbed FONLL-A, FONLL-B, FONLL-C. 
In FONLL-A and FONLL-C the counting is performed as in ACOT schemes, whereas FONLL-B is an intermediate scheme which combines $\mathcal O(\as^2)$ massive contributions with $\mathcal O(\as)$ massless contributions. 
The FONLL-B counting is somewhat similar to the perturbative counting used in BPT, where the heavy quark PDF is formally treated as a quantity of order $\as$.

\subsection{The curious case of the charm quark}\label{sec:ic} 

Among heavy quarks, the charm plays a special role, 
%
since the value of the charm mass is rather close to the scale where one would expect the non-perturbative behaviour to kick in.
For this reason, the assumption that the charm is perturbatively generated at threshold may not be accurate, as it is possible that a non-vanishing charm distribution is generated by non-perturbative effects.
Over the years, various models of non-zero --- `intrinsic' --- charm distributions have been proposed (for a recent review, see~\cite{Brodsky:2015fna}) and several attempts have been made to obtain an empirical determination~\cite{Aubert:1981ix,Harris:1995jx,Dulat:2013hea,Jimenez-Delgado:2014zga,Ball:2016neh,Hou:2017khm}.

Ideally, one would like to allow for the possibility of an initial charm PDF at threshold, which then evolves perturbatively to higher scales.
The initial charm PDF would then be determined by fitting to data, like the light quark and the gluon PDFs.
As we discussed in the previous section, the various VFNSs typically assume that the heavy quark PDF is perturbatively generated above threshold.
Therefore, if one wishes to include the charm PDF alongside the other PDFs, the VFNS should be adequately modified.
In this section we will focus on the FONLL scheme and we will discuss how it can be extended to incorporate intrinsic heavy quark effects. 

In the FONLL scheme, a generic structure function can be constructed as
\begin{equation}
	F_{\text{FONLL}}(Q^2, m_h^2) = F^{[4]}(Q^2, 0) +  F^{[3]}(Q^2, m_h^2) - F^{[3,0]}(Q^2, m_h^2) 	,
\end{equation}
where the last term which subtracts the double counting is defined in term of the so-called massless limit of the 3FS coefficient functions introduced in eq.~\eqref{eq:masslesslimitcf}:
\begin{equation}
	F^{[3,0]}(Q^2, m_h^2) = \sum_{i=g,q,\bar q, h, \bar h} C_i^{[3,0]} \left(\frac{m_h^2}{Q^2}\right) \otimes f_i^{[3]}(Q^2).
\end{equation}
The result is usually conveniently recast in a formula expressed in terms of the 4FS scheme PDFs. 
In its original formulation, the assumption that the heavy quark PDF vanishes at threshold allows one to write
\begin{equation}\label{eq:pdfmatching1}
	f_i^{[3]} (Q^2) = \sum_{j=g, q, \bar q}  \tilde K^{-1}_{ij} \left( \frac{m_h^2}{Q^2} \right) \otimes f_j^{[4]} (Q^2),
\end{equation}
where $\tilde K_{ij}$ is the inverse of the matching matrix defined in~\eqref{eq:pdfmatching} in the subspace of the light partons.
In this case, the FONLL result can thus be written as (`zIC', standing for zero intrinsic charm, stresses the assumption about the absence of an intrinsic component)
\begin{align}\label{eq:FONLLzic}
		F_{\text{FONLL}}(Q^2, m_h^2)\Big|_{\text{ zIC}} =& \sum_{j=g, q, \bar q}\left\{ \sum_{i=g, q, \bar q} \left[C_i^{[3]} \left( \frac{m_h^2}{Q^2} \right) - C_i^{[3,0]}  \left( \frac{m_h^2}{Q^2} \right)  \right]\otimes  \tilde K^{-1}_{ij}  \left( \frac{m_h^2}{Q^2} \right) \right\} \otimes  f_j^{[4]} (Q^2) \nonumber\\
		&  +  \sum_{j=g, q, \bar q} C_j^{[4]} \left(0\right) \otimes  f_j^{[4]} (Q^2)+  \sum_{j=h, \bar h} C_j^{[4]} \left(0\right) \otimes  f_j^{[4]} (Q^2).
\end{align}

If the assumption that the heavy quark PDF is perturbatively generated at threshold is no longer valid, one needs to construct the 3FS PDFs using the whole inverse of the matching matrix:
\begin{equation}\label{eq:pdfmatching2}
	f_i^{[3]} (Q^2) = \sum_{j=g, q, \bar q, h, \bar h} K^{-1}_{ij} \left( \frac{m_h^2}{Q^2} \right) \otimes f_j^{[4]} (Q^2).
\end{equation}
and the FONLL result is
\begin{align}
			F_{\text{FONLL}}(Q^2, m_h^2) =& \sum_{i,j=g, q, \bar q, h, \bar h} \left[C_i^{[3]} \left( \frac{m_h^2}{Q^2} \right) - C_i^{[3,0]}  \left( \frac{m_h^2}{Q^2} \right)  \right]\otimes  K^{-1}_{ij}  \left( \frac{m_h^2}{Q^2} \right) \otimes  f_j^{[4]} (Q^2) \nonumber\\
		& +  \sum_{i=g, q, \bar q h, \bar h} C_i^{[4]} \left(0\right) \otimes  f_j^{[4]} (Q^2).
\end{align}
However, as $ C_i^{[4]} $ and $C_i^{[3,0]}  $ are related by eq.~\eqref{eq:masslesslimitcf}, the difference term vanishes identically, and the expression is simply
\begin{align}\label{eq:FONLLic}
			F_{\text{FONLL}}(Q^2, m_h^2) &= \sum_{i,j=g, q, \bar q, h, \bar h} C_i^{[3]} \left( \frac{m_h^2}{Q^2} \right) \otimes  K^{-1}_{ij}  \left( \frac{m_h^2}{Q^2} \right) \otimes  f_j^{[4]} (Q^2)\nonumber\\
			&\equiv  \sum_{j=g, q, \bar q, h, \bar h} C_j^{[4]} \left( \frac{m_h^2}{Q^2} \right)  \otimes  f_j^{[4]} (Q^2).
\end{align}
The coefficient functions $C_i^{[4]} (m_h^2/Q^2)$ can be identified with the coefficient functions defined in the ACOT scheme, since the matching conditions $K^{-1}_{ij}$ subtract order by order the unresummed collinear logarithms such that the convolution with the PDFs has a well-behaved perturbative expansion.

The difference between the result in eq.~\eqref{eq:FONLLic} and the original FONLL formulation~\eqref{eq:FONLLzic} can be cast in a particular compact form~\cite{Ball:2015dpa}
\begin{equation}\label{eq:deltaFONLL}
	\Delta F_{\text{IC}}(Q^2, m_h^2) = \sum_{h, \bar h} \left [C_i^{[4]} \left( \frac{m_h^2}{Q^2} \right)-C_i^{[4]} (0)  \right]\otimes\left[f_i^{[4]}- \sum_{k,l=g, q,\bar q} K_{ik}\otimes K_{kl}^{-1}\otimes f_l^{[4]} \right],
\end{equation}
which shows that the missing mass corrections in eq.~\eqref{eq:FONLLzic} are entirely contained in the mass dependence of the 4FS heavy quark-initiated contributions.
Since $\Delta F_{\text{IC}}$ becomes zero identically if one sets $C_{h, \bar h}^{[4]} ( m_h^2/Q^2) \rightarrow C_{h, \bar h}^{[4]} (0) $, one can also see explicitly that the original FONLL formulation of ref.~\cite{Forte:2010ta} is equivalent order by order to S-ACOT~\cite{Ball:2015dpa}. 

Eq.~\eqref{eq:deltaFONLL} shows manifestly that $\Delta F_{\text{IC}}$ is suppressed as  $m_h^2/Q^2$ at $Q^2 \gg m_h^2$.
Nevertheless, $\Delta F_{\text{IC}}$ is generally $\mathcal O (1)$ at threshold, unless the heavy quark distribution is purely perturbative. 
In that case, at the initial scale $Q_0$ the difference term is identically zero as eq.~\eqref{eq:pdfmatching1} can be used, whilst for $Q > Q_0$, and with the matching matrix truncated at order $\as^j$, we can see that
\begin{equation}
	f_i^{[4]}(Q^2) - \sum_{k,l=g, q,\bar q} K_{ik}\otimes K_{kl}^{-1}\otimes f_l^{[4]} (Q^2) \sim \mathcal O \left(\as^{j+1} \ln^{j+1 } \frac{m_h^2}{Q^2}\right),
\end{equation}
and so in absence of intrinsic heavy quark distributions $\Delta F_{\text{IC}}$ is subleading at $Q\sim m_h$.

Whilst in the case of top and bottom it is likely safe to assume that the heavy quark PDF is purely perturbative, it is possible that in presence of intrinsic charm one should in principle compute massive coefficient functions with incoming charm, unless some assumptions are made on the nature of the intrinsic charm density.
For instance, if one assumes that the intrinsic contribution is $\mathcal O (\LambdaQ^2/m_h^2)$, then  $\Delta F_{\text{IC}}$ is formally suppressed by $\LambdaQ^2 /Q^2$ and is therefore a higher-twist effect, beyond the accuracy of the factorization theorem. 
Recent analyses have suggested that the impact of $\Delta F_{\text{IC}}$ is indeed small~\cite{Ball:2016neh}, though there is no conclusive evidence on the nature of intrinsic charm. 

\begin{figure}[t]
\centering
   \includegraphics[width=0.49\textwidth,page=1]{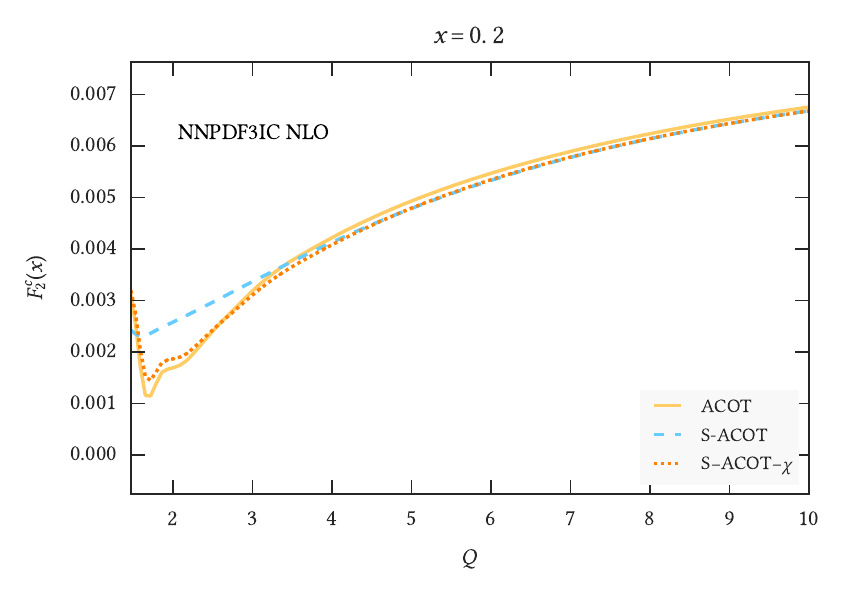}
 \includegraphics[width=0.49\textwidth,page=3]{figures/NLOIC_thesis.pdf}
  \caption{ The charm structure function $F_2^c$ in different heavy quark schemes at NLO accuracy for two different values of $x$, computed with PDFs from ref.~\cite{Ball:2016neh}}
 \label{fig:ACOTvsSACOT}
\end{figure}

From a purely intuitive point of view, however, one could argue that the best accuracy is obtained by VFNSs which respect particle kinematics, like the ACOT scheme.
Since the calculation of massive coefficient functions is however rather cumbersome (currently, only the $ \mathcal O (\as)$ diagrams with incoming heavy quarks are known~\cite{Hoffmann:1983ah,Kretzer:1998ju}), the accuracy of calculations in the presence of intrinsic PDFs can be improved by restoring the massive kinematics in contributions computed in the massless limit.
In fig.~\ref{fig:ACOTvsSACOT}, we compare the charm structure function $F_2^c $ (the part of the structure function in which the struck quark is a charm) at order $\as$, using as input a PDF set where the charm density is fitted to data and therefore can take non-zero values below the matching scale. 
We observe that there are some differences at large-$x$ between S-ACOT and ACOT when $Q\sim m_c$,  $m_c$ being the charm mass, whereas S-ACOT-$\chi$ is always very close to ACOT\footnote{There are several prescriptions for $\chi$-rescaling in the literature. Here we restore the massive kinematics by replacing the integrals of the form $x \int_x^1 dz f(z) C(x/z)$ which appear in the computation of the heavy-quark initiated contribution in the massless scheme with $\chi \int_\chi^1 dz f(z) C(\chi/z)$, where $\chi=x (1+4m_c^2/Q^2)$. With this prescription, S-ACOT-$\chi$ and ACOT coincide at order $\as^{0}$. }. 

\begin{mccorrection}
Whether one should adopt an empirical procedure, and thus determine the heavy quark PDF from a fit to experimental data, or a theoretical prejudice, by setting the heavy quark distribution in the massive scheme to zero, depends on the heavy quark mass and on the precision of the existing data.
The charm mass is sufficiently close to $\Lambda_{\text{QCD}}$ and the experimental data are sufficiently precise that the empirical approach may prove necessary.
Since the differences between schemes which respect particle kinematics and those which do not are non-negligible, the use of a VFNS which allows for possible effects due to an intrinsic charm component in the proton may be preferable in view of applications to PDF determination. 
\end{mccorrection}

\begin{savequote}[8cm]

\textlatin{\`E la somma che fa il totale.}
\qauthor{--- Tot\`o, \textit{Tot\`o, Fabrizi e i giovani d'oggi}}

 
\end{savequote}

\chapter{\label{ch-res}Resummation(s)} 


\lettrine[lines=2]{F}{ixed-order} perturbation theory is based on a simple idea: theoretical predictions require the calculation of a finite number of terms in the perturbative expansion, since higher-order terms get progressively smaller and can be neglected once the desired accuracy is reached\footnote{This idea has an underlying assumption, that is the convergence of the series.
Nevertheless, it is well known that in quantum field theory perturbative expansions are generally divergent~\cite{Dyson:1952tj,tHooft:1977xjm}.
In particular, in QCD the perturbative coefficients display a factorial growth (renormalons); as a consequence, non-perturbative corrections are needed to define the theory unambiguosly.
See e.g. ref.~\cite{Beneke:1998ui} for a thorough review on renormalon ambiguity.}. 
Sometimes, however, (logarithmically)-enhanced contributions appear at all orders and spoil the convergence of the perturbative series.
When this happens, it is necessary to reorganize the perturbative expansion and perform an all-order summation of the enhanced terms. 
This procedure is called resummation. 

Resummed calculations are typically needed in processes which depend on more than one scale.
In such cases, the perturbative coefficients present a single, or double, logarithmic dependence on the ratio $R$ of the scales involved, i.e. terms of the form $(\as \ln R)^n$ or $(\as \ln^2 R)^n$ appear at all orders.
If $\ln R$ gets sufficiently large, these contributions are of order one and must be systematically summed to all orders to obtain reliable predictions.
We have already discussed such a typical large logarithmic enhancement: collinear logarithms of the ratio $\muF/Q_0$ appear when one considers DGLAP evolution from an initial scale $Q_0$ to the factorization scale $\muF$. 
These single-logarithmic contributions do not appear explicitly in perturbative calculations, as they are resummed to all orders in the parton densities by DGLAP evolution.

Different large logarithms are associated to the brehmsstrahlung spectrum of soft gluons. 
Soft gluons can be radiated collinearly and lead to a double-logarithmic enhancement in the partonic cross section. 
As we will see in sect.~\ref{sec:soft-resum}, these soft (or {\itshape Sudakov}) logarithms appear because in spite of the KLN theorem real and virtual contributions can become highly unbalanced. 
Another class of logarithms appears when the hard scale of the collision $Q$ is much smaller than the energy of the centre of mass of the collision $\sqrt{s}$.
In this case, the physical mechanism responsible for the single-logarithmic enhancement is the multiple gluon radiation over the wide rapidity gap which is present at high energy. 
Resummation of small-$x$ ($x = Q/\sqrt{s}$), or high-energy, logarithms will be discussed in sect.~\ref{sec:small-xres}.

\section{Sudakov resummation}\label{sec:soft-resum}

In chapter~\ref{ch-qcd} we have seen how cancellation theorems guarantee the cancellation of soft gluon singularities for IRC safe observables.
These theorems state in formal terms that since particle detectors have a finite energy resolution, (inclusive enough) observables are not sensitive to arbitrary soft radiation in the final state; this undetected real gluon emission cancels exactly the singularities which appear in virtual contributions. 
However, despite the explicit cancellation of singularities, real and virtual contributions can become highly unbalanced in processes where the real radiation is strongly constrained by kinematics, and large logarithms appear as a left-over of the real-virtual cancellation of IRC divergences. 
Typically, this happens for processes at the exclusive boundary of the phase space, or when restrictive cuts are applied, for instance to increase the sensitivity in experimental searches. 
To rescue the predictive power of perturbation theory, these logarithmically-enhanced terms must be resummed to all orders.



\subsection{Sudakov double logarithms}

Let us illustrate the situation with an example. 
Consider a generic infrared and collinear safe observable $v$ and assume without loss of generality that it is dimensionless and that $0 < v < 1$. 
We want to assess how the emission of soft-gluon radiation affects a distribution or cross section $\sigma^{(n)}$ at the next perturbative order.
If $1-z$ denotes the fraction of energy carried by unobserved soft particles in the final state, the real gluon emission probability can be written as (see e.g.~\cite{Catani:1997xc})
\begin{equation}\label{eq:softradreal}
	\frac{d w_{\text{real}} (z)}{dz} = 2C \frac{\as}{\pi} \frac{1}{1-z} \ln \frac{1}{1-z} \Theta(1-z-\eta),
\end{equation} 
where $C$ is a (colour) coefficient which depends on the process. 
Eq.~\eqref{eq:softradreal} is valid in the soft and collinear limit, and we can recognize the double logarithmic divergence discussed in sect.~\ref{sec: IRC safety} due to the combination of the brehmsstrahlung spectrum $dE_g/E_g$ and the angular distribution $d\theta^2/\theta^2$ for collinear radiation.
An (unphysical) lower cut-off $\eta$ has been introduced to regularize the divergence.

One has to include also contributions from virtual emissions, whose probability can be written as 
\begin{equation}
		\frac{d w_{\text{virt}} (z)}{dz} = - 2C \frac{\as}{\pi} \delta(1-z) \int_0^{1-\eta} \frac{d \zeta }{1-\zeta} \ln \frac{1}{1-\zeta}.
\end{equation}
If taken separately, real and virtual contributions are divergent. 
However, in case of infrared and collinear safe observables the total emission probability is finite and one can take the limit $\eta \rightarrow 0$:
\begin{equation}\label{eq:softradtot}
	\frac{d w (z)}{dz}  = \lim_{\eta\rightarrow 0}\left[\frac{d w_{\text{real}} (z)}{dz} + \frac{d w_{\text{virt}} (z)}{dz} \right]=   2C \frac{\as}{\pi}\left(\frac{1}{1-z}\ln \frac{1}{1-z }\right)_+.
\end{equation}
In particular, if the emission probability is integrated from $z=0$ up to $z=1$ we have that
\begin{equation}
	\int_0^1 dz \frac{d w (z)}{dz} = 0. 
\end{equation} 

In physical processes, however, kinematic constraints may unbalance virtual and real contributions, for instance if the real emission becomes strongly suppressed.
As a consequence, the coefficients of the perturbative expansion can develop singularities.
Let us consider the case of an observable $v$ which increases in the presence of soft radiation, such that the emission of a soft gluon with energy fraction $1-z$ leads to an enhancement $\delta v =\mathcal O(1-z)$. 
In the soft limit, $\sigma^{(n+1)}$ can be written as
\begin{equation}
	\sigma^{(n+1)} (v) = \int_0^v dy\,  \sigma^{(n)} (v-y) \left( \frac{d w (z)}{dz} \right)_{z=1-y} + \ldots \ . 
\end{equation}  
For an observable which increases if soft partons are radiated, real emission is hampered in the region $v \rightarrow 0$. 
In this limit the partonic cross section therefore develops a divergent contribution:
\begin{equation}
	\sigma^{(n+1)} (v)   \underset{v \rightarrow 0}{\sim}  -C \frac{\as}{\pi} \sigma^{(n)} (v) \ln^2 v + \ldots \ .
\end{equation}
The double-logarithmic behaviour is known as Sudakov \mccorrect{suppression}~\cite{Sudakov:1954sw} (soft-non collinear and hard-collinear contributions develop at most a single-logarithmic contribution).
When $v \rightarrow 0$, these logarithms can become large and the cross section can become negative even in the perturbative regime $\as \ll 1$.
This example shows that, despite the cancellation theorems of IR singularities, soft gluon effects can still be large if real and virtual terms are kinematically unbalanced\footnote{Though here we have discussed singularities at the exclusive boundary of phase-space, singularities due to soft-gluon emission can also appear {\it inside} the physical region of phase space and give rise to so-called Sudakov shoulders~\cite{Catani:1997xc}.}.
As a consequence, fixed-order predictions are marred by infrared logarithms and all-order calculations are necessary to get reliable results.

Sudakov logarithms occur in a variety of observables, such as event-shape distributions in $e^+ e^-$ annihilation, like thrust $1-T$, $C$-parameter or heavy jet mass $\rho_H$.
Fairly obviously, soft-gluon radiation also affects observables in hadron collisions.
Primary examples are the production of a system of high invariant mass close to threshold (where $1-v \sim M/\sqrt{s} $) and the cumulative cross section
\begin{equation}\label{eq:cumulative}
  \Sigma(v) \equiv \int_0^v dv' \frac{d \sigma(v')}{d v'}
\end{equation}
for the distribution of the transverse momentum $p_t$ in colour-singlet production, where $v\sim p_T/M$.

The first studies for the all-order resummation of Sudakov logarithms for transverse-momentum distributions and for hard processes near threshold started about four decades ago~\cite{Bassetto:1984ik,Dokshitzer:1978hw}, and various techniques to achieve the resummation of the logarithmically-enhanced terms have since been proposed.
A categorization of the different techniques is open to more than one interpretation and necessarily bears some elements of subjectivity.  
In a broad sense, one can recognize a chiefly European/Russian school and a chiefly American school.
The former places the emphasis on the properties of the QCD matrix elements and of gluon radiation (angular ordering and coherence), whereas the latter exploits the factorization properties of the observable in the singular regions of phase space and the associated RG evolution equation to obtain resummed expressions to any logarithmic order. 
In both cases, however, the all-order resummation of infrared logarithms is based on dynamic and kinematic factorization properties of the observable.
Thanks to gauge invariance and unitarity, multi-gluons amplitudes in the soft limit can be factorized into universal contributions. 
However, kinematic factorization is process-dependent and \mccorrect{is usually performed} separately for each observable.
Factorization is typically obtained in a suitable conjugate space, where the phase-space constraints can be tamed.

A more recent approach to resummation, partially related to the American school, is based on methods of Soft-Collinear Effective Theory~\cite{Bauer:2000ew,Bauer:2000yr,Bauer:2000yr,Bauer:2001ct,Beneke:2002ni,Beneke:2002ph,Hill:2002vw} (SCET), an effective theory of QCD constructed to describe in detail soft and collinear radiation (see~\cite{Becher:2014oda} for a recent review).
Similarly to the American school initiated by Sterman et al., the starting point is the derivation of a factorization theorem in terms of soft, collinear, and hard contributions.
The associated RGEs are also in this case usually solved in a conjugate space where the factorization is manifest. 
Unlike Sterman's approach, however, \mccorrect{the factorization of the relevant modes is assumed already at the level of an effective Lagrangian}. 

Here we use the examples of threshold and transverse-momentum resummation to introduce and compare some features of these three approaches.
Finally, we will introduce an alternative approach to resummation, which is based on the coherent branching formalism introduced by Catani, Marchesini and Webber~\cite{Catani:1990rr} and overcomes the need for observable-dependent factorization theorems by performing the resummation in direct space.

\subsection{Logarithmic accuracy}\label{sec:logaccuracy}

Resummation is a reorganization of the perturbative expansion in which classes of logarithms are summed at all orders.
The accuracy of a resummed calculation is therefore determined by the number of terms which are correctly predicted.
In the literature several conventions are being used to define the logarithmic accuracy, which may sometimes be potentially confusing~\cite{Almeida:2014uva}.
We briefly review some common definitions in this section.

For definiteness, we consider a generic \mccorrect{(dimensionless)} observable $\tilde \sigma$ in the conjugate space where it factorizes, and we write its all-order structure as
\begin{equation}\label{eq:sigmanonlog}
	\tilde \sigma (\nu ) = 1 + \sum_{n=1}^\infty \as^n \sum_{m=1}^{2n} c_{nm} \ln^m (\nu ) +\ldots\, ,
\end{equation}
where $\nu$ is the conjugate variable of $v$ and the dots indicate non-singular terms.
As we shall see below, for many observables the dominant class of logarithms exponentiates, $\tilde \sigma (\nu) \approx e^{\mathcal O\left( \as \ln^2 (\nu)\right) }$.
In these cases, it is therefore natural to write the observable in exponential form
\begin{equation}\label{eq:sigmalog}
	\tilde \sigma (\nu ) = g_0 (\as)  \exp\left[ \sum_{n=1}^\infty \as^n \sum_{m=1}^{n+1} b_{nm} \ln^m (\nu ) \right] + \ldots \, ,
\end{equation}
and to perform the counting at the level of the exponent.
Here the dots denote the non-singular terms and the function $g_0$ has an expansion in $\as$ independent of logarithms:
\begin{equation}
 g_0 (\as) = 1 +  g_{01} \as + g_{02} \as^2 + \mathcal O(\as^3).
\end{equation}

The counting at the exponent is widely used in the literature and we shall use it throughout this thesis when discussing Sudakov resummation.
The accuracy is determined by the classes of terms which are correctly predicted in eq.~\eqref{eq:sigmalog}. 
Leading logarithmic (LL) accuracy is achieved if the tower of terms $b_{nm}$ with $m=n+1$ is correctly predicted; next-to-leading logarithmic (NLL) accuracy if also the terms with $m=n$ are under control; generally, at N$^k$LL accuracy all the terms with $n-k+1 \le m \le n+1$ are known. 
This counting is appropriate when the logarithmic terms dominate the perturbative expansion and extends the region of applicability of perturbative QCD to the region $\as \ln (\nu) \lesssim 1 $ (whereas a naive expansion in eq.~\eqref{eq:sigmanonlog} is accurate in the more restricted region $\as \ln^2 (\nu) \lesssim 1 $).
This way, the LL terms are order $1/\as$, NLL are of order 1, NNLL are of order $\as$, and so on.

Within this counting, the order $\as$ term of the constant function $ g_0$ is formally NNLL.
However, since the organization of the constant terms differs from the exponentiated ones, it is possible to perform an alternative counting --- the so-called `primed' counting --- in which $g_{01}$ enters already at NLL$^\prime$.
By comparing eq.~\eqref{eq:sigmanonlog} and~\eqref{eq:sigmalog} one can see that in the primed counting an additional tower of logarithms is included at the level of $\tilde \sigma(\nu)$ (see table~\ref{tab:count}), which increases the logarithmic accuracy by `half' a logarithmic order (for instance, the coefficient of $\as^2 L^2$ is correctly predicted only if $ g_{01}$ is included).
The primed counting is more appropriate in the transition region where the non-singular terms become of the same order of the logarithms and the inclusion of the next order of $ g_0$ can capture the bulk of the next logarithmic order accuracy. 
If the primed counting is used, the resummed expression at N$^k$LL$^\prime$ accuracy contains ingredients at the same order of N$^k$LO computation; therefore, it is natural to match N$^k$LL$^\prime$ results to N$^k$LO, whereas the natural matching for the unprimed counting is N$^k$LL to N$^{k-1}$LO\footnote{\mccorrect{Note that this statement is true as long as an additive matching is performed; in multiplicative schemes}~\cite{Banfi:2012yh,Banfi:2012jm,Banfi:2015pju}, \mccorrect{N$^k$LO ingredients are automatically retrieved in the matching procedure and the natural matching is N$^k$LO+N$^k$LL.}}.

\begin{table*}[t]
\begin{center}
\begin{tabular}{llccl}
  Notation$^\prime$ & Notation* & $c_{nm}$ & $b_{nm}$ & Natural matching \\
  \hline
  LL        &   LL      &\multicolumn{1}{r}{$m= 2n$} & \multicolumn{1}{r}{$m= n+1 $} & LO \\
  NLL       &   NLL*    & $2n-1\le m\le 2n$ & \multicolumn{1}{r}{$n \le m\le n+1$} & LO \\
  NLL$^\prime$    &   NLL     & $2n-2\le m\le 2n$ & \multicolumn{1}{r}{$n \le m\le n+1$} & NLO \\
  NNLL      &   NNLL*   &  $2n-3\le m\le 2n$  & $n-1 \le m\le n+1$ & NLO  \\
  NNLL$^\prime$   &   NNLL    & $2n-4\le m\le 2n$ & $n-1 \le m\le n+1$ & NNLO \\
  N$^3$LL   &   N$^3$LL*& $2n-5\le m\le 2n$ & $n-2 \le m\le n+1$ & NNLO \\
  N$^3$LL$^\prime$ &   N$^3$LL & $2n-6\le m\le 2n$ & $n-2 \le m\le n+1$ & N$^3$LO
\end{tabular}
\caption{Orders of logarithmic approximations in resummed computations. We report in the first two columns the logarithmic accuracy in the primed and in the starred notation, corresponding to the inclusion in the coefficient function of terms as given in the third and fourth column.}
\label{tab:count}
\end{center}
\end{table*}

The notation of `primed' and `unprimed' accuracy, however, was only rather recently introduced in the literature.
In particular, the unprimed notation has been used --- and is still widely used --- to denote resummation performed both at primed and unprimed accuracy.
This lead some authors to introduce a parallel notation, where the unprimed notation is denoted by a star, which we report it in table~\ref{tab:count} for completeness. 
Unless otherwise stated, we shall adopt the primed-unprimed notation.

Finally, let us notice that subleading ambiguities can affect calculations performed at the same formal logarithmic accuracy.
For instance, since the logarithmic accuracy is typically defined in conjugate space, the inverse transform may introduce subleading terms, which in some cases can have a non-negligible impact on phenomenological predictions.
We will briefly discuss this issue in the next section, and we will come back to it in chapter~\ref{ch-respheno}.

\subsection{Threshold resummation}\label{sec:threshold}

Let us consider the inclusive production of a system with high invariant mass $M$. 
Though soft gluons have generally a small effect as they do not significantly modify the kinematics of the process, they become important near the production threshold $M^2/s \rightarrow 1$, where $\sqrt{s}$ is the hadronic centre-of-mass energy.
In this region gluon emission is strongly suppressed, as even a small amount of energy spent in soft radiation can significantly reduce the available energy. 
\mccorrect{Since radiation effects are already included in the PDFs when solving the DGLAP equation, the sign of the correction due to soft radiation is generally scheme-dependent; in the $\MSbar$ scheme, gluon emission does not modify the evolution of the parton densities and \textit{enhances} the coefficient functions}~\cite{Korchemsky:1988si,Albino:2000cp,Forte:2002ni}.
In particular, each extra emission contributes with a double logarithm of the form $\as \ln^{2} (1-z)$, enhancing the fixed-order perturbative calculation.
These contributions become progressively large approaching the kinematic threshold and must be resummed to all orders. 

It is important to observe that these resummed contributions affect the coefficient functions rather than the hadronic cross section.
As a consequence, soft contributions can be sizeable also far from the hadronic threshold.
Since the parton densities are strongly suppressed at large $x$, in the convolution integral eq.~\eqref{eq:hadronfact2} the coefficient functions are typically evaluated much closer to partonic threshold than the hadronic cross section as the dominant value of $\hat s = x_1 x_2 s $ is considerably smaller than $s$~\cite{Bonvini:2012an}.

Threshold resummation is a classic problem, which has been extensively studied in the past decades.
Resummation formul\ae\ were first obtained using factorization techniques by Sterman~\cite{Sterman:1986aj} or a direct diagrammatic approach by Catani and Trentadue~\cite{Catani:1989ne} which lead to equivalent results~\cite{Catani:1990rp,Forte:2002ni}.
Since then, several resummation techniques have been developed and a large variety of observables has been resummed to high logarithmic accuracy.
Threshold resummation is currently known up to N$^3$LL$^\prime$ accuracy (with the exception of the four-loop cusp anomalous dimension, see below) for DIS~\cite{Moch:2005ba}, Higgs~\cite{Bonvini:2014joa,Catani:2014uta,Bonvini:2014tea,Schmidt:2015cea}, Drell-Yan~\cite{Moch:2005ky,Catani:2014uta,Laenen:2005uz,Ahmed:2014cla}, and several other processes such as $t \bar t$ production are known up to NNLL$^\prime$ accuracy~\cite{Czakon:2009zw,Cacciari:2011hy}.
Most of these resummations were performed with traditional techniques, in what has been sometimes called a direct QCD (dQCD) approach.
More recently, alternative formulations based on SCET methods have been derived, which have been shown to be analytically equivalent with the results obtained with the more traditional formalisms~\cite{Bonvini:2012az,Bonvini:2013td,Sterman:2013nya,Bonvini:2014qga,Almeida:2014uva}, despite the different treatment of the subleading terms in the two approaches may lead to appreciable differences at the level of phenomenological predictions~\cite{Bonvini:2014qga}.
Here we shall briefly review the Catani-Trentadue approach and the one by Sterman, considering the case of colour-singlet production close to threshold.
We will finally compare the traditional formulation with the resummation performed in the SCET framework of refs.~\cite{Becher:2007ty,Ahrens:2008qu,Ahrens:2008nc}.

Let us start by discussing the resummation within the Catani-Trentadue approach. 
We follow a simplified argument~\cite{Catani:1996rb} which nevertheless captures the original argument of ref.~\cite{Catani:1989ne} based on properties of QCD dynamics. 
The starting point is the cross section for the production of a colourless system of mass $M$ which according to eq.~\eqref{eq:hadronfact2} can be written as
\begin{equation}\label{eq:factdimless}
	\tilde \sigma(\tau, M^2) = \frac{1}{\tau \sigma_0} \sigma(\tau, M^2) =  \int_\tau^1 \frac{dz}{z} \mathcal L \left (\frac{\tau}{z}\right) C(z,M^2),
\end{equation}
where for the sake of simplicity we omit the sum over partons and we take resummation and factorization scales to be equal to $M$.
In eq.~\eqref{eq:factdimless} we define a dimensionless cross section $\tilde \sigma(\tau, M^2)$ in terms of the Born cross section $\sigma_0$ and $\tau= M^2/s$, where $\sqrt{s}$ is the centre-of-mass energy. 
In the threshold limit $\tau \rightarrow 1 $ the convolution integral is constrained by energy conservation towards $z=1$.
Therefore, at the lowest logarithmic order the coefficient function $C(z,M^2)$ is dominated by the soft-gluon emission probability eq.~\eqref{eq:softradtot} and 
\begin{equation}
	C (z, M^2) = \delta(1-z)  + \sum_{n=1}^\infty \int_0^1 dz_1 \ldots dz_n \frac{d w_n(z_1,\ldots z_n)}{dz_1 \cdots dz_n} \Theta_{\text{PS}}^{(n)} (z; z_1, \cdots z_n) ,
\end{equation}
where $d w_n$ is the multi-gluon emission probability and $ \Theta_{\text{PS}}^{(n)}$ generically denotes the available phase-space region for the process under consideration.

In the soft and collinear limit, due to properties of gauge invariance and unitarity, the most singular part of the multi-gluon emission probability factorizes (see e.g.~\cite{Luisoni:2015xha})
\begin{equation}\label{eq:eikonalfact}
	\frac{d w_n(z_1,\ldots z_n)}{dz_1 \cdots dz_n} \simeq \frac{1}{n!} \prod_{i=1}^n \frac{d w (z_i)}{dz_i}.
\end{equation}
In a non-abelian theory like QCD gluons carry colour charge; gluon correlations are present and can spoil the picture.
The {\it eikonal} approximation eq.~\eqref{eq:eikonalfact} nevertheless allows one to compute exactly the leading contributions, since at this accuracy higher-order gluon correlations for real and virtual corrections cancel by gauge invariance~\cite{Catani:1983bz,Catani:1984dp}.
In case of total cross sections, the phase space is constrained by longitudinal momentum conservation
\begin{equation}
	\Theta_{\text{PS}}^{(n)} (z; z_1, \cdots z_n)  = \delta(z-z_1\cdots z_n).
\end{equation}
This kinematic constraint is factorized exactly by considering the Mellin transform
\begin{equation}\label{eq:phasespacefact}
\int_0^1dz\, \, z^{N-1} \delta(z-z_1\cdots z_n)  = z_1^{N-1} \cdots z_n^{N-1};
\end{equation}
as a consequence, using eq.~\eqref{eq:eikonalfact} and eq.~\eqref{eq:phasespacefact},  the coefficient function exponentiates in $N$ space:
\begin{align}
		{\bm  C}_{\text{res}}(N,M^2) &= \exp\left[ \int_0^1 dz\, z^{N-1} \frac{d w(z)}{dz} \right]\nonumber\\
		&= \exp\left[ 2 C \frac{\as}{\pi} \int_0^1 dz\, z^{N-1} \left(\frac{1}{1-z}\ln \frac{1}{1-z }\right)_+  \right] \underset{N \gg 1}{\simeq} \exp\left[- C \frac{\as}{\pi} \ln^2 N + \mathcal O(\as \ln N)\right].
\end{align}
In our derivation we have neglected running coupling effects; these can be taken into account by observing that the transverse momenta of the gluons fix the proper scale at which $\as$ should be evaluated, thereby extending the accuracy to LL at the exponent
\begin{equation}\label{eq:clandau}
			{\bm C}_{\text{res}}(N,M^2) =  \exp\left[ \frac{2 C}{\pi} \int_0^1 dz\, z^{N-1} \left(\frac{1}{1-z}  \int_{M^2}^{M^2(1-z)} \frac{d \mu^2}{\mu^2}\as(\mu^2)\right)_ + \right]. 
\end{equation}
We notice that eq.~\eqref{eq:clandau} is ill-defined, since by integrating up to $z=1$ one hits the Landau pole at $z= 1-\frac{\LambdaQ^2}{M^2}$. 
However, the integrand can be expanded in powers of $\as(M^2)$ such that for each term the Mellin transform is well defined.
The resulting series is divergent; however, by keeping only the leading-power contributions in the large-$N$ limit one obtains a summable series, which leads to a finite result in $N$-space.

The result we have obtained is valid up to LL accuracy; at higher logarithmic orders the coefficient function takes the form 
\begin{align}\label{eq:CataniTh1}
	{\bm C}_{\text{res}}(N,M^2) &= \bar g_0  (\as (M^2))\nonumber\\
	 &\quad  \times \exp \left\{  \int_0^1dz\,z^{N-1} \left[ \frac1{1-z} \left( \int_{M^2}^{M^2(1-z)^2} \frac{d\mu^2}{\mu^2} 2A_{\text{cusp}}\left(\as(\mu^2)\right) + D\left(\as([1-z]^2M^2)\right) \right)\right]_+  \right\}\nonumber \\
	 &= \bar g_0 (\as(M^2)\exp  \bar {\mathcal S} (\as (M^2) , N) ,
\end{align}
which correctly includes also subleading gluon-correlation effects~\cite{Catani:1989ne}. 
The functions $\bar g_0$, $A_{\text{cusp}}$, and $D$ admit a perturbative expansion in $\as$, with $\bar g_0 (0) = 1$ and $A_{\rm cusp}(0)=D(0)=0$.
The function $A_{\text{cusp}}$ (cusp anomalous dimension) is the numerator of the divergent part of the relevant diagonal Altarelli-Parisi splitting function ($P_{gg}$ for Higgs boson production, or $P_{qq}$ for Drell-Yan pair production)
\begin{equation}\label{eq:splittinglimit}
	P(z,\as) = \frac{A_{\rm cusp}(\as)}{(1-z)_+} + B (\as) \delta(1-z) +\mathcal O  \( (1-z)^0\) ,
\end{equation}
and embodies the effect of soft-collinear radiation, whilst the process-dependent function $D$ contains the effects due to soft non-collinear radiation. 
The knowledge of the coefficient $A_{\text{cusp}}^{(1)}$ leads to the resummation of the LL terms; knowledge of $\{ A_{\text{cusp}}^{(1)}, A_{\text{cusp}}^{(2)}, D^{(1)} \} $ allows one to resum the NLL terms, etc. 
The function $A_{\text{cusp}}$ and the function $D$ for a variety of processes are currently known analytically at three loops (see for instance~\cite{Moch:2005ba,Moch:2005ky,Laenen:2005uz}), whereas the four-loop cusp anomalous dimension has only recently been computed numerically~\cite{Moch:2017uml,Moch:2018wjh}; to achieve N$^3$LL accuracy, the coefficient $A_{\text{cusp}}^{(4)}$ has been sometimes estimated with a Pad\'e approximant~\cite{Moch:2005ba}.

The Mellin integrals in eq.~\eqref{eq:CataniTh1} are frequently evaluated in the large-$N$ limit, thereby keeping only those contributions which do not vanish at large $N$ and behave as powers of $\ln N$.
In this limit, the resummed coefficient function can be written as 
\begin{align}\label{eq:CataniTh3}
	{\bm C}_{\text{res}}(N,M^2) =\hat g_0  (\as (M^2)) \exp \left\{   \int_{M^2}^{M^2/\bar{N^2}} \frac{d\mu^2}{\mu^2} \left[A_{\text{cusp}}\left(\as(\mu^2)\right) \ln\frac{M^2}{\mu^2 \bar N^2} + \hat D\left(\as(M^2)\right)\right] \right\},
\end{align}
where $\bar N = N e^{\gamma_E}$, and the relation between the perturbative expansions of the functions in eqs.~(\ref{eq:CataniTh1},\ref{eq:CataniTh3}) can be computed perturbatively~\cite{Bonvini:2013td,Bonvini:2014qga,Bonvini:2015pxa}.
Usually, eq.~\eqref{eq:CataniTh3} is rewritten as~\cite{Catani:1989ne} 
\begin{align}
\label{eq:CataniTh2}
{\bm C}_{ N{\text{-soft}}} &=g_0(\as) \left[\frac{1}{\as} g_1(\as\ln N)+ g_2(\as\ln N)+ \as g_3(\as\ln N)+\dots\right],
\end{align}
where the functions $g_i(\as\ln N)$ with $i\geq1$ resum $\as^n \ln^n N$ contributions to all orders in perturbation theory and can be derived by computing the integrals as an expansion in powers of $\as$ at fixed $\as\ln N$.
The functions $g_i, \, i = 1, 2, 3, 4$ can be found for several processes in ref.~\cite{Moch:2005ba}.
We note that the two expressions eqs.~\eqref{eq:CataniTh2} and~\eqref{eq:CataniTh1} differ by subleading terms, as in eq.~\eqref{eq:CataniTh2} all $N$-independent terms are included in $g_0$, whereas in eq.~\eqref{eq:CataniTh1} some constant terms are exponentiated.
The relation between $g_0$ and $\bar g_0$ is detailed for instance in ref.~\cite{Ball:2013bra}.
In eq.~\eqref{eq:CataniTh2} the subscript $N$-soft stresses that the result is obtained in the large-$N$ limit.
Other resummation schemes, which are equivalent to the $N$-soft scheme in the $N\rightarrow\infty$ limit, but which preserve the analytic structure of the fixed-order coefficient functions at finite $N$, have also been considered~\cite{Bonvini:2014joa,Ball:2013bra,Bonvini:2014jma,Muselli:2015kba} and will be introduced in chapter~\ref{ch-respheno}, where we also discuss the exponentiation of the constants.

The result eq.~\eqref{eq:CataniTh2} corresponds to the result found by Sterman in~\cite{Sterman:1986aj} using factorization and RGE arguments (for a review of Sterman's method, see~\cite{Sterman:1995fz,Laenen:2004pm}).
Sterman's derivation starts from a {\it re-factorization} of the partonic cross section in the limit $1-z\rightarrow 0$, which in this limit can be written as 
\begin{equation}\label{eq:Sterman1}
	C(z,M^2,\muF^2) = H (\muF^2)\,  \tilde S\left(z, \muF^2, \ln \frac{M^2 (1-z) }{ \muF^2}\right) + \mathcal O ((1-z)^0)
\end{equation}
where $\muF$ is a factorization scale, $H$ is an infrared-safe function which does not depend on $z$, and the {\it soft} function $S$, whose definition requires the concept of Wilson lines\footnote{These are theoretical objects which encode the coupling of soft gluons to a single parton neglecting recoil effects, and therefore provide a natural building block for the definition of cross sections in the soft (or eikonal) limit (see e.g.~\cite{White:2015wha}).}, has a perturbative expansion in $\as$ and contains large logarithms of $\ln \frac{M^2 (1-z) }{ \muF^2}$.

We learned in chapter~\ref{ch-qcd} that whenever there is factorization there is evolution, and wherever there is evolution, there is resummation.
The Mellin transform of the soft function $\tilde {\bm S} \left(N,\muF^2,L\right) $, where $L$ is $\ln \frac{M^2}{\muF^2 \bar N^2}$, obeys the RGE
\begin{equation}\label{eq:RGsoft}
	\frac{d}{d \ln \mu} \tilde {\bm S} \left(N,\mu^2,L\right)  = -2 A_{\text{cusp}} (\as(\mu^2)) \ln \frac{M^2}{\bar N^2 \mu^2} - 2 \gamma_W (\as (\mu^2))
\end{equation}
where $\gamma_W$ is a process-dependent anomalous dimension. 
The resummation formula can be obtained by solving the RGE~\eqref{eq:RGsoft} and the associated RGE for the parton densities in the large-$N$ limit
\begin{equation}\label{eq:RGPDF}
	\mu \frac{\partial}{\partial \mu} \ln {\bm f} (N,\mu) = -2 A_{\text{cusp}} (\as(\mu^2)) \ln \bar N + B (\as (\mu^2)),
\end{equation} 
where $B$ is the coefficient of the $\delta (1-x)$ term of the relevant Altarelli-Parisi splitting functions, see eq.~\eqref{eq:splittinglimit}. 
The large logarithms in the soft function can be resummed to all orders by solving the RGE \eqref{eq:RGsoft} from a scale $M/\bar N$ to the factorization scale $\muF$: 
\begin{equation}\label{eq:RGsoftsol}
\tilde {\bm S} \left(N, \muF^2,L\right) = \tilde {\bm S} (N, M^2/\bar N^2,0) \exp \left\{ \int_{M/\bar N}^{\muF} \frac{d\mu'}{\mu'} \left(2 A_{\text{cusp}} \ln \frac{{\mu'}^2 \bar N^2}{M^2}-2\gamma_W(\as({\mu'}^2))\right)\right\},
\end{equation}
whereas the solution of eq.~\eqref{eq:RGPDF} is
\begin{equation}
	{ \bm f} (N, \muF) ={\bm f} (N, M) \exp\left[\int_{M}^{\muF} \frac{d \mu'}{\mu'} (-2 A_{\text{cusp}}(\as ({\mu'}^2))\ln \bar N + B (\as({\mu'}^2)))\right]. 
\end{equation}
By combining the two RGEs and absorbing the $\muF$ dependence associated to $B$ into the short-distance function $H$, the $N$-space partonic coefficient function can be reduced to the large-$N$ result eq.~\eqref{eq:CataniTh2}~\cite{Catani:1990rp}.

To obtain the resummed cross section in direct space one has to compute the inverse Mellin transform of the resummed Mellin space expression
\begin{equation}\label{eq:MellinInverse}
	\tilde \sigma (\tau ,M^2) = \frac{1}{2 \pi i} \int_{c-i \infty}^{c+i\infty} dN\, \tau^{-N}\mathcal {\bm  L} (N) \, {\bm  C}_{ N{\text{-soft}}} (N,M^2).
\end{equation}
However, such an inverse transform does not exist.
Indeed, the functions $g_i$ in eq.~\eqref{eq:CataniTh2} have a branch cut in the complex $N$ plane along the real axis for $N \ge N_L$, where $N_L = \exp \frac{1}{2 \as \beta_0}$. 
As Mellin transforms always have a convergence abscissa, ${\bm  C}_{ N{\text{-soft}}} (N,M^2)$ cannot be the Mellin transform of any function.
A simple solution is provided by the minimal prescription (MP)~\cite{Catani:1996yz}. 
The MP defines the resummed cross section in direct space by computing the integral in eq.~\eqref{eq:MellinInverse} with an abscissa $c_{\text{MP}}$ which lies between the rightmost singularity in eq.~\eqref{eq:MellinInverse}, but to the left of the branch cut. 
The slope of the integration path is then bent to ensure \mccorrect{numerical stability}.
The resulting cross section is finite and converges asymptotically to the sum of the divergent series obtained by expanding the coefficient function in $N$-space in powers of $\as$ and performing a Mellin transform of each term.
Alternatively, one can consider the Borel sum of the divergent series~\cite{Forte:2006mi}.

A different approach to threshold resummation, which provides an alternative solution to the Landau pole problem, is based on the framework of soft-collinear effective theories.
Within the SCET formalism, hard, collinear and soft modes are integrated out in a series of matching steps.
This allows one to write the coefficient function in a factorized form in terms of a soft and a hard function, in complete analogy to eq.~\eqref{eq:Sterman1}.
As we have discussed above, these functions obey a RGE which can be solved in a closed form. 
However, the procedure to solve the RGE and the choice of scales differ from the traditional dQCD formalism.
In particular, the soft and the hard function are evaluated at a soft scale $\muS$ and a hard scale $\muH$, respectively, and the coefficient function eq.~\eqref{eq:Sterman1} is supplemented with an evolution factor $U(\muH^2,\muS^2,\muF^2)$ to the common factorization scale $\muF$. 

Thanks to the introduction of the soft scale $\muS$ the resummed coefficient function can be written in $z$ space, without hitting the Landau pole:
\begin{align}\label{eq:scetfact2}
&C(z,M^2,\muF^2) = H(\muH^2) U(\muH^2,\muS^2,\muF^2)\tilde s\left(\ln\frac{M^2}{\muS^2}+\partial_\eta,\muS^2\right)
\frac{z^{-\eta}}{(1-z)^{1-2\eta}}  \frac{e^{2\gamma\eta}}{\Gamma(2\eta)},\nonumber
\end{align}
where
\begin{align}
\eta &= - 2 \int_{\muS^2}^{\muF^2}  \frac{d \mu^2}{\mu^2} A_{\text{cusp}}\(\alpha(\mu^2)\),
\end{align}
and $\tilde s$ is a process-dependent soft function, related to $\tilde S$ in eq.~\eqref{eq:Sterman1}, which admits a perturbative expansion in powers of $\as$.
The evolution factor $U$ depends on several anomalous dimensions, but takes a simple form if the hard scale $\muH$ and the factorization scale $\muF$ are set to $M$:
\begin{equation}
	U(M^2,\muS^2) = \exp \left\{ -\int_{M^2}^{\muS^2} \frac{d\mu^2}{\mu^2} \left[ A_{\text{cusp}}(\as(\mu^2))\ln \frac{\mu^2}{M^2} -\gamma_W (\as(\mu^2)) \right] \right\}.
\end{equation}

In actual computations the soft scale has to be set to a particular value to avoid a reappearance of the Landau pole.
For instance, if $\muS \propto M (1-z)$ the Landau pole is present when computing the convolution integral~\cite{Beneke:2011mq}.
In the approach of refs.~\cite{Becher:2007ty,Ahrens:2008qu,Ahrens:2008nc} the soft scale is instead proportional to a hadronic scale,  $\muS \propto M (1-\tau)$. 
With this choice, the coefficient function is free from the Landau pole and the integral over $z$ can be performed, leading to a well-defined result.
The hadronic cross section, however, no longer factorizes when taking a Mellin transform, though the impact of these factorization-violating terms are negligible~\cite{Bonvini:2014qga}.
This is perhaps not surprising, as not even the MP is a convolution, since it includes terms exponentially suppressed in $\Lambda^2_{\text{QCD}}/M^2$ in the $z\geq 1$ region~\cite{Bonvini:2010tp,Bonvini:2012sh}. 
This confirms that any finite definition of the all-order cross section requires the introduction of non-perturbative effects to regularize the Landau pole.

The SCET result can be recovered (up to subleading and power-suppressed terms) by manipulating the dQCD result eq.~\eqref{eq:CataniTh3} and inserting a dependence on an extra scale $\muS$~\cite{Bonvini:2014qga}.
However, these contributions can be important if resummation predictions are used to improve the accuracy of perturbative calculations far from threshold, as in Higgs production in gluon fusion~\cite{Bonvini:2014qga,Bonvini:2014tea}.
Subleading terms can therefore have a non-negligible impact in phenomenological applications.
Their effect can be used to assess more robustly the perturbative uncertainty from missing higher orders, as we shall discuss in chapter~\ref{ch-respheno}.

\subsection{Transverse-momentum resummation}\label{sec:tmomres}

Another classical example of \mccorrect{an} observable which is affected by large Sudakov logarithms is the transverse-momentum distribution of systems with a high invariant mass $M \gg p_t$, where the transverse momentum $p_t$ vanishes at the Born level.
In these systems -- for instance, Higgs boson or Drell-Yan pair production --- the LO transverse-momentum distribution is strongly peaked at $p_t = 0$; therefore, if the heavy system is produced with a transverse momentum much smaller than $M$ the emission of real radiation is strongly suppressed and cannot balance the virtual contributions.  
Double logarithms of $p_t/M$ appear at all orders and the convergence of the series is spoiled at small $p_t$.

The resummation for transverse-momentum distributions is particularly delicate as $p_t$ is a vectorial quantity.
A naive exponentiation of the most leading logarithmic contributions at small $p_t$  leads to a cross section which is exponentially suppressed in the limit $p_t \rightarrow 0 $ (the so-called DDT formula~\cite{Dokshitzer:1978hw}).
This formula was obtained by considering the leading soft and collinear contributions from a ensemble of $n$ gluons whose transverse momenta $k_{t,i}$ are {\it strongly ordered}:
\begin{equation}
	k^2_{t,n } \ll \cdots \ll  k^2_{t,2 } \ll  k^2_{t,1 } \lesssim p_t^2 \ll s.
\end{equation}
As a consequence, the cross section becomes naturally suppressed if $p_t \ll M$ as there is no phase space left for soft gluon production.

This is not the only mechanism, however, which leads to a system with small transverse momentum.
Indeed, the only requirement for having a system with $p_t \sim 0$ is that the vectorial sum $\sum_{i=1}^n \vec k_{t,i}$ is small.
In a seminal work, Parisi and Petronzio showed that around and below the peak of the distribution, kinematic cancellations become predominant, and the spectrum vanishes as $d \sigma/d p_t\sim p_t$ rather than exponentially~\cite{Parisi:1979se}.
In their paper, Parisi and Petronzio suggested to perform the resummation in the impact-parameter ($b$) space where the two competing effects leading to a vanishing $p_t$ are correctly handled through a Fourier transform.
In Fourier space, the phase-space constraint factorizes,
\begin{equation}\label{eq:fourier}
	\Theta_{\text{PS}} (\vec p_t, \vec k_{t,1},\ldots \vec k_{t,n} )=\delta^{(2)} \left(\vec p_t - \sum_{i=1}^n \vec k_{t,i} \right) = \int d^2 b \frac{1}{4 \pi^2} e^{i \vec b \cdot \vec p_t  } \prod_{i=1}^n e^{-i \vec b \cdot \vec k_{t,i} },
\end{equation}
and transverse-momentum conservation is respected.

Using the $b$-space formulation, Collins, Soper and Sterman (CSS) established a formalism to resum the transverse momentum in Drell-Yan pair production in ref.~\cite{Collins:1984kg}.
As with the case of threshold resummation in the Sterman's approach, the starting point is a re-factorization of the differential cross section, to identify the main integration regions for the parton momenta.
In the formalism of ref.~\cite{Collins:1984kg}, the partonic cross section is written as a convolution of parton-in-parton distributions ${\mathcal P}_{i/j}(x,{\vec k})$, at momentum fraction $x$ and transverse momentum ${\vec k}$, with an additional eikonal function $U$ which describes coherent soft-gluon emission~\cite{Collins:1981mv,Collins:1981tt,Collins:1982wa,Collins:1983ju},
\begin{align}
\frac{d \sigma_{ab\to F}}{dM^2 d{ p}^2_T}
\simeq &  \sum_{c}\ \hat \sigma_{c\bar c \to F}^{({0)}}(M^2) \;
H_{c \bar c}(M)\;
\int dx_a d^2{\vec k}_a\,{\mathcal P}_{c/a}(x_a,{\vec k}_a,M)
\int dx_b d^2{\vec k}_b\, {\mathcal P}_{\bar c/b}(x_b,{\vec k}_b,M)
\nonumber\\
& \times
\int  d^2{\vec q}\ U_{c\bar c}({\vec q})\ \delta(M^2-x_ax_bs)\
\delta^{(2)} \left(\vec p_t + {\vec q}- {\vec k}_a-{\vec k}_b\right)  ,
\label{eq:CSSfact}
\end{align}
where $\hat \sigma^{(0)}_{c \bar c \to F}$ is the LO cross section for the process $c \bar c \rightarrow F$, where $F$ can be a electroweak boson (Drell-Yan pair production, $c,\bar c = q, \bar q$) or a Higgs boson ($c,\bar c = g, g$).
The factor $H_{c\bar c} (M)=1 + {\cal O}(\alpha_s(M))$ absorbs hard-gluon corrections and is computable in perturbation theory.
Under Fourier transform, the cross section factorizes and RGEs are developed for the separate pieces and solved.
In $b$-space, the logarithms of $p_t/M$ become logarithms of $bM$ and exponentiate. 
The final result is usually written as (here we use the notation of ref.~\cite{Catani:2000vq}) 
\begin{align}\label{eq:CSSCatani}
	\frac{d \sigma_{p p \to F}}{dM^2 dp_t} =& \sum_{a,b} \int_0^1 dx_1 dx_2\, f_{a/h_1} (x_1, \muF^2)  f_{b/h_2} (x_2, \muF^2) \int_0^\infty db\, b p_t J_0(p_t b) \nonumber \\
	  & \sum_{c} \int_0^1 dz_1 dz_2\, C_{c a} (\as(b^2_0/b^2),z_1) C_{\bar c b} (\as(b^2_0/b^2),z_2)  \delta(M^2-z_1 z_2 x_1 x_2 s)\notag \\
	  &\hat \sigma^{(0)}_{c \bar c \to F} (M) H_{\text{CSS}} (M) \exp\left\{ - R_{\text{CSS},c}(b) \right\},
\end{align}
where the Bessel function $J_0$ descends from the integration over the azimuthal angles in eq.~\eqref{eq:fourier} and $b_0 = 2e^{-\gamma_E}$.
The Sudakov form factor $R_{\text{CSS}}(b)$ is defined as
\begin{align}
R_{\text{CSS},c}(b) &= \int_{b_0^2/b^2}^{M^2}\frac{d k_t^2}{k_t^2}
  {R}_{{\text{CSS}},c}'\left(k_{t}\right)=\int_{b_0^2/b^2}^{M^2}\frac{d
                    k^2_{t}}{k^2_t}
                    \left(A_{{\text{CSS}},c}(\as(k^2_t))\ln\frac{M^2}{k_t^2} +
                    B_{{\text{CSS}},c}(\as(k^2_t))\right).
\end{align}
The anomalous dimensions $A_{{\text{CSS}},c}$ and $B_{{\text{CSS}},{c}}$, the coefficient functions $C_{ab} (\as, z)$, and the process-dependent hard function $H_{\text{CSS}}$ admit an expansion in the strong coupling 
\begin{align}
&A_{{\CSS},c}(\as)=\sum_{n=1}^{4}\left(\frac{\as}{\pi}\right)^nA^{(n)}_{{\CSS},c},\,\quad
B_{{\CSS},c}(\as)=\sum_{n=1}^{3}\left(\frac{\as}{\pi}\right)^nB^{(n)}_{{\CSS},c},\nonumber \\
&C_{ab} (\as,z) = \delta_{ab} \delta(1-z) + \sum_{n=1}^{\infty} \left(\frac{\as}{\pi}\right) ^nC_{ab}^{(n)} (z), \quad
H_{{\CSS}}(M)=1 + \sum_{n=1}^{\infty}\left(\frac{\as(M^2)}{\pi}\right)^nH_{{\CSS}}^{(n)}(M).
\end{align}
Currently, all the ingredients to perform resummation up to N$^3$LL accuracy are known analytically\footnote{At sufficiently high accuracy one needs to include also the contribution from the gluon collinear correlation functions $G$, which gives a contribution analogous to the one in eq.~\eqref{eq:CSSCatani} with the coefficient functions $C$ replaced by $G$. \mccorrect{These contributions have a quantum-mechanical origin and are related to spin correlations in the gluon fusion hard-scattering subprocess}~\cite{Catani:2010pd}.}~\cite{Catani:2011kr,Catani:2012qa,Gehrmann:2014yya,Li:2016ctv,Vladimirov:2016dll}, \mccorrect{except} for the cusp anomalous dimension $A_{\text{cusp}}^{(4)}$ which enters in the expression for $A_{\text{CSS}}^{(4)}$ and is known only numerically.
Using the $b$-space formulation, the $p_t$ spectrum for colour-singlet production has been resummed up to NNLL accuracy both for Higgs~\cite{Bozzi:2005wk} and for Drell-Yan pair production~\cite{Bozzi:2010xn,Banfi:2012du}.
Results with the same accuracy have been obtained also in refs.~\cite{Becher:2012yn,Neill:2015roa,Becher:2010tm} where the factorization theorem for the $p_t$ spectrum has been re-derived within a SCET approach.
Recently, N$^3$LL results for the Higgs transverse momentum have been obtained in the SCET framework in ref.~\cite{Chen:2018pzu}.
Combined threshold and transverse-momentum resummation have been considered in refs.~\cite{Laenen:2000ij,Kulesza:2002rh,deFlorian:2005fzc} and more recently in refs.~\cite{Lustermans:2016nvk,Marzani:2016smx,Muselli:2017bad}.

To obtain theoretical predictions, one has to compute the integral over the impact parameter $b$ in eq.~\eqref{eq:CSSCatani}.
However, analogously to the threshold resummation case, the integral hits the Landau pole at large values of $b$.
As a consequence, it is necessary to introduce a prescription to regularize the integral if one were to compute it for arbitrary values of $p_T$. 
Several solutions have been proposed in the literature. 
In the $b_*$-prescription~\cite{Collins:1984kg} the impact parameter is `frozen' at a value 
\begin{equation}
	b_* = \frac{b}{\sqrt{1+(b/b_{\text{lim}})^2}}, 	\qquad b_* < b_{\text{lim}},
\end{equation} 
where the separation between perturbative and non-perturbative regimes is set by the parameter $b_{\text{lim}} \sim 1/\LambdaQ$.
The $b_*$ prescription is typically supplemented by an additional factor $\sim e^{-gb^2}$ to model the non-perturbative region, where the parameter $g$ is tuned to data.
However, this procedure typically leads to numerical instabilities when the matching to fixed order is performed.
To overcome these instabilities, other prescriptions have been proposed.
A minimal prescription, constructed along the lines discussed in the threshold resummation case, was suggested in~\cite{Kulesza:2002rh}.
Alternatively, one can resort to Borel summation~\cite{Bonvini:2008ei} or adopt scales which depend on the hadronic variable $p_t$ within the SCET approach of refs.~\cite{Becher:2011xn,Becher:2012yn}.

Another possible solution has been proposed in ref.~\cite{Ellis:1997ii}, where the resummed expression is constructed as an extension of the DDT formula in $p_t$ (direct) space, by computing the Fourier inversion integral of the expanded resummed result in powers of $\as$ and retaining only the leading terms.
This procedure, however, is not stable with respect to the inclusion of subleading corrections, which are necessary to reproduce the correct behaviour in the $p_t\rightarrow 0$ limit due to the vectorial nature of the observable.
Beyond LL, it is indeed not possible to construct a closed analytic expression for the resummed distributions which is simultaneously free of singularities at finite $p_t$ values and of logarithmically subleading corrections~\cite{Frixione:1998dw}.

The problem of resummation of transverse-momentum distributions in $p_t$ space has received further attention~\cite{Kulesza:1999sg,Kulesza:1999gm,Kulesza:2001jc} but was only recently solved.
A solution to the problem has been presented in refs.~\cite{Monni:2016ktx,Bizon:2017rah}, where the Higgs transverse momentum has been resummed up to N$^3$LL accuracy and matched to NNLO.
The problem has been addressed also in ref.~\cite{Ebert:2016gcn} using a SCET approach, where the RG evolution is solved in momentum space. 
An alternative technique to relate analytically the impact parameter space and the momentum space results has been proposed in ref.~\cite{Kang:2017cjk}.
The approach of refs.~\cite{Monni:2016ktx,Bizon:2017rah} is summarized in the next section, where we will use it as a prototype to introduce a formalism which performs resummation in direct space.

\subsection{Towards automation: resummation in direct space}\label{sec:radish}

Resummations based on factorization properties of the observables lead to calculations with high accuracy for various distributions and cross sections. 
However, all these approaches share some limitations.
The main disadvantage is that only observables for which a factorization theorem is known can be resummed.
Moreover, since factorization is usually achieved in a conjugate space, one has to compute an inverse transform, which sometime causes numerical instabilities. 
Therefore, one may ask if it is possible to achieve resummation in a more observable-independent way, without the need to separately establish factorization properties on a case-by-case basis. 

A general approach to evaluate higher-order contributions to partonic cross section is based on Monte Carlo (MC) parton showers (for a recent review, see~\cite{Hoche:2014rga}), which provide an all-order approximation of the cross section in the soft and collinear limit.
Parton showers can treat exactly multi-parton kinematics and often include models of non-perturbative effects to provide a complete description of the hard-scattering event at the hadron level.
However, they give a probabilistic description of the event based on a probability distribution which approximates the {\it square} of the matrix elements.
Whilst the kinematic part of the matrix element factorizes, quantum-interference effects due to colour charge are present and spoil the independent-emission picture.
To overcome this problem, parton showers exploit QCD coherence through angular ordering (or in some other kinematic variable) to achieve leading-logarithmic accuracy.
However, parton showers typically contain some subleading contributions and might be NLL accurate only for a few inclusive-enough observables (see~\cite{Dasgupta:2018nvj} for a very recent study).

In ref.~\cite{Catani:1990rr}, Catani, Marchesini and Webber (CMW) took advantage of the factorization properties of QCD radiation to introduce an algorithm in the context of DIS and Drell-Yan processes.
In this method, a cutoff scale $q_0$ is introduced and the radiation above and below the resolution scale is treated differently.
The infrared safety of the observable guarantees that there is no residual logarithmic dependence on the regularization scale $q_0$, such that the limit $q_0\rightarrow 0$ can be taken.
At leading-logarithmic accuracy, the CMW method reduces to a conventional parton shower; however, it can be in principle improved to reach arbitrary logarithmic accuracy.
The CMW method has been successfully applied to a series of observables measured at LEP at NLL accuracy (see e.g~\cite{Catani:1991kz}); however, a systematic extension beyond NLL requires the description of additional effects, such as double-soft radiation at wide angle, which proves to be not trivial within the CMW framework.

Methods based on the the CMW theoretical framework have however an advantage, as they do not rely on observable-dependent factorization theorems.
An algorithm based on the branching formalism of ref.~\cite{Catani:1990rr,Catani:1991kz}, but with a different implementation, was proposed in refs.~\cite{Banfi:2001bz,Banfi:2003je,Banfi:2004yd} and initially applied to event-shape resummation in $e^+ e^-$ collisions~\cite{Banfi:2004nk} (\texttt{ARES/CAESAR} framework).
The idea of this method is to translate the resummability of the observable into properties of the observable in the presence of multiple radiation.
In particular, resummation can be performed semi-analytically for observables which are continuously global and recursive infrared collinear (rIRC) safe~\cite{Banfi:2004yd}\footnote{In words, an observable is rIRC safe if a) in the presence of multiple soft and/or collinear emissions it has the same scaling properties as with just one of them and b) there exists a scale $q_0$, independent of the observable, such that emissions below $q_0$ do not contribute significantly to the observable's value. An observable is continuously global if its scaling with respect to the transverse momentum of a soft and/or collinear emission is the same in the whole phase space. We refer the reader to the original paper~\cite{Banfi:2004yd} for a more precise definition.}.
The rIRC safety of the observable allows one to introduce a resolution scale $q_0$ such that unresolved emission (that is, below $q_0$) can be treated as totally uncorrelated, making the need of a conjugate space to achieve factorization unnecessary. 
In the resolved region, the emissions can be treated exclusively with efficient MC methods, directly in momentum space.
The method was shown to be general enough to resum a wide class of observables at NLL accuracy~\cite{Banfi:2004yd} and was later extended to NNLL in ref.~\cite{Banfi:2014sua,Banfi:2016zlc}.

A review of the \texttt{CAESAR} method has been presented in ref.~\cite{Luisoni:2015xha} and very recently in a concise, yet pedagogical way in ref.~\cite{Bauer:2018svx}, where a compelling connection between automated resummation in direct QCD and resummation in the SCET approach has been devised. 
The basic idea of the \texttt{CAESAR} algorithm is to relate the observable which one wants to resum to a simpler observable, whose resummation can be carried analytically, and to compute numerically the difference between the two observables.
In formul\ae, if one knows the cumulative distribution~\eqref{eq:cumulative} of the simple observable $\Sigma_s (v_s)/v_s$, the cumulative distribution of the more complex observable can be computed through the convolution
\begin{equation}
	\Sigma(v) = v \int \frac{d v_s}{v_s	} \Sigma_s (v_s) \mathcal F (v, v_s).
\end{equation}
Choosing a simpler observable which shares the same (double) leading logarithms of the resummed observable allows for a much easier implementation of the all-order result.
Indeed, if the simpler observable is known at N$^{k}$LL accuracy, the function $\mathcal F $ relating the two observables only requires ingredients at N$^{k-1}$LL accuracy, since rIRC safety guarantees that the leading logarithmic behaviour is fully captured by the simpler observable.
 A natural choice for the simpler observable, for instance, is the value of observable in its soft-collinear approximation~\cite{Banfi:2001bz,Banfi:2004yd,Banfi:2014sua,Banfi:2016zlc}.

In this section, we present an extension of the \texttt{CAESAR} method to observables affected by azimuthal cancellations, such as the transverse-momentum distribution in the small $p_t$ limit~\cite{Monni:2016ktx,Bizon:2017rah}.
As we will see below, in the original \texttt{CAESAR} approach some approximations are made to dispose of subleading effects once the desired logarithmic accuracy has been achieved.
However, as we discussed in sect.~\ref{sec:tmomres}, subleading effects are essential to reproduce the correct scaling of the distribution at small $p_t$~\cite{Frixione:1998dw}; as a consequence, the approximations which lead to the \texttt{CAESAR} result cannot be trivially performed.

We start by discussing a general formalism for the resummation of a generic rIRC safe observable $V$ in the reaction $pp\rightarrow B$, where $B$ is a colourless system of invariant mass $M$, but we will then specialize to the specific case of the transverse momentum $p_t$ at NLL accuracy, assuming that the parton densities are independent on the scale and setting them to one for the sake of simplicity.
We will then discuss the correct treatment of parton luminosities and show how the method can be extended up to N$^3$LL accuracy, referring the reader to ref.~\cite{Bizon:2017rah} for additional technical details of the derivation.

The quantity at the centre of our discussion is the cumulative cross section $\Sigma(v)$~\eqref{eq:cumulative} for $V$ smaller than some value $v$.
At Born level, the final state is a colour singlet of mass $M$, whereas beyond Born level radiation takes place and the final state consists of the colour singlet and $n$ partons with outgoing momenta $k_1,\dots,k_n$.
The observable $V$ we wish to resum is a function of all momenta, $V=V(k_1,\ldots k_n)$.

In the case of transverse-momentum resummation, $v$ is equal to $p_t/M$.
{\it Transverse} observables do not depend on the rapidity of the radiation; in particular, for the specific case of $p_t$, $V$ obeys the following parameterization in the presence of a single soft emission $k$ collinear to the leg $\ell$:
\begin{equation}
	V(k ) = \frac{k_t}{M}, 
\end{equation}  
where $k_t$ is the transverse momentum with respect to the beam axis.
%
The transverse momentum is moreover an {\it inclusive} observable, as it depends only upon the total momentum of the radiation: $V(k_1,\dots, k_n) = V(k_1+\dots +k_n)$.

In the IRC limit, $\Sigma(v)$ receives contributions both from virtual corrections and from soft and/or collinear radiation. 
The IRC divergences of the virtual corrections exponentiate at all orders~\cite{Dixon:2008gr} and we denote them with ${\cal V}$.
Therefore we can write
\begin{equation}
  \label{eq:Sigma-2}
  \Sigma(v) = \int d\Phi_B {\cal V} \sum_{n=0}^{\infty}
  \int\prod_{i=1}^n [dk_i]
  |\M(k_1,\dots ,k_n)|^2\,\Theta\left(v-V(k_1,\dots,k_n)\right)\,,
\end{equation}
where $\M$ is the matrix element for $n$ real emissions (which reduces to the Born matrix element for $n=0$), $[dk_i]$ denotes the phase space for the emission $k_i$, and $\Phi_B$ denotes the Born phase space. 
The $\Theta$ function represents the measurement function for the observable under consideration, which in the case of $p_t$ reads $\Theta\left(v-V(k_1,\dots,k_n)\right) = \Theta (p_t/M - | \sum_i \vec k_{i}/M | )$.

To obtain a resummed expression for $ \Sigma(v)$, one needs to establish an explicit logarithmic counting for the squared matrix element.
To do so, it is convenient to define the $n$-particle correlated ($n$PC) matrix element as
\begin{align}
|\tilde{\M}(k_a)|^2 =&~\frac{|\M(k_a)|^2}{|\M_B|^2}=~|\M(k_a)|^2,\notag\\
|\tilde{\M}(k_a,k_b)|^2 =&~\frac{|\M(k_a,k_b)|^2}{|\M_B|^2}-\frac1{2!}|\M(k_a)|^2|\M(k_b)|^2,\notag\\
|\tilde{\M}(k_a,k_b,k_c)|^2 =&~\frac{|\M(k_a,k_b,k_c)|^2}{|\M_B|^2}-\frac1{3!}|\M(k_a)|^2|\M(k_b)|^2|\M(k_c)|^2\notag\\
&~-|\tilde{\M}(k_a,k_b)|^2|\M(k_c)|^2-|\tilde{\M}(k_a,k_c)|^2|\M(k_b)|^2-|\tilde{\M}(k_b,k_c)|^2|\M(k_a)|^2,\nonumber \\
\vdots& 
\end{align}
For $n\geq 2$, these represent the contributions to the $n$-particle squared matrix element that vanish in strongly-ordered kinematic configurations, which cannot be factorized in terms of lower-multiplicity squared amplitudes, i.e. the fully correlated part.
With these definitions, the renormalized squared amplitude for the emission of $n$ real gluons can be decomposed as\footnote{Here the multiplicity factors are taken into account explicitly.}
\begin{align}
\label{eq:nPC}
&\frac{|\M(k_1,\dots ,k_n)|^2}{ |\M_B|^2} =\left\{\left(\frac{1}{n!}\prod_{{\substack{i=1\\\phantom{x}}}}^{n} \left|\M(k_i)\right|^2\right) \right. + \left.  \left[\sum_{a > b}\frac{1}{(n-2)!}\left(\prod_{\substack{i=1\\ i\neq a,b}}^{n} \left|\M(k_i)\right|^2 \right)\left|\tilde{\M}(k_a, k_b)\right|^2+\right.\right. \notag\\
&\left.\left.\sum_{a > b}\sum_{\substack{ c > d\\ c,d\neq a,b}}\frac{1}{(n-4)!2!}\left(\prod_{\substack{i=1\\ i\neq a,b,c,d}}^{n} \left|\M(k_i)\right|^2 \right)\left|\tilde{\M}(k_a, k_b)\right|^2 \left|\tilde{\M}(k_c, k_d)\right|^2\right. + \dots \right]\notag\\
&\left. + \left[\sum_{a > b > c}\frac{1}{(n-3)!}\left(\prod_{\substack{i=1\\ i\neq a,b,c}}^{n} \left|\M(k_i)\right|^2 \right)\left|\tilde{\M}(k_a, k_b,k_c)\right|^2 +\dots\right]+\dots\right\}.
\end{align}
Each of the $n$PC matrix element admits an expansion in powers of $\as$,
\begin{equation}
\label{eq:nPC-def}
|\tilde{\M}(k_a, \dots,k_n)|^2~\equiv~ \sum_{j=0}^{\infty}\left(\frac{\alpha_s}{2\pi}\right)^{n+j}n\mbox{PC}^{(j)}(k_a, \dots,k_n),
\end{equation}
where $j$ denotes the order of virtual corrections to the squared amplitude with $n$ real emissions.
We further decompose the $n$PC blocks by isolating the leading (soft-collinear) term (henceforth denoted by a subscript `$\text{sc}$'), obtained by taking the soft and collinear limit of all emissions, from the less-singular part (denoted by `$\cancel{\text{sc}}$').

As $p_t$ is an inclusive observable, we can integrate the $n$PC blocks for $n > 1$ prior to evaluating the observable. 
This amounts to making the replacement
\begin{align}
\label{eq:nPC-inclusive}
\sum_{n=0}^{\infty}|\M(&k_1,\dots ,k_n)|^2 \longrightarrow |\M_{B}|^2\notag\\ & \times
\sum_{n=0}^{\infty}\frac{1}{n!}\left\{\prod_{i=1}^{n} \left(|\M(k_i)|^2+\int [d k_a][d k_b]|\tilde{\M}(k_a,k_b)|^2\delta^{(2)}(\vec{k}_{t,a}+\vec{k}_{t,b}-\vec{k}_{t,i})\delta(Y_{ab}-Y_i)\right.\right.\notag\\ &\left.\left. +
     \int [d k_a][d k_b][d k_c]|\tilde{\M}(k_a,k_b,k_c)|^2\delta^{(2)}(\vec{k}_{t,a}+\vec{k}_{t,b}+\vec{k}_{t,c}-\vec{k}_{t,i})\delta(Y_{abc}-Y_i)
     + \dots\right.\bigg)\right.\Bigg\}\notag\\
& \equiv |\M_{B}|^2\sum_{n=0}^{\infty}\frac{1}{n!}\prod_{i=1}^{n} |\M(k_i)|_{\text{inc}}^2,
\end{align}
where $Y_{abc...}$ is the rapidity of the $k_a+k_b+k_c+\dots$ system
in the centre-of-mass frame of the collision.
With the above notation, we can rewrite eq.~\eqref{eq:Sigma-2} as
\begin{equation}
  \label{eq:Sigma-start-1}
  \Sigma(v) = \int d\Phi_B |\M_{B}|^2{\cal V} \sum_{n=0}^{\infty}
  \frac{1}{n!}  \int\prod_{i=1}^n [dk_i]
  |\M(k_i)|_{\text{inc}}^2\,\Theta\left(v-V(k_1,\dots,k_n)\right)\,,
\end{equation}
where $|\M(k_i)|_{\text{inc}}^2$ is defined in
Eq.~\eqref{eq:nPC-inclusive}.

We can now exploit the fact that the observable is global and rIRC safe.
On the one hand, this allows us to establish a hierarchy between the different blocks in the decomposition~\eqref{eq:nPC}, since correlated blocks with $n$ particles start contributing at one logarithmic order higher than correlated blocks with $n-1$ particles.
Indeed, in the soft and collinear limit $\prod_{i=1}^{n} |\M(k_i)|^2$ comes with a factor $\as^n \ln(v)^{2n}$, whereas correlated blocks with $n$ emissions $|\tilde \M(k_1, \ldots k_n)|^2$ (and similarly the contributions from the virtual corrections eq.~\eqref{eq:nPC-def}) contribute at most with a factor of $\as^n \ln (v) ^{n+1}$.
This guarantees that the knowledge of a finite number of correlated blocks is enough to construct an all-order amplitude at a given logarithmic accuracy.
Indeed, upon integration over the full phase space, the expansion in terms of the correlated blocks of the squared amplitude eq.~\eqref{eq:nPC} (or equivalently eq.~\eqref{eq:nPC-inclusive}) with $n$ emissions can be put in a one to one correspondence with the logarithmic structure
\begin{equation}
	|\M(k_1,\dots ,k_n)|^2 \rightarrow \mathcal O (\as^n \ln(v)^{2n}) +\mathcal O( \as^n \ln(v)^{2n-1} ) + \mathcal O ( \as^n \ln(v)^{2n-2} )\ldots\, ,
\end{equation}
thus providing a recipe to include systematically terms up to the desired logarithmic accuracy (see table~\ref{tab:log-order}).
For instance, the knowledge of the 1PC$^{(0)}_{ \cancel {\text {sc}}}$, 1PC$^{(1)}_{\text {sc}}$ and  2PC$^{(0)}_{\text {sc}}$ is enough to reach NLL accuracy as it predicts all the terms of order $\as^n \ln(v)^{n}$ in $\ln \Sigma(v)$ (that is, all the terms of order $\as^n \ln(v)^{2n-1}$ in $\Sigma(v)$, cfr. table~\ref{tab:count})\footnote{A careful reader might object that to reach NLL accuracy in the \textit{logarithm} of $\Sigma$ one should also include other contributions, such as the 3PC$^{(0)}$ block in its soft and collinear limit; nonetheless, as we shall see below, for the class of observables discussed here these contributions can be included with a suitable choice of the scale in the running coupling~\cite{Banfi:2004yd}.}.

\begin{table}[t]
\begin{center}
\begin{tabular}{ccc}
Logarithmic order & \multicolumn{2}{c}{Blocks required} \\
 & $n$PC$^{(j)}_{\text{sc}}$ & $n$PC$^{(j)}_{ \cancel {\text {sc}}}$ \\
\midrule
  LL       & $n+j \leq 1$ &  ---  \\
  NLL      & $n+j \leq 2 $ & $n+j \leq 1$ \\
$\vdots$ & $\vdots$ & $\vdots$ \\
  N$^k$LL  &  $n+j \leq k+1 $ & $n+j \leq k$ \\
\end{tabular}
\caption{Blocks to be included in the squared-amplitude decomposition
  at a given logarithmic order.}
\label{tab:log-order}
\end{center}
\end{table}

On the other hand, rIRC safety can be used to single out the IRC singularities of the real matrix element to achieve the cancellation of the exponentiated divergences of virtual origin.
To this end, we consider a configuration where the radiation corresponding to the first (hardest) block $|\M(k_1)|_{\text {inc}}^2$ has occurred (the contribution with $n=0$ in eq.~\eqref{eq:Sigma-start-1} vanishes since it is infinitely suppressed by the pure virtual corrections ${\cal V}$).
We then order the inclusive blocks described by $|\M(k_i)|_{\text{inc}}^2$ according to their contribution to the observable $V(k_i)$ i.e. $k_{t,1} > k_{t,2} > \dots > k_{t,n}$.
The rIRC safety of the observable allows us to introduce a slicing parameter $\epsilon \ll 1$  such that all inclusive blocks with $k_{t,i} < \epsilon k_{t,1}$ can be neglected in the computation of the observable. 
We classify inclusive blocks $k$ as {\it resolved} if $k_{t}> \epsilon k_{t,1}$ and as {\it unresolved} if $k_{t}< \epsilon k_{t,1}$. 
With this separation eq.~\eqref{eq:Sigma-start-1} becomes
\begin{align}
\label{eq:Sigma-start-2}
\Sigma(v) &= \int d\Phi_B |\M_{B}|^2{\cal V}\notag\\
& \quad \times\int [dk_1]
|\M(k_1)|_{\text{inc}}^2 \left(\sum_{l=0}^{\infty}
\frac{1}{l!}  \int\prod_{j=2}^{l+1} [dk_j]
|\M(k_j)|_{\text{inc}}^2\,\Theta(\epsilon \Vsc(k_1)- \Vsc(k_j))\right)\notag\\
& \quad \times\left(\sum_{m=0}^{\infty}
\frac{1}{m!}  \int\prod_{i=2}^{m+1} [dk_i]
|\M(k_i)|_{\text{inc}}^2\,\Theta(\Vsc(k_i)-\epsilon \Vsc(k_1))\Theta\left(v-V(k_1,\dots,k_{m+1})\right)\right)\,.
\end{align}
The phase space of the unresolved real ensemble is now solely constrained by the upper resolution scale, since it does not contribute to the evaluation of the observable. 
Therefore, it can be exponentiated directly in eq.~\eqref{eq:Sigma-start-2} and employed to cancel the divergences of the virtual corrections ${\cal V}$.

We can now proceed to evaluate eq.~\eqref{eq:Sigma-start-2} at NLL accuracy.
Double logarithmic terms of the form $\as^n \ln^{2n} (1/v)$ entirely arise from the 1PC$^{(0)}_{\text{sc}}$ block.
However, if one wants to control all the leading-logarithmic terms of order $\alpha_s^n \ln^{n+1}(1/v)$ in the {\it logarithm} of $\Sigma(v)$ (see discussion in sect.~\ref{sec:logaccuracy}), then also the leading (soft-collinear) term of the 1PC$^{(1)}$ and 2PC$^{(0)}$ blocks must be included.
Within the inclusive approximation we introduced above, we find
\begin{align}
\label{eq:2PC-inclusive}
|\M(k)|_{\text{inc}}^2&\simeq|\M(k)|^2 + \int [dk_a][dk_b]|\tilde{\M}(k_a,k_b)|^2\delta^{(2)}(\vec{k}_{t,a}+\vec{k}_{t,b}-\vec{k}_{t})\delta(Y_{ab}-Y)\notag\\
&= \frac{\alpha_s(\mu)}{2\pi}1{\text{PC}}^{(0)}(k)\left(1+\alpha_s(\mu)\left(\beta_0\ln\frac{k_{t}^2}{\mu^2} + \frac{K}{2\pi}\right)+\dots\right),
\end{align}
where 
\begin{equation}
\label{eq:K}
K = \left(\frac{67}{18}-\frac{\pi^2}{6}\right)C_A - \frac{5}{9} n_f\,
\end{equation}
encodes the contribution from the one-loop cusp anomalous dimension and is a pure NLL contribution.
At this accuracy, one can integrate inclusively over the invariant mass of the 2PC$^{(0)}$ block, while keeping the bounds on the rapidity $Y \leq \ln M/k_t$ as computed from the massless kinematics~\cite{Bizon:2017rah}. 
Indeed, as one can see from table~\ref{tab:log-order}, the exclusive treatment of the 2PC$^{(0)}$ enters only at NNLL. 
Up to LL accuracy, we can absorb the contribution of the $\beta_0$ term in the running  of the coupling of the 1PC$^{(0)}_{\text{sc}}$ block by setting the scale $\mu$ at which it is evaluated to the scale $k_t$ of each emission $k$ in the parametrization~\cite{Catani:1992ua,Catani:1989ne} as we discussed in sect.~\ref{sec:threshold}.
Therefore, we find that the inclusive matrix element square and the phase space which controls all the LL terms is
\begin{equation}
  \label{eq:sc-amplitude}
[dk] | \M(k)|_{\text{inc}}^2\simeq[dk]\M^2_{\text{sc}}(k)= \sum_{\ell=1,2} 2 C_\ell \frac{\alpha_s(k_{t})}{\pi}\frac{dk_{t}}{k_{t}} 
 \frac{d z^{(\ell)}}{1-z^{(\ell)}}\, \Theta\left((1-z^{(\ell)}) - k_{t}/M\right) \Theta(z^{(\ell)})
\frac{d\phi}{2\pi}\,,
\end{equation}
where $\M_{\text{sc}}(k)$ denotes the amplitude in the soft approximation, $C_{\ell}$ is the Casimir of the emitting leg ($C_{\ell}=C_A $ for gluons, $C_{\ell}=C_F$ for quarks), and we denote by $1-z^{(\ell)}$ the fraction of the momentum (entering in the emission vertex) which is carried by the emitted parton.

To reach NLL accuracy, we need to include also the 1PC$^{(1)}$ and 2PC$^{(0)}$ in their soft and collinear limit, that is the term proportional to $K$, as well as the contributions from the hard-collinear limit of the 1PC$^{(0}$ block, which we have so far ignored.
We obtain
\begin{align}
  \label{eq:single-emsn}
  [dk] | \M(k)|_{\text{inc}}^2&= [dk]\M^2_{\text{CMW}}(k)
     \notag \\ &\quad +
    \sum_{\ell=1,2}
    \frac{dk_t^2}{k_t^2}\frac{dz^{(\ell)}}{1-z^{(\ell)}}\frac{d\phi}{2\pi}\frac{\alpha_s(k_t)}{2\pi}\left(
    (1-z^{(\ell)})\tilde  P^{(0)}(z^{(\ell)})-\lim_{z^{(\ell)}\to 1}\left[(1-z^{(\ell)})\tilde  P^{(0)}(z^{(\ell)})\right]
    \right)\,,
\end{align}
where we employed the so-called Catani-Marchesini-Webber (CMW) scheme for the running coupling~\cite{Catani:1990rr} and we defined
\begin{equation}
  \label{eq:sc-amplitude-CMW}
[dk]\M^2_{\text{CMW}}(k)= \sum_{\ell=1,2} 2 C_\ell
\frac{\alpha_s(k_{t})}{\pi}\left (1 +  \frac{\alpha_s(k_t)}{2\pi}
  K\right)\frac{dk_{t}}{k_{t}} 
\frac{d z^{(\ell)}}{1-z^{(\ell)}}\, \Theta\left((1-z^{(\ell)}) - k_{t}/M\right) \Theta(z^{(\ell)})
 \frac{d\phi}{2\pi}\,.
\end{equation}
In eq.~\eqref{eq:single-emsn}, $\tilde  P^{(0)}$ denotes the leading-order unregularized splitting function (see appendix~\ref{app:sudakov-radiator}).
At NLL accuracy, we can treat the above hard-collinear contribution by neglecting recoil effects both in the phase-space boundaries of other emissions and in the observable, since these effects enter at NNLL.

We can now insert~\eqref{eq:single-emsn} into eq.~\eqref{eq:Sigma-start-2}.
At NLL all constant terms and virtual corrections can be neglected and the singular structure of the virtual corrections depends only on the invariant mass of the singlet
\begin{equation}
{\cal V} \simeq {\cal V}(M^2) = \exp\left\{-\int [dk]|\M(k)|_{\text{inc}}^2\right\} \,\,\,\text{at NLL},
\end{equation}
and can be combined with the unresolved contributions to give rise to a Sudakov suppression factor
\begin{align}
{\cal V}(M^2)&\exp\left\{\int [dk]|\M(k)|_{\text{inc}}^2\,\Theta(\epsilon
  \Vsc(k_1)- \Vsc(k))\right\}\notag\\
\simeq &\exp\left\{-\int [dk]|\M(k)|_{\text{inc}}^2\,\Theta(\Vsc(k) -\epsilon
  \Vsc(k_1))\right\} = e^{-R(\epsilon \Vsc(k_1))},
\end{align}
where the Sudakov radiator at this order reads~\cite{Banfi:2004yd,Banfi:2014sua}
\begin{equation}
  \label{eq:radiator}
  \begin{split}
    R(v) \simeq \int [dk] \M_{\text{CMW}}^2(k)\Theta\left(\ln
      \left(\frac{k_t}{M}\right)^{a} -\ln
      v\right) + \sum_{\ell=1,2}C_\ell B_\ell\int
    \frac{dk_t^2}{k_t^2}\frac{\alpha_s(k_t)}{2\pi}\Theta\left(\ln\left(\frac{k_t}{M}\right)^{a}-\ln v
    \right)\,,
  \end{split}
\end{equation}
having defined
\begin{equation}
  C_\ell B_\ell = \int_0^1 \frac{d z^{(\ell)}}{1-z^{(\ell)}}\left((1-z^{(\ell)}) P^{(0)}(z^{(\ell)}) - \lim_{z^{(\ell)}\to 1}\left[(1-z^{(\ell)}) P^{(0)} (z^{(\ell)})\right]\right)\,.
\end{equation}

To find the final result at NLL accuracy, we need to treat the resolved real blocks for which $V(k_i) > \epsilon V(k_1)$.
To this end, it is necessary to treat correctly the kinematics and the phase space in the presence of additional radiation.
At NLL accuracy, however, the real radiation can be approximated with its soft limit, and the phase space of each emission becomes independent of the remaining radiation in the event.
Upon inclusive integration, the inclusive squared amplitude and its phase space can be parametrized by introducing a function $R'$
\begin{equation}
  \label{eq:single-emsn-rp}
  \begin{split}
    [dk_i] |\M(k_i)|^2_{\text{inc}} & = \frac{dv_i}{v_i}
    \frac{d\phi_i}{2\pi}
    R'(v_i)
    = \frac{d\zeta_i}{\zeta_i} \frac{d\phi_i}{2\pi}
    R'\left(\zeta_i v_1\right) \,,
  \end{split}
\end{equation}
where $v_i= V(k_i)$ and $\zeta_i = V(k_i)/V(k_1)$.
We thus obtain that eq.~\eqref{eq:Sigma-start-2} becomes
\begin{align}
\label{eq:master-NLL}
  \Sigma(v) &= \sigma^{(0)}\int \frac{d v_1}{v_1} \int_0^{2\pi} \frac{d\phi_1}{2\pi}
              e^{-R(\epsilon v_1)}R'\left(v_1\right)   \notag\\ &\quad \times \sum_{n=0}^{\infty}\frac{1}{n!}
                                                           \prod_{i=2}^{n+1}
                                                           \int_{\epsilon}^{1}\frac{d\zeta_i}{\zeta_i}\int_0^{2\pi}
                                                           \frac{d\phi_i}{2\pi} 
                                                           R'\left(\zeta_i v_1\right)\, \Theta\left(v-V(k_1,\dots, k_{n+1})\right)\,,
\end{align}
where we introduced the total Born cross section
\begin{equation}
\sigma^{(0)}=\int d\Phi_B |\M_{B}|^2.
\end{equation}

This formula can be directly evaluated using MC techniques, as it is finite in four dimensions since the $\epsilon$ dependence cancels exactly.
However, it contains logarithmically subleading effects with respect to the formal NLL accuracy, which in the original \texttt{CAESAR} approach are discarded by a series of approximations.
These approximations, which are legitimate for observables which vanish only in the Sudakov limit, cannot be trivially performed for observables which feature kinematic cancellations.
Let us now discuss this problem in some detail.
Ref.~\cite{Banfi:2004yd} suggests to perform the following Taylor expansion around $v$ in eq.~\eqref{eq:master-NLL}
\begin{align}
R(\epsilon v_1) &=  R(v) + \frac{dR(v)}{d\ln(1/v)}\ln\frac{v}{\epsilon v_1} + {\cal O}\left(\ln^2\frac{v}{\epsilon v_1}\right),\notag\\
R'\left(v_i\right) &=  R'(v) + {\cal O}\left(\ln\frac{v }{v_i}\right),
\end{align}
which is motivated by the fact that at NLL the resolved emissions are such that $v_i \sim v_1 \sim v$ and hence the terms neglected are at most NNLL. 
We can identify $R'$ with $d R(v) / d \ln (1/v) = R'(v)$ (i.e. the real radiator is fully parametrized by the Sudakov radiator) and we obtain
\begin{align}
\label{eq:master-NLL-CAESAR}
  \Sigma(v) &\simeq \sigma^{(0)}\int \frac{d v_1}{v_1} \int_0^{2\pi} \frac{d\phi_1}{2\pi}
              e^{-R(v)}e^{- R'(v)\ln\frac{v}{\epsilon v_1}} R'\left(v\right) \notag\\ &\quad \times \sum_{n=0}^{\infty}\frac{1}{n!}
                                                           \prod_{i=2}^{n+1}
                                                           \int_{\epsilon}^{1}\frac{d\zeta_i}{\zeta_i}\int_0^{2\pi}
                                                           \frac{d\phi_i}{2\pi}                                                           R'\left(v\right)\, \Theta\left(v-V(k_1,\dots, k_{n+1})\right)\,.
\end{align}
The integration over $v_1$ can now be performed analytically and~\eqref{eq:master-NLL-CAESAR} reduces to the \texttt{CAESAR} formulation of ref.~\cite{Banfi:2004yd}. 

However, an expansion about the observable's value $v$ is valid only if the ratio $v_i/v$ remains of order one in the whole emission phase space.
Since rIRC safety ensures that emissions with $v_i \ll v$ exponentiates, the condition $v_i /v \sim 1$ is satisfied if there are no configurations such that $v_i \gg v$.
However, this is not true for observables which feature kinematic cancellations; in particular, eq.~\eqref{eq:master-NLL-CAESAR} geometrically diverges in $ R' (v) \sim 2$~\cite{Monni:2016ktx}. 

Nevertheless, though eq.~\eqref{eq:master-NLL} can be directly evaluated using Monte-Carlo (MC) techniques, it is convenient~\cite{Monni:2016ktx,Bizon:2017rah} to perform an alternative expansion about the observable's value of the hardest block $v_1$
\begin{align}
\label{eq:expansion-right}
R(\epsilon v_1) &=  R(v_1) + \frac{dR(v_1)}{d\ln(1/v_1)}\ln\frac{1}{\epsilon} + {\cal O}\left(\ln^2\frac{1}{\epsilon}\right),\notag\\
R'\left(v_i\right) &=  R'(v_1) + {\cal O}\left(\ln\frac{v_1 }{v_i}\right).
\end{align}
Indeed, rIRC safety guarantees that $v_i \sim v_1$ $(\zeta_i \sim 1 )$ such that the terms neglected are at most NNLL.
However, a class of higher-order terms still remains through the dependence on $v_1$ and ensures the absence of divergences at small $v$.
With this choice (let us recall that in our case $v_1 = k_{t,1}/M$), eq.~\eqref{eq:master-NLL} reads
\begin{align}
\label{eq:master-NLL-kt}
  \Sigma(v) &= \sigma^{(0)}\int \frac{d k_{t,1}}{k_{t,1}} \int_0^{2\pi} \frac{d\phi_1}{2\pi}
              e^{-R(k_{t,1})} \epsilon^{R'(k_{t,1})}  R'\left(k_{t,1}\right) \notag\\ & \quad \times \sum_{n=0}^{\infty}\frac{1}{n!}
                                                           \prod_{i=2}^{n+1}
                                                           \int_{\epsilon}^{1}\frac{d\zeta_i}{\zeta_i}\int_0^{2\pi}
                                                           \frac{d\phi_i}{2\pi} 
                                                           R'\left(\zeta_i k_{t,1}\right)\, \Theta\left(v-V(k_1,\dots, k_{n+1})\right)\,.
\end{align}
The above approximations make the evaluation of eq.~\eqref{eq:master-NLL-kt}  considerably simpler than its original form, as discussed in ref.~\cite{Bizon:2017rah}.
\mccorrect{With this choice, the logarithmic accuracy is effectively defined in terms of logarithms of $k_{t,1}/M $}.

\begin{figure}[t]
\begin{center}
\includegraphics[width=0.50\textwidth]{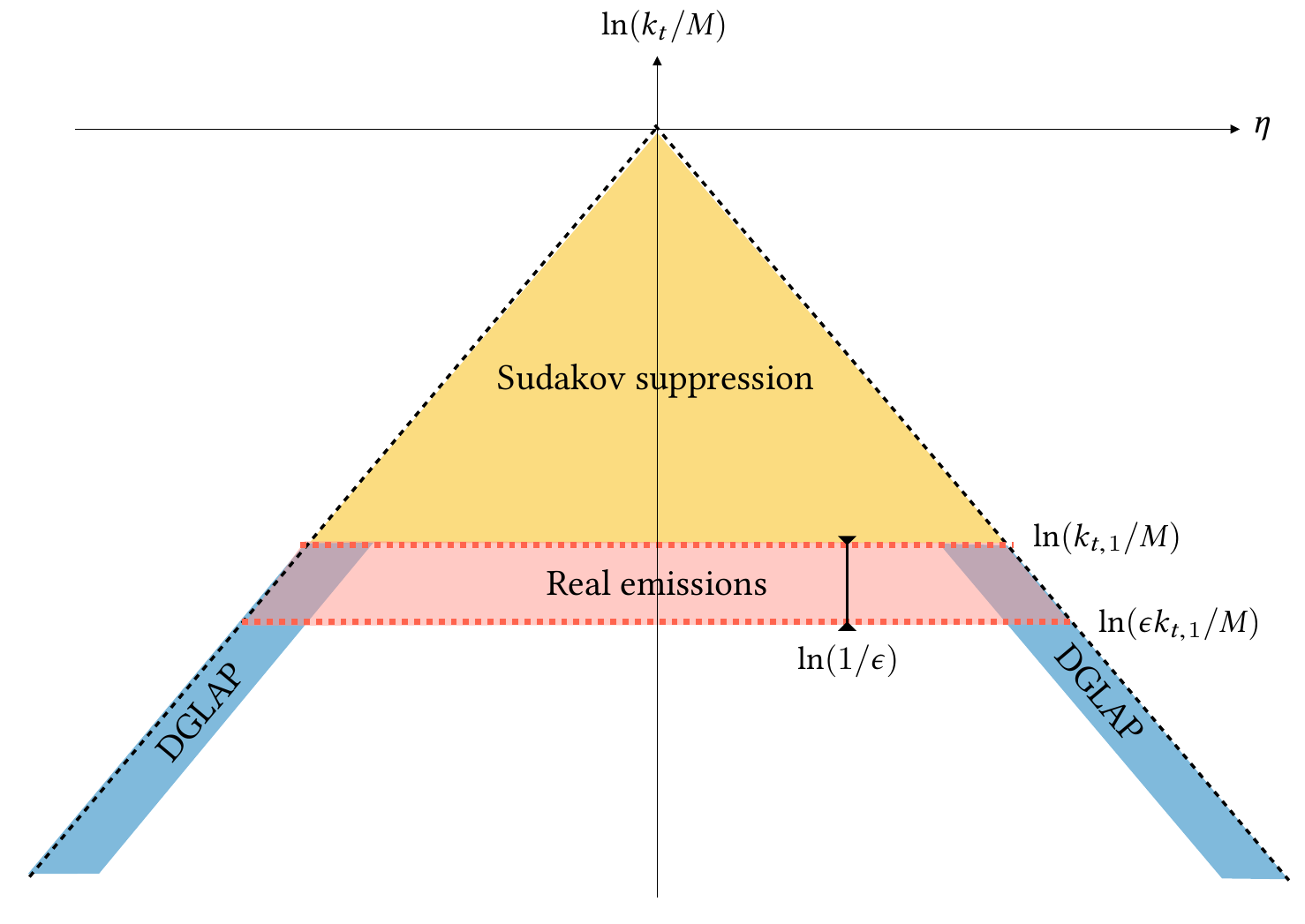}
 \caption{Lund diagram representing the phase space for a secondary real emission.}
  \label{fig:lund}	
\end{center}
 \end{figure}

To complete our treatment of the NLL result, we need to include the parton densities, which we have neglected so far.
To show how they can be accounted for, let us again consider configurations in which the emissions are ordered in $k_{t,i}$ and the hardest resolved emission $k_{t,1}$ has already occurred.
The phase space for any secondary emission can be depicted in the $\ln (k_{t}/M) - \eta$ (Lund) plane, see fig.~\ref{fig:lund}.
Here $\eta $ is the rapidity in the centre-of-mass frame of the incoming partons, which are extracted from the colliding hadrons at a factorization scale $\mu_0$.
In this plane, the resolved real radiation lives in a strip of width $\ln (1/\epsilon)$ between $\epsilon k_{t,1}$ and $ k_{t,1}$.
The remaining unresolved real emissions are combined with the virtual corrections to give rise to the Sudakov form factor which inhibits secondary emission in the yellow region labelled `Sudakov suppression'.

DGLAP evolution governs the radiation in the strictly collinear limit (corresponding to the blue strips labelled `DGLAP').
Since rIRC safety ensures that emissions in the unresolved region do not contribute significantly to the observable, DGLAP evolution can be performed inclusively up to $\epsilon k_{t,1}$. 
In the overlapping region, however, hard-collinear emissions modify the observable's value and the evolution should be performed exclusively, that is unintegrated in $k_t$. 
However, as we discussed above, at NLL accuracy the real radiation can be approximated with its soft limit: emissions in the hard-collinear edge of phase space enter at NNLL accuracy.
Therefore, we can perform DGLAP evolution inclusively up to $k_{t,1}$, which corresponds to evaluating the parton densities at a scale $\muF = k_{t,1}$.
At NLL accuracy we can thus write the cumulative cross section differential over the Born phase space $d \Phi_B$ as
\begin{align}
\label{eq:master-NLL-kt-with-Lumi}
  \frac{d\Sigma(v)}{d\Phi_B} &=\int \frac{d k_{t,1}}{k_{t,1}} \int_0^{2\pi} \frac{d\phi_1}{2\pi}
              e^{-R(k_{t,1})} \epsilon^{R'(k_{t,1})}  {\cal L}_{\text{NLL}}(k_{t,1})  R'\left(k_{t,1}\right) \notag\\ &\quad \times \sum_{n=0}^{\infty}\frac{1}{n!}
                                                           \prod_{i=2}^{n+1}
                                                           \int_{\epsilon}^{1}\frac{d\zeta_i}{\zeta_i}\int_0^{2\pi}
                                                           \frac{d\phi_i}{2\pi} 
                                                           R'\left(\zeta_i k_{t,1}\right)\, \Theta\left(v-V(k_1,\dots, k_{n+1})\right)\,,
\end{align}
where we defined
\begin{equation}
	{\cal L}_{\text{NLL}}(k_{t,1}) = \sum_{c}\frac{d|M_{B}|_{c\bar c}^2}{d\Phi_B} f_c\!\left(x_1,k^2_{t,1}\right)f_{\bar c}\!\left(x_2,k^2_{t,1}\right).
\end{equation}
At this logarithmic accuracy, the Sudakov radiator reads
\begin{align}
	R(k_{t,1}) = - L g_1 \(\as (\muR^2) L\) - g_2 \(\as (\muR^2) L\), \qquad L \equiv \frac{M}{k_{t,1}},
\end{align}
and the functions $g_1$ and $g_2$ are collected for instance in the appendix of ref.~\cite{Bizon:2017rah}.

This completes our derivation for the resummation of the transverse-momentum spectrum in direct space at NLL accuracy. 
To reach higher logarithmic accuracy, we need to relax some of the approximations that we have made so far.
First, we need to include systematically the correlated blocks necessary to achieve the desired logarithmic accuracy as summarized in table~\ref{tab:log-order}.
Furthermore, at NNLL and beyond one has to consider the exact rapidity bounds (see eq.~\eqref{eq:2PC-inclusive}) which give rise to subleading corrections neglected at NLL. 
These corrections can be taken into account by including additional terms in the expansion~\eqref{eq:expansion-right}~\cite{Banfi:2014sua}.
Finally, one needs to specify a complete treatment for hard-collinear radiation.
Indeed, at NLL the only hard-collinear contribution comes from the 1PC$^{(0)}$ block eq.~\eqref{eq:single-emsn}, which has been treated in the soft approximation and neglecting recoil effects.

Let us now briefly discuss how to include the evolution of the hard-collinear radiation.
To repeat the procedure that led to eq.~\eqref{eq:master-NLL} at higher logarithmic accuracy, we need to handle the phase space in the multiple-emission kinematic region.
In the NLL case, indeed, all resolved real emissions are soft and collinear and do not modify each other's phase space.
Starting at NNLL one or more real emissions can be hard and collinear to the emitting leg.
Therefore, the available phase space for subsequent real emissions changes.
In particular, at NNLL one needs to work out the corrections due to a single hard-collinear resolved emission within an ensemble of soft-collinear radiation; similarly, at N$^3$LL one has to consider up to two resolved hard-collinear emissions embedded in an ensemble of soft-collinear radiation. 

A correct treatment of the hard-collinear emissions can be achieved by identifying two different contributions in the real matrix element, one of which is fully analogous to the one giving rise to $R'$ in eq.~\eqref{eq:master-NLL}, and another contribution which corresponds to an exclusive step of DGLAP evolution~\cite{Bizon:2017rah}.
This separation allows one to disentangle the $R'$ contribution, whose kinematics is soft by construction, from that of the exclusive DGLAP, which by construction is hard and collinear.
As a result, the cross section at all orders can be constructed by splitting the contribution to the cross section from each inclusive block into a $R'$-contribution and an exclusive DGLAP step.
This amounts to performing the last step of DGLAP evolution in fig.~\ref{fig:lund} unintegrated in $k_{t}$.

In addition to the parton densities, at NNLL one needs to include the collinear coefficient functions which emerge from their renormalization and originate from emissions that occur at the edges of the phase space in fig.~\ref{fig:lund} (cfr. eq.~\eqref{eq:CSSCatani}).
The coefficient functions contribute to the logarithmic structure only through the scale of their running coupling, which is the transverse momentum of the emission(s) with which they are associated. 
Analogously to the parton densities, one can evolve them inclusively up to the resolution scale $\epsilon k_{t,1}$, whereas their evolution must be instead treated exclusively in the resolved strip. 
At sufficiently higher orders, one must finally include also the contribution from the collinear coefficient functions $G$, which describe the azimuthal correlations with the initial-state gluons.
This contribution starts at N$^3$LL for gluon-induced processes.

Therefore, higher-order logarithmic corrections can be included simply by adding higher-order correlated blocks: this corresponds to incorporating higher-order logarithmic corrections to the radiator $R$ and its derivative $R'$, as well as in the anomalous dimensions which drive the evolution of the parton densities and coefficient functions.
In ref.~\cite{Bizon:2017rah} this construction was used to compute the resummed transverse-momentum spectrum at N$^3$LL accuracy.
The result is equivalent to the CSS formulation in Fourier space barring a change of scheme and reproduces correctly the Parisi-Petronzio scaling at small values of $p_t$~\cite{Bizon:2017rah}.
We will discuss phenomenological applications of the formalism introduced here in chapter~\ref{ch-respheno}, where we will present matched predictions for the transverse-momentum in Higgs production at the LHC at NNLO+N$^3$LL accuracy.

\section{High-energy resummation}\label{sec:small-xres}

Another class of logarithmic corrections which mar perturbative computations is that of high-energy logarithms. 
As the centre-of-mass energy $\sqrt{s}$ of the collision increases, so does the phase space available for parton radiation, which is enhanced by large logarithms of $x=Q^2/s$, where $Q$ is the hard scale of the process. 
As an example, let us consider how high-energy logarithms arise in deep-inelastic scattering.
Here, $x$ represents the longitudinal momentum fraction of the proton carried by the struck parton. 
When the centre-of-mass energy squared of the collision $W^2 = Q^2 (1-x)/x \simeq Q^2/x $ (see sect.~\ref{sec:DISkin}) is very large, only a small fraction of it can be put on shell by the virtual photon.
This leaves a large phase space for the emission of a cascade of partons, each of which has a very \mccorrect{large fraction of the} longitudinal momentum of the parent's parton, such that the final \mccorrect{struck} parton at the end of the cascade carries only a very small fraction of the original longitudinal momentum of the proton.
Similarly to what we discussed in the previous section, each parton emission is logarithmically enhanced; however, the soft parton is now the one which is struck by the virtual photon, and each emission thus occurs with a {\it single} logarithmic enhancement. 

These single logarithms affect generically higher order corrections to both splitting functions and coefficient functions\footnote{Notably, the heavy top quark limit induces a double-logarithmic enhancement at small $z$ in the coefficient function for Higgs production in gluon fusion~\cite{Catani:1990xk,Catani:1990eg,Hautmann:2002tu}.}. 
Specifically, leading logarithms of $1/x$ appear as $1/x \ln(1/x)^{n-1}$ contributions to the gluon splitting functions at $n$-loop. 
Small-$x$ logarithms mostly affect the singlet sector; (double) logarithms of $x$ appear also in the nonsinglet sector but are suppressed by an extra power of $x$ (see for instance~\cite{Blumlein:1995jp}).
Coefficient functions can also contain small-$x$ logarithms.
For gluon-induced processes like Higgs or top production they affect already the LO cross section and are thus a LL$x$ effect.
In quark-induced processes, the need of a gluon-to-quark conversion to develop small-$x$ logarithms makes it a NLL$x$ effect. 
Nevertheless, in either cases at small-$x$ and at high-energy $\as \ln (1/x)$ can become large, and reliable predictions require the resummation of the large logarithms in both splitting and coefficient functions.
We refer the reader to ref.~\cite{Caola:2018xxx} for a recent review.

The inclusion of high-energy resummation into DGLAP evolution and into the coefficient functions to produce consistent predictions has proved far from trivial and sufficient progress has only very recently been made to make this endeavour possible~\cite{Bonvini:2016wki,Bonvini:2017ogt,Ball:2017otu}.
High-energy resummation is indeed a \texttt{HELL} of a challenge, as an aptly named code which resums small-$x$ logarithms suggests.
Different implementations have been proposed, each distinguished by its pros and cons but all sharing kindred complexities in their implementation.  
Here we limit ourselves to a short discussion of small-$x$ resummation in the Altarelli-Ball-Forte (ABF) approach~\cite{Ball:1997vf,Altarelli:1999vw,Altarelli:2001ji,Altarelli:2003hk,Altarelli:2005ni,Altarelli:2008aj}.
Our treatment of high-energy resummation will be rather qualitative and aimed at introducing the ingredients necessary for PDFs with small-$x$ resummation, which we will discuss in chapter~\ref{ch-respdfs}.
After a brief introduction to $k_t$-factorization and to the BFKL equation, we will show how the BFKL kernel can be used to resum high-energy logarithms.
We will then summarize the main features of the ABF approach and discuss how it has been revived and improved to allow for an effective implementation in the context of PDF fits.

\subsection{$k_t$-factorization \amper the BFKL equation}

\begin{figure}[t]
\begin{center}
\includegraphics[width=0.6\textwidth]{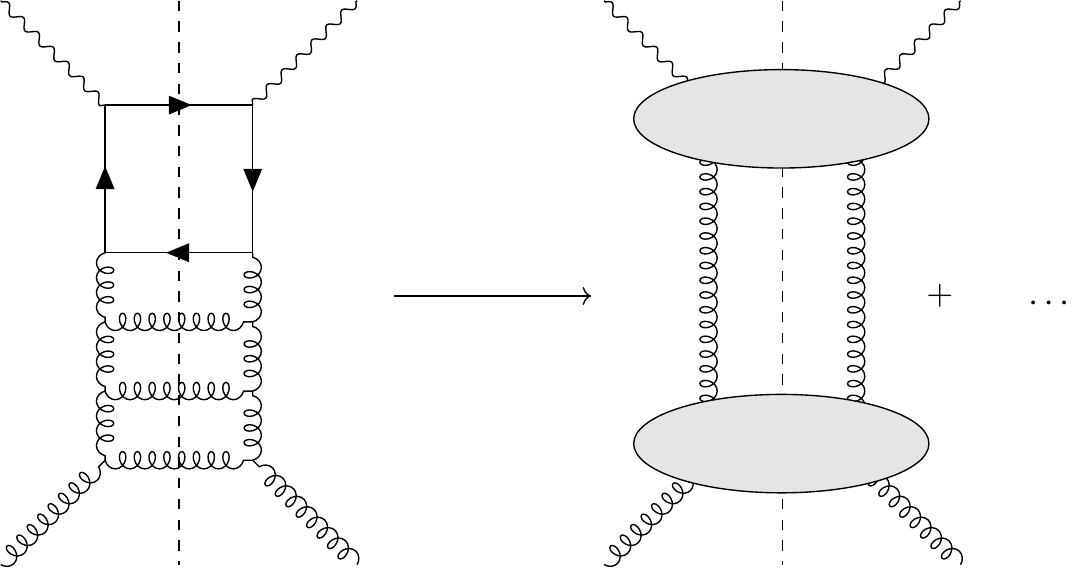}
 \caption{Representative ladder diagram for DIS process in the gluon channel (left) and schematic representation in $k_t$ factorization of the two-gluon reducible contribution in the high-energy limit (right).}
  \label{fig:ladder}	
\end{center}
 \end{figure}

Let us consider again the simple DIS example we just introduced, where a proton of mass $M_p \equiv Q_0$ is struck by a photon of virtuality $Q^2 \gg Q_0^2 $.
The correct treatment of such collision requires the resummation of large collinear logarithms $\as \ln Q^2$, which is performed by the DGLAP equation~\eqref{eq:AP}.
In an axial gauge, these logarithms are generated by so-called `ladder' diagrams (see fig.~\ref{fig:ladder}), in which the emitted partons with transverse momentum $k_{t,i}$ and (fractional) longitudinal momentum $z_i$ follows a strong ordering in $k_{t,i}$ 
\begin{equation}
	Q^2 \gg k_{t,n}^2 \gg k_{t,n-1}^2 \gg \ldots k_{t,1}^2 \gg Q_0^2,
\end{equation}
such that the LL solution of DGLAP equation can be interpreted as the all-order sum of the ladder diagrams: the diagram with $n$ rugs corresponds to the $(\as \ln Q^2)^n$ contribution.

When the centre-of-mass energy squared $s$ is much larger than $ Q^2$, smaller values of $x = Q^2/s$ become accessible.
An inspection of the splitting functions~\eqref{eq:APsplittingLO} shows that the dominant ladder in this regime is the gluon ladder (with gluon rugs), since the splitting function $P^{(0)}_{gg}(z)$ grows as $1/z$, as opposed to $P^{(0)}_{qg}(z)$ and/or $P^{(0)}_{qq}(z)$ which are not singular. 
Therefore, at small-$x$ gluon ladders with repeated iterations of $P_{gg}$ become dominant, and there is also a strong ordering in the longitudinal momentum $z_i$
\begin{equation}
		z_n \ll z_{n-1}  \ll \ldots \ll z_1.
\end{equation}
In this kinematic regime, logarithms of $1/x$ associated with the emission of soft partons down the ladder become increasingly large and must be resummed to all orders.

Though this analysis is rather qualitative, it can indeed be proven~\cite{Lipatov:1976zz,Fadin:1975cb,Kuraev:1976ge,Kuraev:1977fs,Balitsky:1978ic} that the dominant contribution (which gives rise to small-$x$ enhanced terms) in the high-energy regime comes from diagrams which are at least two-gluon reducible in the $t$-channel.
In this regime, predictions for the structure functions are made using the so-called $k_t$ factorization (see e.g.~\cite{Catani:1990xk,Catani:1990eg}),
\begin{align}\label{eq:kt-fact}
	F (x,Q^2) = \sum_j \int d k_t^2 \int \frac{dz}{z} \mathcal C_j\left(\frac{x}{z},\frac{k_t^2}{Q^2}, \as\right) \mathcal F_j (z, k_t^2),
\end{align}
where $\mathcal F$ is an unintegrated distribution function, which encodes all-order gluon emission, and $\mathcal C$ is an off-shell partonic cross section, which describes the cross section in the $t$-channel for the process of interest, computed with off-shell partons.
Since in the high-energy limit the dominant contribution is given by gluons, we will drop the summation over the index $j$ and we shall consider only the gluon unintegrated distribution $\mathcal G$ in the following discussion.

The $k_t$-factorization formula eq.~\eqref{eq:kt-fact} generalizes the collinear factorization formula of eq.~\eqref{eq:AP} to the regime $s \gg Q^2 \gg Q_0^2$, regardless of the size of the transverse momentum $k_t^2$, which is allowed to take all the values in the $k_t^2$ integral.
On the contrary, the validity of collinear factorization is restricted to the region  $Q^2 \sim s$, where the integration over $k_t$ is important only in the region $k_t^2 \ll Q^2$.

The $k_t$-factorization formula eq.~\eqref{eq:kt-fact} is diagonalized by a double Mellin transform,
\begin{equation}
	F(N,M) = \int_0^1 dx \, x^{N-1} \int_0^\infty \frac{dQ^2}{Q^2} \left(\frac{Q^2}{\mu^2} \right)^{-M} F(x,Q^2),
\end{equation}
where $\mu$ is a reference scale.
In double Mellin space the structure function is simply given by the product
\begin{equation}
	\bm F(N,M) = \bm {\mathcal C}(N, M, \as)  \bm{ \mathcal  G}(N, M),
\end{equation}
with
\begin{align}
	\bm{\mathcal C}(N,M,\as) &= \int_0^1 dz \, z^{N-1} \int_0^\infty \frac{d\rho}{\rho} \rho^{-M} {\mathcal C}(z, \rho, \as),\nonumber \\
	 \bm{ \mathcal  G} (N, M) &=  \int_0^1 dz \, z^{N-1} \int d k_t^2 \left(\frac{k_t^2}{\mu^2}\right)^{-M} \mathcal G(z, k_t^2).
\end{align}

The Mellin transform in $x$ maps the $x \rightarrow 0$ region to the $N \rightarrow 0$ one; in particular, it maps the ($\ln x$)-enhanced terms to poles in $N=0$. 
In non-pathological theories, one can prove that the coefficient function $\bm {\mathcal C}$ is regular in $N=0$ and that the leading small-$x$ terms are encoded in $\bm {\mathcal G}$.
As we will see below, to resum the leading $\ln x$ terms it is sufficient to solve RG equations for the unintegrated gluon distribution function $\bm{ \mathcal G}$.

The first equation that the unintegrated distribution function satisfies is the renowned Balitsky-Fadin-Kuraev-Lipatov (BFKL) equation~\cite{Lipatov:1976zz,Fadin:1975cb,Kuraev:1976ge,Kuraev:1977fs,Balitsky:1978ic}, which reads 
\begin{equation}\label{eq:BFKLdirect}
	\frac{d}{d \xi} \mathcal G (\xi, Q^2) = \int_0^\infty \frac{d k^2}{k^2} K\left(\as, \frac{Q^2}{k^2}\right) \mathcal G (\xi, k^2),
\end{equation}
where we have defined $\xi = \ln 1/z$.
In Mellin space the BFKL equation is
\begin{equation}\label{eq:BFKLmellin}
	\frac{d}{d \xi}  \bm{ \mathcal  G}(\xi, M) =   \chi (\as, M)  \bm{ \mathcal  G} ( \xi, M),
\end{equation}
where we have defined the BFKL kernel in Mellin space
\begin{equation}\label{eq:chidefinition}
	\chi (\as, M) = \int_0^\infty \frac{dQ^2}{Q^2} \left(\frac{Q^2}{k^2} \right)^{-M} K \left(\as, \frac{Q^2}{k^2}\right),
\end{equation}
which has a perturbative expansion in powers of $\as$, $\chi (\as, M) = \as \chi_0 (M) + \as^2 \chi_1 (M) + \ldots $ and has been known for quite some time at NLO accuracy~\cite{Fadin:1998py}.
The BFKL equation describes the evolution of the (unintegrated) parton densities along the direction of the variable $\xi$ and resums large-$\xi$, i.e. small-$x$ logarithms, which under $N$-Mellin are mapped in the $N=0$ region.
Besides BFKL equation, we would also like to exploit DGLAP equation, which controls the collinear region $M\rightarrow 0$.
However, the function $\mathcal G$ is an {\it unintegrated} distribution, which depends on the transverse momentum $k_t$, and which therefore is not necessarily described by DGLAP.
Nevertheless, we can define an {\it integrated} parton distribution
\begin{equation}
	G(x, Q^2) \equiv \int^{Q^2} d k_t^2\,  \mathcal G (x, k_t^2),
\end{equation}
which does not depend on the transverse momentum and which can be identified with  $x g (x,Q^2)$, such that its evolution in Mellin space is described by the DGLAP equation
\begin{equation}
	\frac{d}{d t} \bm G (N, Q^2) = \gamma (\as (Q^2) , N) \bm G (N, Q^2), 
\end{equation}
where $t=\ln Q^2/\mu^2$.

We can now use the BFKL equation to study the asymptotic behaviour at small-$x$ of the gluon distribution $G$. 
At LO, the BFKL kernel reads
\begin{equation}\label{eq:BFKLLO}
	\chi_0 (M) = \frac{C_A}{\pi} [2 \psi(1) - \psi (M) - \psi (1-M)],
\end{equation}
where $\psi $ is the digamma function $\psi(x) = \Gamma'(x)/\Gamma(x)$.
The solution of the BFKL equation for $\bm G$ is therefore
\begin{equation}
	G(x, Q^2) = \int_{c-i\infty}^{c+i\infty} \frac{dM}{2 \pi i} \tilde{ \bm G}_0 (M) e^{\as \chi_0 (M) \xi + Mt}. 
\end{equation}
For sufficiently small $x$ and large $Q^2$, we can use a saddle point approximation to capture the asymptotic behaviour of the gluon distribution. 
For fixed values of $Q^2$ and in the $x\rightarrow 0$ limit, the saddle is determined by the minimum of the BFKL kernel,
\begin{equation}\label{eq:pomeron}
	G(x, Q^2) \sim e^{\as \chi_0(1/2) \ln \frac{1}{x}} = x^{\alpha_g}. 
\end{equation}
Eq.~\eqref{eq:pomeron} is the famous pomeron solution. 
For values of $Q \lesssim 10$ GeV, where $\as \sim 0.2$, the intercept for $x g(x, Q^2)$ is $\alpha_g \simeq -0.5$: nevertheless, such steep rise for the gluon PDF has not been observed in experiments at HERA collider, which probe values of $x$ down to $x \sim 10^{-5}$.
This analysis, however, is based on the BFKL at LO.
It turns out that the NLO corrections to the BFKL kernel are very large and that the NLO kernel is dramatically different from the LO result.
The failure of LO BFKL to successfully predict the small-$x$ behaviour of the gluon distribution and the perturbative instabilities of the BFKL kernel stimulated a series of studies by several groups: the aforementioned Altarelli, Ball and Forte (ABF), Ciafaloni, Colferai, Salam and Stasto (CSSS)~\cite{Salam:1998tj,Ciafaloni:1998iv,Ciafaloni:1999yw,Ciafaloni:2002xf,Ciafaloni:2003ek,Ciafaloni:2003rd,Ciafaloni:2003kd,Ciafaloni:2007gf}, and Thorne and White (TW)~\cite{Thorne:1999sg,Thorne:1999sg,Thorne:2001nr,White:2006yh}.
These studies showed that a simultaneous resummation of collinear and high-energy logarithms can be obtained if one consistently combines the DGLAP and BFKL equations.
In this section, we will mainly focus on the ABF approach; the theoretical ingredients used by the various groups were nevertheless similar and lead to compatible results (see refs.~\cite{Forte:2009wh,Dittmar:2005ed} for a detailed comparison).

\subsection{From factorization to resummation: duality \amper the double-leading approximation}

Reliable results in the high-energy region can be obtained by exploiting an important relation between the DGLAP and the BFKL kernels, which is more transparent if one writes the two equations satisfied by the gluon distribution $G$ in double Mellin space.
Here, the relation between the integrated and the unintegrated parton distribution takes the simple form
\begin{equation}
	\bm G(N,M) = \frac{1}{M} \bm {\mathcal G }(N,M),
\end{equation}
and the evolution equations for the integrated parton density $\bm G(N,M)$ can be written as algebraic relations
\begin{align}\label{eq:duality1}
	N \bm G(N, M) &= \chi (M, \as) \bm G (N, M) + \bar{ \bm G}_0 (M),\nonumber \\
	M \bm G(N, M) &= \gamma (N, \as) \bm  G (N, M) + \bm G_0 (N),
\end{align}
where $\bm G_0(N)$ and $\bar {\bm G}_0 (M)$ are non-perturbative initial conditions.
The symmetric form of eq.~\eqref{eq:duality1} is rather intriguing, as it suggests a relation between the kernels of the DGLAP and the BFKL equation.
This relation, called duality~\cite{Ball:1997vf,Jaroszewicz:1982gr}, is the key for high-energy resummation of splitting functions.

The solutions to the two equations~\eqref{eq:duality1} are
\begin{align}\label{eq:duality2}
 &\bm G(N,M) = \frac{\bm G_0(N)}{M - \gamma (\as, N) },  &\bm G(N,M) = \frac{\bar {\bm G}_0(M)}{N - \chi (\as, M ) }.
\end{align}
We can now find the result in $t$-space by computing the inverse Mellin transform.
Up to subleading power corrections, we find that
\begin{align}\label{eq:dualityrel}
	\bm G(N,t) &= \int_{c-i\infty}^{c+i \infty} \frac{dM }{2 \pi i} e^{Mt} \frac{\bar {\bm {G}}_0(M)}{N-\chi (\as, M)} = \frac{\bm {\bar G}_0 (M^*)}{- \chi'(\as, M^*)} e^{ M^* t}\nonumber ,\\
	\bm G(N,t) &= \int_{c-i\infty}^{c+i \infty} \frac{dM }{2 \pi i} e^{Mt} \frac{\bm G_0(N)}{M-\gamma (\as, N)} = \bm G_0 (N) e^{\gamma(\as, N) t},
\end{align}
where $M^*$ is the position of the rightmost pole in eq.~\eqref{eq:duality2} and is implicitly defined by the equation
\begin{equation}
	\chi (\as, M^*) = N.
\end{equation}
By requiring the leading twist consistency of the $x$ and $Q^2$ RGE which the gluon distribution $G$ obeys we arrive to the duality relation~\cite{Ball:1997vf}
\begin{align}\label{eq:duality}
	&\chi (\as, \gamma (\as, N)) = N,  &\gamma (\as, \chi (\as, M)) = M,
\end{align}
dictated by the position of the pole.
The consistency also relates the boundary conditions which appear in eq.~\eqref{eq:dualityrel}~\cite{Bonvini:2012sh}.

The duality relations eq.~\eqref{eq:duality} allow one to extract the all-order behaviour both from $\chi$ or $\gamma$ if the fixed-order behaviour of the other kernel is known~\cite{Altarelli:1999vw}.
Let us consider an expansion of the anomalous dimension $\gamma$ and the BFKL kernel $\chi$ in powers of $\as$ and at fixed $\as/N$, $\as/M$
\begin{align}
	\gamma (\as, N) &= \gamma_s (\as/N) + \as \gamma_{ss} (\as/N) + \ldots\\
	\chi (\as, M) &= \chi_s (\as/M) + \as \chi_{ss} (\as/M) + \ldots \, ,
\end{align}
where $\gamma_s$ and $\gamma_{ss}$ ($\chi_s$ and $\chi_{ss}$) capture the leading and next to leading logarithms of $1/x$ ($Q^2$).  
We can now relate the coefficients of the fixed-order expansion of $\chi$ to the resummed expressions for $\gamma_s$ and $\gamma_{ss}$ (or vice versa). 
Using the duality relation we have
\begin{align}
	\chi_0 \left( \gamma_s \left(\frac{\as}{N} \right) \right) + \as \left[ \chi_0' \left( \gamma_s \left(\frac{\as}{N} \right) \right) \gamma_{ss} \left(\frac{\as}{N} \right) +  \chi_1 \left( \gamma_s \left(\frac{\as}{N} \right) \right) \right]+ \mathcal O \left(\as^{k+2}/N^k\right)= \frac{N}{\as},
\end{align}
and therefore we obtain
\begin{align}
	& \gamma_s  \left(\frac{\as}{N} \right)  = \chi_0^{-1} \left( \frac{N}{\as} \right) ,
	&\gamma_{ss}  \left(\frac{\as}{N} \right) = - \frac{\chi_1 (\gamma_s (\as/N) )}{\chi_0' (\gamma_s (\as/N) )}.
\end{align}
An equivalent result can be obtained for $\chi_s$ and $\chi_{ss}$, which can be expressed as in terms of the LO and the NLO DGLAP splitting functions $\gamma_0$ and $\gamma_1$.

\begin{figure}[t]
\begin{center}
\includegraphics[width=0.6\textwidth]{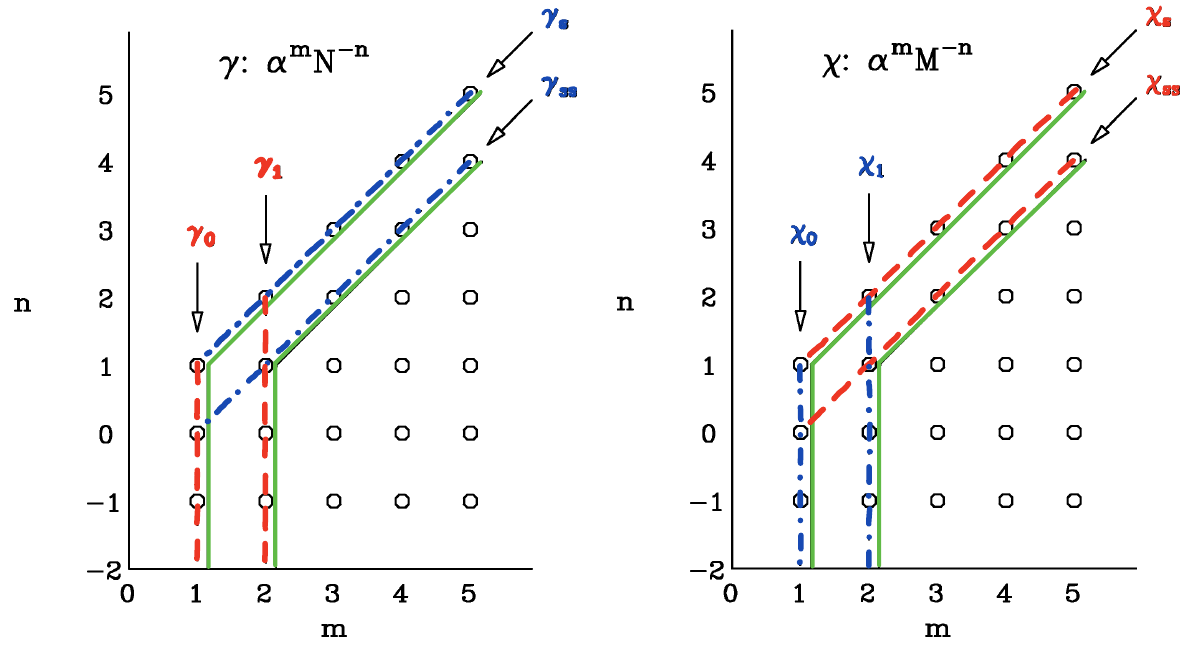}
 \caption{Structure of the DL approximation for the anomalous dimension $\gamma$ and the BFKL kernel $\chi$. Vertical lines correspond to terms of the same fixed order in $\as$, whereas diagonal lines correspond to the same order in $\as$ at fixed $\as/N$ (left) or $\as/M$ (right). The sum of the terms in vertical lines in the $\chi$ plot is related by duality to the sum of terms in a diagonal line in the $\gamma$ plot, and vice versa. Green solid lines denote terms of the same order in the DL expansion (fig. adapted from ref.~\cite{Altarelli:1999vw}).}
  \label{fig:duality}	
\end{center}
 \end{figure}
 
The two expressions which we have constructed describe the leading and the next-to-leading $\ln 1/x$ effects, but do not capture the resummation of the $\ln Q^2$ terms which are not enhanced at small-$x$. 
This resummation can be obtained by performing a {\it double leading} (DL) expansion of the DGLAP anomalous dimension:  
\begin{align}\label{eq:DLexp}
	\gamma_{\text{DL}} (\as, N) =& \left[ \as \gamma_0 (N) + \gamma_s\left(\frac{\as}{N}\right) - \text{double counting} \right] \nonumber	 \\
	& + \left[ \as^2 \gamma_1 (N) + \as \gamma_{ss} \left(\frac{\as}{N}\right) - \text{double counting} \right] + \mathcal O \left(\as^3, \as^{k+2}/N^k\right) .
\end{align} 
This result contains information about both the all-order small-$x$ behaviour (encoded in $\gamma_s$, $\gamma_{ss}$) and the large-$Q^2$ behaviour (encoded in $\gamma_0$, $\gamma_1$), and explicitly resums the leading and next-to leading logarithms of $x$ in the DGLAP splitting functions.
One usually refers to the first line as LO DL, to the second as NLO DL, et cetera (see fig.~\ref{fig:duality}). 

By construction, the DL expansion eq.~\eqref{eq:DLexp} is close to the DGLAP result in the $M\rightarrow 0$ limit and to (the dual of) the BFKL result when $N \rightarrow 0$.
To identify how the double-leading result describes the small-$x$ behaviour of $G$, one needs to determine the rightmost $N$-space singularity of $\gamma(\as,N)$, which governs the leading behaviour at small-$x$.
It turns out that the LO DL anomalous dimension is dramatically different from the NLO DL one. 
Indeed, the former has a square-root branch-point at some small positive value of $N$, whereas the latter is not even singular.
Moreover, both results differ from the fixed-order anomalous dimension, which has a simple pole at $N=0$ both at LO and at NLO.

The bad perturbative behaviour of the resummed result at $N=0$ is related to the poor perturbative stability of the BFKL kernel.
In particular, the behaviour of the BFKL kernel at $M=0$ is unstable: the LO contribution to the BFKL eq.~\eqref{eq:BFKLLO} has a simple pole at $M=0$, whereas the NLO contribution has a double pole in $M=0$ with the opposite sign~\cite{Fadin:1998py}.
Both results are different from the all-order behaviour, which is instead determined by duality: the momentum conservation equation for the parton distribution implies $\gamma (\as, 1) = 0 $ to all orders and thus $\chi (\as, M=0) =1$. 

It is therefore convenient to use the additional information contained in the one and two loop anomalous dimension $\gamma_0$ and $\gamma_1$ to first construct a DL expansion of the BFKL kernel, analogously to eq.~\eqref{eq:DLexp}, with the roles of $\gamma$ and $\chi$ interchanged.
The resummed DGLAP anomalous dimension is then determined by duality.
It can be seen that the perturbative instabilities at $M=0$ are indeed less dramatic if one considers the double leading result $\chi_{\text{DL}} (\as, M)$.
The LO DL kernel satisfies the momentum constraint at $M=0$, whereas at NLO there is a small violation which can be removed by adding subleading terms to the DL result. 
However, the DL result is still perturbatively unstable for larger values of $M$ due to singularities in $M=1$.
Therefore, one must also resum the singularities in $M = 1$ to obtain a stable DL expansion.

The solution to this problem comes from a symmetry argument.
Since the three gluon vertex is symmetric if the radiated and radiating gluons are interchanged, the BFKL kernel should satisfy
\begin{equation}
	\frac{1}{Q^2} K \left(\as,\frac{Q^2}{k^2} \right) = 	\frac{1}{k^2} K \left(\as,\frac{k^2}{Q^2} \right),
\end{equation}
which in Mellin space translates into the symmetry of the BFKL kernel upon the interchange $M \leftrightarrow 1-M$. 
Though this symmetry is manifest in the LO kernel eq.~\eqref{eq:BFKLLO}, it is broken beyond LO by running coupling corrections and by the choice of the DIS kinematics.
However, the terms which break the symmetry can be computed exactly, allowing one to construct a symmetrized version of the DL expansion~\cite{Altarelli:2005ni} which completely removes the large-$M$ instabilities.

By combining symmetrization and momentum conservation one can therefore construct a perturbatively stable BFKL kernel, which, by duality, allows one to obtain resummed DGLAP anomalous dimensions.
The final result for the resummed BFKL kernel has a minimum at a finite value of $M$, which would therefore translate into a branch cut for the DGLAP anomalous dimension by duality.
However, in our discussion we have so far included perturbatively the corrections due to the running of the strong coupling. 
Though additional running coupling effects are subleading at small-$x$, it turns out that reliable results require the resummation of the running coupling corrections at all orders by solving the running-coupling BFKL equation~\cite{Altarelli:2001ji,Altarelli:2005ni,Ciafaloni:1999yw,Ciafaloni:2002xf,Ciafaloni:2003rd}.
The resummation of running coupling effects substantially changes the small-$x$ behaviour of the anomalous dimension: the branch cut singularity of the anomalous dimension is removed and is replaced by a pole (as in the fixed-order case) which is however shifted from $N=0$ to a finite value $N_0> 0$. 
The net result is a softer growth at small-$x$ of the gluon distribution in the small-$x$ region.
For instance, in the kinematic region probed by HERA, with $x \lesssim 10^{-3}$ and $Q \lesssim 10$~GeV, where $\as \simeq 0.2$, one obtains $G(x, Q^2 ) \sim x^{-0.2}$, which is fully compatible with the experimental data (see for instance~\cite{Altarelli:2001ji,Altarelli:2005ni}).

\subsection{To \texttt{HELL} \amper back}\label{sect:HELLandback}

The discussion in the previous section shows that by combining the physical content of the BFKL and of the DGLAP equations one obtains perturbatively stable anomalous dimensions for small-$x$ evolution, which include resummation effects and describe the correct behaviour.
However, all the necessary ingredients to perform a realistic global PDF fit with resummation of small-$x$ logarithms have only recently been implemented in a public code named \texttt{HELL} (\texttt{High-Energy Large Logarithms}) capable of resumming both splitting function and partonic coefficient functions.
Crucially, the code also allows for the matching to NNLO fixed order.
Indeed, whereas the HERA dataset \mccorrect{can be satisfactorily described by NLO theory}~\cite{Ball:2017otu}, there is some indication of a tension between the data and NNLO predictions~\cite{Caola:2009iy,Caola:2010cy,Abramowicz:2015mha,Harland-Lang:2016yfn} \mccorrect{(this tension is already present at NLO in refs.}~\cite{Abramowicz:2015mha,Harland-Lang:2016yfn}), and the need for small-$x$ resummation at (N)NLO and beyond has been long advocated (see e.g.~\cite{Forte:2005mw}).
While following the same general approach as ABF, the \texttt{HELL} implementation incorporates a number of technical improvements which makes the numerical implementation more robust, as well as some new developments; a detailed discussion and comparison is given in ref.~\cite{Bonvini:2016wki}.

Let us now briefly sketch how all-order high-energy effects can be included in a PDF fitting framework to achieve consistent small-$x$ resummed phenomenology. 
First, let us notice that though for the sake of simplicity we have considered small-$x$ resummation only for the gluon, DGLAP evolution couples the gluon to the singlet.
Therefore, one should rather consider resummation in the whole singlet sector.
As a consequence, the resummed anomalous dimension $\gamma$ corresponds to the largest eigenvalue at small-$x$ (that is, the one which is enhanced) $\gamma_+$ of the DGLAP evolution matrix
\begin{equation}
	\Gamma (N, \as) \equiv \left( 
	\begin{array}{cc}
		\gamma_{gg} & \gamma_{gq} \\
		\gamma_{qg} & \gamma_{qq}
	\end{array}
	\right),
\end{equation}
and analogously the distribution $\bm G$ should be identified with the `plus' eigenvector $\bm f_+$ in the basis where DGLAP evolution in the singlet sector is diagonal\footnote{To make contact with the \texttt{HELL} papers, we have absorbed a factor of $x$ in the definition of the parton density $\bm f_+$.}.
The resummation of the plus eigenvalue with all the above ingredients was originally achieved by ABF in ref.~\cite{Altarelli:2005ni}, while the inclusion of the quark contributions and hence the rotation to the physical basis of the singlet sector was completed in ref.~\cite{Altarelli:2008aj}.
This procedure is rather cumbersome, as it relies on the relation between the  $qg$ anomalous dimension and the largest eigenvalue $\gamma_+$, which however is not not known in a closed form in the $\overline{\text{MS}}$ scheme (or in the $Q_0\overline{\text{MS}}$ scheme, commonly used in the context of small-$x$ resummation since it avoids large cancellations between evolution and coefficient functions, see e.g.~\cite{Ciafaloni:2005cg}).
Nevertheless, the rotation can be performed and resummed splitting functions can be obtained by computing the inverse Mellin transform of the anomalous dimension in the physical basis.
The resummed splitting functions take the generic form
\begin{equation}\label{eq:HELLsplitting}
	P_{ij}^{\text{N}^k\text{LO}+\text{N}^h\text{LL$x$}}(x,\as) = P_{ij}^{\text{N}^k\text{LO}}(x,\as)+\Delta_k P_{ij}^{ \text{N}^h\text{LL$x$}}(x,\as),
\end{equation}
where the first contribution is the splitting function computed to fixed-order $k$ and the second term is the resummed contribution, computed to either LL$x$ ($h=0$) or NLL$x$ ($h = 1$), minus its expansion to the fixed-order $k$ to avoid double counting.

\begin{figure}[t]
\centering
  \includegraphics[width=0.49\textwidth, page=1]{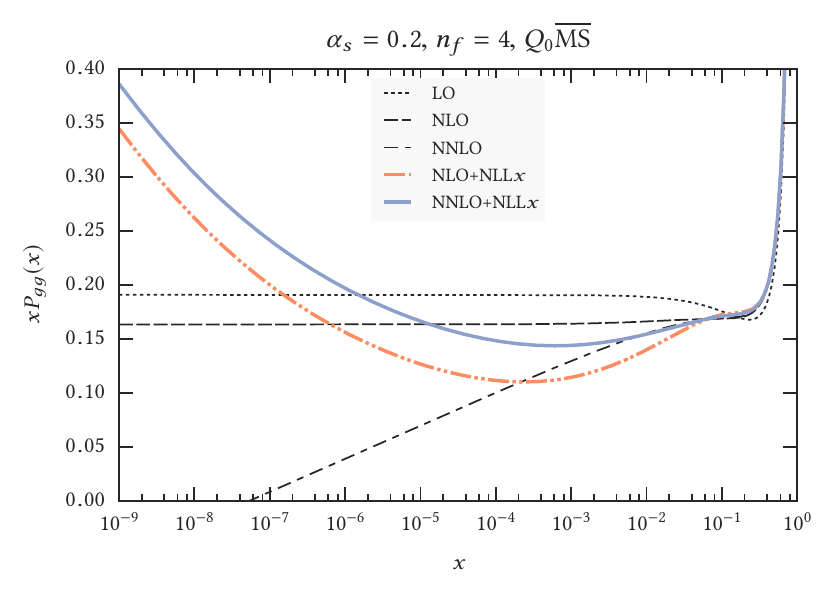}
  \includegraphics[width=0.49\textwidth, page=2]{figures/Pgg_and_Pqg.pdf}
  \caption{Comparison of the fixed-order gluon-gluon
    $xP_{gg}(x,\as)$ (left) and the quark-gluon $xP_{qg}(x,\as)$ (right)
    splitting functions with the NLO+NLL$x$ and NNLO+NLL$x$ results including small-$x$ resummation.
  }
  \label{fig:splittingfunctions}
\end{figure}

We show in fig.~\ref{fig:splittingfunctions} a comparison of the fixed-order gluon-gluon $x P_{gg} (x, \as)$ (left panel) and the quark-gluon $x P_{qg} (x,\as)$ (right panel) splitting functions with the matched results at NLO+NLL$x$ and NNLO+NLL$x$ accuracy.
The comparison is performed in the $Q_0 \overline{\text{MS}}$ scheme, with $n_f=4$ and with $\as=0.2$. 
Since the scheme change between the $Q_0 \overline{\text{MS}}$ scheme and $\overline{\text{MS}}$ scheme is of order $\as^3$, NLL$x$ resummation can be matched directly to the usual fixed-order NNLO  $\overline{\text{MS}}$ scheme calculation.
The effect of resummation is more important at NNLO than at NLO, due to the perturbative instabilities at small-$x$, which are visible by comparing the NLO and the NNLO curves in fig.~\ref{fig:splittingfunctions}.
In the gluon-gluon splitting function case, we see that the NLO evolution is closer to the all-order result (NNLO+NLL$x$) than NNLO evolution.
The situation is similar in the quark-gluon splitting function case, with the resummed results closer to NLO and NNLO for $10^{-5} \lesssim x \lesssim 10^{-1}$.
At N$^3$LO the instabilities would be larger, due to the appearance of two extra-powers of small-$x$ logarithms (since the leading logarithms at NLO and NNLO are accidentally zero).
This would translate into an even more dramatic impact of resummation, thus making the inclusion of small-$x$ resummation effects at N$^3$LO mandatory.

Once the resummed splitting functions have been constructed, the last remaining ingredient to perform a fit with small-$x$ resummation is the resummation of the coefficient functions.
The general formalism is based on $k_t$-factorization and has been applied to various processes at the lowest non-trivial order (note that in DIS the first non-trivial order is NLL$x$, since there are no LL$x$ contributions in the partonic coefficient functions).
In the ABF approach, the resummation of coefficient function was first developed in ref.~\cite{Altarelli:2008aj} and is performed in double Mellin space.
The \texttt{HELL} approach~\cite{Bonvini:2016wki,Bonvini:2017ogt,Bonvini:2018xvt,Bonvini:2018iwt}, despite being based on the same ingredients, performs the resummation directly in momentum space, avoiding technical complications related to the use of a conjugate space. 
In a nutshell, the idea is to use the relation between $k_t$ factorization and standard collinear factorization to construct resummed coefficient functions.
In particular, in the high-energy limit the unintegrated gluon density $\bm{ \mathcal G} (N, k_t^2)$ is related to the standard resummed PDF by a function $ \bm {\mathcal U}$
\begin{equation}
	\bm { \mathcal G} (N, k_t^2) = \bm {\mathcal U} \left( N, \frac{k_t^2}{Q^2} \right) \bm f_{+} (N, Q^2).
\end{equation}
The comparison between the $k_t$-factorization formula eq.~\eqref{eq:kt-fact} and the high-energy contribution to a generic cross section
\begin{equation}
	\bm \sigma(N, Q^2) \sim \bm C_+ (N, \as) \bm f_+ (N, Q^2)
\end{equation}
allows one to write
\begin{equation}
	\bm C_+ (N,\as) = \int dk_t^2 \bm{ \mathcal C} \left(N, \frac{k_t^2}{Q^2},\as \right) \bm {\mathcal U} \left( N, \frac{k_t^2}{Q^2} \right).
\end{equation}
Therefore, to construct the resummed coefficient functions one has to convolute the off-shell coefficient function $\bm {\mathcal C}$ with the evolution factor, which in the aforementioned $Q_0\overline{\text{MS}}$ scheme takes the form
\begin{align}
	\bm {\mathcal U} \left( N, \frac{k_t^2}{Q^2} \right) = \frac{d}{dk_t^2} \exp \left[ \int_{Q^2}^{k_t^2} \frac{dq_1^2}{q_1^2} \gamma_+ (N, \as (q_1^2))	 \right],
\end{align}
which contains the DGLAP evolution from the scale $Q^2$ to the scale $k_t^2$ and the conversion from unintegrated to integrated PDF. 

Once the resummation of $C_+$ has been achieved, one needs to rotate back to the physical basis to obtain an expression for $C_q$ and $C_g$.
Like the previous rotation, the procedure is not trivial and it is further complicated by the presence of running coupling effects which have to be handled with a certain care~\cite{Ball:2007ra,Bonvini:2016wki}.
Eventually, the resummed coefficient functions are computed analogously to \eqref{eq:HELLsplitting}:
\begin{equation}
	C_{i}^{\text{N}^k\text{LO}+\text{N}^h\text{LL$x$}}(x,\as) = C_{i}^{\text{N}^k\text{LO}}(x,\as)+\Delta_k C_{i}^{ \text{N}^h\text{LL$x$}}(x,\as).
\end{equation}
Resummed coefficient functions for DIS structure functions, including mass effects, have been recently implemented in \texttt{HELL2.0}~\cite{Bonvini:2017ogt}.

As we discussed in sect.~\ref{sec:factheavyquarks}, since a consistent PDF fit spans several orders of magnitude in $Q^2$, one has to consider a different number of active flavours at different energies to resum large collinear logarithms due to massive quarks. 
It turns out that the matching conditions eq.~\eqref{eq:pdfmatching} which relate the PDFs above and below threshold also contain small-$x$ logarithmic enhancements, which one needs to consistently resum: their resummation is available in \texttt{HELL2.0}. 
This last ingredient allows one to implement a resummation of the FONLL variable flavour number scheme used in the NNPDF fits.


\begin{savequote}[8cm]
\textlatin{Consistency is the last refuge of the unimaginative.}
  \qauthor{--- Oscar Wilde, \textit{The Relation of Dress to Art}}
\end{savequote}

\chapter{\label{ch-respdfs}Resummed parton distribution functions} 


\lettrine[lines=2]{S}{tate-of-the-art} global PDF sets are extracted from a variety of data, collected in very different environments over a period of time which spans more than two decades. 
The kinematic coverage of the data is typically parametrized with a dimensionful scale $Q$ (the hard scale of the process) and a dimensionless variable $x=Q^2/s$, where $\sqrt{s}$ is the centre-of-mass energy of the experiment. 
The use of a dataset which extends over a wide range in the ($x$,$Q^2$)-plane therefore provides a large number of constraints and allows for a precise extraction of PDFs. 

To make the most of the precise data available, it is necessary to supplement the measurements with accurate theoretical predictions. 
Currently, partonic cross sections are computed up to next-to-next-to leading order (NNLO) accuracy in fixed-order theory and DGLAP evolution is consistently included up to next-to-next-to-leading logarithmic (NNLL) accuracy. 
Nevertheless, as we discussed in chapter~\ref{ch-res}, in the large-$x$ (threshold) and in the small-$x$ (high-energy) regions partonic cross sections and DGLAP splitting functions are affected by logarithmic enhancement at all orders in perturbation theory.
Both regions are probed by experiments at the LHC, which cover a vast kinematic range in $x$ and $Q^2$.

As a consequence, it becomes important to assess the role of the logarithmically-enhanced contributions in these kinematic regions. 
Consistent calculations require resummation of the partonic cross sections, and should in principle be computed with PDFs extracted using a consistent theory.
Though attempts to include threshold (or large-$x$)~\cite{Corcella:2005us} and high-energy (or small-$x$) resummation~\cite{White:2006yh} in PDF fits were made more than a decade ago, state-of-the-art global PDF sets with resummation effects have only recently been determined.
We discuss these sets in this chapter: in sect.~\ref{sec:pdflarge} we consider a PDF set with large-$x$ resummation and we present a set which includes small-$x$ resummation effects in sect.~\ref{sec:pdfsmall}.
Finally, we discuss the prospects for PDF sets which include both large-$x$ and small-$x$ resummation in sect.~\ref{sec:jointlargesmall}.


\section{Parton distribution functions with threshold resummation}\label{sec:pdflarge}

In this section we present NLO+NLL$^\prime$ and NNLO+NNLL$^\prime$ global fits, constructed as variants of the NNPDF3.0 global fit~\cite{Ball:2014uwa}.
As we discussed in chapter~\ref{ch-res}, the foundations for threshold resummation were laid about thirty years ago and resummed calculations at high perturbative accuracy for processes relevant for PDF determination, such as Drell-Yan (DY) and DIS, have been available for quite some time.
However, to obtain a truly global resummed fit one needs to include a larger number of processes such as top production data, as well as jet production and $W$ production at the level of the measured lepton distributions.
Nevertheless, for the last two processes threshold resummation is not yet available in an amenable form.
Though resummed calculations for inclusive jets were used to construct approximate expressions for the NNLO contributions (see e.g.~\cite{deFlorian:2013qia}), there are no codes which are publicly available.
For $W$ production, resummation is available only at the level of the reconstructed $W$.

Therefore, the fit we discuss here is based on neutral and charged current deep-inelastic structure functions (both for fixed-target and collider experiments), fixed-target and collider neutral current DY production, and inclusive top-quark pair production. 
Though a DY+DIS+top fit is a first step towards a consistent resummed phenomenology at the LHC, it is however affected by larger uncertainties than the NNPDF3.0 global fit on which it is based, since the missing experiments affect in particular the constraints on the gluon.
Therefore, it will be important to produce updated resummed fits which include these processes as soon as the calculations become available; as we discuss below, the use of fixed-order PDFs with resummed matrix elements can lead to misleading results, particularly at NLO.

This section is organized as follows.
We first review the theoretical framework and the practical implementation in sect.~\ref{sec:large-implementation}.
In sect.~\ref{sec:large-settings} we discuss the fit settings, which are based on the NNPDF3.0 global fit.
We then present PDFs with threshold resummation effects in sect.~\ref{sec:large-PDFs} and we finally explore some implications for LHC phenomenology in sect.~\ref{sec:large-pheno}.
\mccorrect{Throughout this section, we shall use the shorthand `resummed PDFs' to indicate fits performed including threshold resummation in the coefficient functions: we emphasize that no resummation is included at the level of PDF evolution, since in the $\overline{\text{MS}}$ scheme all effects are contained in the coefficient functions.}

\subsection{Theoretical framework \amper implementation}\label{sec:large-implementation}

We have already discussed the theoretical formalism for threshold resummation in sect.~\ref{sec:threshold}, where we considered for simplicity the colour-singlet case.
The global fit we present here includes also processes where coloured particles appear in the final state, such as DIS and top production.
Before discussing the implementation, let us briefly review how the formalism presented in sect.~\ref{sec:threshold} can be extended to deal with these processes.

The starting point is a generic cross section for a hadronic process, which we can write in Mellin space as (for the sake of simplicity, the factorization and renormalization scales are set to $Q$)
\begin{equation}
	\bm \sigma (N, Q^2) = \sum_{a,b} \mathcal L_{ab} (N,Q^2)  \hat {\bm  \sigma}_{ab} (N, Q^2, \as),
\end{equation}
where $a, b $ run over the parton flavours, $Q^2$ is the hard scale of the process and $N$ is the conjugate variable to $x$. 
Here we consider three processes: DIS, DY production and top-anti-top production ($ t\bar t$).
In DIS, $Q$ is the off-shellness of the exchanged boson $Q^2 = -q^2$ and $x = Q^2/2 p\cdot q$, where $p$ is the hadron momentum (see sect.~\ref{sec:DISkin}); in DY, $Q$ is the invariant mass of the lepton pair and $x = Q^2/s$; for $t \bar t$, $Q ^2= 4m^2_t$ and $x = Q^2/s$ . 
Finally, $\mathcal L_{ab} (N,Q^2)$ is the Mellin transform of a parton luminosity (defined in eq.~\eqref{eq:partonlumi}) in the hadron-hadron collision case, or of a single PDF in the DIS case.

In Mellin space, the resummed partonic cross section can be written as the product of a Born contribution and an all-order coefficient function
\begin{equation}
	  \hat {\bm \sigma}_{ab}^{\text{(res)}} (N, Q^2, \as) = \bm \sigma_{ab}^{\rm Born} (N, Q^2, \as)\, \bm C_{ab}^{\rm res} (N, \as), 
\end{equation}
where, analogously to eq.~\eqref{eq:CataniTh1}, 
\begin{align}\label{eq:gen_resum}
\bm C^{(\text{res})}_{ab}(N,\as)&=\sum_{\bm I} \bar g_{ab}^{({\bm I})}(\as)\exp \bar{\mathcal{S}}^{({\bm I})}(N,\as),
\nonumber \\
 \bar{\mathcal{S}}^{({\bm I})}(N,\as)&= \ln \Delta_a+\ln \Delta_b+\ln J_c+\ln J_d+\ln \Delta^{({\bm I})}_{ab\to c d}.
\end{align}
We use the notation $ab\to c d$ to accommodate all the processes that enter the fit.
For $t \bar t$ production, we have to consider the resummation of two Born-level processes, namely $q \bar q \to t \bar t $ and $g g \to t \bar t$; for DIS instead we have $V^* q \to q$ and for DY $q \bar q \to V^*$.
Moreover, while in DIS and DY we have one colour structure, in the $t \bar t$ case we have two contributions, i.e.\ $\bm{I}= \text{singlet, octet}$.

Let us now examine the different contributions to the resummed exponent.
If $i$ is a colour-singlet, then $\Delta_i=J_i=1$. 
For each initial-state QCD parton, we have an initial-state jet function
\begin{equation} \label{eq:beam_func}
\ln \Delta_i=\int_0^1 d z \frac{z^{N-1}-1}{1-z}\int_{Q^2}^{(1-z)^2Q^2} \frac{d q^2}{q^2}A_{\text{cusp},i} \(\as \(q^2 \) \), \quad i=a,b.
\end{equation}
For each \emph{massless} final-state QCD parton we have a final-state jet function
\begin{equation} \label{eq:massless_jet_func}
\ln J_i=\int_0^1 d z \frac{z^{N-1}-1}{1-z} \[ \int_{(1-z)^2Q^2}^{(1-z)Q^2} \frac{d q^2}{q^2}A_{\text{cusp},i}  \(\as \(q^2 \) \)+ \frac{1}{2}B_i  \(\as \(Q^2(1-z) \) \) \], \quad i=c,d,
\end{equation}
while there is no jet-function for $t$ or $\bar t$. 

Finally, we also need to consider large-angle soft contributions, which depend in principle on both the process and the color flow: 
\begin{equation} \label{eq:soft1}
\ln \Delta^{({\mathbf I})}_{ab\to c d}=\int_0^1 d z \frac{z^{N-1}-1}{1-z} D^{({\bm I})}_{ab\to c d} \(\as \(Q^2(1-z)^2 \) \).
\end{equation}
The functions $A_{\text{cusp},i} \(\as \)$, $B_i \(\as \)$, $D^{({\bm I})}_i \(\as \)$, and $\bar g_{ab}^{({\bm I})} (\as)$ are free of large logarithms and can be computed in fixed-order perturbation theory.
As discussed in sect.~\ref{sec:threshold}, the accuracy of their determination fixes the logarithmic accuracy of the resummation.
In particular, (N)NLL$^\prime$ requires $A_{\text{cusp},i}$ to second (third) order in the strong coupling $\as$ and $B_i $, $D^{({\bm I})}_i $, and $\bar g_{ab}$ to first (second) order.
If one evaluates the Mellin integrals in the resummed expression eq.~\eqref{eq:gen_resum} in the $N\to \infty$ limit, the resummed coefficient function takes the form eq.~\eqref{eq:CataniTh3}; it is understood that all the modifications we discuss are applied to each partonic subprocess and each colour flow.

To produce resummed predictions for DY differential in rapidity, threshold resummation should be extended to rapidity distributions.
Here we follow the approach of ref.~\cite{Bonvini:2010tp}, which is based on the observation that the resummed partonic rapidity distribution coincides with the rapidity integrated one up to terms which are power-suppressed in the threshold limit. 
This means that to obtain the hadron-level resummed rapidity distribution one needs to modify only the parton luminosity. 

We can now move on to discuss the numerical implementation of the $N$-soft threshold resummation which we just described.
For DIS and DY processes, threshold resummation has been implemented in the public code \texttt{TROLL}~\cite{TROLL,Bonvini:2014joa,Bonvini:2014tea}, standing for \texttt{TROLL Resums Only Large-x Logarithms}, whereas we use the public code \texttt{Top++}~\cite{Czakon:2013goa} for top-pair production.
The contribution from resummation is encoded into $\Delta_j K_{\text{N$^k$LL$^\prime$}}$, defined as the difference between a resummed $K$-factor at N$^j$LO+N$^k$LL$^\prime$ and the fixed-order $K$-factor at N$^j$LO, such that
\begin{equation}\label{eq:deltaK}
\sigma_{\text{N$^j$LO+N$^k$LL$^\prime$}} = \sigma_{\text{N$^j$LO}} + \sigma_{\text{LO}}\times \Delta_jK_{\text{N$^k$LL$^\prime$}}\, ,
\end{equation}
where all the cross sections are evaluated with a common N$^j$LO+N$^k$LL$^\prime$ PDF set.
The fixed-order calculation is obtained from a separate code (specifically, the internal \texttt{FKgenerator} code used for the fits of the NNPDF3.0 family).

\begin{figure}[t]
  \centering
  \includegraphics[width=0.41\textwidth,page=1]{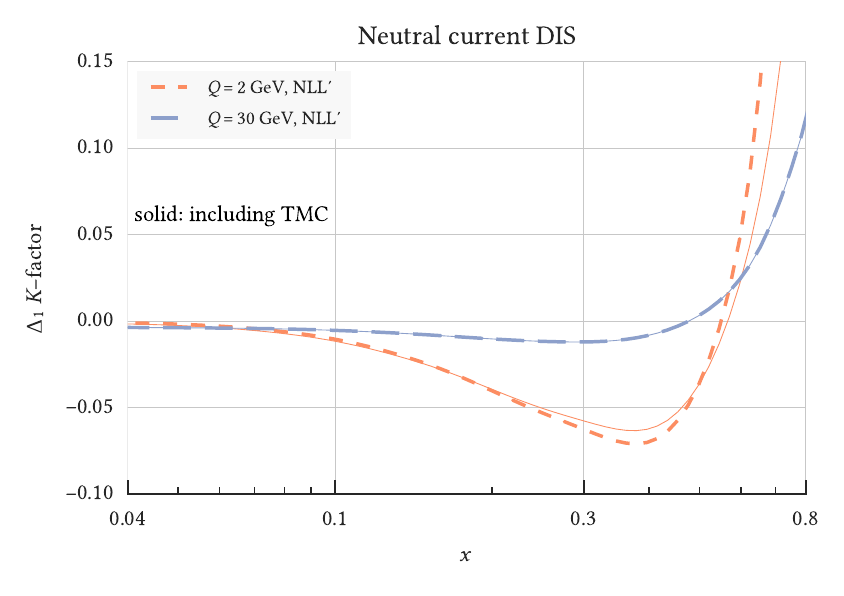}\qquad \qquad
  \includegraphics[width=0.41\textwidth,page=2]{figures/DIS_thesis.pdf}
  \caption{$\Delta K$-factors for the neutral current DIS structure function $F_2(x,Q)$,
    as a function of $x$, for $Q=2$~GeV and $Q=30$~GeV.
    The plot on the left (right) corresponds to $j=1$, $k=1$ ($j=2$, $k=2$) in eq.~\eqref{eq:deltaK}.
    The effect of TMCs is shown as a thin solid line.}
  \label{fig:DIS-Kfactor}
\end{figure}

\begin{figure}[t]
  \centering
  \includegraphics[width=0.41\textwidth,page=1]{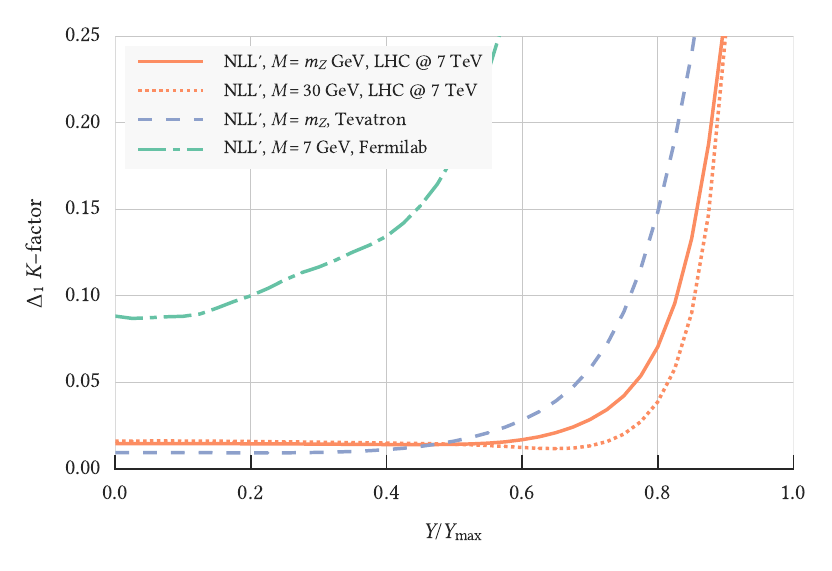}\qquad \qquad
  \includegraphics[width=0.41\textwidth,page=2]{figures/DY_thesis.pdf}
  \caption{Same as fig.~\ref{fig:DIS-Kfactor} for the neutral-current
    DY rapidity distribution, for different experiments and values of the lepton invariant mass, which allows one to probe different kinematic regimes: \mccorrect{close to threshold (Fermilab's fixed-target Drell-Yan experiments at $M = 7$ GeV), an intermediate region (Tevatron and LHC at the $Z$ pole) and away from threshold (low mass DY at LHC).} }
  \label{fig:DY-Kfactor_Rap}
\end{figure}

The $\Delta K$-factors for the neutral current DIS structure function $F_2 (x,Q)$ as a function of $x$ for two hard scales are shown in fig.~\ref{fig:DIS-Kfactor}. 
In the left plot, we show $\Delta_1 K_{\text{NLL$^\prime$}}$, namely NLL$^\prime$ matched to NLO, whereas in the right plot we show $\Delta_2 K_{\text{NNLL$^\prime$}}$ (NNLL$^\prime$ matched to NNLO).
The net effect of the resummation is an enhancement at large $x$, whereas the effect at small-$x$ is very small, as expected.
There is also a dip in the intermediate region of $x$, which is also present in fixed-order calculations~\cite{Vermaseren:2005qc}.
We also show the effect of Target Mass Corrections (TMCs) at next-to-leading twist, which are included in the resummation according to the same prescription used in the NNPDF fitting code~\cite{Ball:2008by}, where the Mellin transform of the partonic coefficient functions is multiplied by an $N$-dependent factor.
Their effect is non-negligible only at small scales, where they partially reduce the effect of resummation.

In fig.~\ref{fig:DY-Kfactor_Rap} we show the $\Delta K$ factors for the lepton pair rapidity distribution in DY processes, as a function of the ratio between the rapidity $Y$ and the maximum rapidity $Y_{\text{max}} = \frac{1}{2} \ln (s/M^2) $ allowed by kinematics.
In particular, we show the effect of resummation on experiments which probe different kinematic regimes.
We observe that the effect of threshold resummation is relevant also at central rapidities for fixed-target kinematics at low invariant masses.

Finally, in table~\ref{tab:top} we collect the $K$-factors for $t \bar t$ production at the LHC at 7 and 8 TeV, calculated as the ratio between the (N)NLO+(N)NLL$^\prime$ cross section and the (N)NLO cross section, using the same (N)NLO PDFs to compute the numerator and denominator.
We observe that the effect of resummation is almost $10\%$ at NLO+NLL$^\prime$ and is comparable to the NNLO correction.

\begin{table}[t]
  \centering
  \footnotesize
  \begin{tabular}{ccc}
    & LHC $7$ TeV &  LHC $8$ TeV \\
    \midrule
    $\sigma_{\text{NLO+NLL$^\prime$}}/\sigma_{\text{NLO}}$ & 1.086 & 1.081 \\
    $\sigma_{\text{NNLO+NNLL$^\prime$}}/\sigma_{\text{NNLO}}$ & 1.031 & 1.029 \\
    $\sigma_{\text{NNLO}}/\sigma_{\text{NLO}}$ & 1.123 & 1.122
  \end{tabular}
  \caption{$K$-factors for $t\bar t$ production at LHC at $7$ and $8$~TeV.}
  \label{tab:top}
\end{table}

\subsection{Fit settings}\label{sec:large-settings}

Le us now discuss the settings used to produce the resummed fit.
These are constructed as a variant of the NNPDF3.0 global fit and share the same fitting methodology and input parameters.
As stated above, the experimental dataset used in the resummed fits is a subset of that of the NNPDF3.0 paper.
We first review the experimental dataset and then we explain the procedure used to construct the resummed $K$-factors to include threshold resummation in the PDF fit.

\paragraph{\itshape\mdseries Experimental data.}

The list of datasets used in this analysis and in NNPDF3.0 is presented in table~\ref{tab:completedataset}.
This analysis contains all the neutral and charged current DIS data, neutral current DY production and top quark production data included in NNPDF3.0, whereas it does not contain jet data and charged-current DY datasets for the reasons discussed above.
By comparing the datasets included in the two PDF analyses, one can see that the analysis based on the reduced dataset loses experimental constraints on the medium and large-$x$ gluon (due to the missing jet data) and on the quark flavour separation (due to the missing $W$ data). 
Nevertheless, as we will discuss below, the loss of accuracy is not dramatic, since the resummed fit includes more than 3000 data points.

\begin{table}[t]
\footnotesize
\begin{centering}
\begin{tabular}{ccccc}
    {Experiment} & Observable & {Ref.} &
NNPDF3.0 global  & NNPDF3.0 DIS+DY+top\\
&&& (N)NLO & (N)NLO [+(N)NLL$^\prime$]
    \tabularnewline 
\midrule
NMC & $\sigma^{\text{NC}}_{\text{DIS}},F^d_2/F_2^p$ & \cite{Arneodo:1996kd,Arneodo:1996qe}&  Yes  &  Yes\\[0.06cm]
BCDMS & $F^d_2,F_2^p$ & \cite{Benvenuti:1989rh,Benvenuti:1989fm}&  Yes  &  Yes \\[0.06cm]
  SLAC & $F^d_2,F_2^p$  & \cite{Whitlow:1991uw}  & Yes & Yes \\[0.06cm]
  CHORUS & $\sigma^{\text{CC}}_{\nu N}$ &  \cite{Onengut:2005kv}       & Yes & Yes \\[0.06cm]
  NuTeV &  $\sigma^{\text{CC},charm}_{\nu N}$ & \cite{Goncharov:2001qe}  & Yes & Yes \\[0.20cm]    

  HERA-I & $\sigma^{\text{NC}}_{\text{DIS}},\sigma^{\text{CC}}_{\text{DIS}}$ & \cite{Aaron:2009aa} & Yes & Yes \\[0.06cm]
ZEUS HERA-II & $\sigma^{\text{NC}}_{\text{DIS}},\sigma^{\text{CC}}_{\text{DIS}}$ & \cite{Chekanov:2009gm,Chekanov:2008aa,Abramowicz:2012bx,Collaboration:2010xc}
& Yes & Yes \\[0.06cm]
H1 HERA-II & $\sigma^{\text{NC}}_{\text{DIS}},\sigma^{\text{CC}}_{\text{DIS}}$ & \cite{Aaron:2012qi,Collaboration:2010ry}  & Yes & Yes \\[0.06cm]
HERA charm & $\sigma^{\text{NC},charm}_{\text{DIS}}$ &  \cite{Abramowicz:1900rp}& Yes & Yes \\[0.20cm]

  DY E866 & $\sigma^{\text{NC}}_{\text{DY},p},\sigma^{\text{NC}}_{\text{DY},d}/\sigma^{\text{NC}}_{\text{DY},p}$  & \cite{Towell:2001nh,Webb:2003ps,Webb:2003bj}& Yes & Yes \\[0.06cm]
  DY E605 &  $\sigma^{\text{NC}}_{\text{DY},p}$ &  \cite{Moreno:1990sf} & Yes & Yes \\[0.20cm]

 CDF $Z$ rap & $\sigma^{\text{NC}}_{\text{DY},p}$ & \cite{Aaltonen:2010zza} & Yes & Yes \\[0.06cm]
 CDF Run-II $k_t$ jets & $\sigma_{\text{jet}}$ & \cite{Abulencia:2007ez}& Yes & No \\[0.06cm]
 D0 $Z$ rap & $\sigma^{\text{NC}}_{\text{DY},p}$ &  \cite{Abazov:2007jy} & Yes & Yes \\[0.20cm]

ATLAS $Z$ 2010 & $\sigma^{\text{NC}}_{\text{DY},p}$  & \cite{Aad:2011dm}  & Yes & Yes \\[0.06cm]
 ATLAS $W$ 2010 & $\sigma^{\text{CC}}_{\text{DY},p}$ & \cite{Aad:2011dm}  & Yes & No \\[0.06cm]
    ATLAS 7 TeV jets 2010 & $\sigma_{\text{jet}}$&  \cite{Aad:2011fc} & Yes & No \\[0.06cm]
    ATLAS 2.76 TeV jets  &  $\sigma_{\text{jet}}$ &\cite{Aad:2013lpa} & Yes & No \\[0.06cm]
       ATLAS high-mass DY  & $\sigma^{\text{NC}}_{\text{DY},p}$  & \cite{Aad:2013iua} & Yes & Yes \\[0.06cm]
       ATLAS $W$ $p_T$  &  $\sigma^{\text{CC}}_{\text{DY},p}$  & \cite{Aad:2011fp} & Yes & No \\[0.20cm]

 CMS $W$ electron asy &  $\sigma^{\text{CC}}_{\text{DY},p}$& \cite{Chatrchyan:2012xt} & Yes & No \\[0.06cm]
   CMS $W$ muon asy  & $\sigma^{\text{CC}}_{\text{DY},p}$  & \cite{Chatrchyan:2013mza}& Yes & No \\[0.06cm]
   CMS jets 2011     &   $\sigma_{\text{DIS}}$ & \cite{Chatrchyan:2012bja}& Yes & No \\[0.06cm]
   CMS $W+c$ total  & $\sigma^{\text{NC},charm}_{\text{jet},p}$  & \cite{Chatrchyan:2013uja}& Yes & No \\[0.06cm]
   CMS 2D DY 2011   &  $\sigma^{\text{DY}}_{\text{DY},p}$ & \cite{Chatrchyan:2013tia}  & Yes & Yes \\[0.20cm]

    LHCb $W$ rapidity  & $\sigma^{\text{NC}}_{\text{DY},p}$  &\cite{Aaij:2012vn} & Yes & No \\[0.06cm]
      LHCb $Z$ rapidity & $\sigma^{\text{NC}}_{\text{DY},p}$  & \cite{Aaij:2012mda} &  Yes & Yes \\[0.20cm]

ATLAS CMS top prod &  $\sigma(t\bar{t})$  &  \cite{ATLAS:2012aa,ATLAS:2011xha,TheATLAScollaboration:2013dja,Chatrchyan:2013faa,Chatrchyan:2012bra,Chatrchyan:2012ria}&
Yes & Yes \\[0.06cm]
\end{tabular}
\par\end{centering}
\caption{List of all the experiments used in the NNPDF3.0 global analysis and whether or not
  they are now included in the present analysis.
  For each dataset we also provide the type
  of cross section that has been measured and 
  the corresponding
  publication reference(s).
}
\label{tab:completedataset}
\end{table}

We use a similar set of kinematic cuts as those applied in the NNPDF3.0 analysis.
For DIS data, the minimum value of $Q^2$ is set to $3.5$ GeV$^2$; to reduce the dependence on higher twists at large $x$, a cut $W^2 \geq 12.5 $ GeV$^2$ is applied.
We performed a study of the stability for fixed-target DY predictions, and we found that they become unstable if the data get too close to the production threshold, either because the lepton pair mass $M_{ll}$ becomes too large, or because $Y$ is too close to $Y_{\text{max}}$.
Therefore, we changed the kinematic cuts with respect to those of NNPDF3.0; we take $ \tau\le 0.08 $ and $|Y|/Y_{\text{max}}\le 0.663$.
For the collider DY data, the cuts are the same as in NNPDF3.0.


\paragraph{\itshape\mdseries Calculation of the resummed $K$-factors.}

As we discussed in sect.~\ref{sec:threshold}, the effects of threshold resummation in the $\overline{\text{MS}}$ scheme are encoded in the partonic cross sections; therefore, the theoretical settings in the resummed fits can be the same as those of the NNPDF3.0 analysis, apart from the modification of the hard-scattering cross sections.

In the NNPDF3.0 analysis the hadronic observables are calculated using fast NLO calculations~\cite{Carli:2010rw,Wobisch:2011ij}, which are supplemented by NNLO/NLO $K$-factors at NNLO when required.
These $K$-factors are computed as ratios of the NNLO over the NLO calculations, using a common PDF luminosity computed using NNLO PDFs.
To include the effect of the resummation in the hard cross sections here we follow the same procedure, by including the effect of resummation supplementing the fixed-order computation with a resummed $K$-factor.

As we discussed in sect.~\ref{sec:large-implementation}, the resummed contributions are encoded in the form of $\Delta K$-factors. 
For DIS processes, we incorporate them into $K$-factors using
\begin{equation}
\label{eq:cfact1}
K^{\text{N$^k$LO+N$^k$LL$^\prime$}}_{\text{DIS}} \equiv \frac{\sigma^{\text{N$^k$LO+N$^k$LL$^\prime$}}}{\sigma^{\text{N$^k$LO}}}
    = 1+ \Delta_{k} K_{\text{N$^k$LL$^\prime$}} \times \frac{\sigma^{\text{LO}}}{\sigma^{\text{N$^k$LO}}},
\end{equation}
with $k=1,2$ for NLO+NLL$^\prime$ and NNLO+NNLL$^\prime$ respectively. 
Here the NNLO calculation is implemented exactly in the NNPDF code.
For hadronic processes we use a similar expression,
\begin{align}
\label{eq:cfact3}
K^{\text{NLO+NLL$^\prime$}}_{\text{hadr}} &\equiv \frac{\sigma^{\text{NLO+NLL$^\prime$}}}{\sigma^{\text{NLO}}}
= 1+ \Delta_{1} K_{\text{NLL$^\prime$}} \times \frac{\sigma^{\text{LO}}}{\sigma^{\text{NLO}}}\, , \\
K^{\text{NNLO+NNLL$^\prime$}}_{\text{hadr}} &\equiv \frac{\sigma^{\text{NNLO+NNLL$^\prime$}}}{\sigma^{\text{NLO}}}
= K^{\text{NNLO}} + \Delta_{2} K_{\text{NNLL$^\prime$}} \times \frac{\sigma^{\text{LO}}}{\sigma^{\text{NLO}}} \, ,
\label{eq:cfact2}
\end{align}
where $K^{\text{NNLO}}=\sigma^{\text{NNLO}}/\sigma^{\text{NLO}}$ is the NNLO/NLO $K$-factor.
All contributions are meant to be computed with the same N$^k$LO+N$^k$LL$^\prime$ PDF set.
We computed the LO cross sections combining LO coefficient functions with PDFs with NLO and NNLO evolution using \texttt{APFEL}~\cite{Bertone:2013vaa}.

Since the $K$-factors are computed externally using a fixed set of PDFs, the contributions in the above equations should in principle be recomputed for several iterations of the fit until convergence has been reached.
In practice, however, the computation of $K^{\text{NNLO}}$ is time-consuming, so for that contribution we use the same fixed value used in the NNPDF3.0 fits.
The only exception are the $K$-factors for the fixed-target DY experiments, which we have recomputed using \texttt{Vrap}~\cite{Anastasiou:2003ds} and NNPDF3.0 as input PDF set.
We found that two iterations of the fit are enough to ensure a satisfactory convergence of the resummed $K$-factors, namely the $K$-factors are essentially identical if we use resummed PDFs from the last or the penultimate iteration of the fit. 

Let us now illustrate the effect of threshold resummation for some of the datasets used in the fit by plotting the resummed $K$-factors for some representative experimental dataset, using the same kinematics as for the data-points which we use in the fit.
To compute the $K$-factors we have consistently used the NNPDF3.0 DIS+DY+top NLO+NLL$^\prime$ and NNLO+NNLL$^\prime$ PDF sets which we will present in the next section, with $\alpha_s(m_Z^2)=0.118$, in both the fixed-order and resummed cross sections.
For hadronic experiments, we factor out $K^{\text{NNLO}}$ in eq.~\eqref{eq:cfact2} to isolate the effect of the resummation.

\begin{figure}[t]
\begin{center}
\includegraphics[width=0.41\textwidth]{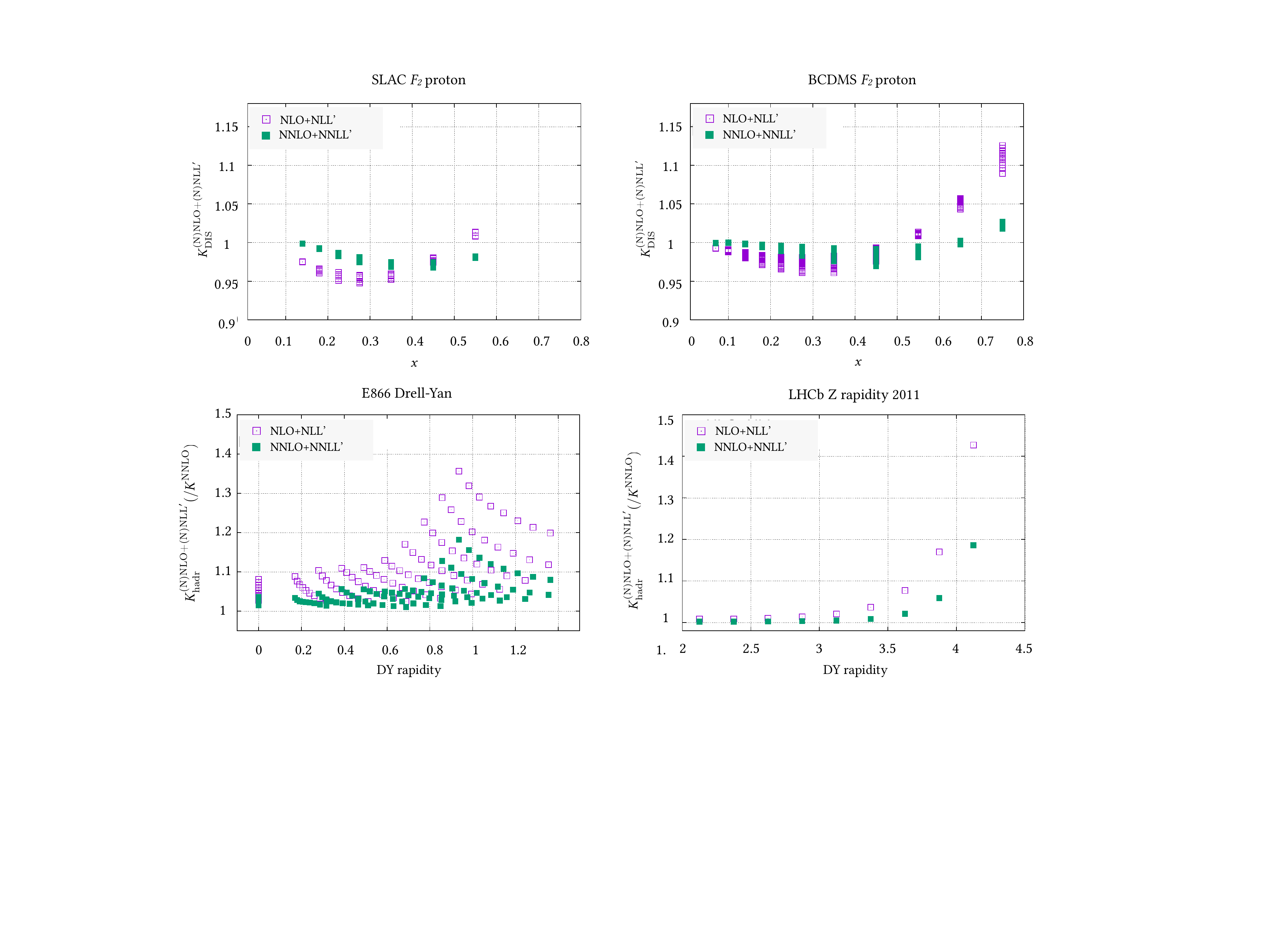}\qquad \qquad
\includegraphics[width=0.41\textwidth]{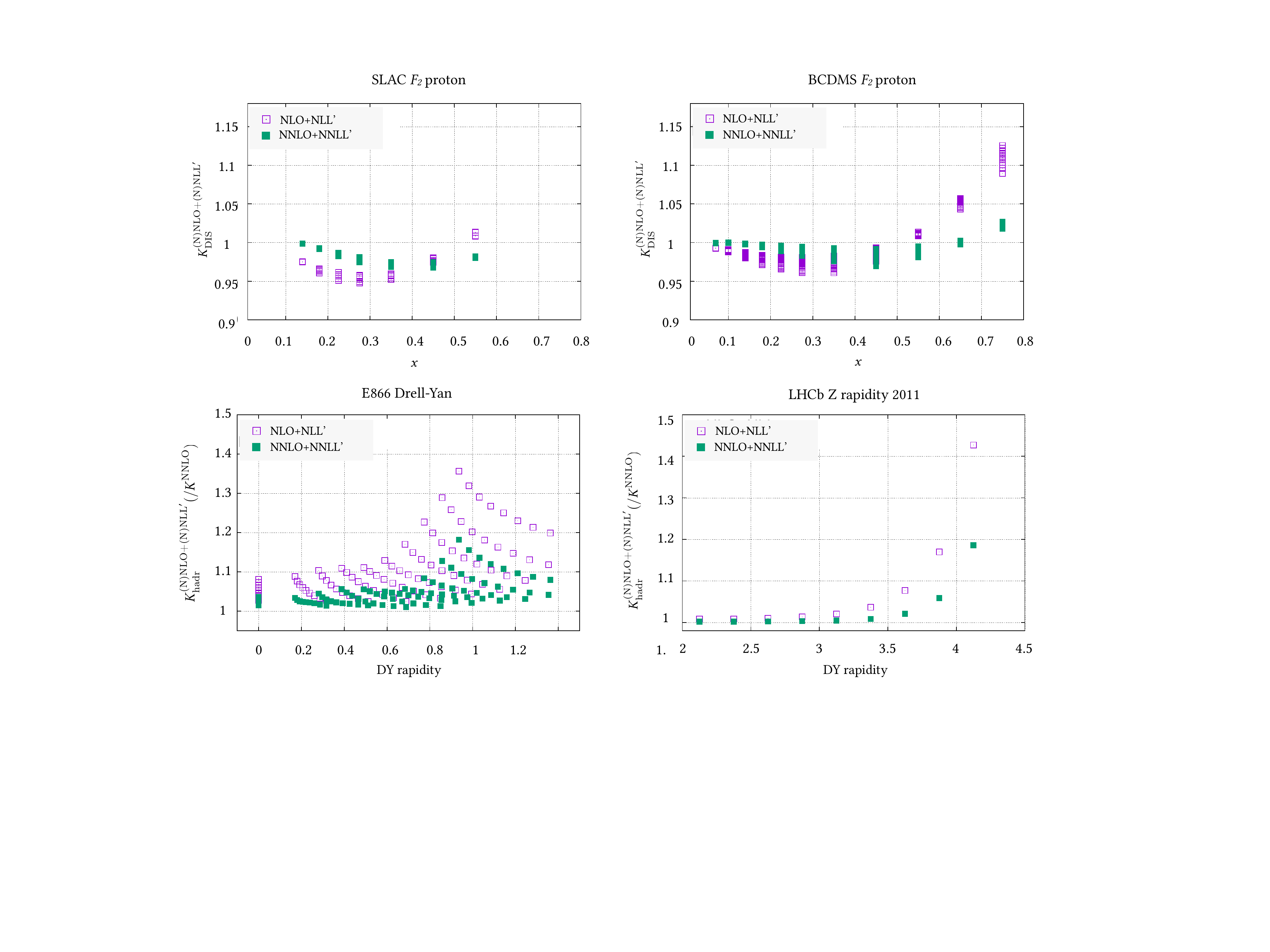}
\caption{The resummed $K$-factors for DIS, eq.~\eqref{eq:cfact1}, for a representative subset
  of the experiments included in the resummed fit.
  We show both the results corresponding to NLL$^\prime$ and to NNLL$^\prime$ resummation.
  }
   \label{fig:nnpdf_cfact_1nsoft}
  \end{center}
\end{figure}

We show the results for the DIS case in fig.~\ref{fig:nnpdf_cfact_1nsoft} for two representative datasets.
For each experiment we show both the NLL$^\prime$ and the NNLL$^\prime$ $K$-factors.
We use the DIS kinematics $(x,Q^2,y)$ used in the experimental data, therefore for each value of $x$ there are measurements at different values of $Q^2$ and $y$. 
Here we choose not to show results for the HERA datasets, since resummation turns out to be negligible as the data are at large scales and at small values of $x$.
We observe that the effect of resummation is most relevant for BCDMS data, where NLL$^\prime$ resummation can reach $15\%$ at the largest value of $x$; the effect is reduced to a few percent in the NNLL$^\prime$ calculation. 
The effect of resummation is milder for the other datasets; here we show the SLAC data, where the effects of resummation is always below $5\%$.
The situation is similar for the other datasets included.

\begin{figure}[t]
  \begin{center}
    \includegraphics[width=0.41\textwidth]{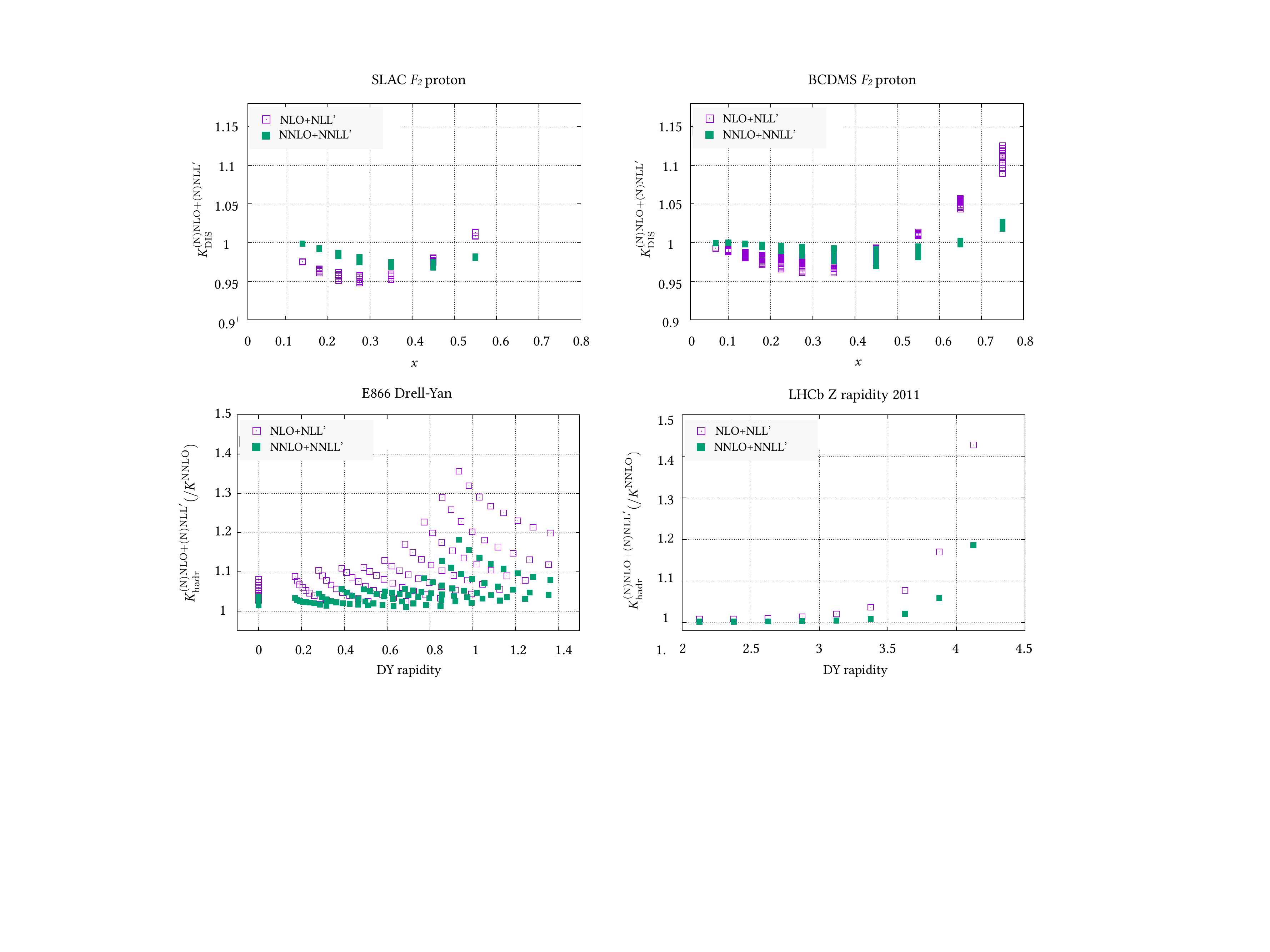}\qquad \qquad
    \includegraphics[width=0.41\textwidth]{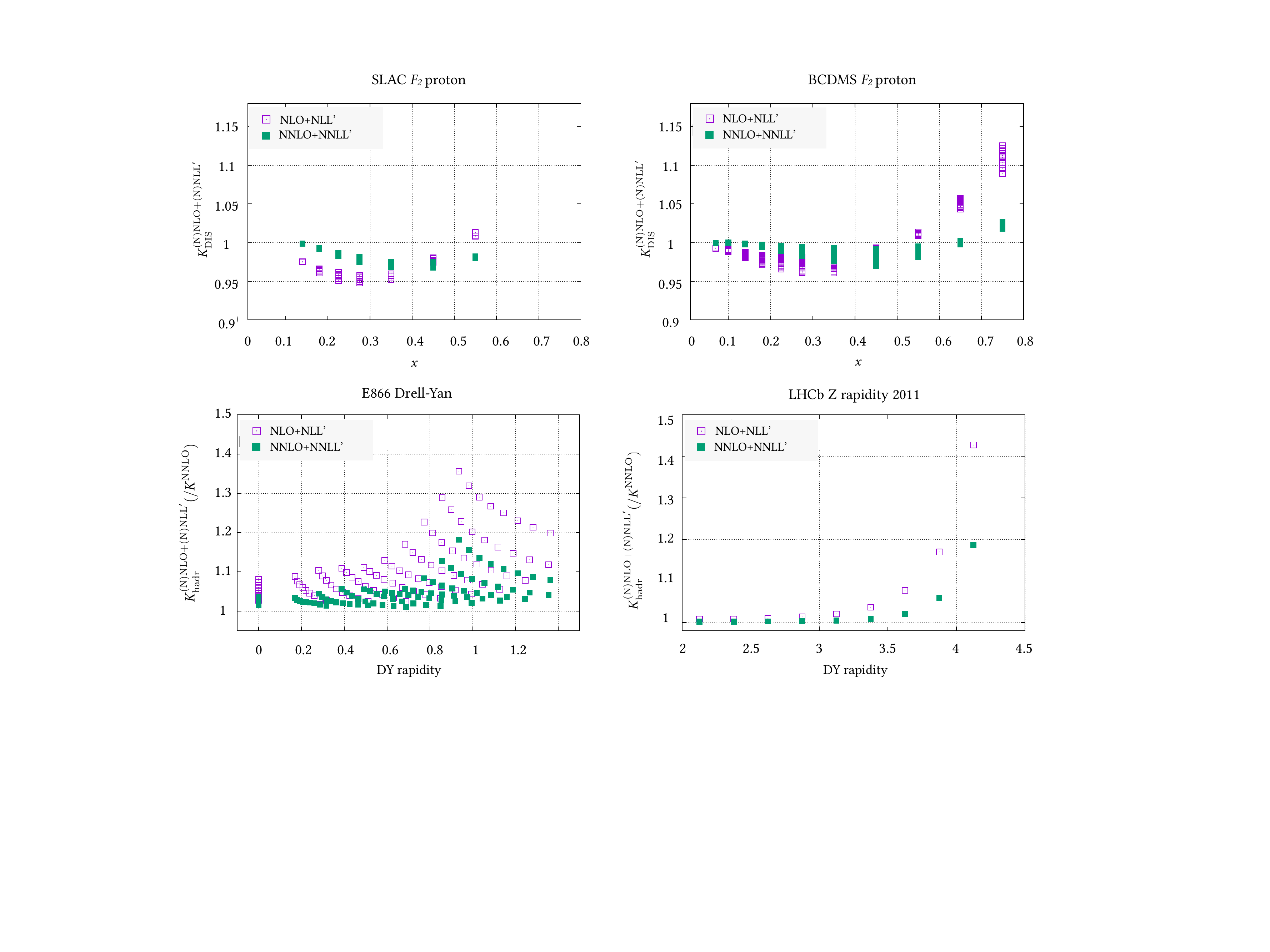}
\caption{Same as Fig.~\ref{fig:nnpdf_cfact_1nsoft} for selected Drell-Yan experimental datasets included in the fit.
The resummed $K$-factors are now those defined in eqs.~\eqref{eq:cfact3} and~\eqref{eq:cfact2},
but to isolate the effect of resummation from that of the fixed-order
NNLO corrections, in the NNLL$^\prime$ case we divide eq.~\eqref{eq:cfact2} by $K^{\text{NNLO}}$.
In the left plot data points differ by the values of the rapidity and the invariant mass of the pair,
but only the dependence on the rapidity is shown.
} 
    \label{fig:nnpdf_cfact_2nsoft}
\end{center}
\end{figure}

In fig.~\ref{fig:nnpdf_cfact_2nsoft} we show the analogous result for DY experiments.
We take one fixed target dataset, namely the DY E866 $pp$ cross section, and one collider dataset, the LHCb $Z\rightarrow \mu \mu$ rapidity distribution.
We observe that the impact of resummation is always more important at NLO+NLL$^\prime$ than at NNLO+NNLL$^\prime$ and grows with the rapidity of the lepton pair, as the kinematic threshold is approached.
The effect at LHCb can be as large as $50\%$ at NLL$^\prime$, whereas it is still about $20\%$ at NNLL$^\prime$ for the points at larger rapidity. 
In the case of the E866 dataset, the effect of resummation is as large as $35\%$ at NLL$^\prime$ and $20\%$ at NNLL$^\prime$.

\subsection{Parton distributions with large-$x$ resummation}\label{sec:large-PDFs}

We are now ready to present the results of the NNPDF3.0 fits with threshold resummation.
We start by comparing our baseline fit with the NNPDF3.0 global fits to quantify the impact of the missing datasets on the PDF accuracy. 
We then quantify the impact of threshold resummation on the fit quality and we compare the resummed sets with our baseline, both at the level of PDF and of $\chi^2$. 

\paragraph{\itshape\mdseries Baseline fits.}

As discussed above, our baseline DIS+DY+top set differs from the NNPDF3.0 global set since it is based on a reduced dataset.
Nevertheless, it is based on the exact same methodology and settings and we thus expect the two sets to be consistent, with the DIS+DY+top fit affected by larger PDF uncertainties.
We focus only on the NLO case, since the impact of the dataset is approximately independent of the perturbative order.
%

\begin{figure}[t]
  \begin{center}
    \includegraphics[width=0.8\textwidth]{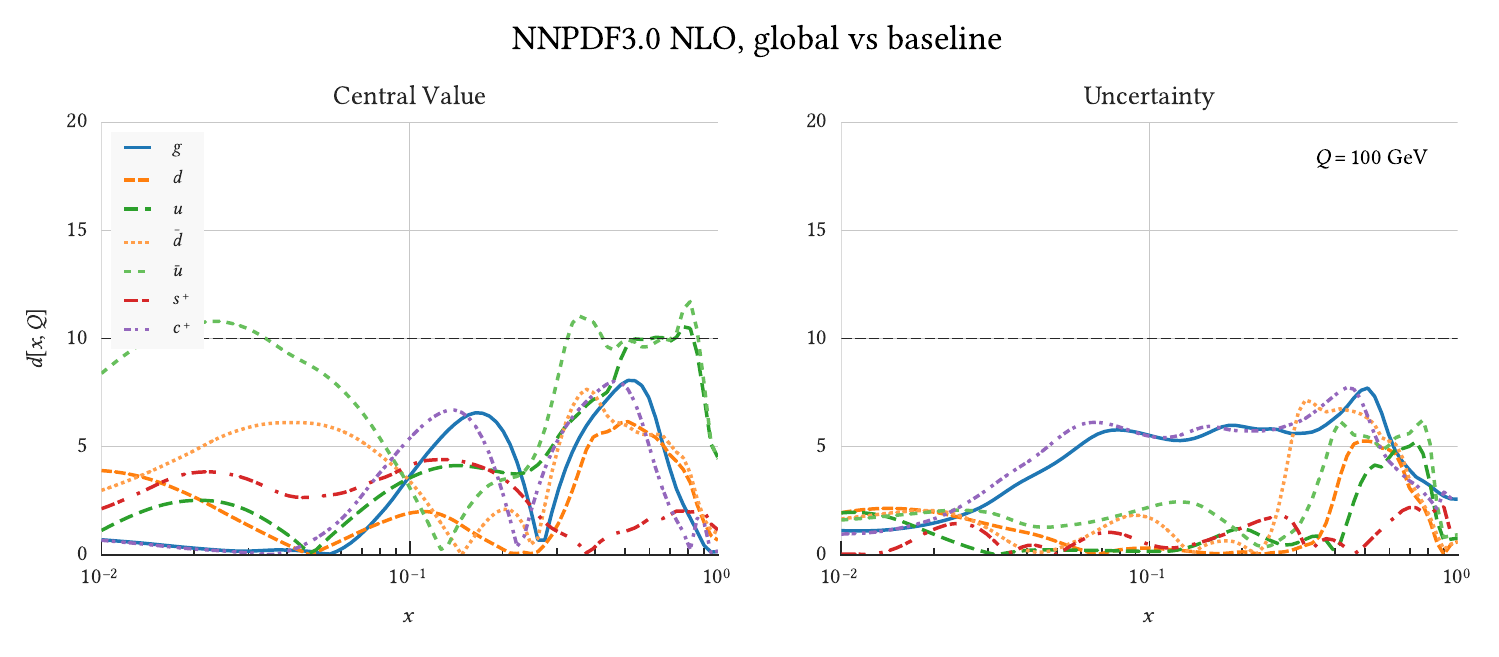}
\caption{The statistical distances between the central values (left) and the PDF uncertainties (right plot) of the NNPDF3.0 NLO and the DIS+DY+top fits at $Q = 100$ GeV in the flavour basis.} 
    \label{fig:baseline-vs-global-nlo-distances}
\end{center}
\end{figure}

To quantify the differences between the two PDF sets it is convenient to use a distance estimator (for a definition, see e.g ref.~\cite{Ball:2010de,Ball:2014uwa}) which allows us to represent in a concise way how two PDF fits differ among themselves, both at the level of central values and of PDF uncertainties. 
We show these distances between the NNPDF3.0 NLO and the DIS+DY+top fits at $Q = 100$~GeV in fig.~\ref{fig:baseline-vs-global-nlo-distances}.
Since both the fits have $N_{\text{rep}}=100$ replicas, a distance of $d\sim 10$ corresponds to a variation of one-sigma of the central values or the PDF uncertainties in units of the corresponding standard deviation.
We observe that there is a reasonable agreement for most of the PDF flavours, with somewhat larger differences in the central values for $u$ and $\bar u$, whereas the uncertainties which are effected the most are those of the gluon and of the charm quark (which is generated perturbatively in the NNPDF3.0 analysis).
Overall, the PDFs in the DIS+DY+top fit are fairly similar to the ones in the global fit; it follows that the calculations performed with the reduced dataset should be comparable to those performed with the global PDFs.

\begin{figure}[t]
  \begin{center}
    \includegraphics[width=0.41\textwidth,page=2]{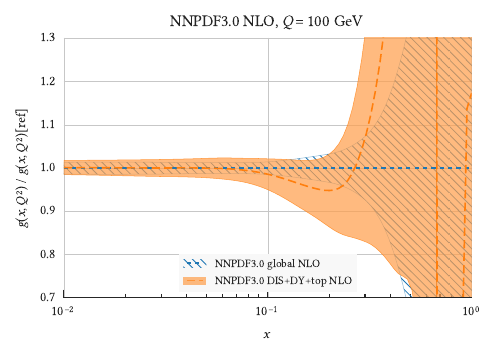}\qquad \qquad
    \includegraphics[width=0.41\textwidth,page=4]{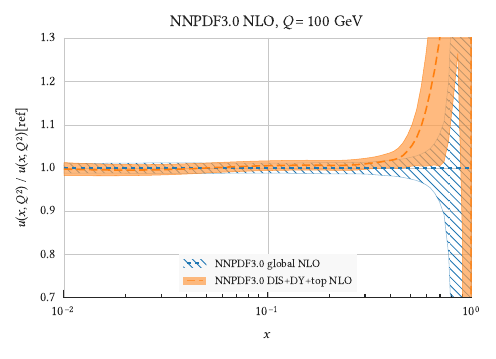}	
\caption{Comparison of the fixed-order NNPDF3.0 NLO fits based on global and DIS+DY+top dataset, for $ \as(m^2_Z) = 0.118$, at a typical LHC scale of $Q= 100$ GeV. Results are normalized to the central prediction of the NNPDF3.0 NLO global fit. Left: gluon; right: up quark.
} 
    \label{fig:baseline-vs-global-nlo-pdfs}
\end{center}
\end{figure}

We compare the NNPDF3.0 and the DIS+DY+top PDF at the level of PDFs in fig~\ref{fig:baseline-vs-global-nlo-pdfs}.
We show results for the gluon and the up quark, normalized to the central prediction of the global fit at a scale $Q=100$ GeV. 
We observe a substantial increase in the uncertainty associated to the gluon PDF in the DIS+DY+top case; the situation is only partly mitigated by the constraint provided by the total $t \bar t$ pair production dataset, which offers a handle on the large-$x$ gluon PDF.
In the up quark case, the PDF uncertainties are similar in the global and in the DIS+DY+top case, though the bands only partially overlap at very large values of $x$.   

This concludes our analysis of the impact of the reduced dataset on the baseline fits.
We can now study the effect of resummation by comparing our baseline fits with fits based on the same datasets but where the fixed-order calculations are supplemented with resummation as described above.

\paragraph{\itshape\mdseries DIS+DY+top resummed PDFs.}

To quantify the impact of threshold resummation, we start by comparing the fit quality of our baseline fit with that of the threshold resummed fit. 
In table~\ref{tab:chi2disdyfit} we present the $\chi^2/N_{\text{dat}}$ of the NLO+NLL$^\prime$ and of the NNLO+NNLL$^\prime$ fits alongside those of the corresponding fixed-order fits.

In the case of the NLO+NLL$^\prime$ fit, the fit quality of most of the hadron collider experiments improves with respect to the fixed-order NLO fit; this is particularly marked in the case of the LHCb $Z$ rapidity dataset and top quark pair production.
We also observe an improvement in the fit quality of SLAC data, as well as a marginal improvement in the quality of the CHORUS neutrino structure functions.
For the other dataset the fit quality is very similar, with the exception of the fixed target DY data, where threshold resummation makes the $\chi^2$ worse.
\mccorrect{To further study this issue, we have performed a variant of the fit based on a reduced dataset, which includes HERA data only and fixed-target DY data.
It appears that whereas in the global case the fit with resummation struggles to give a simultaneously good description of the DY and other datasets, in the reduced fit the description of the DY data is fine and $\chi^2\sim 1$ for the DY data also at the resummed level.}
The overall fit quality is however very similar, with the resummed $\chi^2$ being slightly worse, since the improvement in the description of some datasets is compensated by the deterioration of others.

The effect of resummation is more moderate for the NNLO+NNLL$^\prime$ fits.
For many datasets, the fit quality is very similar, with some improvement in the description of the ATLAS high-mass DY dataset and of the LHCb $Z$ rapidity dataset.
Again, we observe a marked deterioration of the fixed-target DY data, especially in the case of the E866 dataset.
Overall, however, the fit quality is essentially identical in the two cases.
 
\begin{table}[t]
\footnotesize
\begin{centering}
\begin{tabular}{c C{1.9cm} C{1.9cm} C{1.9cm} c}
  Experiment & 
    \multicolumn{4}{c}{NNPDF3.0 DIS+DY+top} \\
    & NLO & NNLO &  NLO+NLL$^\prime$ &  NNLO+NNLL$^\prime$ \\
\midrule
    NMC            &  1.39 & 1.34 &   1.36    &  1.30     \\
      SLAC         &  1.17 & 0.91 &   1.02    &  0.92     \\ 
BCDMS              &  1.20 & 1.25 &   1.23    &  1.28     \\ 
  CHORUS           &  1.13 & 1.11 &   1.10    &  1.09     \\ 
  NuTeV            &  0.52 & 0.52 &   0.54    &  0.44     \\[0.20cm] 

  HERA-I           &  1.05 & 1.06 &   1.06    &  1.06     \\
ZEUS HERA-II       &  1.42 & 1.46 &   1.45    &  1.48     \\
H1 HERA-II         & 1.70  & 1.79 &   1.70    &  1.78     \\
HERA charm         & 1.26  & 1.28 &   1.30    &  1.28     \\[0.20cm] 

 DY E866           & 1.08 & 1.39  &   1.68    &  1.68     \\
 DY E605           & 0.92 & 1.14  &   1.12    &  1.21     \\[0.20cm] 

 CDF $Z$ rap       & 1.21 & 1.38  &   1.10    &  1.33     \\
 D0 $Z$ rap        & 0.57 & 0.62  &   0.67    &  0.66     \\[0.20cm] 

ATLAS $Z$ 2010     & 0.98 & 1.21  &   1.02    &  1.28     \\
ATLAS high-mass DY & 1.85 & 1.27  &   1.59    &  1.21     \\[0.20cm] 

CMS 2D DY 2011     & 1.22 & 1.39  &   1.22    &  1.41      \\[0.20cm] 

LHCb $Z$ rapidity  & 0.83 & 1.30  &   0.51    &  1.25      \\[0.20cm] 

ATLAS CMS top prod & 1.23 & 0.55  &   0.61    &  0.40      \\[0.20cm] 
\midrule
Total              &  1.233  &  1.264   &  1.246  &   1.269    \\
\end{tabular}
\par\end{centering}
\caption{  The  $\chi^2$ per data point
  for all experiments
  included in the DIS+DY+top fits threshold resummed fits, at NLO and
  NNLO, compared with their resummed counterparts.}
\label{tab:chi2disdyfit}
\end{table}

\begin{figure}[t]
  \begin{center}
    \includegraphics[width=0.8\textwidth]{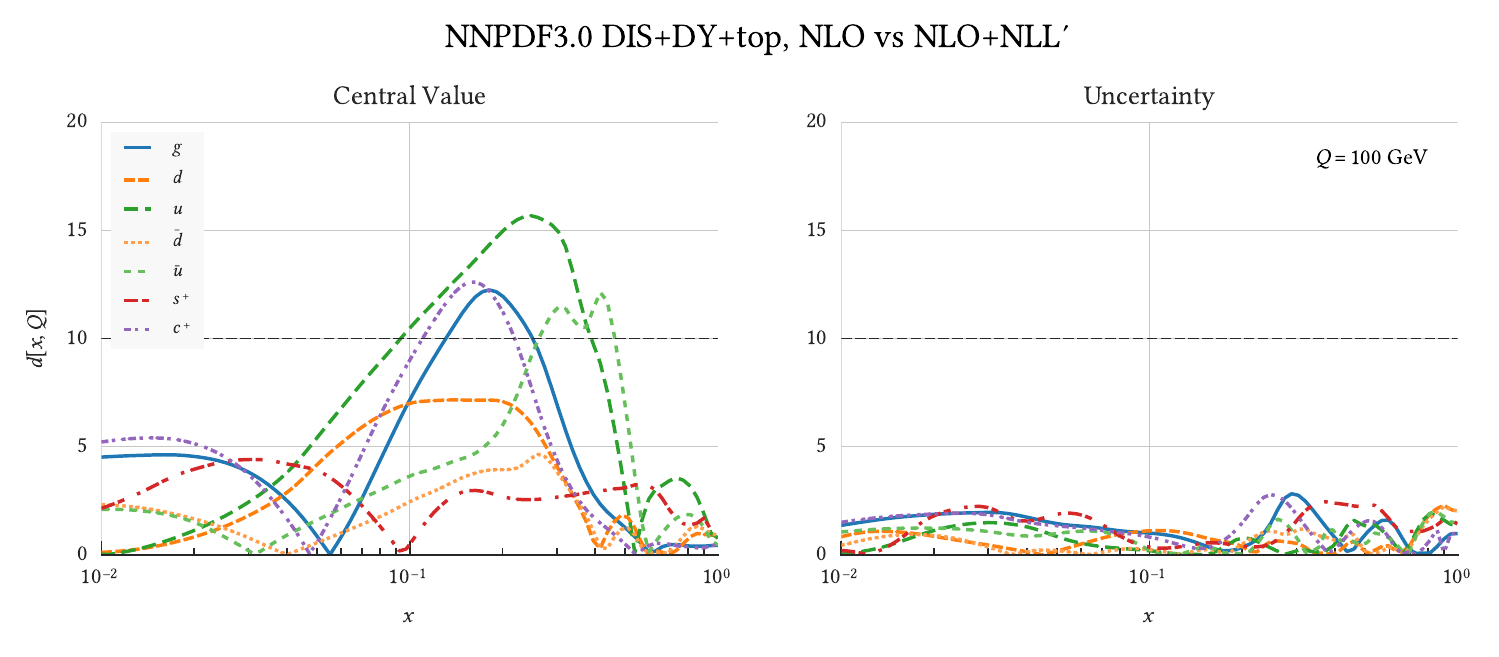}
    \includegraphics[width=0.8\textwidth]{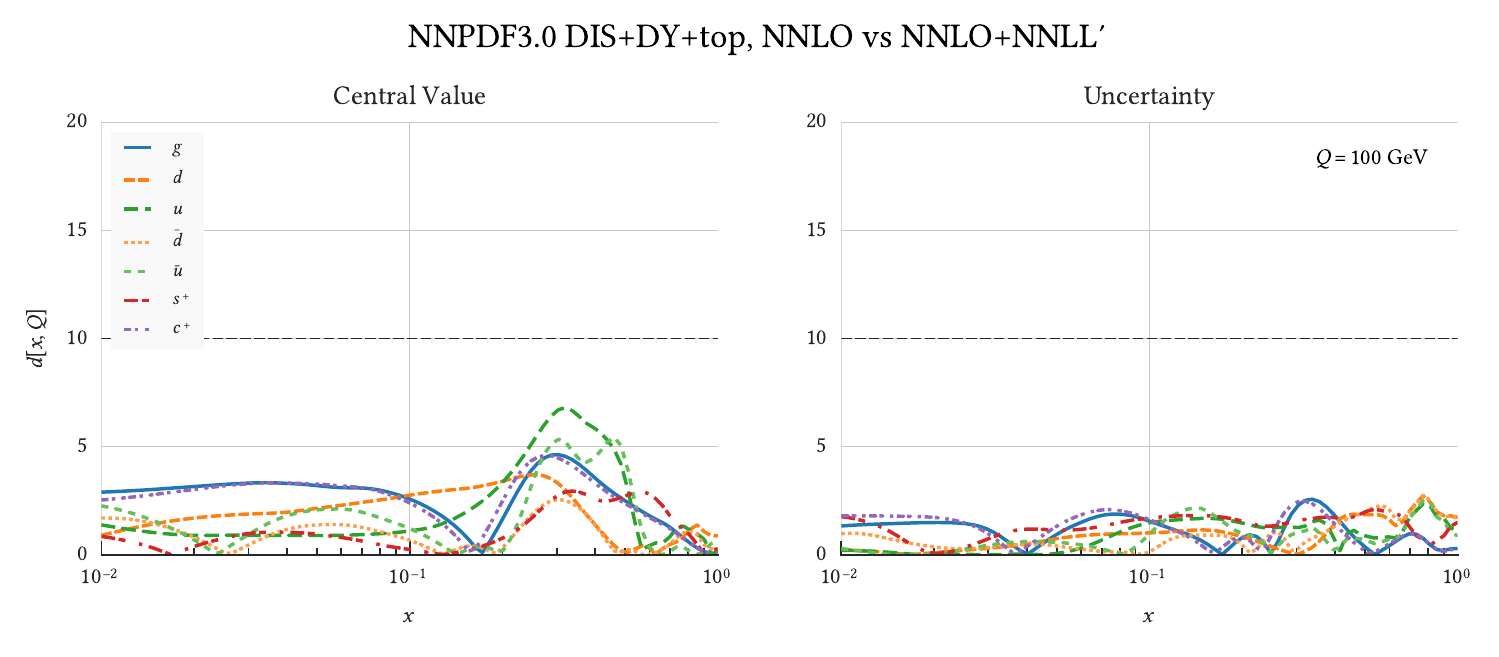}
\caption{Same as fig.~\ref{fig:baseline-vs-global-nlo-distances}, now for the NLO and NLO+NLL$^\prime$ DIS+DY+top fits (upper panel) and for the NNLO and NNLO+NNLL$^\prime$ (lower panel).} 
    \label{fig:nlo-vs-nlonll-distances}
\end{center}
\end{figure}

In fig.~\ref{fig:nlo-vs-nlonll-distances} we show the distances between the NLO and the NLO+NLL$^\prime$ fits (top panel) and the corresponding distances in the case of NNLO and NNLO+NNLL$^\prime$ (lower panel).
We see that at NLO the effect of resummation is particularly pronounced for the gluon, the up and anti-up quark, as well as for the charm quark and the down quark. 
The effect is less pronounced at NNLO, where the distances are smaller and only the up and anti-up are partially affected.
In both cases, the uncertainties are, as expected, essentially equivalent, as the PDF sets are based on the same dataset.

\begin{figure}[t]
  \begin{center}
    \includegraphics[width=0.41\textwidth,page=2]{figures/nn30Xnlonll100_ratio.pdf}\qquad \qquad
    \includegraphics[width=0.41\textwidth,page=1]{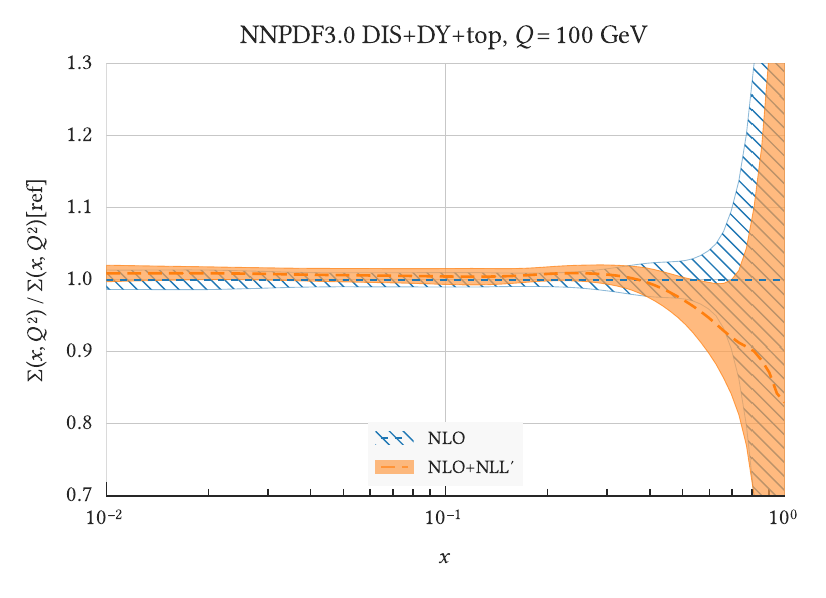}	
      \includegraphics[width=0.41\textwidth,page=7]{figures/nn30Xnlonll100_ratio.pdf}\qquad \qquad
    \includegraphics[width=0.41\textwidth,page=6]{figures/nn30Xnlonll100_ratio.pdf}	
\caption{Comparison between the NNPDF3.0 DIS+DY+top dataset NLO and NLO+NLL$^\prime$, for $ \as(m^2_Z) = 0.118$, at a typical LHC scale of $Q= 100$ GeV. 
} 
    \label{fig:nlo-vs-nll-pdfs}
\end{center}
\end{figure}

\begin{figure}[htp]
  \begin{center}
    \includegraphics[width=0.41\textwidth,page=2]{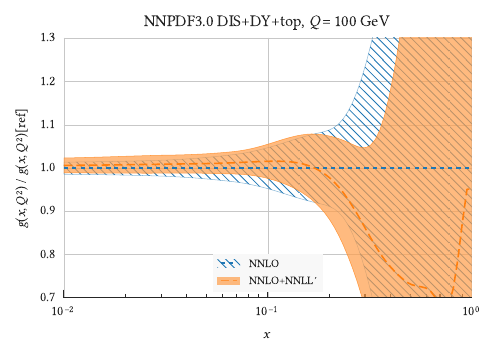}\qquad \qquad
      \includegraphics[width=0.41\textwidth,page=7]{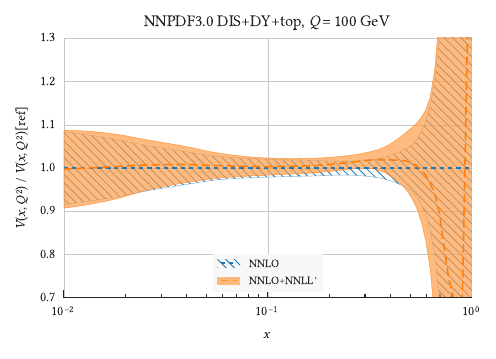}
\caption{Comparison between the NNPDF3.0 DIS+DY+top dataset NNLO and NNLO+NNLL$^\prime$, for $ \as(m^2_Z) = 0.118$, at a typical LHC scale of $Q= 100$ GeV. 
} 
    \label{fig:nnlo-vs-nnll-pdfs}
\end{center}
\end{figure}

We show the comparison of the PDFs between the NLO and NLO+NLL$^\prime$ fits in fig.~\ref{fig:nlo-vs-nll-pdfs}, as well as those in the NNLO+NNLL$^\prime$ fits in fig.~\ref{fig:nnlo-vs-nnll-pdfs}.
In the NLO case, we show the gluon PDF, the valence, the singlet and the strangeness.
We observe that the effect of resummation for the quarks is a suppression at large $x$ and an enhancement of the valence at smaller values of $x$, presumably due to a compensation for the suppression at large $x$ through the sum rule.
For the singlet, the shift of the central value is about $5\%$ at $x \sim 0.5$; the uncertainty bands of the PDFs are departing from each other, though they still overlap.
For the valence, the difference between the central values is about $5\%$ for $x$ between $0.1$ and $0.3$, and the uncertainty bands barely overlap.
As a consequence, we expect resummation to have a phenomenological impact for the calculation of quark-initiated heavy production processes in BSM scenarios, where large values of $x$ are probed.
The large-$x$ gluon is also suppressed, though the PDF uncertainty gets larger than the shift for $x \gtrsim 0.3 $.
The effect of resummation is negligible at small-$x$, as expected.

In the NNLO case, we show only the gluon and the valence, since the impact of resummation is considerably smaller.
The PDF uncertainties are very similar and they overlap in the entire range.
Significant shifts in the central values are present only at very large $x$, where the PDF uncertainties are however very large.
For instance, at $x\sim 0.3$ the central value of the resummed gluon is $\sim 20\%$ smaller than the fixed-order one, but the shift is smaller than the PDF uncertainties.

\subsection{Implications for new physics searches at the LHC}\label{sec:large-pheno}

In this section we explore some of the phenomenological implications of the threshold resummed fits at the LHC.
We start by considering parton luminosities and then we estimate the effects of threshold resummation in DY pair production.
We finally discuss implications for heavy-resonance searches at the LHC.

\paragraph{\itshape\mdseries Parton luminosities.}

To provide a first insight on the impact of threshold resummation it is useful to consider its effect on parton luminosities.
We assume the production of a hypothetical final state of mass $M_X$ such that $x=M_X^2/s$, where $\sqrt{s}=13$~TeV is the centre of mass energy of the collision at the LHC, with $\mu_F= M_X$. 
This comparison provides direct information on how the cross sections for a final state with mass $M_X$ are affected by resummation; however, consistent calculations require the inclusion of resummation effects also in the partonic matrix elements, which could compensate the impact of resummation in the PDFs.
These consistent comparisons will be discussed below.
Here we focus on the comparison between NLO and NLO+NLL$^\prime$ fits, since we have seen in the previous section that at NNLO the impact of resummation is smaller \mccorrect{and below the level of PDF uncertainties}.

\begin{figure}[t]
  \begin{center}
    \includegraphics[width=0.41\textwidth,page=1]{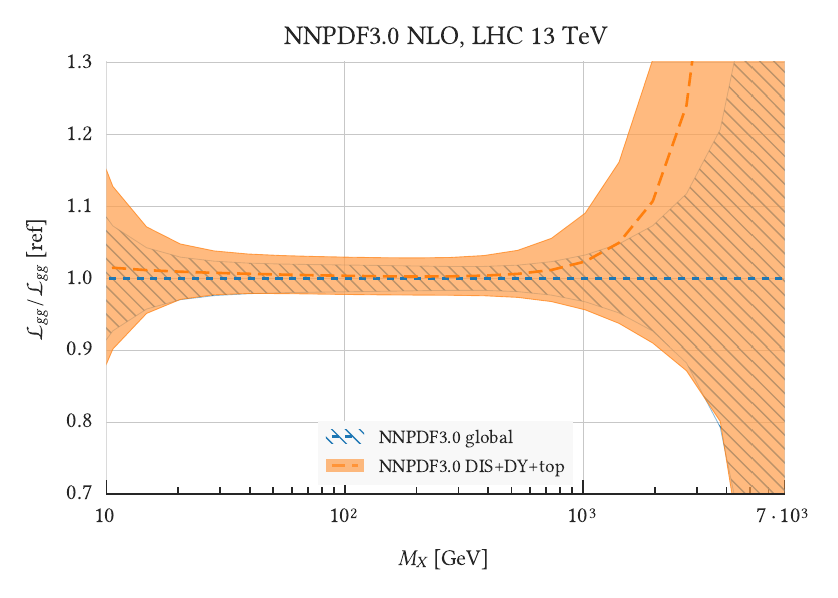}\qquad \qquad
    \includegraphics[width=0.41\textwidth,page=2]{figures/NNPDF30vsbaseline_1D_13.pdf}	
      \includegraphics[width=0.41\textwidth,page=3]{figures/NNPDF30vsbaseline_1D_13.pdf}\qquad \qquad
    \includegraphics[width=0.41\textwidth,page=4]{figures/NNPDF30vsbaseline_1D_13.pdf}	
\caption{Comparison of the NNPDF3.0 NLO partonic luminosities in the global fit and in the DIS+DY+top fit which is used as fixed-order baseline for the resummed fits.
} 
    \label{fig:baseline-vs-global-nlo-lumi}
\end{center}
\end{figure}

We start by estimating the effect on PDF luminosities comparing our baseline fit with the NNPDF3.0 global fit with $\as= 0.118$.
In fig.~\ref{fig:baseline-vs-global-nlo-lumi} we show the gluon-gluon, quark-gluon, quark-antiquark and quark-quark parton luminosities.
We see some important differences between the global fit and our baseline fits.
The missing jet data have a significant impact on the $gg$ and the $qg$ luminosities; for instance, the $gg$ luminosity increases by more than a factor of two for $M_X$ larger than $0.5$~TeV.
The effect is smaller for the $qq$ and the $q \bar q$ luminosity, where the differences are sizeable only at large $M_X$; for example, the $q \bar q$ luminosities differ by about $10\%$ at $M_X \sim 3$~TeV. 
Therefore, to assess the impact of resummation consistently one should compare the resummed fit to the baseline DIS+DY+top fits, rather than with the NNPDF3.0 global fit.

\begin{figure}[t]
  \begin{center}
    \includegraphics[width=0.41\textwidth,page=1]{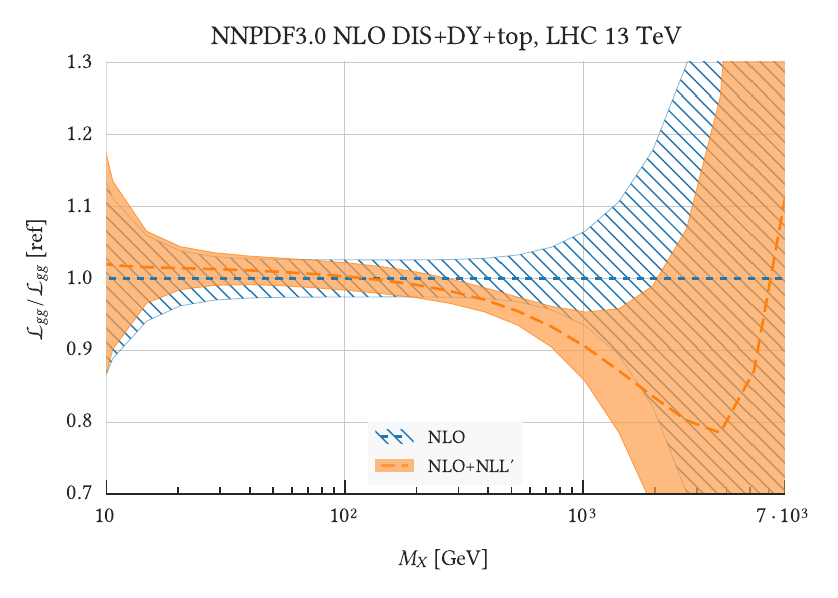}\qquad \qquad
    \includegraphics[width=0.41\textwidth,page=2]{figures/NNPDF30nlo_1D_13.pdf}	
      \includegraphics[width=0.41\textwidth,page=3]{figures/NNPDF30nlo_1D_13.pdf}\qquad \qquad
    \includegraphics[width=0.41\textwidth,page=4]{figures/NNPDF30nlo_1D_13.pdf}	
\caption{Same as fig.~\ref{fig:baseline-vs-global-nlo-lumi}, now comparing the DIS+DY+top NLO fit to the NLO+NLL$^\prime$ fit.
} 
    \label{fig:nlo-vs-nll-lumi}
\end{center}
\end{figure}

The comparison between the resummed and the fixed-order DIS+DY+top fits at NLL$^\prime$ is shown in fig.~\ref{fig:nlo-vs-nll-lumi}.
We see that in all cases the fixed-order and the resummed fits agree at a level of one sigma.
The $q q$ and the $q \bar q$ luminosities are enhanced for values of $M_X$ smaller than 1~TeV by about one sigma and they are suppressed at larger values of $M_X$.
Their behaviour follows closely those of the quark PDFs which we observed above. 
In particular, the $q\bar q$ luminosity at NLO+NLL$^\prime$ is reduced by about $15 \%$. 
In the $qq$ case, however, the suppression is smaller until very large values of invariant masses.
The $gg$ and $gq$ luminosities are also suppressed at large $M_X$, though they are still consistent with the fixed-order fit within the large PDF uncertainties.

\paragraph{\itshape\mdseries High-Mass Drell-Yan dilepton mass distributions.}

High-mass Drell-Yan is one of the most important processes at the LHC in many new physics searches, for example for $Z'$ boson production.
Therefore, it is interesting to assess the effect of consistently including threshold resummation both in the PDF and in matrix element and comparing these results with those obtained by considering resummation only in the matrix element.

\begin{figure}[t]
  \centering
  \includegraphics[width=0.41\textwidth,page=2]{figures/DY_vrap_thesis.pdf}\qquad \qquad
  \includegraphics[width=0.41\textwidth,page=1]{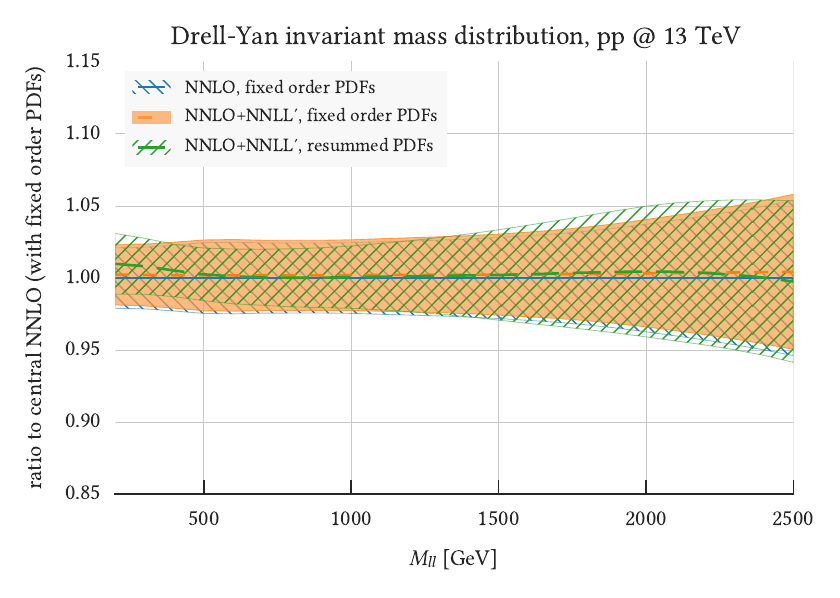}
  \caption{ Left: dilepton invariant mass distribution for
  high-mass neutral current Drell-Yan production at the LHC 13 TeV.
  Results are shown normalized to the
  central prediction of the fixed-order NLO calculation.
  Right: the same comparison at NNLO.
}
\label{fig:DYnnll}
\end{figure}

In fig.~\ref{fig:DYnnll} we show the dilepton invariant mass distribution for neutral current Drell-Yan production at the LHC 13 TeV, comparing the predictions of fixed-order and resummed calculations. 
To compute the fixed-order predictions we have used \texttt{Vrap}, and we have used \texttt{TROLL} to compute the resummation effects.
We show the predictions for the invariant mass distribution of the lepton pair, comparing the (N)NLO fixed-order with the resummed (N)NLO+(N)NLL$^\prime$ calculations, using either fixed-order or resummed PDFs.
We observe that the effect of resummation is rather moderate: even at NLO and at large invariant masses the effect of resummation is at the level of a few percent and within the PDF uncertainty.
The effect is further reduced if resummed PDFs are used; in the whole range $M_{ll}\in [1.5,2.5]$ TeV the resummed calculation and the fixed-order one differ by less than $1\%$.
The effect of resummation is completely negligible at NNLO.

\paragraph{\itshape\mdseries Supersymmetric particle production.}

Theoretical predictions for high-mass supersymmetric pair production at hadron colliders are currently done at NLO for several processes, supplemented with NLL$^\prime$ or NNLL$^\prime$ resummation (though approximate calculations at higher accuracy have been performed; NNLO$_{\text{appr}}$+NNLL$^\prime$ calculations have been presented in e.g.~\cite{Beenakker:2016lwe} and even N$^3$LO$_{\text{appr}}$+N$^3$LL$^\prime$ calculations have been made~\cite{Ahmed:2016otz}).
\mccorrect{Calculations appearing in analyses prior to this work use resummation only in the computation of the partonic matrix elements}.
This creates a potential mismatch especially at high masses --- the crucial region for supersymmetric searches --- since possible effects due to compensations are not taken into account.
The resummed PDFs we have presented, barring their limitations, offer now the possibility to compute predictions where resummation is consistently taken into account both in the matrix elements and in the luminosities.

\begin{figure}[t]
\begin{center}
\includegraphics[width=0.41\textwidth]{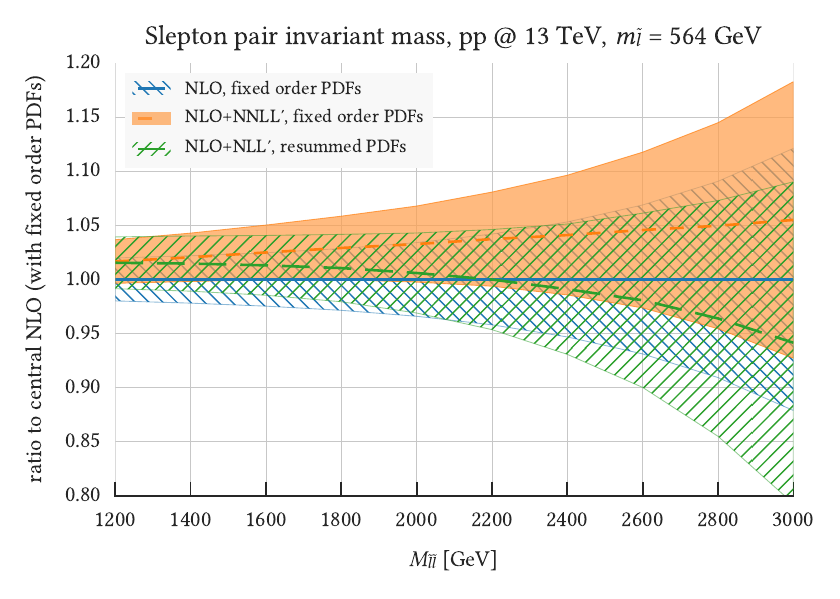}\qquad \qquad
\includegraphics[width=0.41\textwidth]{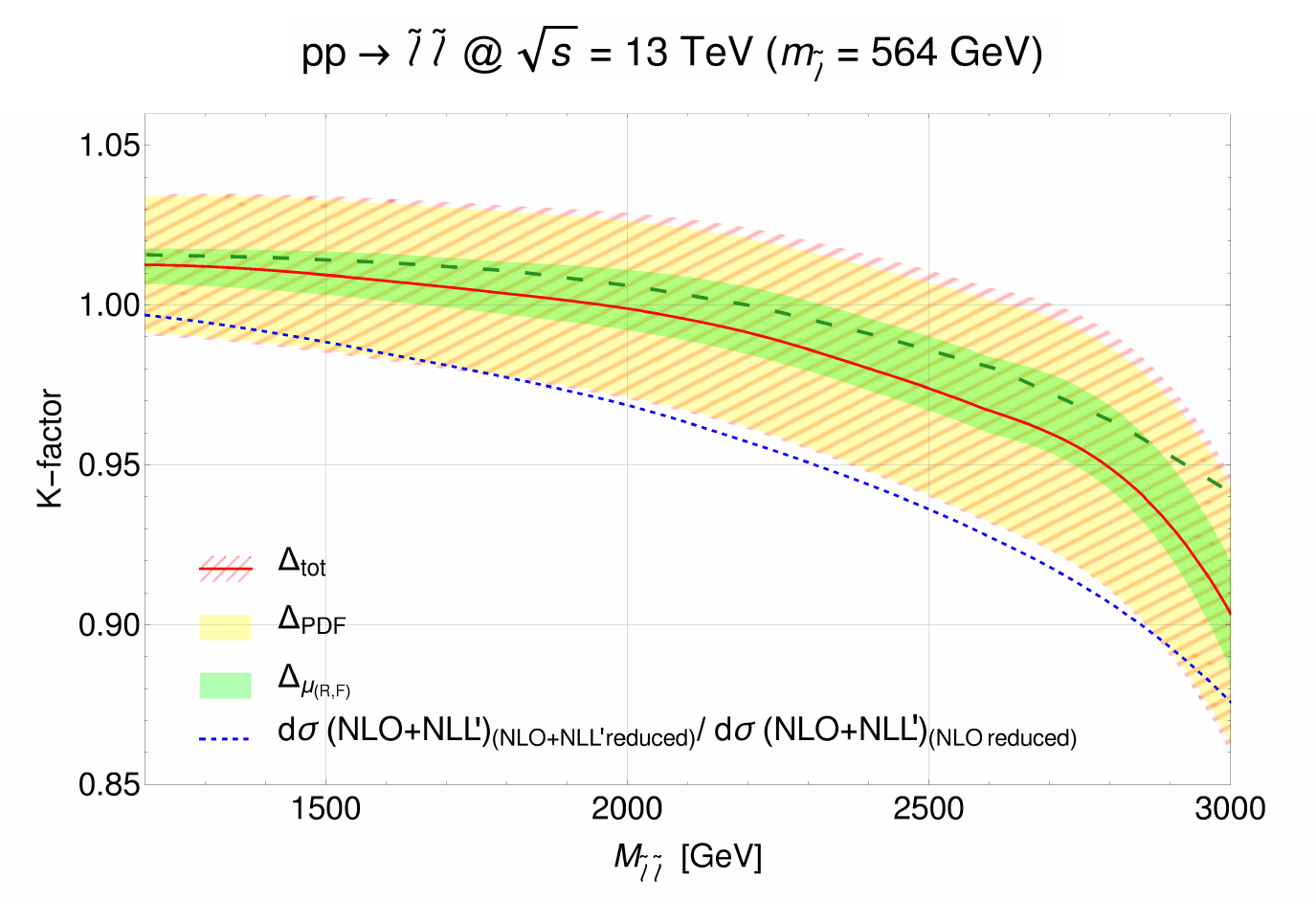}
\end{center}
\vspace{-0.3cm}
\caption{ Left panel: The NLO+NLL$^\prime$  calculation of the invariant mass distribution for 
  slepton pair production at the LHC 13 TeV using the {\tt Resummino} program,
  using both the NLO and NLO+NLL$^\prime$ NNPDF3.0 DIS+DY+top PDFs
  as input normalized to the NLO calculation. 
  Right panel: The $K$-factor eq.~\eqref{eq:Kfactkulesza} as a function of $M_{\tilde l \tilde l}$. We superimposed the green curve of the left panel to facilitate the comparison (figure adapted from ref.~\cite{Fiaschi:2018xdm}; the DIS+DY+top fit is labelled `reduced' in the figure). 
  }
\label{fig:resummino}
\end{figure}

To examine the effect of resummed PDFs in the context of supersymmetric particle production we show the NLO+NLL$^\prime$ predictions for left-handed slepton pair production using the public code \texttt{Resummino}~\cite{Fuks:2013vua,Fuks:2013lya,Bozzi:2007qr}.
The production of sleptons is mostly sensitive to the $q\bar q$ luminosity; other processes, like squark and gluino pair production, would instead be sensitive at $qg$ and $gg$ channels.
The results are shown in the left panel of fig.~\ref{fig:resummino}, where we show the invariant mass distribution as a function of the mass of the slepton pair $M_{\tilde l \tilde l}$ at the LHC 13 TeV, using the standard settings of \texttt{Resummino} and a slepton mass of $m_{\tilde l}= 564$~GeV.
All results are normalized to the NLO curve, computed with NLO baseline PDFs.
We observe that if the resummation is included only in the matrix element the cross section is enhanced by several percent, up to $5\%$ at $M_{\tilde l \tilde l} \sim 3$~TeV, though within the NLO uncertainty bands.
On the other hand, if the resummation is included consistently both in the PDFs and in the matrix elements there is only a marginal increase in the cross section for $M_{\tilde l \tilde l} \sim 1.2 $~TeV, whereas the NLO+NLL$^\prime$ calculation is even suppressed by a factor of $5\%$ at very high $M_{\tilde l \tilde l}$, albeit within the large PDF uncertainties.

To overcome the limitations due to the reduced experimental dataset, the authors of ref.~\cite{Beenakker:2015rna} suggested the introduction of a $K$-factor, defined as
\begin{equation}\label{eq:Kfactkulesza}
	K = \frac{\sigma_{\text{NLO+NLL$^\prime$,NLO global}}}{\sigma_{\text{NLO,NLO global}}} \times \frac{\sigma_{\text{NLO+NLL$^\prime$,NLO+NLL$^\prime$ DIS+DY+top}}}{\sigma_{\text{NLO+NLL$^\prime$,NLO DIS+DY+top}}},
\end{equation}
where the first subscript refers to the accuracy of the computation and the second to the PDF to be used.
This allows one to reduce the impact of the larger uncertainties by computing the resummed NLO+NLL$^\prime$ cross section as
\begin{equation}
	\sigma_{\text{NLO+NLL$^\prime$}} = K \times \sigma_{\text{NLO,NLO global}}
\end{equation}
and using the PDF error associated to the global PDF as an estimate of the uncertainties due to the PDFs.
The authors of ref.~\cite{Beenakker:2015rna} used this prescription to calculate squark and gluino production cross sections at LHC 13 TeV.
Their recipe has been recently used to compute the invariant mass distributions for the pair production of left-handed and right-handed sleptons in ref.~\cite{Fiaschi:2018xdm}.
The right panel of fig.~\ref{fig:resummino} shows the $K$-factor eq.~\eqref{eq:Kfactkulesza} for left-handed slepton pair production.
The yellow, green and red bands correspond to the PDF, scale and total uncertainties, respectively, whereas the blue line corresponds to the change of PDFs alone (namely, the second factor appearing in the RHS of eq.~\eqref{eq:Kfactkulesza}).
The results of ref.~\cite{Fiaschi:2018xdm} are in good agreement with the green curve in the left plot, which is affected by the reduced dataset used.

These results show that the use of resummation only in the matrix element can lead to inaccurate results, because it may overestimate cross sections and invariant mass distributions. 
This is more likely to happen in NLO+NLL$^\prime$ calculations, since at NNLO+NNLL$^\prime$ the effect of the resummation is much smaller.
In conclusion, threshold resummed matrix elements should be used in conjunction with resummed PDFs.
The prescription eq.~\eqref{eq:Kfactkulesza} allows one to produce consistently resummed calculations with threshold resummation effects with a PDF error comparable to state-of-the-art PDF determinations, pending the availability of resummed PDFs based on a wider dataset.

\section{Parton distribution functions with high-energy resummation}\label{sec:pdfsmall}

Since in the $\overline{\text{MS}}$ scheme threshold resummation affects only the coefficient functions, there are few obstacles preventing its inclusion in PDF fits, as long as resummed calculations are available.
The situation is instead more complex for high-energy resummation, since in general both coefficient functions and splitting functions receive single-logarithmic contributions to all orders in perturbation theory, as we discussed in sect.~\ref{sec:small-xres} .
Despite the fact that the formalism for resumming small-$x$ logarithms has existed for quite some time, the number of phenomenological analyses has indeed been very limited, due to the rather convoluted technical details of high-energy resummation.
The recent theoretical insight obtained by reviving the ABF approach discussed in sect.~\ref{sect:HELLandback} and, crucially, the implementation of the intricate results of high energy resummation in the \texttt{HELL} public code~\cite{Bonvini:2016wki,Bonvini:2017ogt,Bonvini:2018iwt,Bonvini:2018xvt} have finally made small-$x$ resummation available for phenomenological applications.

In this section, we present a state-of-the-art PDF determination in which NLO and NNLO fixed-order perturbation theory is matched to NLL$x$ small-$x$ resummation.
%
This will be done by supplementing the recent NNPDF3.1 PDF determination~\cite{Ball:2017nwa} with high-energy resummation of DGLAP evolution and DIS coefficient functions using \texttt{HELL}, thereby leading to resummed PDF sets.
This will allow us to assess the impact of small-$x$ resummation in a PDF fit and to demonstrate that the tension which some groups have observed in the description of the HERA data in the small-$x$ and small-$Q^2$ region disappears if resummation effects are included.
We will show that the inclusion of small-$x$ resummation improves quantitatively the description of the HERA data, in particular at NNLO.
The results we show fulfill a program that was initiated more than two decades ago, when the first measurements of the $F_2$ structure function at HERA stimulated the first studies on the inclusion of small-$x$ resummation in perturbative evolution.

We start by discussing how resummation affects PDF evolution and DIS structure functions in sect.~\ref{sec:small-implementation}.
We then present the settings of the fits, which we dub NNPDF3.1sx, in sect.~\ref{sec:small-settings}, and we discuss the results in sect.~\ref{sec:small-PDFs}.
We finally show evidence of the onset of resummation effects in the HERA data in sect.~\ref{sec:small-HERA} and we discuss some implications for small-$x$ phenomenology at the LHC and beyond in sect.~\ref{sec:smallLHC}.

\subsection{Implementation of small-$x$ resummation}\label{sec:small-implementation}

To facilitate the use of small-$x$ resummation, we have interfaced the code \texttt{HELL} with the evolution library \texttt{APFEL},  thus providing a framework for the systematic inclusion of small-$x$ resummation in PDF fits. 
The interface allows for a straightforward inclusion of resummation effects in the PDF evolution and in DIS structure functions. 
We use the so-called `exact' solution of the DGLAP evolution, rather then the `truncated' solution which is used in ABF and is routinely used in NNPDF fits, in which one expands out systematically the subleading corrections~\cite{Ball:2008by}. 
The difference between the two solutions becomes smaller at higher perturbative orders, so the choice will not affect significantly our best NNLO+NLL$x$ result.

\begin{figure}[t]
\begin{center}
 \includegraphics[width=0.41\textwidth,page=1]{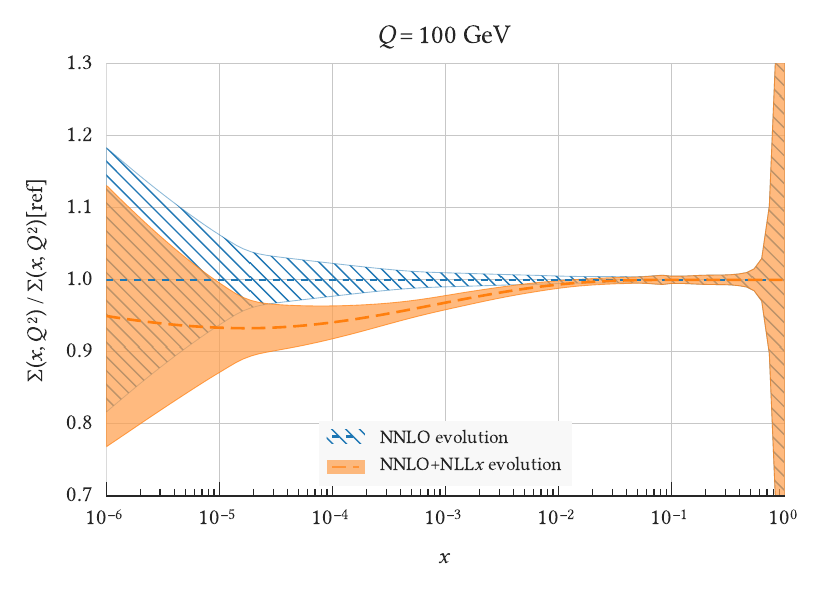}\qquad \qquad
 \includegraphics[width=0.41\textwidth,page=2]{figures/nn31x_apfelevol100_ratio.pdf}
 \caption{ The ratio of the quark singlet (left panel)
     and gluons (right panel) for the evolution from
     a fixed boundary condition at $Q_0=1.65$ GeV up to $Q=100$ GeV
     using either NNLO fixed-order theory or NNLO+NLL$x$
     resummed theory for the DGLAP evolution.
  }
  \label{fig:resummedPDFs}
\end{center}
\end{figure}

The effect of resummation on PDF evolution can be investigated by evolving the PDFs with resummed splitting kernels rather than with standard fixed-order DGLAP splitting functions.
To illustrate this effect, we take a given boundary condition at a low scale $Q_0$ and we evolve it upward using fixed-order or resummed evolution.
Though this comparison allows us to qualitatively study the differences induced at high scales by resummation, its physical meaning is however limited, since in a PDF fit with small-$x$ resummation the initial conditions are likely to change significantly.
\mccorrect{Moreover, since parton densities themselves are not observables, the inclusion of resummation in the coefficient function may compensate for the effect seen at the level of PDFs.}
We show the result of this comparison in fig.~\ref{fig:resummedPDFs}, where we show the ratio of the gluon and of the quark singlet at NNLO and at NNLO+NLL$x$ as a function of $x$ at $Q=100$ GeV.
To obtain the curves we have taken as a boundary condition NNPDF3.1 NNLO at $Q_0=1.65$ GeV.
We observe that the effects of PDF evolution are negligible at large and medium $x$, but can reach a few percent for $x \lesssim 10^{-4}$, a region which is probed by the HERA structure functions data.
Though this study is only illustrative, it shows that small-$x$ resummation has a sizeable impact on PDF evolution, which will affect the determination of PDFs at small $x$.

Before discussing the fit strategy, we can estimate the impact of small-$x$ resummation on DIS structure functions by comparing theoretical predictions at fixed order with predictions which include resummation.
The \texttt{HELL} code contains all ingredients to implement resummation in the FONLL variable flavour number scheme which is used in NNPDF fits.
Since the kinematic region where resummation effects are important (small $x$ and small $Q^2$) is rather close to the charm threshold, a careful treatment of the charm is essential.
In this analysis we fit the initial charm distribution, as in the NNPDF3.1 analysis.
As we discussed in chapter~\ref{ch-qcd}, when the FONLL scheme is extended to fitted charm, it receives an extra contribution $\Delta F_{\text{IC}}$, which is currently known at $\mathcal O(\as)$.
In this analysis we do not include the NNLO and the small-$x$ resummation corrections to $\Delta F_{\text{IC}}$; however, since the $\mathcal O(\as)$ contribution is a small correction, we expect these additional corrections to be practically insignificant.

\begin{figure}[t]
\begin{center}
  \includegraphics[width=0.49\textwidth,page=5]{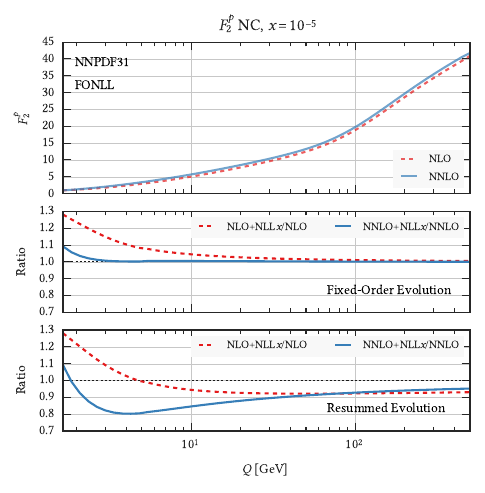}
  \includegraphics[width=0.49\textwidth,page=5]{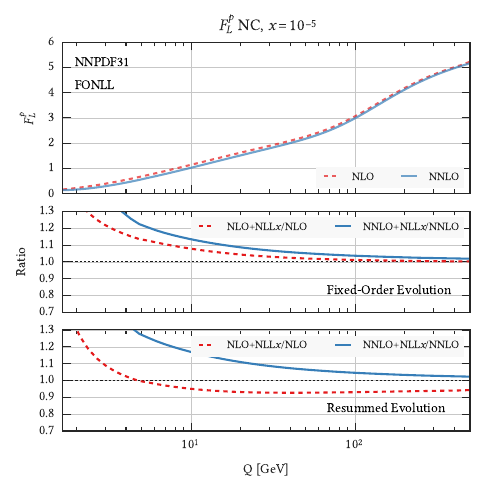}
  \caption{
   The proton NC structure function $F_2(x,Q^2)$ (left plot) and the longitudinal structure function $F_L(x,Q^2)$ as a function of $Q$ using different calculational schemes for $x=10^{-5}$.
  }
  \label{fig:resummedSFfixedPDF}
\end{center}
\end{figure}

We show the results for the proton structure function $F_2(x,Q^2)$  and $F_L(x,Q^2)$ in neutral-current (NC) DIS as a function of $Q$ and for $x=10^{-5}$ in fig.~\ref{fig:resummedSFfixedPDF}.
To disentangle the effect of resummation on PDF evolution from that in the coefficient functions, the effect of resummation is taken into account in two steps.
In the first step, we compute the structure functions using the same input PDF and we include resummation effects only in the coefficient functions.
In the second step, we include resummation both in the coefficient functions and in the evolution.
In the upper panel, we show the results at NLO and at NNLO, including heavy quark effects using the FONLL-B and FONLL-C schemes respectively.
In the middle panel, we show the ratio between the resummed (N)NLO+NLL$x$ results and the correspondent fixed-order one, including resummation only in the splitting functions.
In the lower panel, we show the same ratio now including resummation also in the evolution. 
As a boundary condition, we take NNPDF3.1 at (N)NLO at $Q_0=1.65$ GeV and we perform all calculations with $\as(m_Z^2)=0.118$ and a (pole) charm mass $m_c=1.51$ GeV.

We observe that if resummation is included only in the coefficient functions the effect is rather mild for $F_2$, especially at NNLO, even for small values of $Q^2$.
However, the situation changes if resummation is included also in the PDF evolution, showing that most of the impact of small-$x$ resummation in this case arises from DGLAP evolution. 
The effects are now smaller at NLO than at NNLO, where they can reach $20\%$.
The effects in the coefficient functions are substantially more pronounced for $F_L$.
In this case, the effect is larger at NNLO than at NLO already at the level of the coefficient functions.
When resummation effects are included in PDF evolution, their impact is somewhat reduced at NLO, while it is further enhanced at NNLO where it can reach $30\%$. 
This discussion suggests that we expect few differences between fixed-order and resummed PDFs at NLO and more significant differences at NNLO.

\subsection{Fitting strategy}\label{sec:small-settings}

We can now move on to discuss the settings of the NNPDF3.1 fits with small-$x$ resummation, as well as the fixed-order fits which we use as a baseline.
We denote these fits as NNPDF3.1sx.
We briefly present the input dataset and we review the theoretical treatment of the data used in the fit, as well as the strategy used to choose the kinematic cuts for both DIS and hadronic processes.

\paragraph{\itshape\mdseries Fit settings.}

The settings of the fits follow rather closely those of the NNPDF3.1 global analysis.
The same input dataset is used, which includes fixed-target~\cite{Arneodo:1996kd,Arneodo:1996qe,Benvenuti:1989rh,Benvenuti:1989fm,Whitlow:1991uw,Onengut:2005kv,  Goncharov:2001qe,Mason:2006qa} and HERA~\cite{Abramowicz:2015mha} DIS inclusive structure functions; charm and bottom cross sections from HERA~\cite{Abramowicz:1900rp}; fixed-target DY production~\cite{Webb:2003ps,Webb:2003bj,Towell:2001nh,Moreno:1990sf}; gauge boson and inclusive jet production from the Tevatron~\cite{Aaltonen:2010zza,Abazov:2007jy,Aaltonen:2008eq,Abazov:2013rja,D0:2014kma}; and electroweak boson production, inclusive jet, $Z$ $p_T$ distributions, and $t\bar{t}$ total and differential measurements from  ATLAS~\cite{Aad:2011dm,Aad:2013iua,Aad:2011fp,Aad:2011fc,Aad:2013lpa, ATLAS:2012aa,ATLAS:2011xha,TheATLAScollaboration:2013dja,Aad:2015auj,Aaboud:2016btc,Aad:2014kva,Aaboud:2016pbd,Aad:2015mbv,Aad:2014qja,Aad:2014xaa}, CMS~\cite{Chatrchyan:2012xt,Chatrchyan:2013mza,Chatrchyan:2013tia,Chatrchyan:2013uja,Chatrchyan:2013faa,Chatrchyan:2012bra,Chatrchyan:2012ria,Khachatryan:2016pev,Khachatryan:2015luy,Khachatryan:2016mqs,Khachatryan:2015oqa,Khachatryan:2015oaa} and LHCb~\cite{Aaij:2012vn,Aaij:2012mda,Chatrchyan:2012bja,Aaij:2015gna,Aaij:2015zlq} at $\sqrt{s}=7$ and 8 TeV.
We produce fits at NLO+NLL$x$ and at NNLO+NLL$x$, as well as their fixed-order counterparts. 

As in the NNPDF3.1 analysis, the charm PDF is fitted alongside the light quark and the gluon PDFs.
We use heavy quark pole masses, whose values are the same as in the NNPDF3.1 analysis.
All the results presented here are produced taking $\as(m_Z^2)=0.118$.
As we mentioned above, the fits are produced by using the exact solution of the DGLAP evolution, rather than the truncated solution which has been used in NNPDF3.1.
The initial scale $Q_0$ at which the PDFs are parametrized is chosen to be $Q_0 = 1.64$ GeV ($Q_0^2=2.69$ GeV$^2$), which is slightly smaller than the scale used in NNPDF3.1, which was $1.65$ GeV.
We chose this scale since this way we are able to include the $Q^2=2.7$ GeV$^2$ bin in the HERA structure function data, which we expect to be particularly sensitive to small-$x$ resummation.

We use the same settings as in NNPDF3.1 for the evaluation of the hadronic hard-scattering matrix element.
Theoretical predictions for Drell-Yan fixed-target and Tevatron and LHC cross sections are obtained using either fixed-order or resummed evolution for (N)NLO and (N)NLO+NLL$x$ fits, respectively.
However, the partonic cross sections for these processes are always computed at fixed-order accuracy, since the implementation of the hadronic processes in \texttt{HELL} is still a work in progress.
To account for this limitation, we cut all the data in kinematic regions where we expect small-$x$ corrections to the matrix element to be significant, as we explain below.
We also produced DIS-only fits where small-$x$ resummation is included consistently in both evolution and coefficient functions for all the data points included in the fit.
Whilst PDF uncertainties are much larger due to the lack of hadronic data, the constraints from HERA data are still the dominant ones in the small-$x$ region.

\paragraph{\itshape\mdseries Kinematic cuts.}

The kinematic cuts used in this analysis are the same as those used in the NNPDF3.1 fit, with some differences. 
As discussed above, we lower the $Q^2$ cut to $Q^2_{\text{min}}=2.69$ GeV$^2$ to allow for the inclusion of a further bin of the HERA inclusive cross section dataset.
This way, the kinematic coverage of the small-$x$ region extends down to $x_{\text{min}} \simeq 3 \times 10^{-5}$.
Furthermore, we do not include any additional cuts to the HERA charm production cross sections, contrary to NNPDF3.1 where the data with $Q^2 \leq 8$ GeV$^2$ were excluded from the NNLO fit.
The inclusion of the extra points does not affect the PDFs, though the $\chi^2$ of the charm data somewhat deteriorates at NNLO.
Finally, we apply a kinematic cut on collider processes for which resummation is included only in the PDF evolution, lest the fit be biased.
Therefore, we include in the fit only data for which the impact of small-$x$ resummation on the coefficient function can be assumed to be negligible.

Quantifying the effect of small-$x$ resummation on the partonic coefficient functions would require the knowledge of such resummation.
Here we turn to a more qualitative argument, based on the observation that since in a generic factorization scheme logarithms affect both the evolution kernels and the coefficient function they are expected to be similar in size.

To implement the cuts, we resort to a parametrization of the resummation in the $(x,Q^2)$ plane.
Since small-$x$ logarithms should be resummed when $\as(Q^2) \ln (1/x) $ approaches unity, we define our kinematic cut such as to remove those data points for which
\begin{equation}
	\as(Q^2) \ln \frac{1}{x} \geq H_{\text{cut}},
\end{equation}
where $H_{\text{cut}} \lesssim 1$ is a fixed parameter; the smaller its value, the more data are removed.
Assuming one-loop running for the strong coupling constant, the cut becomes a straight line in the $(x,Q^2)$ plane, determined by the equation
\begin{equation}\label{eq:hadrcut}
	\ln \frac{1}{x} \geq \beta_0 H_{\text{cut}} \ln \frac{Q^2}{\LambdaQ^2},
\end{equation}
where $\LambdaQ \simeq 88$ MeV is the QCD Landau pole for $n_f=5$ and $\beta_0 \simeq 0.61$.
The value of $x$ in the definition of the cut is determined using leading-order kinematics.
%
%
The value of $H_{\text{cut}}$ must be conservative enough to guarantee that the effect of the unresummed logarithms is under control.
However, it would be ideal if the dataset is large enough such that the NNPDF3.1sx fits are competitive with NNPDF3.1.
Here we choose $H_{\text{cut}}=0.6$, which has been shown to satisfy both criteria in ref.~\cite{Ball:2017otu}.

\begin{figure}[t]
\begin{center}
  \includegraphics[width=0.8\textwidth]{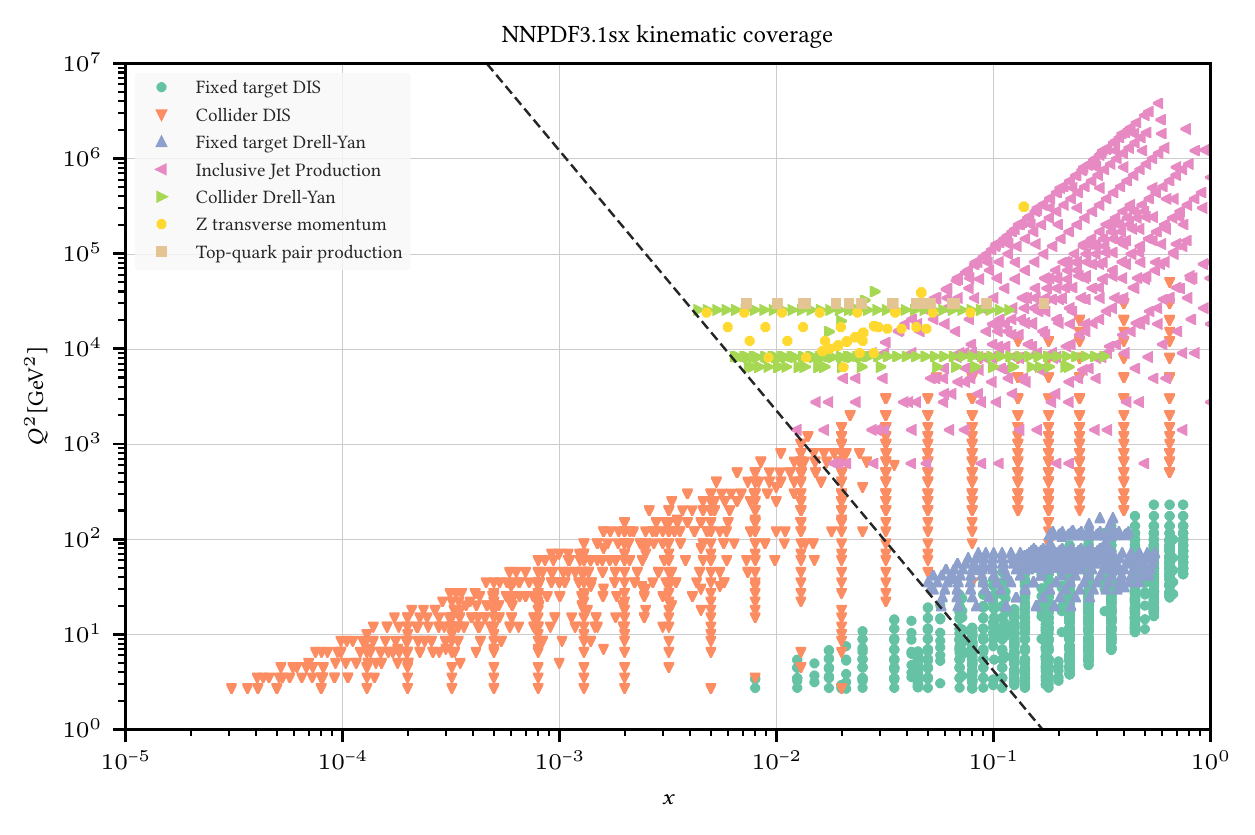}
  \caption{
 The kinematic coverage in the $(x,Q^2)$ plane of the data included in the NNPDF3.1sx fit with the default value of the kinematic cut to the hadronic data, $H_{\text{cut}}=0.6$. The diagonal line indicates the value of the cut eq.~\eqref{eq:hadrcut}, below which the hadronic data are excluded from the fit. \mccorrect{The reader can compare this figure with fig.}~\ref{fig:kinplot31} \mccorrect{to get an idea of the number of points excluded}.
  }
  \label{fig:nnpdf31sxkin}
\end{center}
\end{figure}

The kinematic coverage of the NNPDF3.1sx fits in the $(x,Q^2)$ plane for the default value of $H_{\text{cut}}=0.6$ is shown in fig.~\ref{fig:nnpdf31sxkin}.
The diagonal line indicates the region below which the hadronic data are removed by the cut eq.~\eqref{eq:hadrcut}.
As a consequence, the hadronic dataset is restricted to the large-$Q^2$ and medium- and large-$x$ region.
\mccorrect{We find that} most of the ATLAS and CMS measurements included in NNPDF3.1 are included in the NNPDF3.1sx fits, though with a reduced number of datapoints, the LHCb measurements are removed altogether, highlighting the sensitivity of forward $W,\, Z$ production data to the small-$x$ region.

\subsection{Parton distributions with small-$x$ resummation}\label{sec:small-PDFs}

We can now present the NNPDF3.1sx fits with small-$x$ resummation.
We will start by discussing the results of the DIS-only fits.
We then focus the discussion on the global fits based on the dataset described in sect.~\ref{sec:small-settings} with the default choice of $H_{\text{cut}}$ and we will compare them to the DIS-only fits.
%

\paragraph{\itshape\mdseries DIS-only fits.}

Let us start by discussing the results of the DIS-only fits, where resummation is included consistently both in the evolution and in the coefficient functions.
We collect the values for $\chi^2/\Ndat$ obtained in the NLO, NLO+NLL$x$, NNLO and NNLO+NLL$x$ fits in table~\ref{tab:chi2tab_dis}, computed using the experimental definition of the covariance matrix.
To better quantify the differences between the resummed and the fixed-order results for each experiment, we also show the difference in $\chi^2$ between the resummed and fixed-order results
\begin{equation}
\label{eq:deltachi2def}
\Delta\chi^2_{\text{(N)NLO}} \equiv \chi^2_{{\text{(N)NLO+NLL}}x}-\chi^2_{\text{(N)NLO}} \, .
\end{equation}

\begin{table}[t]
\centering
   \footnotesize
   \renewcommand{\arraystretch}{1.10}
   \begin{tabular}{l C{1.7cm}C{1.7cm}C{1.3cm}|C{1.7cm}C{2.2cm}C{1.3cm}}
     & \multicolumn{2}{c}{$\chi^2/N_{\text{dat}}$}  & $\Delta\chi^2$  & \multicolumn{2}{c}{$\chi^2/N_{\text{dat}}$}  & $\Delta\chi^2$ \\
 & NLO &  NLO+NLL$x$ &  &   NNLO & NNLO+NLL$x$   &  \\
 \midrule
NMC    &    1.31   &   1.32   &  +5  &   1.31   &   1.32   & +4 \\    
SLAC    &    1.25   &   1.28   &  +2  &  1.12   &   1.02    & $-8$ \\    
BCDMS    &    1.15   &   1.16   & +7   &  1.13   &   1.16     &  +14 \\    
CHORUS    &    1.00   &   1.01   &  +9  &   1.00   &   1.03    & +26  \\    
NuTeV dimuon    &    0.66   &   0.56      & $-8$ &      0.80   &   0.75   &   $-4$   \\[0.20cm]  
HERA I+II incl. NC    &   1.13      &  1.13    &  $+6$   &  1.16    & 1.12       & $-47$    \\
HERA I+II incl. CC    &    1.11    &  1.09   &  $-1$   &  1.11    &   1.11      &  -    \\    
HERA $\sigma_c^{\text{NC}}$    &    1.44   &   1.35   & $-5$   &   2.45   &   1.24   & $-57$ \\    
HERA $F_2^b$       &    1.06   &   1.14   & +2    &  1.12  &   1.17    & +2  \\    
\midrule
    { Total}
    &    1.113   &   1.119   &  +17 &  1.139  &   1.117  &  $ -70$ \\    
  \end{tabular}
  \vspace{0.5cm}
\caption{ The values of $\chi^2/\Ndat$ for the total and the individual datasets included
  in the DIS-only NNPDF3.1sx NLO, NLO+NLL$x$, NNLO and NNLO+NLL$x$
  fits.
  We also indicate the absolute difference $\Delta\chi^2$ between the resummed
  and fixed-order results, eq.~\eqref{eq:deltachi2def}.
  We indicate with a dash the case $|\Delta\chi^2| < 0.5$.
}
\label{tab:chi2tab_dis}
\end{table}

The $\chi^2/\Ndat$ is very similar for the NLO, the NLO+NLL$x$ and the NNLO+NLL$x$ fits, whereas the NNLO fit gives the highest $\chi^2/\Ndat$.
We observe a significant improvement of the $\chi^2/\Ndat$ between the NNLO result and the NNLO+NLL$x$ result.
On the contrary, the effect of resummation is very mild at NLO; this is not surprising, as we have discussed previously how resummation is expected to cure instabilities in the fixed-order perturbative expansion which are more important at NNLO than at NLO.

The bulk of the difference in the fit quality between the NNLO and the NNLO+NLL$x$ fit arises from the HERA inclusive neutral current and charm dataset, whose $\chi^2/\Ndat$ decreases from 1.16 to 1.12 ($\Delta \chi^2=-47$) and from 2.54 to 1.24 ($\Delta \chi^2 = -57$), respectively.
The description of the fixed-target DIS experiments, which is less sensitive to the small-$x$ region than the HERA data, is not significantly affected by the inclusion of small-$x$ resummation.
The only exception is a decrease in the fit quality between NNLO and NNLO+NLL$x$ for BCDMS and CHORUS.
We show below that these differences are reduced in the global fit, as the inclusion of the collider dataset stabilizes the PDFs at large $x$.

We note that the fit quality of the charm data at NNLO is somewhat high.
Indeed, the description of the charm data can be rather sensitive to the details of the heavy quark scheme\footnote{In a similar analysis by the xFitter collaboration~\cite{Abdolmaleki:2018jln}, where the charm PDF is perturbatively generated and the details of the heavy quark scheme differ from those used in the present analysis, the fit quality of the charm data is very similar at NNLO and at NNLO+NLL$x$; see in particular the discussion in sect.~4.3 of ref.~\cite{Abdolmaleki:2018jln}.}.
For instance, one may set to zero the $\Delta F_{\text{IC}}$ term and introduce a phenomenological damping to dispose of formally subleading terms close to the charm threshold (though this manipulation is in general not legitimate in the presence of intrinsic charm, as discussed in sect.~\ref{sec:ic}).
When this damping is included, the $\chi^2/\Ndat$ becomes 1.10 at NNLO, whereas it remains pretty stable at NNLO+NLL$x$ (1.23).
The deterioration of the $\chi^2$ is mostly driven by a poor description of the low-$x$ and low-$Q^2$ bins; indeed, by applying the more restrictive cut used in NNPDF3.1 ($Q^2 \geq 8$ GeV$^2$) the $\chi^2/\Ndat$ is essentially identical at the fixed-order and at the resummed level (1.38 and 1.35, respectively).
This region is rather sensitive to the treatment of the subleading terms, which is ultimately driven by phenomenological reasons; therefore, it is possible that by tuning the subleading terms one might achieve a better description at the fixed-order level.
Since at NLO the description of the charm data is satisfactory for all the datapoints both at the fixed-order and at the resummed level, here we decide to use the same theory settings of the NNPDF3.1 paper and interpret the more marked dependence on the subleading terms as a limitation of the fixed-order theory at NNLO.
It will however be interesting to study to what extent this picture is modified if the charm dataset is replaced by the recently released final HERA combination of charm and beauty structure functions~\cite{H1:2018flt}.

\begin{figure}[t]
\centering
   \includegraphics[width=0.41\textwidth,page=2]{figures/nn31x_nlonlldis100_ratio.pdf}\qquad \qquad
   \includegraphics[width=0.41\textwidth,page=1]{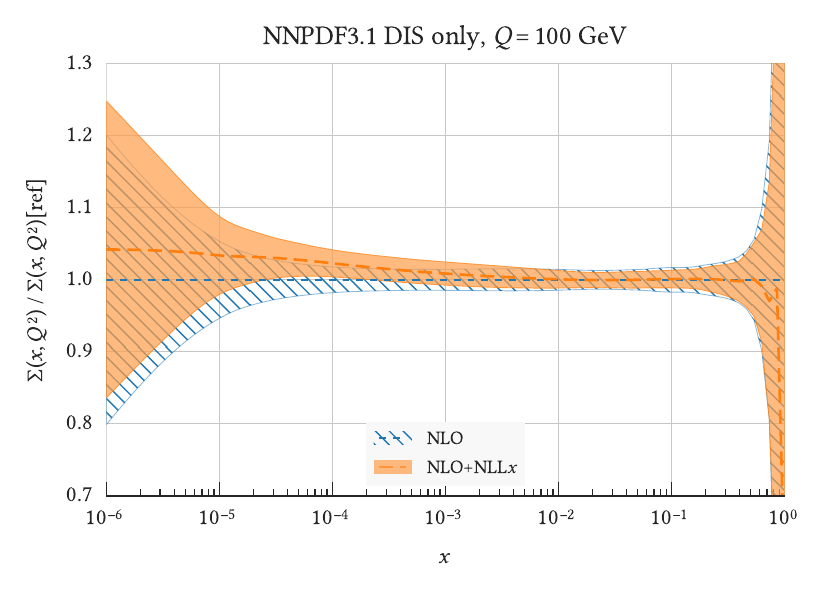}
   \includegraphics[width=0.41\textwidth,page=2]{figures/nn31x_nnlonlldis100_ratio.pdf}\qquad \qquad
   \includegraphics[width=0.41\textwidth,page=1]{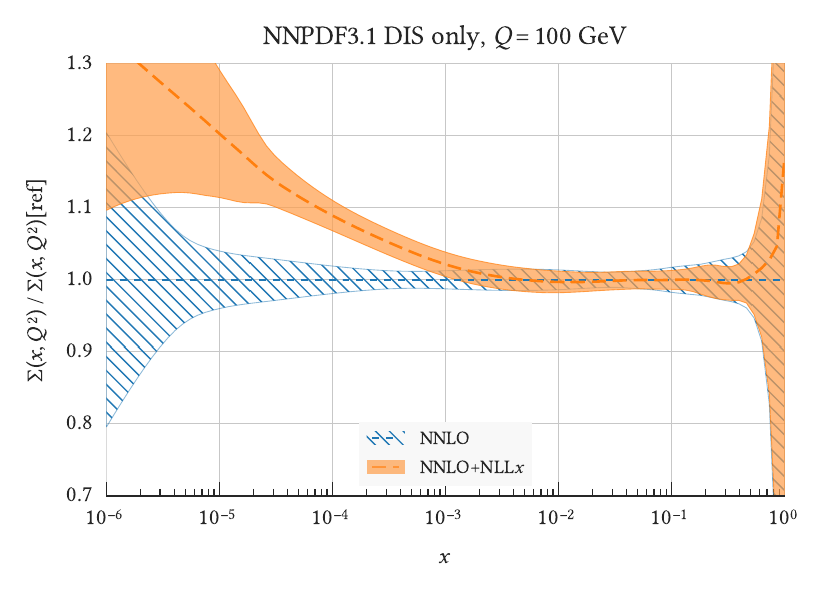}
  \caption{ Comparison between the gluon (left) and the total quark singlet (right plots) from the NLO and NLO+NLL$x$ (upper plots) and from the NNLO and NNLO+NLL$x$ DIS-only fits (lower plots).
}
  \label{fig:PDFfit_results_singlet_gluon}
\end{figure}

Let us now discuss how small-$x$ resummation affects the PDFs and their uncertainties. 
In fig.~\ref{fig:PDFfit_results_singlet_gluon} we show the ratio between the gluon (left) and the quark singlet (right) at $Q=100$ GeV in the NLO+NLL$x$ fit as compared to the baseline in the upper plots and in the NNLO+NLL$x$ fit as compared to NNLO in the lower plots.
The effect of resummation is moderate at NLO+NLL$x$; the gluon PDF is enhanced between $x=10^{-5} $ and $x=10^{-2}$, where the uncertainty bands only partially overlap, whereas the central value of the quark singlet is within the uncertainty bands.
This is consistent with what we discussed in section~\ref{sec:small-implementation}, where we found that NLO theory is a reasonable approximation to the resummed theory in the small-$x$ region.  
The picture changes significantly at NNLO+NLL$x$.
In this case, we see that the resummed gluons and singlet PDFs are systematically harder than their fixed-order counterparts, by an amount which can be as large as $20\%$ for $x\sim 10^{-5}$ and with a shift well outside the uncertainty bands.
Note that this comparison is performed at a scale significant for LHC phenomenology, where the effect of resummation in DGLAP evolution is combined with the change of the PDFs at the fitting scale.
This comparison therefore shows that for LHC observables which probe the small-$x$ region, such as DY production in the forward region, the use of resummed PDFs is likely necessary to obtain reliable predictions.

\begin{figure}[t]
\centering
   \includegraphics[width=0.8\textwidth]{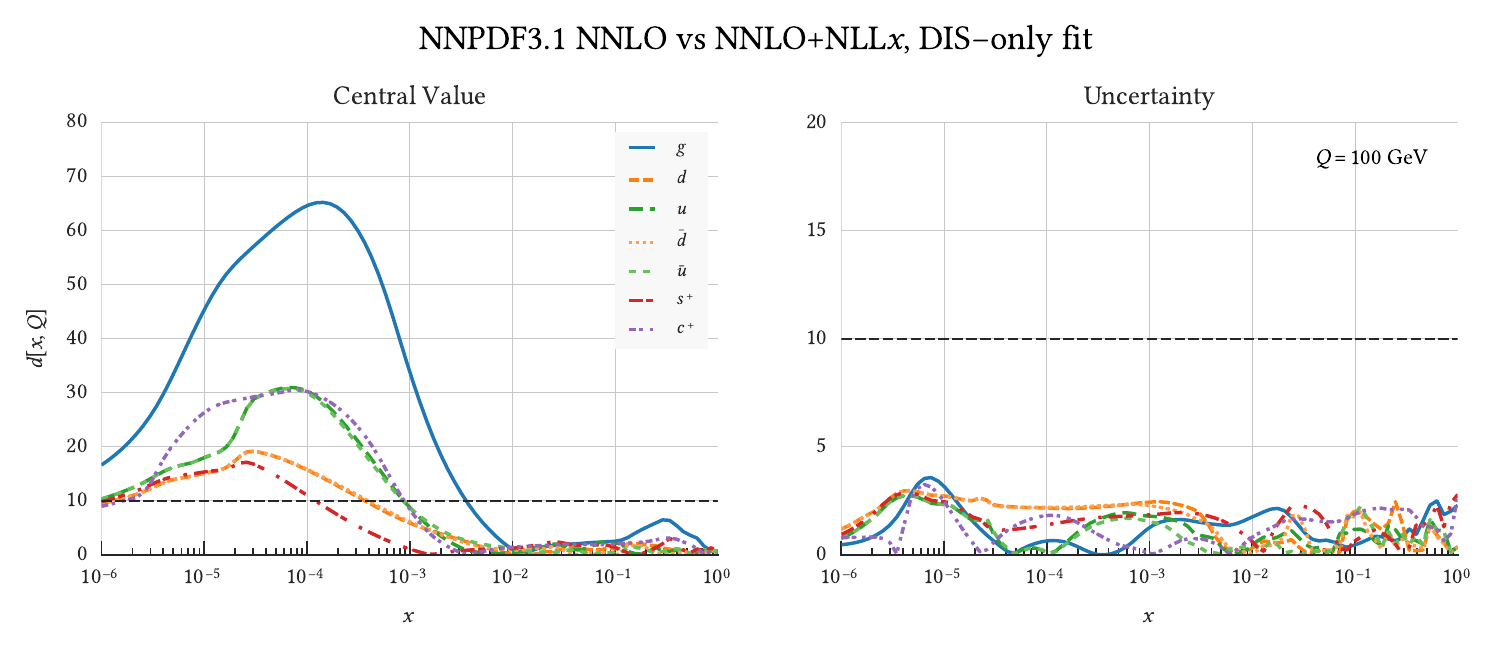}
  \caption{ The statistical distances between the
    central values (left) and the PDF uncertainties (right plot)
    of the NNPDF3.1sx NNLO and NNLO+NLL$x$ fits at $Q=100$~GeV
    in the flavour basis.}
  \label{fig:distance-sx-dis}
\end{figure}

So far we focused on the gluon and quark singlet, as small-$x$ resummation affects PDFs in the singlet sector.
Before moving to the global fits, we can also quantify the impact of small-$x$ resummation in the physical basis \mccorrect{using the statistical distances which we already used in sect.}~\ref{sec:large-PDFs}.
We show the distances between the central values (left) and the PDF uncertainties (right) of the NNLO and NNLO+NLL$x$ fits at $Q=100$ GeV in fig.~\ref{fig:distance-sx-dis}.
Also in this case, a distance of $d\sim 10$ corresponds to one-sigma variation of the central values or of the PDF uncertainties in units of the corresponding standard deviation.
We see that the impact of resummation peaks between $x\sim 10^{-3}$ and $x \sim 10^{-5}$, where $d \gtrsim 30$ for most flavours, corresponding to a shift in central value more than three times larger than the corresponding PDF uncertainty.
The most affected PDF is the gluon, where $d$ can be as large as $60$, followed by the charm and the light quark PDFs.
As expected, the impact of the resummation is negligible for the PDF uncertainties, as the experimental information used in the two fits is the same.

\paragraph{\itshape\mdseries Global fits.}

We now move our attention towards the global fits.
In table~\ref{tab:chi2tab_pertorder} we collect the fit quality of the global NNPDF3.1sx fits at NLO, NLO+NLL$x$, NNLO, NNLO+NLL$x$ for our default choice $H_{\text{cut}} = 0.6$.
Besides the value of $\chi^2/\Ndat$ we also include the absolute difference between the resummed and the fixed-order result eq.~\eqref{eq:deltachi2def}.
We observe that the NNPDF3.1sx fit based on NNLO+NLL$x$ theory leads to the best total fit quality $\chi^2/\Ndat= 1.100$.
The NNLO fit has again the highest value of $\chi^2/\Ndat= 1.130$, such that the overall improvement obtained by moving from NNLO to NNLO+NLL$x$ is $\Delta \chi^2 = -121$.
The effect of resummation at NLO is very mild, similarly to what we observed in the DIS-only fits, with the resummed $\chi^2/\Ndat$ slightly worse than the fixed-order one ($\Delta \chi^2 = +11$).

\begin{table}[tp]
\centering
  \scriptsize
  \renewcommand{\arraystretch}{1.10}
   \begin{tabular}{l C{1.5cm}C{1.5cm}C{1.3cm}|C{1.5cm}C{1.7cm}C{1.3cm}}
     & \multicolumn{2}{c}{$\chi^2/N_{\text{dat}}$}  & $\Delta\chi^2$  & \multicolumn{2}{c}{$\chi^2/N_{\text{dat}}$}  & $\Delta\chi^2$ \\
     & NLO &  NLO+NLL$x$ &  &   NNLO & NNLO+NLL$x$   &  \\
\midrule
NMC     & 1.35   &   1.35   & +1    & 1.30   &   1.33  &  +9  \\    
SLAC      &  1.16  &   1.14   & $-$1  &   0.92   &   0.95  & +2 \\    
BCDMS      & 1.13   &   1.15   & +12  &   1.18   &   1.18   & +3  \\    
CHORUS     & 1.07   &   1.10   & +20   &  1.07   &   1.07   &  $-$2 \\    
NuTeV dimuon     &  0.90   &   0.84   & $-$5   & 0.97   &   0.88  & $-$7  \\[0.20cm]  
HERA I+II incl. NC    &  1.12      &  1.12   & -2    &   1.17   &   1.11    &  $-62$  \\
HERA I+II incl. CC    &    1.24   &   1.24   & -    &   1.25   &  1.24      &  $-1$   \\
HERA $\sigma_c^{\text{NC}}$     &  1.21   &   1.19   & $-$1   &   2.33   &   1.14  & $-$56 \\    
HERA $F_2^b$                &  1.07   &   1.16   & +3   &  1.11   &   1.17    & +2  \\[0.20cm]  
DY E866 $\sigma^d_{\text{DY}}/\sigma^p_{\text{DY}}$     &  0.37   &   0.37   & -    &  0.32   &   0.30    & -  \\    
DY E886 $\sigma^p$                              &  1.06   &   1.10   & +3    &  1.31   &   1.32  & -  \\    
DY E605  $\sigma^p$                             &  0.89   &   0.92   &  +3   &  1.10   &   1.10  & -  \\    
CDF $Z$ rap                                     &  1.28   &   1.30   &  -   &  1.24   &   1.23   &  -  \\    
CDF Run II $k_t$ jets                           &  0.89   &   0.87   &  $-$2   &  0.85   &   0.80  & $-$4  \\    
D0 $Z$ rap                                      &  0.54   &   0.53   &  -   &  0.54   &   0.53  & -  \\    
D0 $W\to e\nu$  asy                             &  1.45   &   1.47   & -    &  3.00   &   3.10    & +1  \\    
D0 $W\to \mu\nu$  asy                           &  1.46   &   1.42   &  -   & 1.59   &   1.56    &  -  \\[0.20cm]  
ATLAS total                                 &  1.18   &   1.16   &  $-$7    &   0.99   &   0.98    &  $-2$   \\    
ATLAS $W,Z$ 7 TeV 2010                      &  1.52   &   1.47   &  -    &  1.36   &   1.21     & $-1$   \\    
ATLAS HM DY 7 TeV                           &  2.02   &   1.99   &   -   &   1.70   &   1.70     & - \\    
ATLAS $W,Z$ 7 TeV 2011                      &  3.80   &   3.73   &   $-$1   &   1.43   &   1.29  & $-1$  \\    
ATLAS jets 2010 7 TeV                       &  0.92   &   0.87   &  $-$4   &  0.86   &   0.83  &  $-2$   \\    
ATLAS jets 2.76 TeV                         &  1.07   &   0.96   &  $-$6  &  0.96   &   0.96    &  -  \\    
ATLAS jets 2011 7 TeV                       &  1.17   &   1.18   & -  &  1.10   &   1.09  & $-1$   \\    
ATLAS $Z$ $p_T$ 8 TeV $(p_T^{ll},M_{ll})$    &  1.21   &   1.24   & +2    &  0.94   &   0.98  &  +2  \\    
ATLAS $Z$ $p_T$ 8 TeV $(p_T^{ll},y_{ll})$    &  3.89   &   4.26   & +2   &  0.79   &   1.07    & +2  \\    
ATLAS $\sigma_{tt}^{tot}$                    &  2.11   &   2.79   & +2      &  0.85   &   1.15  & +1 \\    
ATLAS $t\bar{t}$ rap                        &  1.48   &   1.49   & -    &  1.61   &   1.64    & -  \\[0.20cm]  
CMS total                               &  0.97   &   0.92   &  $-$13  &  0.86   &   0.85    & $-3$ \\    
CMS Drell-Yan 2D 2011                   &  0.77   &   0.77   & -    & 0.58   &   0.57  &  - \\    
CMS jets 7 TeV 2011                     &  0.88   &   0.82   &  $-$9  & 0.84   &   0.81   & $-3$ \\    
CMS jets 2.76 TeV                       &  1.07   &   0.98   &  $-$7 &  1.00   &   1.00    &  - \\    
CMS $Z$ $p_T$ 8 TeV $(p_T^{ll},y_{ll})$  &  1.49   &  1.57    &  +1   & 0.73   &   0.77   &  - \\    
CMS $\sigma_{tt}^{\text{tot}}$                  &  0.74   &   1.28   & +2    & 0.23   &   0.24   &  - \\    
CMS $t\bar{t}$ rap                      &  1.16   &   1.19   &  -   &   1.08   &   1.10   &  - \\    
\midrule
{ Total }    &     1.117   &   1.120   &  +11   &     1.130   &  1.100   &   $-121$   \\    
\end{tabular}
\vspace{0.5cm}
\caption{Same as table~\ref{tab:chi2tab_dis}, now for the global NNPDF3.1sx NLO, NLO+NLL$x$,
  NNLO and NNLO+NLL$x$ fits, using the default value of $H_{\text{cut}}=0.6$ for the cut to the hadronic data.
}
\label{tab:chi2tab_pertorder}
\end{table}
 
We note that the improvement at NNLO+NLL$x$ is mainly due to the better description of the HERA charm and neutral current structure function data, where $\Delta \chi^2 = -62$ and $\Delta \chi^2 = -56$, respectively.
For all the other datasets the changes in the $\chi^2$ are instead less significant and compatible with statistical fluctuations.

We find that the NNLO theory achieves a better description of the ATLAS and CMS measurements than NLO theory.
This is particularly evident in the case of the high-precision data such as the ATLAS $W, \, Z$ 2011 rapidity distributions and the ATLAS and CMS 8 TeV $Z$ $p_T$ distributions.
Thanks to the improved description of these datasets the $\chi^2/\Ndat$ total values for ATLAS and CMS decreases from 1.18 (1.16) and 0.97 (0.92) at NLO(+NLL$x$) to 0.99 (0.98) and 0.86 (0.85) at NNLO(+NLL$x$).
Despite the improvement in the description of the large-$Q^2$ collider data, however, the total NNLO $\chi^2$ remains higher than the NLO one.
This was not the case in the NNPDF3.1 analysis, where instead the fit quality at NNLO was markedly better than the NLO one.
Indeed, the weight of the the high-precision Drell-Yan and $Z$ $p_T$ points --- poorly described by NLO theory --- is diminished since the $H_{\text{cut}}$ partly removes some of the points.
The main reason for the large $\chi^2$ at NNLO is the poor description of the HERA inclusive and charm dataset, which contains almost one third of the number of points included in the fit (specifically, $1209$ out of $3930$).

By comparing the fit quality of the global fit to those of the DIS-only fit we note that in the global fit the improvement in the description of the HERA inclusive data at NNLO+NLL$x$ is larger than in the global fit, so that $\Delta \chi^2$ decreases from $-47$ to $-62$.
We also observe that the description of the NuTeV dimuon data worsen in the global fit, irrespectively of resummation, due to tensions with the LHC data relative to the strange content of the proton, as observed already in~\cite{Ball:2017nwa}.

\begin{figure}[t]
\centering
  \includegraphics[width=0.8\textwidth]{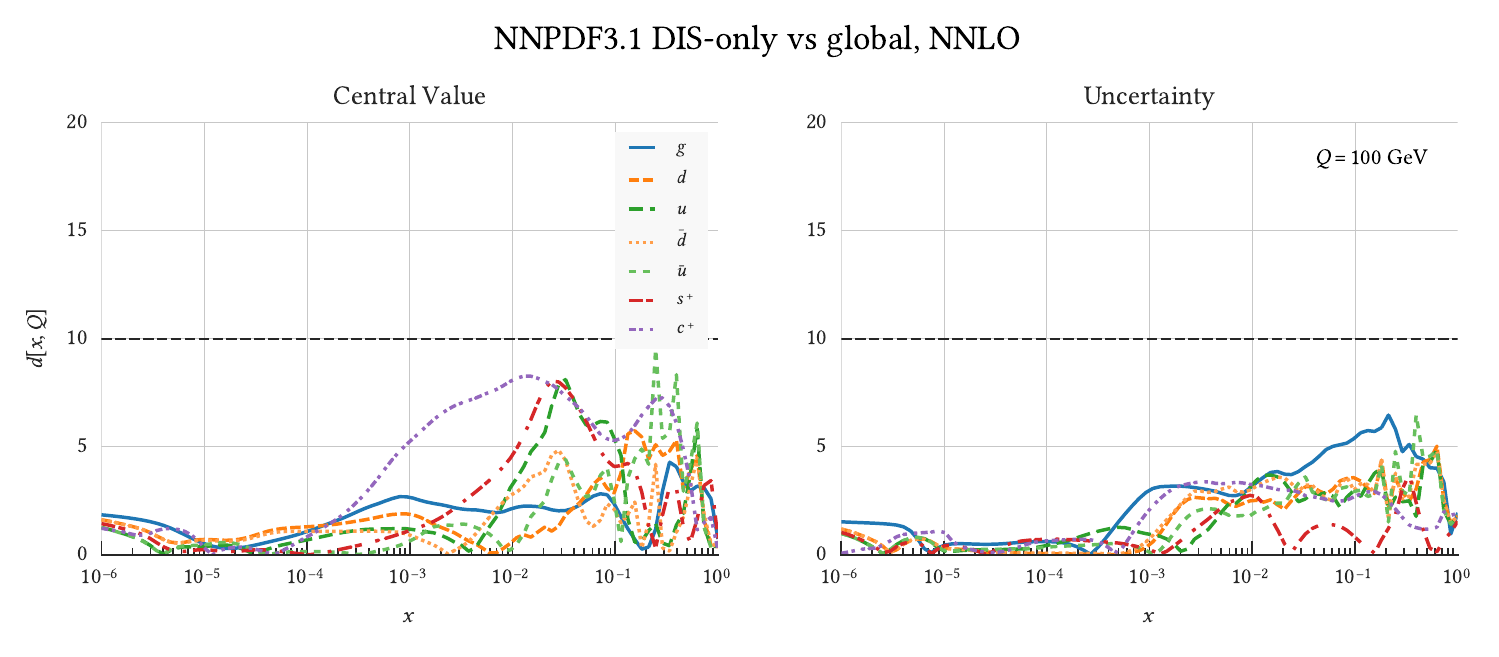}
  \caption{ Same as Fig.~\ref{fig:distance-sx-dis}
    for the comparison between the fixed-order NNLO NNPDF3.1sx
    DIS-only and global fits.}
  \label{fig:distance-sx-global-dis}
\end{figure}

To quantify the differences between the DIS-only and the global fit, we show in fig.~\ref{fig:distance-sx-global-dis} the distance estimator for the NNLO fits.
Since the conservative cut applied on the hadronic data removes points in the small-$x$ region, the constraints in the small-$x$ region are essentially the same.
Indeed, we see that the differences between the global and DIS-only fits are mainly localized at medium and at large $x$.
The PDF flavours which are most affected are the charm and the strange PDFs, whose distance is about 10 for $x\sim 10^{-2}$.
We also observe a visible decrease in the PDF uncertainty at medium and at large $x$, especially for the gluon PDF, which is only indirectly constrained in a DIS-only fit.

This comparison makes us confident that the use of a global fit is beneficial from the point of view of PDF uncertainties and at the same time does not spoil the results in the small-$x$ region obtained in a DIS-only fit.
Therefore, we will henceforth focus our attention on the global fit and in particular on the NNLO+NLL$x$ fits, since we have seen that the effect of resummation is less significant at NLO.

\begin{figure}[t]
\centering
  \includegraphics[width=0.8\textwidth]{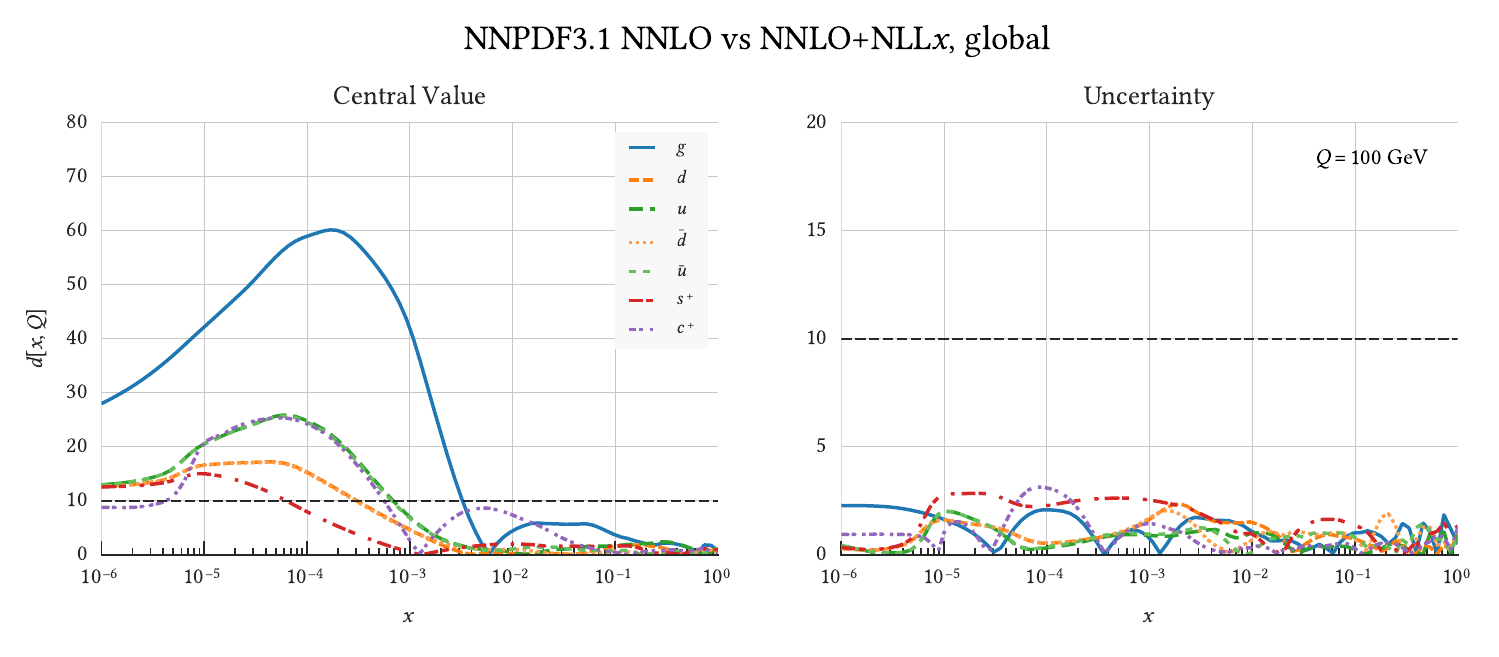}
   \caption{ Same as Fig.~\ref{fig:distance-sx-dis}
    for the NNPDF3.1sx global fits.}
    \label{fig:distance-sx-global}
\end{figure}

\begin{figure}[t]
\centering
  \includegraphics[width=0.41\textwidth,page=2]{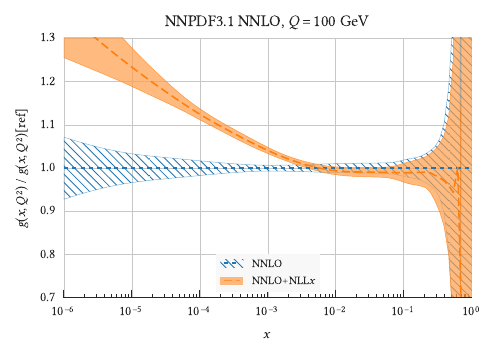}\qquad \qquad
  \includegraphics[width=0.41\textwidth,page=4]{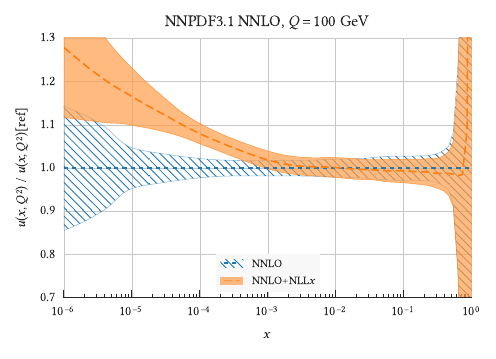}
  \caption{ Comparison of the NNPDF3.1sx  NNLO and NNLO+NLL$x$ global fits
     at $Q=100$~GeV.
    We show the gluon PDF and the up quark PDF,
    normalized to the central value of the baseline NNLO fit.}
  \label{fig:smallx-nnlo-global}
\end{figure}

We start by computing the distances between the NNLO and the NNLO+NLL$x$ fits for the NNPDF3.1sx global fits, which we show in fig.~\ref{fig:distance-sx-global}.
The result is very similar to what we obtained in the DIS-only case, with the distances at large $x$ somewhat larger, due to the reduced PDF uncertainties.
These effects are shown at the PDF level in fig.~\ref{fig:smallx-nnlo-global}, where we show two of the flavours most affected by small-$x$ resummation, namely the gluon and the up quark at a scale $Q=100$ GeV. 
In the up quark case, the impact of resummation is mild at medium and large values of $x$, whereas it increases at small-$x$, though it remains at the level of one or two sigma in units of the PDF uncertainty. 
The effect is much bigger for the gluon, where the difference is larger than $20\%$ at $x\sim 10^{-5}$, way outside the uncertainty band.

\begin{figure}[t]
\centering
   \includegraphics[width=0.41\textwidth,page=2]{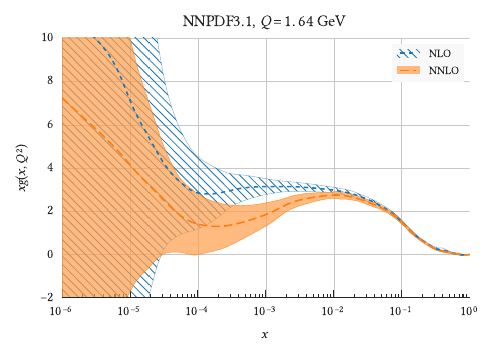}\qquad \qquad
 \includegraphics[width=0.41\textwidth,page=2]{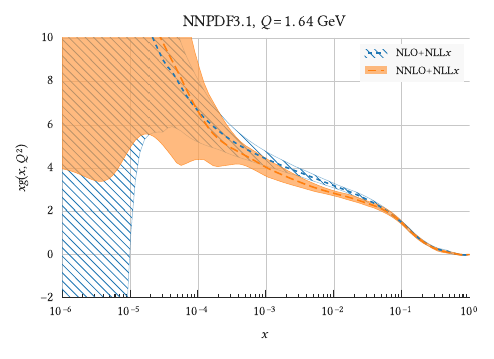} 
 \caption{ Comparison of the NLO and NNLO gluon PDFs and of the NLO+NLL$x$ and NNLO+NLL$x$ fit results (right) at the input parametrization scale of $Q=1.64$~GeV.}
 \label{fig:smallx-nnlo-global-lowQ}
\end{figure}

It is also interesting to consider the PDFs at the parameterization scale $Q_0$, which allows us to appreciate how small-$x$ resummation remedies the perturbative instabilities of the PDFs at the initial scale.
In particular, we show in figure~\ref{fig:smallx-nnlo-global-lowQ} the comparison for the gluon PDF at NLO and at NNLO (left plot) and at NLO+NLL$x$ and at NNLO+NLL$x$ (right plot).
The effect of resummation is rather marked for the fitted gluon.
We observe a drop in the medium- and small-$x$ region at NNLO driven by perturbative instability, which is absent at NNLO+NLL$x$.
%
Indeed, the NNLO+NLL$x$ curve is close to the NLO and the NLO+NLL$x$ ones.
Since the effect of the unresummed logarithms is expected to be larger at higher fixed-order accuracy, high-energy resummation would be even more crucial at N$^3$LO to cure the perturbative instability at small $x$. 

\subsection{Small-$x$ resummation \amper HERA structure functions}\label{sec:small-HERA}

The results we have shown in the previous section show that the inclusion of resummation improves the description of the datasets which represent the best probe of the small-$x$ region, namely the inclusive and the charm HERA structure functions.
In this section we focus on the description of the HERA data in the small-$x$ and small-$Q^2$ region to further quantify the improvement when the fixed order is supplemented by small-$x$ resummation.
We start by comparing the HERA structure function data in the low-$x$ region with the fixed-order and the resummed theoretical predictions, highlighting the improvement in the description when NNLO+NLL$x$ theory is used.
We then study quantitatively the evidence for the onset of small-$x$ resummation introducing a set of estimators building upon the set of diagnostic tools of refs.~\cite{Caola:2010cy,Caola:2009iy}.
%

\paragraph{\itshape\mdseries The HERA data in the small-$x$ region.}

\begin{figure}[t]
\centering
  \includegraphics[width=0.49\textwidth,page=1]{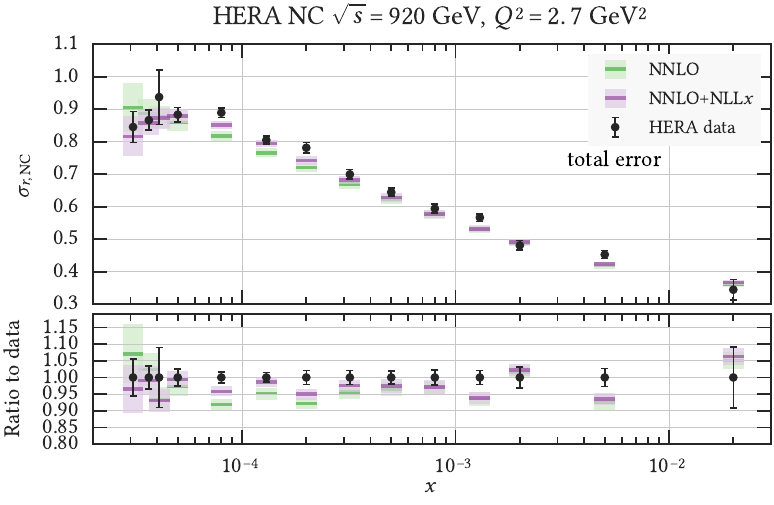}
  \includegraphics[width=0.49\textwidth,page=1]{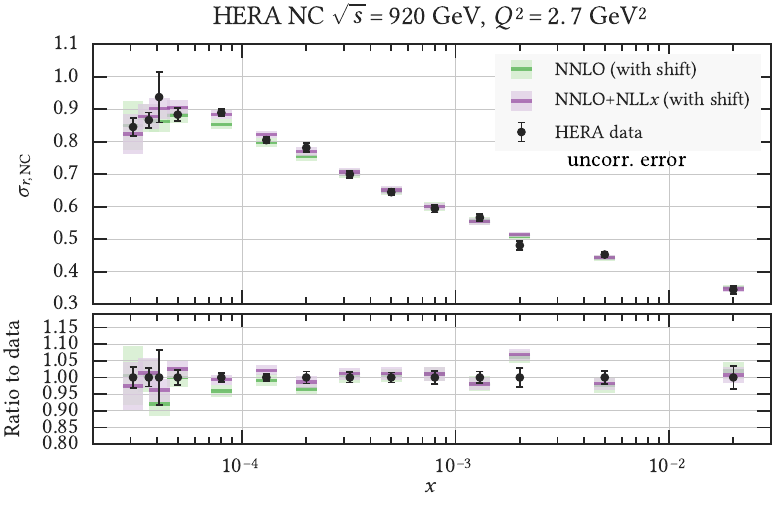}\\
  \includegraphics[width=0.49\textwidth,page=2]{figures/HERA920_samerange_thesis.pdf}
  \includegraphics[width=0.49\textwidth,page=2]{figures/HERA920_samerange_shift_thesis.pdf}  \caption{ Comparison between the HERA NC
   reduced cross section data from the $\sqrt{s}=920$~GeV
   dataset and the results of the NNLO and NNLO+NLL$x$ fits with the
   corresponding PDF uncertainties.    
     We show in the bottom panel the ratio of
    the theoretical predictions to the experimental data.
    The plots on the right show the theoretical prediction including the
    shifts as discussed in the text.}
  \label{fig:datathcomp}
\end{figure}

To investigate the improvement in the description in the HERA data in the low-$x$ and low-$Q^2$ region we first perform a comparison of the theoretical predictions using fixed-order and resummed predictions at NNLO and NNLO+NLL$x$, respectively.
In particular, we show in fig.~\ref{fig:datathcomp} the NC reduced cross section, defined as
\begin{equation}
\label{eq:sigmaRed2}
\sigma_{r,\text{NC}}(x,Q^2,y)\equiv \frac{d^2\sigma_{\text{NC}}}{dx dQ^2}\times
\frac{Q^4x}{2\pi\alpha Y_+}=F_2(x,Q^2)-\frac{y^2}{Y_+}F_L(x,Q^2)\, ,	
\end{equation}
where $Y_+=1+(1-y)^2$ and $y$ is the inelasticity eq.~\eqref{eq:xandy}.
The comparison is shown for the two lowest $Q^2$ bins above the kinematic cut of the $\sqrt{s} = 920$ GeV dataset, corresponding to $Q^2=2.7\, $ and $3.5$ GeV$^2$.
In the left plots, the uncertainty of the experimental data is computed as the sum in quadrature of the correlated and uncorrelated uncertainties, whereas the theoretical predictions include the associated PDF uncertainties.
In the right plots we include only the uncorrelated uncertainties, and the correlations are taken into account via shifts (see discussion in appendix~\ref{sec:qualityandmin}).
While the graphical comparisons provide some useful information, the quantitative agreement between data and theory must be judged from a detailed analysis of the fit quality, which we perform below.

We see that for values of $x$ larger than $5 \times 10^{-4}$ the NNLO and the NNLO+NLL$x$ predictions are essentially identical and that in both cases the predictions in the left plots tend to undershoot the data.
The two predictions start to differ from $x \lesssim 5 \times 10^{-4}$.
Around this value, we also observe a change in the slope of the experimental data: the cross section data stop rising and start decreasing.
As a consequence, the NNLO prediction starts to overshoot the data, whereas the resummed prediction is in reasonable agreement with the data.
We note that the differences between the fixed-order and the resummed predictions are relatively small and are limited to a few points in the small-$x$ region.
The two predictions differ by at most $10\%$ and only for the smallest values of $x$, as one can see the bottom panels of fig.~\ref{fig:datathcomp} where we show the ratio to the experimental data.
Nevertheless, thanks to the precision of the HERA dataset, these differences have a significant impact at the $\chi^2$ level, as we saw previously in sect.~\ref{sec:small-PDFs}.

It is tempting to relate the improvement in the description of the reduced cross section data to the role of the longitudinal structure function $F_L$, since its contribution to the reduced cross section is relevant at high $y$, which correspond to small $x$ and small $Q^2$ in the HERA kinematic case.
Since $F_L$ vanishes at the Born level, it receives gluon-induced contributions already at the first non-trivial order and it is therefore particularly sensitive to deviations from standard DGLAP evolution as we saw in sect.~\ref{sec:small-implementation}.

\begin{figure}[t]
\centering
  \includegraphics[page=1,width=0.6\textwidth]{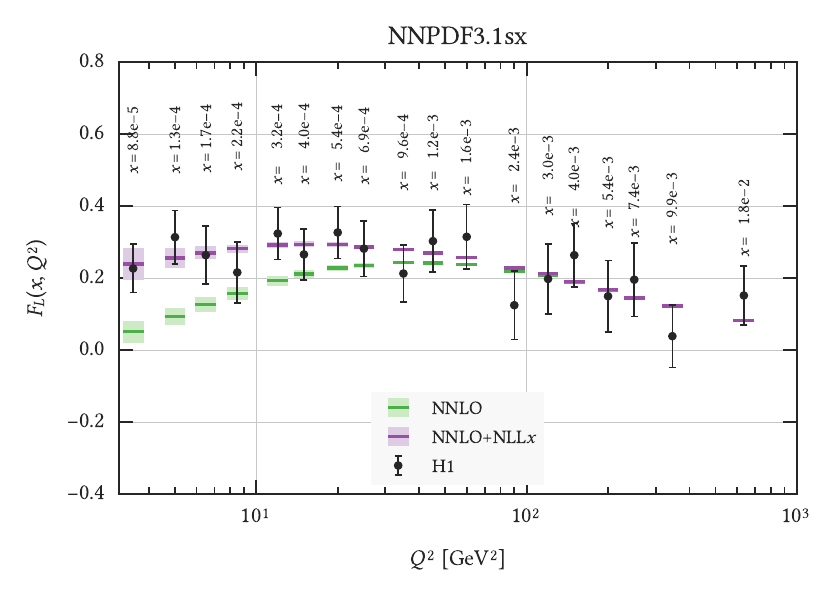}
  \caption{ The longitudinal structure function
    $F_L(x,Q^2)$ as a function of $Q^2$ for different $x$ bins
    for the most recent H1 measurement~\cite{Andreev:2013vha},
    comparing the results of the NNLO and NNLO+NLL$x$ fits.}
  \label{fig:FLres}
\end{figure}

It is therefore interesting to compare theoretical predictions for the longitudinal structure function $F_L$ using NNLO and NNLO+NLL$x$ theory.
We compare the latest measurements of $F_L$ obtained by the H1 collaboration to the NNLO and NNLO+NLL$x$ predictions obtained in fig.~\ref{fig:FLres}.
We show the total experimental uncertainties which have been added in quadrature.
Note that each value of $Q^2$ corresponds to a different $x$ bin as indicated in the plot.
We compute the theoretical predictions down to the lowest value of $Q^2$ for which they are are reliable, which is set by the parametrization scale $Q_0^2$.

The plot shows that there are significant differences between the fixed-order and the resummed predictions for $Q^2 \lesssim 100$ GeV$^2$.
The NNLO+NLL$x$ prediction is larger than the NNLO result by a significant amount; at $10$ GeV$^2$ the fixed-order calculation is more than a factor of two smaller than the resummed result.
At low values of $Q$, in particular, the NNLO+NLL$x$ result exhibits a flat behaviour, whereas the fixed-order bends down and approaches zero.
The flatter behaviour of the resummed curve translates into a smaller reduced cross section at small $x$.
As a consequence, the resummed prediction for the reduced cross sections features a more pronounced slope, which agrees with data better than the fixed-order prediction, which is instead harder at small $x$.

\begin{figure}[t]
\centering
  \includegraphics[width=0.495\textwidth,page=1]{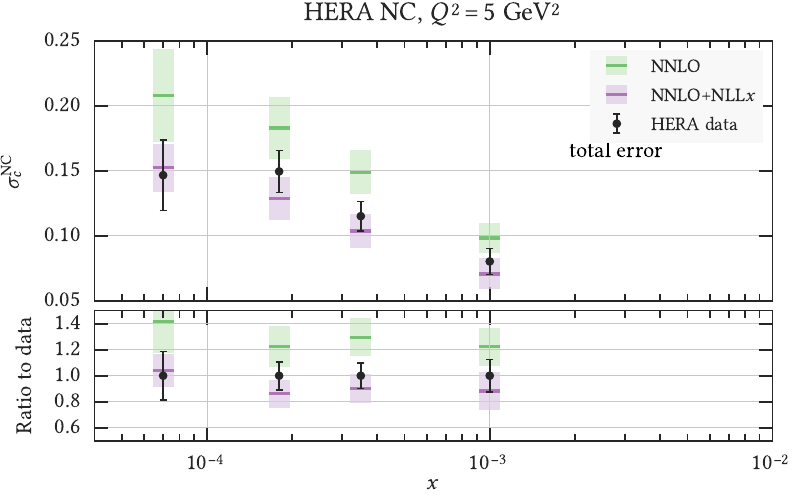}
  \includegraphics[width=0.495\textwidth,page=1]{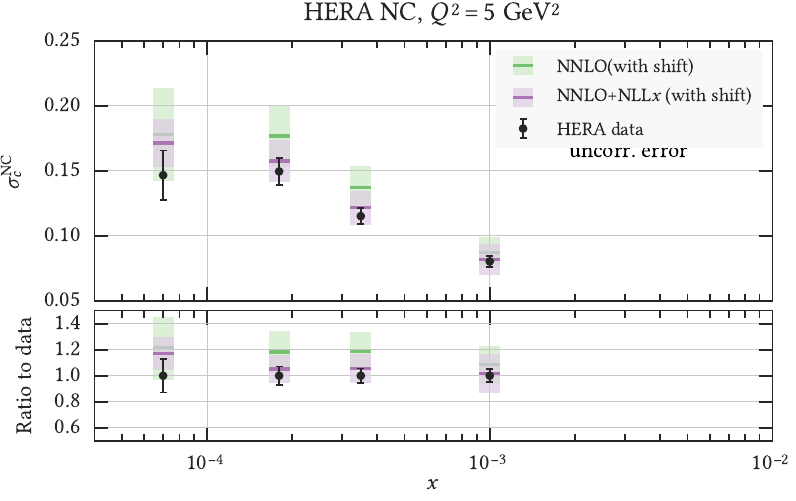}
  \includegraphics[width=0.49\textwidth,page=2]{figures/HERAF2C_samerange_NNLO_thesis.pdf}
  \includegraphics[width=0.49\textwidth,page=2]{figures/HERAF2C_samerange_NNLO_shift_thesis.pdf}
  \caption{ Same as Fig~\ref{fig:datathcomp} for the
  HERA charm production cross sections.}
  \label{fig:datathcompF2c}
\end{figure}

In fig.~\ref{fig:datathcompF2c} we show a similar comparison to fig.~\ref{fig:datathcomp}, now for the HERA charm production reduced cross sections for the two lower bins above the $Q^2_{\text{min}}$ cut.
We find that especially in the $Q^2=5$ GeV$^2$ bin the NNLO+NLL$x$ prediction is in better agreement with the HERA data, whereas the NNLO one overshoots them.
The data are instead somewhat in the middle of the two theoretical predictions for the $Q^2=7$ GeV$^2$ bin.
We note that the HERA charm data are extracted from the fiducial cross section~\cite{Abramowicz:1900rp} by extrapolating to the full phase space using the fixed-order calculation at $\mathcal O(\as)$ based on the fixed-flavour number scheme.
Contrary to this, the inclusive NC structure function measurements are determined from the outgoing lepton kinematics and therefore do not assume any theory input. 
It is therefore possible that a more consistent analysis of the raw data based on an extrapolation using resummed theory might further improve the already good agreement of the charm cross section with the NNLO+NLL$x$ fit.

\paragraph{\itshape\mdseries Quantifying the onset of BFKL dynamics in HERA data.}

To quantify more precisely the onset of BFKL dynamics in the small-$x$ and small-$Q^2$ region of the HERA data we now introduce several statistical estimators.
We start by performing a detailed $\chi^2$ analysis to identify the regions in the $(x,Q^2)$ plane which are responsible for the improvement in the $\chi^2$ observed in sect.~\ref{sec:small-PDFs} when one supplements the NNLO theory with small-$x$ resummation.

\begin{figure}[t]
  \centering
    \includegraphics[width=0.41\textwidth]{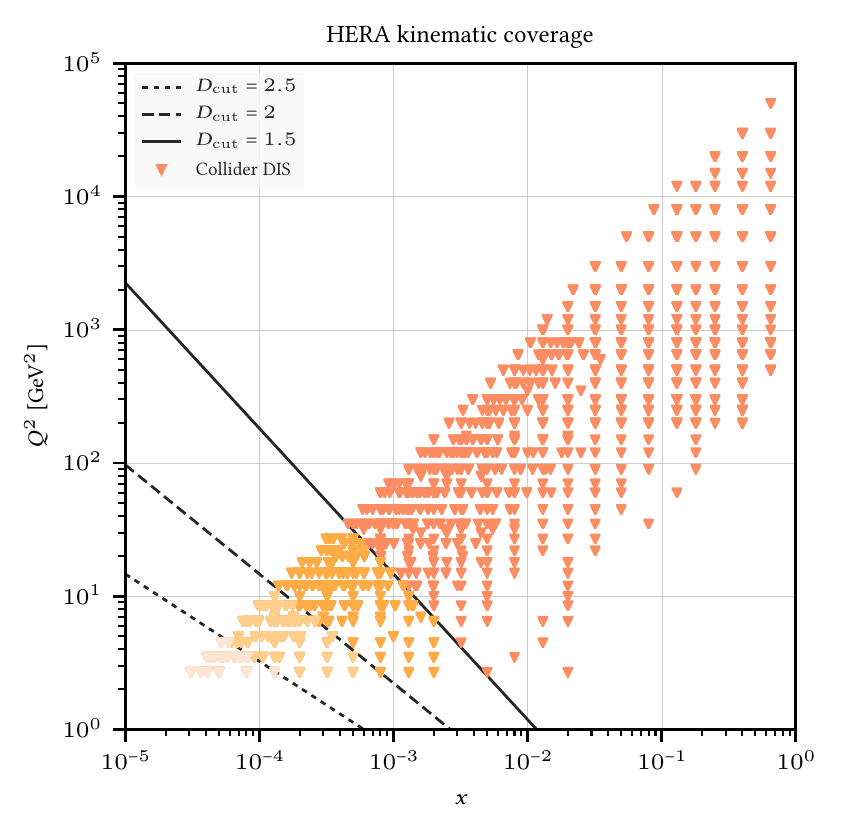}\qquad \qquad
    \includegraphics[width=0.41\textwidth]{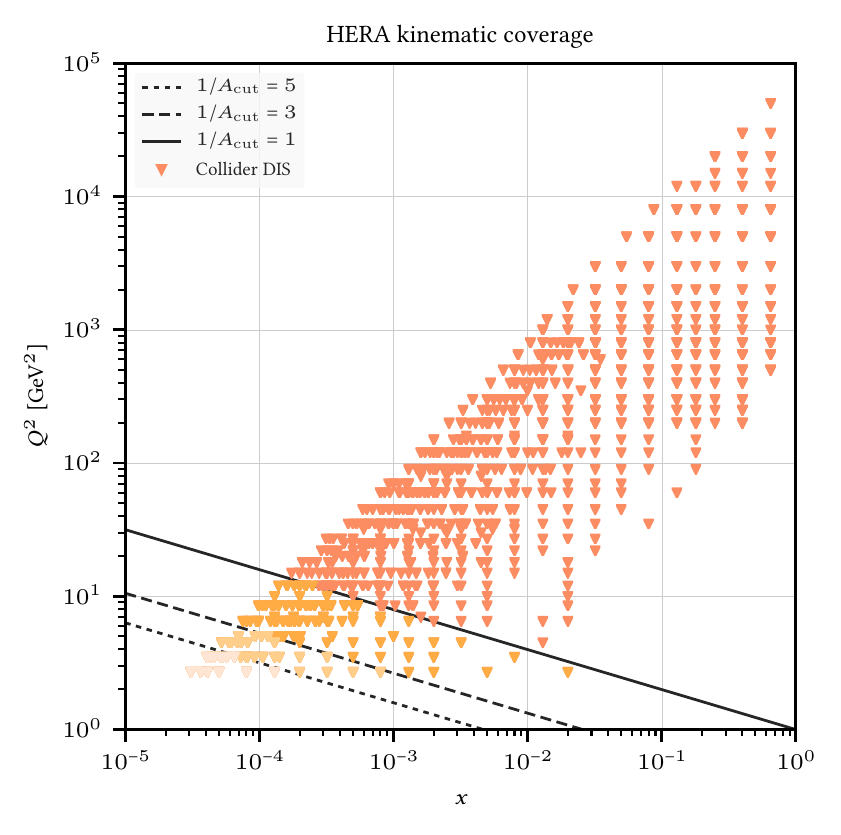}
    \caption{ The kinematic coverage of the
      HERA inclusive structure function data that enters
      the NNPDF3.1sx fits. 
      The tilted lines represent illustrative values of the
      cut to DIS data applied after the fit to study evidence for
      BFKL effects at small-$x$ and small-$Q^2$.
   Left plot: perturbative-inspired cut eq.~(\ref{eq:dcut});
   right plot: saturation-inspired cut
   eq.~(\ref{eq:saturationcut}).
  The data points affected by the various cuts are plotted
   with different shades.
    }
  \label{fig:kincutDIS}
\end{figure}

To this end, we have recomputed the value of $\chi^2/\Ndat$ of the HERA inclusive and charm cross section with the NNPDF3.1sx NLO, NNLO, NLO+NLL$x$ and NNLO+NLL$x$ fits using the default choice $H_{\text{cut}}=0.6$, now removing the data points for which
\begin{align}\label{eq:dcut}
	\as(Q^2) \ln \frac{1}{x} \geq D_{\text{cut}}.
\end{align}
By varying the value of $D_{\text{cut}}$ in eq.~\eqref{eq:dcut} we can vary the number of points excluded in the computation of the $\chi^2/\Ndat$: the smaller its value, the higher is the number of points at small $x$ and $Q^2$ which are cut away. 
We show in fig.~\ref{fig:kincutDIS} the HERA structure function data in the $(x,Q^2)$ plane which are cut for some representative values of $D_{\text{cut}}$ eq.~\eqref{eq:dcut}.
Let us stress that this cut is fundamentally different from the cut $H_{\text{cut}}$ defined in eq.~\eqref{eq:hadrcut}. 
Here the parameter $D_{\text{cut}}$ is applied only to DIS structure functions and is used as an \textit{a posteriori} tool after the fit has been performed.

\begin{figure}[t]
  \centering
    \includegraphics[width=0.41\textwidth]{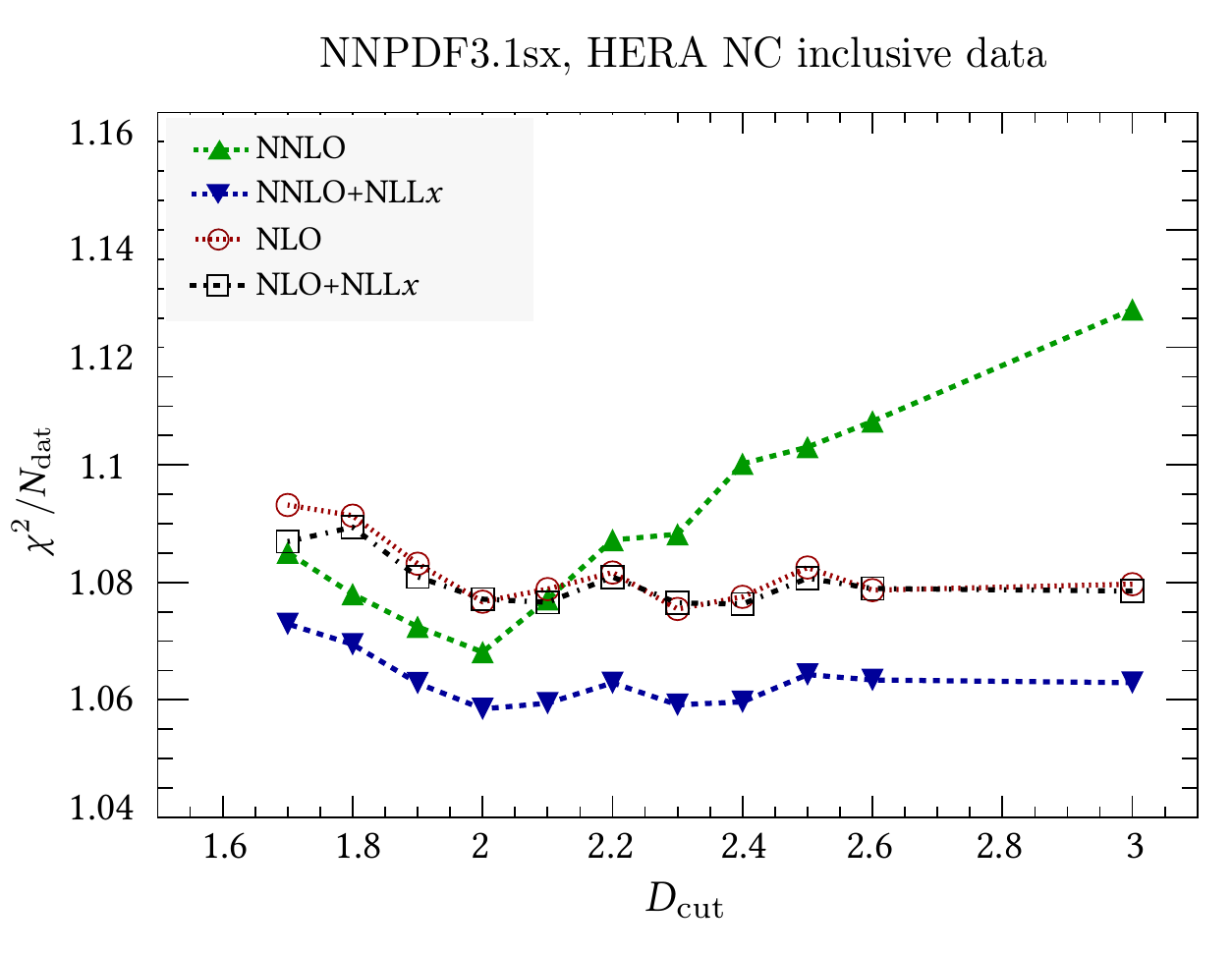}\qquad \qquad
    \includegraphics[width=0.41\textwidth]{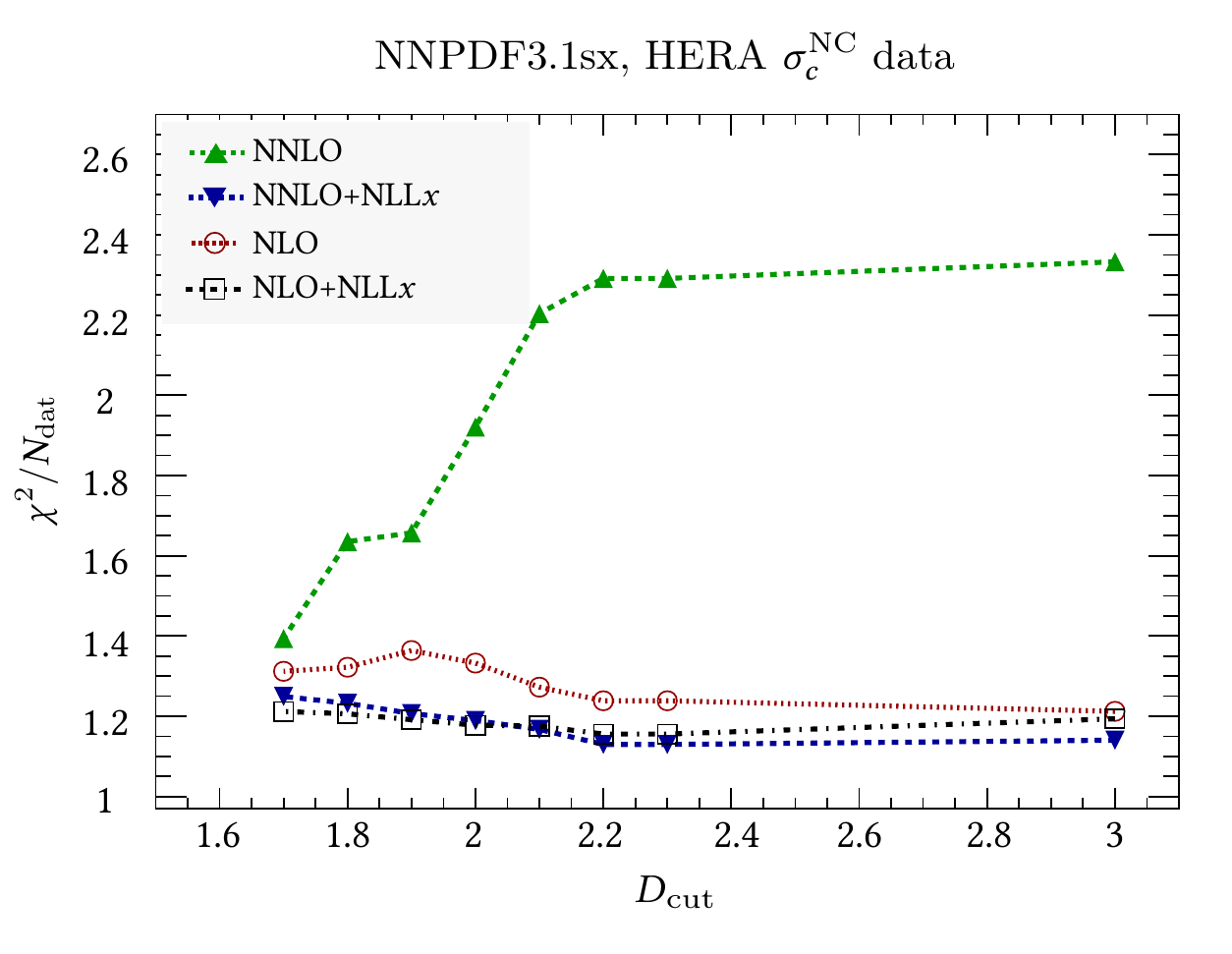}
   \caption{ Upper left: the values of $\chi^2/N_{\text{dat}}$
     in the NNPDF3.1sx global fits
    for the HERA NC inclusive structure function data for different
    values of the cut $D_{\text{cut}}$ eq.~(\ref{eq:dcut}), comparing
     the results of the NLO, NLO+NLL$x$, NNLO,
    and NNLO+NLL$x$ fits.
    Upper right: same as above for the HERA
    charm production data.
   }
   \label{fig:chi2prof}
\end{figure}

We display the values of $\chi^2/\Ndat$ for the HERA NC inclusive and charm data reduced cross sections as a function of $D_{\text{cut}}$ in fig.~\ref{fig:chi2prof}. 
We immediately note that the $\chi^2/\Ndat$  at NNLO increases steadily for values of $D_{\text{cut}} \gtrsim 2 $, whereas it remains flat for the NNLO+NLL$x$, NLO, and NLO+NLL$x$ fits.
We can trace this behaviour with what we observed in sect.~\ref{sec:small-PDFs}, where we saw how the latter theories are rather similar to each other.
The best fit quality is achieved in the NNLO+NLL$x$ case, highlighting the importance of the NNLO corrections to describe the precise HERA data in the medium- and large-$x$ region.
On the other hand, NNLO theory alone is not sufficient to achieve a good description of the HERA data at small $x$ and $Q^2$.
The best description is achieved by supplementing NNLO with NLL$x$ resummation, providing a compelling evidence of the need of small-$x$ resummation to describe the small-$x$ region in a satisfactory manner.

A similar analysis to the one just described was performed in refs.~\cite{Caola:2010cy,Caola:2009iy}, though the cut on the data in the small-$x$ and small-$Q^2$ region was inspired by saturation arguments.
In refs.~\cite{Caola:2010cy,Caola:2009iy} the condition used to discard data points was
\begin{equation}\label{eq:saturationcut}
	Q^2 x^\lambda \geq A_{\text{cut}},
\end{equation}
where $\lambda=0.3$.
The larger the value of $A_{\text{cut}}$, the more points are excluded. 
Whilst the motivations for the two cuts are different, from a practical point of view the result is rather similar, with some differences in the shape of the cut in the $(x, Q^2)$ plane as shown in fig.~\ref{fig:kincutDIS}.
In particular, the trend of the $\chi^2/\Ndat$ values as a function of $1/A_{\text{cut}}$ is similar to what we showed above using  $D_{\text{cut}}$~\cite{Ball:2017otu}.

These results demonstrate the onset of BFKL dynamics for both the inclusive and the charm data in the small-$x$ and $Q^2$ region.
Since the deterioration of the fit quality at NNLO starts at $D_{\text{cut}} \sim 2 $,  the onset of BFKL effects is approximately situated at
\begin{equation}
	\ln \frac{1}{x	} \geq 1.2 \ln \frac{Q^2}{\LambdaQ^2},
\end{equation}
which corresponds to a value of $x \sim 3 \times 10^{-4}$ for $Q^2 = 6.5$ GeV$^2$, consistent with the results presented above in sect.~\ref{sec:small-PDFs}.
Our results offer also useful guidance to estimate the region where small-$x$ resummation becomes phenomenologically relevant at the LHC or in future colliders.

\begin{figure}[t]
  \begin{center}
    \includegraphics[width=0.41\textwidth]{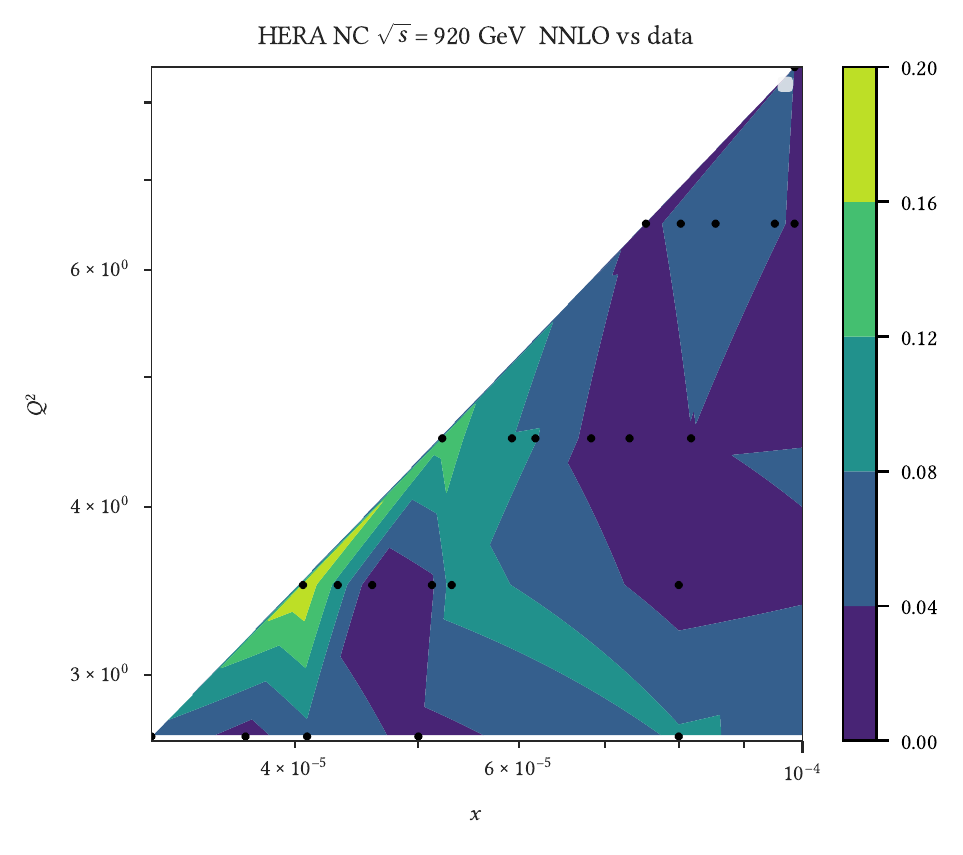}\qquad\qquad
    \includegraphics[width=0.41\textwidth]{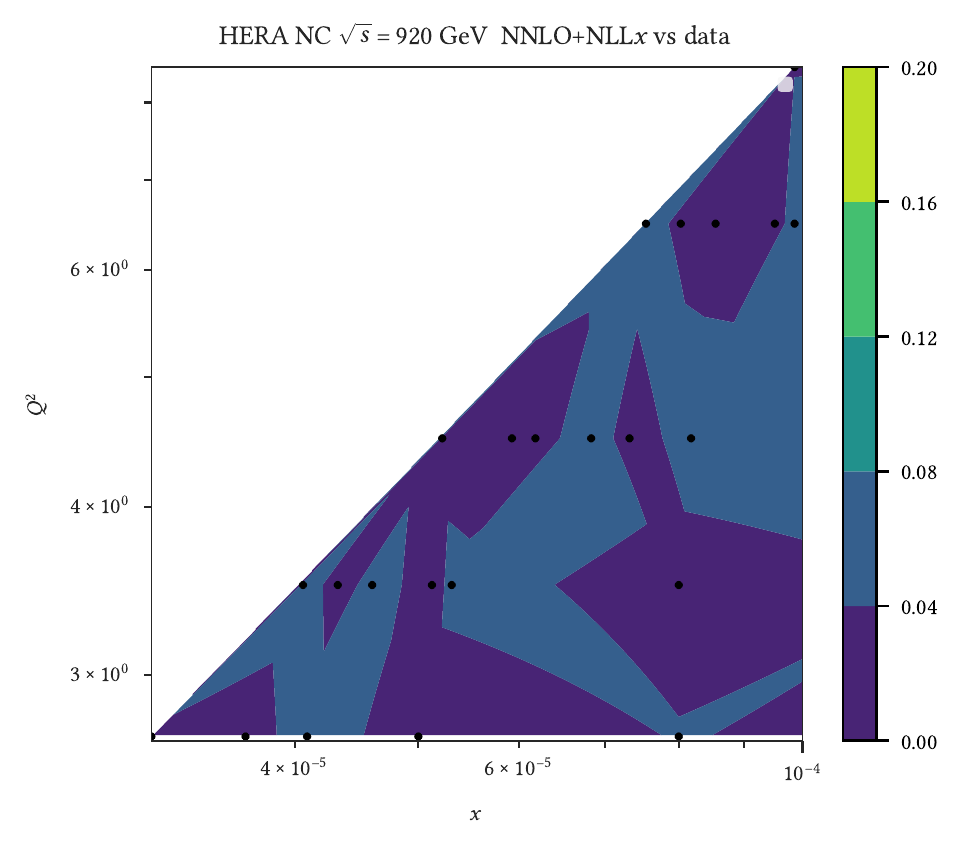}
    \caption{Left panel: interpolated representation of the relative pull
      eq.~(\ref{eq:relativedistances})
      between the HERA NC reduced cross section data at $\sqrt{s}=920$ GeV and the
      NNLO fit, in the small-$x$
      and small-$Q^2$ region.
      Right panel: same as the left panel now
      for the NNLO+NLL$x$ fit.
  }
  \label{fig:reldiff-data}
\end{center}
\end{figure}

A complementary tool to further investigate the onset of novel QCD dynamics in the small-$x$ region is the relative pull between experimental data and theory, defined as
\begin{equation}
	\label{eq:relativedistances}
P_i^{\text{rel}}(x,Q^2)\equiv  \frac{\big|\sigma_{{\text{data}},i}-\sigma_{{\text{th}},i}\big|}{
(\sigma_{{\text{data}},i}+\sigma_{{\text{th}},i})/2 
} \, ,
\end{equation}
where the normalization is given as the average of the central values.
This estimator allows us to quantify the absolute size of the differences between data and theory in units of the cross section.
In fig.~\ref{fig:reldiff-data} we represent the relative pull eq.~\eqref{eq:relativedistances} as a function of $(x,Q^2)$ in the kinematic region relevant for the HERA data.
We show an interpolation of the relative pull for the data points of the $\sqrt{s}=920$ GeV HERA NC dataset for the NNLO (left panel) and the NNLO+NLL$x$ (right panel) fits. 
In the former case, the relative differences between data and theory can be as large as $20\%$ at small $x$ and $Q^2$, whereas they reduce to a few percent at larger values of $x$ and $Q^2$.
In the latter case, the agreement between data and theory is essentially the same in all the kinematic regions considered, with differences always below the $8\%$ level.
This result offers yet another indication that NNLO+NLL$x$ provides a satisfactory description of the whole inclusive HERA dataset.

\subsection{Small-$x$ resummation beyond HERA}\label{sec:smallLHC}

The search for evidence of BFKL dynamics at small-$x$ has been an ongoing enterprise ever since the HERA collider started operations about 25 years ago.
Despite some hints being reported, until now no conclusive evidence had been found in the HERA inclusive deep-inelastic structure functions.
The results discussed in this section demonstrate that including small-$x$ resummation stabilizes the perturbative expansion of the DIS structure functions at small-$x$ and $Q^2$ and thus also of the PDFs extracted from them.
Indeed, the PDFs obtained with small-$x$ resummation using NLO+NLL$x$ and NNLO+NLL$x$ theory are in much closer agreement with each other at medium and small $x$ than the corresponding fixed-order NLO and NNLO PDFs.
This suggests in turn that the theoretical uncertainty due to missing higher order corrections in a NNLO+NLL$x$ resummed calculation is smaller (at least in the small-$x$ region) than that of the corresponding fixed-order NNLO calculation.
This result is reflected in the marked improvement of the quantitative description of the HERA inclusive structure function data at small-$x$ when fixed order is supplemented by resummation: the NNLO+NLL$x$ theory describes the low $Q^2$ and low $x$ bins of the HERA data just as well as it describes the data at higher $Q^2$ and larger $x$.

The relevance of small-$x$ resummation, however, goes beyond the improved description of the HERA data. 
It has been understood for some time that the effect of resummation on the evolution of the PDFs can have a significant impact on the shape of parton luminosities and thus of hadronic cross sections at the LHC~\cite{Ball:2007ra}. 
Indeed, the inclusion of small-$x$ resummation in a PDF fit to low-$Q^2$ data at small $x$ has a sizeable effect on parton luminosities even at high scales~\cite{Ball:2017otu}.
As a consequence, small-$x$ resummation might have significant effects at the LHC, either at low invariant masses or at high rapidities.
The accurate description of processes in these kinematic regions will therefore require small-$x$ resummation.
Conversely, present and future LHC measurements might provide further evidence for the onset of BFKL dynamics, this time in proton-proton collisions.
\mccorrect{Since the improvement in the description of the HERA data is concentrated in the low-$x$, low-$Q^2$ region (high $y$ in DIS kinematics), measurements at the LHC would allow one to investigate the small-$x$ region at relatively high $Q^2$, thus breaking the ($x$-$Q^2$) degeneracy}\footnote{\mccorrect{Due to the kinematic correlations between $x$ and $Q^2$ at fixed and high $y$, in refs.}~\cite{Harland-Lang:2016yfn,Abramowicz:2015mha} \mccorrect{an improvement in the description of the HERA data was found by adding a higher-twist correction, with no $x$ dependence, to $F_L$.}}.
A further probe of BFKL dynamics is provided by the ultra high-energy neutrino-nucleus cross sections, where differences in event rates could be observed by upcoming measurements with neutrino telescopes such as IceCube and KM3NET~\cite{Ball:2017otu,Bertone:2018dse}.

Finally, high-energy resummation also plays a crucial role in shaping the physics case for future high-energy lepton-proton colliders, such as the LHeC and the FCC-eh.
These colliders would extend the coverage of HERA by up to two orders of magnitude deep into the small-$x$ region. 
In this respect, the NNPDF3.1sx fits can be used to improve the accuracy of existing calculations of deep inelastic scattering processes in the kinematic regions which would eventually be probed by these new machines.

\section{Towards parton distribution functions with double resummation}\label{sec:jointlargesmall}

In this chapter we have presented two global PDF analyses which incorporate the effect of large-$x$ and of small-$x$ resummation, respectively.
The main limitation of these analyses is the need to impose stringent cuts to the fitted hadronic data, to ensure that the contamination from unresummed partonic cross sections is kept to a minimum.

On the one hand, it is now well understood how to combine resummation corrections to partonic cross sections with resummed parton luminosities to obtain fully resummed cross sections, even in the more complex case of small-$x$ resummation.
On the other hand, for several processes theoretical calculations are still not available in a format amenable to systematic phenomenology.
Therefore, some effort is still required before they can be used in PDF fits.  
Future work in this direction will allow for the inclusion of a wider range of hadron collider data into a fully consistent resummed global fit by removing the need for hadronic cuts.
This way, it would be possible to obtain resummed PDFs with an accuracy competitive to that of fixed-order mainstream global fits.

As LHC data become ever more precise, the theoretical challenge is to reduce theoretical uncertainties down to the percent level.
At this level of precision, all-order calculations which simultaneously resum small-$x$ and large-$x$ logarithms will likely be required for processes at high rapidity.
Indeed, in such processes, such as forward Drell-Yan and $D$ meson production at LHCb,  one of the partons is at very small $x$, whereas the other is at very large $x$.
\mccorrect{
For consistency, the PDFs to be used in these calculations should be determined using a double-resummed theory.
}
The first calculation which combines small-$x$ and large-$x$ resummation has been recently presented in ref.~\cite{Bonvini:2018ixe} (see also~\cite{Bonvini:2018iwt}) for the inclusive Higgs production in gluon fusion.
The technique presented there can be generalized and applied to other processes more relevant in the context of PDF fits, such as DY or heavy-quark production, \mccorrect{and in particular it can be extended to differential distributions.
In the inclusive case, since the PDFs are depleted at large $x$, the dominant contribution in the convolution integral eq.~}\eqref{eq:hadronfact2} \mccorrect{comes from the region where the coefficient functions are probed at somewhat large values of $z$ and the PDF luminosities (and thus the PDFs) at relatively small values of $x=\tau/z$.
However, distributions differential in rapidity provide an additional set of constraints, thus allowing for a more precise determination both in the small-$x$ and in the large-$x$ region.
In particular, as the number of available datapoints differential in rapidity increases, the impact of threshold resummation on PDF fits, currently moderate at NNLO, will likely become more significant, thanks to the reduction in the PDF uncertainty.
The simultaneous inclusion of high-energy and threshold resummations in the theoretical predictions will in turn allow for PDFs determined with an unprecedented level of accuracy.}
The calculation of ref.~\cite{Bonvini:2018ixe} therefore opens up the possibility of global PDF fits where fixed-order is supplemented by both resummations, blazing a trail towards calculations which consistently combine high-energy and threshold resummations.

\begin{savequote}[8cm]

\textlatin{Science does not aim at establishing immutable truths and eternal dogmas; its aim is to approach the truth by successive approximations, without claiming that at any stage final and complete accuracy has been achieved.}
 \qauthor{--- Bertrand Russell, \textit{ABC of Relativity}}
\end{savequote}

\chapter{\label{ch-respheno}Resummation for Higgs processes} 


\lettrine[lines=2]{T}{he discovery} of the Higgs boson was the major achievement of the LHC run I~\cite{Aad:2012tfa,Chatrchyan:2012xdj} and established the Standard Model as a successful theory of fundamental interactions.
The first measurements performed by the ATLAS and CMS experiments allowed for the determination of the mass of the Higgs boson $m_H$ and its coupling strengths to other SM particles.
Once $m_H$ is known, all the properties of the SM Higgs boson are completely predicted.
Increasingly precise measurements have so far confirmed that all the observed properties of the new particle, such as its quantum numbers and couplings, are consistent with SM expectations.

The start of the LHC run II brought us into the Higgs precision era.
Since the data collected suggest that new physics (NP) will not likely manifest as a direct signal, it becomes progressively important to study the properties of the Higgs boson in detail to measure possible deviations from what the SM predicts.
The measurements of the Higgs production cross section and differential distributions and their comparison with accurate theoretical calculations therefore are of paramount importance.
Moreover, a reliable estimate of the associated theoretical uncertainties becomes essential at this level of precision.

In this chapter we will consider the production of a Higgs boson in gluon-fusion and we will present accurate predictions for its inclusive cross section and its transverse-momentum distribution at small $p_t^H$.
Besides their importance from a purely phenomenological perspective, these studies offer an opportunity to achieve a deeper theoretical understanding.
Indeed, the relative simplicity of these observables allows one to compute predictions which include several orders of perturbative corrections, thus providing an ideal framework to probe many non-trivial features of QCD.

\section{The Higgs inclusive cross section}\label{sec:totalhiggs}

Though it is not directly measurable, the inclusive cross section for Higgs production is an observable of great interest, both experimentally --- being used as normalization for differential distributions --- and theoretically, due to the poor convergence properties of its perturbative series.
The dominant production channel at the LHC is the gluon-fusion mode, where the Higgs boson is generated by the fusion of two gluons through a fermion loop, predominantly top quarks.
The cross section can be written as 
\begin{align}
	&\sigma(\tau, \mH^2, \mt^2) = \sigma_0(\mH^2,\mt^2) \tau \sum_{ij} \int_\tau^1 \frac{dz}{z} \mathcal L_{ij} \left(\frac{\tau}{z},\muF^2 \right) C_{ij} \left(z, \mH^2, \mt^2, \as(\muR^2), \frac{\mH^2}{\muF^2}, \frac{\mH^2}{\muR^2}\right),  &\tau=\frac{\mH^2}{s} ,
\end{align}
where $\mathcal L_{ij} (z, \mu^2)$ is the parton luminosity eq.~\eqref{eq:partonlumi} and the prefactor $\sigma_0$ is chosen such that $C_{gg}$ is normalized to $\delta(1-z)$.
The dimensionless coefficient functions admit an expansion in the strong coupling $\as$ (for the ease of notation, we let the dependence on the factorization scale $\muF$ and renormalization scale $\muR$ be understood)
\begin{align}\label{eq:Cseries}
	C_{ij} (z, \mH^2, \mt^2, \as) &= \delta_{ig}\delta_{jg} \delta(1-z) + \as C_{ij}^{(1)} (z, \mH^2, \mt^2)\nonumber \\
	&\quad + \as^2 C_{ij}^{(2)} (z, \mH^2, \mt^2)+ \as^3 C_{ij}^{(3)} (z, \mH^2, \mt^2)	+\mathcal O (\as^4).
\end{align}
Whilst the NLO coefficient $C_{ij}^{(1)}$ is known exactly~\cite{Spira:1995rr} and the NNLO coefficient is known as an expansion in $\mH^2/\mt^2$, the milestone computation of the third-order coefficient has only recently been performed in the large-$\mt$ effective field theory (EFT)~\cite{Anastasiou:2015ema,Anastasiou:2016cez,Mistlberger:2018etf}.
Indeed, since the top quark is heavier than the SM Higgs, it is possible to integrate out the top quark such that the Higgs couples directly to the gluons through an effective pointlike vertex.
In the EFT, the dependence on the top mass is encoded in a Wilson coefficient squared $W$ and the coefficient function further factorizes
\begin{equation}
	C_{ij} (z, \mH^2, \mt^2, \as) = W(\mH^2,\mt^2) \tilde C_{ij} (z, \as),
\end{equation}
where $ \tilde C_{ij} (z, \as)$ has an expansion in $\as $ analogous to eq.~\eqref{eq:Cseries} and $W=1+\mathcal O(\as)$.
The EFT is usually improved by rescaling the cross section by the ratio of the exact LO over the LO in the EFT, leading to the so-called rescaled effective field theory (rEFT).
The rEFT is known to be a very good approximation for $\mH \lesssim \mt$ for not too high collider energies, since the large-$s$ behaviour of the coefficient functions exhibits a double-logarithmic scaling at high energy~\cite{Hautmann:2002tu}, while the full theory features only a single-logarithmic enhancement~\cite{Marzani:2008az}.

The QCD corrections to the inclusive cross sections are huge.
The NLO correction amounts to more than $100\%$ of the LO cross section, whereas the NNLO correction adds another $\sim 80\%$. 
The N$^3$LO correction appears to be small, thus indicating that the perturbative series finally manifests convergence.
Due to the bad convergence properties of the perturbative series, the estimate of the uncertainty from missing higher order (MHOU) using canonical seven-point scale variations does not appear to be reliable at fixed order.
For these reasons, there has been a complementary effort to improve the convergence of the perturbative expansion by computing higher-order corrections resorting to all-order resummation methods. 
In the case of the inclusive cross section in gluon fusion, the most studied all-order technique is large-$x$ (threshold) resummation, which we introduced in chapter~\ref{ch-res}.
Threshold resummation for Higgs production in gluon fusion is currently known to N$^3$LL$^\prime$ accuracy~\cite{Bonvini:2014joa,Catani:2014uta,Bonvini:2014tea,Schmidt:2015cea}.
In this section we discuss the impact of threshold resummation on the N$^3$LO result.
Furthermore, we show that a robust estimate of the theoretical uncertainties due to MHOU can be obtained by considering various forms of threshold resummation which differ by the treatment of subleading terms.
We finally compare the results obtained in direct QCD to those obtained using a SCET approach.

\subsection{Threshold resummation}

The general formalism for threshold resummation has been introduced in sect.~\ref{sec:threshold}.
Here we analyse the case of Higgs production in gluon fusion and review various treatments of the subleading terms following the discussion of refs.~\cite{Bonvini:2014joa,Bonvini:2014tea,Bonvini:2016frm,Ahmed:2016otz}.
Since soft-gluon resummation affects only the gluon-gluon channel, we will drop the flavour indices and let it be understood that we focus on the $gg$ channel.
Under Mellin transform, the cross section factorizes
\begin{equation}
{\bm \sigma}(N, \mH^2) \equiv 	\int_0^1 d\tau \, \tau^{N-1}\frac{ \sigma(\tau, \mH^2)}{\tau} = \sigma_0 {\bm{\mathcal L}}(N) \bm C(N, \as).
\end{equation}
In Mellin space, the threshold limit $z\rightarrow 1$ corresponds to the $N\rightarrow \infty$ limit, and the $N$-space resummed coefficient function takes the form eq.~\eqref{eq:CataniTh1}
\begin{equation}
	\bm C_{\text{res}} (N, \as) = \bar g_0 (\as) \exp \bar {\mathcal S} (\as, N).
\end{equation}
Let us note that in the full theory the top-mass dependence is in $\bar g_0$ and further factorizes in the rEFT as
\begin{equation}
	\bar g_0 (\as)  = W(\mH^2, \mt^2) \tilde {\bar {g}}_0 (\as)
\end{equation}
where now the function $ \tilde {\bar {g}}_0$ does not depend on the top mass.
Currently, all the ingredients necessary to reach N$^3$LL$^\prime$ accuracy in the large-$\mt$ limit are known~\cite{Moch:2005ba, Moch:2005ky,Laenen:2005uz,Anastasiou:2014vaa} with the only exception of the four-loop contribution to the cusp anomalous dimension which is known only numerically~\cite{Moch:2017uml,Moch:2018wjh}).
Its value here is computed with a Pad\'e estimate~\cite{Moch:2005ba}, though a numerical analysis shows that its impact on the resummed result is negligible.

Since the computation of the integrals in eq.~\eqref{eq:CataniTh1} is rather cumbersome, it is usually performed in the large-$N$ limit.
The resulting integrals are written as a function of $\ln N$ only and the all-order resummed coefficient function results in what we dubbed $N$-soft prescription (cfr. eq.~\eqref{eq:CataniTh2}) in sect.~\ref{sec:threshold}
\begin{equation}
	\bm C_{N-\text{soft}} = g_0 (\as)  \exp {\mathcal S} (\as, N).
\end{equation}
We recall that in standard $N$-soft resummation all the constant terms are removed from the exponent and collected in the function in front (which we denote by $g_0$, without bar). 
Similarly, the function $\bar {\mathcal S} $ changes to a function ${\mathcal S}$ which contains only logarithmic terms (explicit formul\ae\ for ${\mathcal S}$, $\bar g_0$ and $g_0$ are collected in appendix \ref{app:dQCD}). 

Besides $N$-soft, there exist several prescriptions, formally equivalent in the $N\rightarrow \infty$ limit, which differ by either power-suppressed (subdominant) $1/N$ or subleading logarithmic contributions.
We refer the reader to ref.~\cite{Bonvini:2014joa} for a thorough discussion.
Here we consider a variant of the $N$-soft resummation, the so-called $\psi$-soft prescription, based on the simple replacement $\ln N \rightarrow \psi_0 (N)$ in the $N$-soft expression, where $\psi_0(N)$ is the Euler digamma function.
This prescription has the advantage of reproducing the function $\bar {\mathcal S} $ up to $\mathcal O (1/N^2)$ corrections, whilst $N$-soft reproduces $\bar {\mathcal S} $ up to $\mathcal O (1/N)$ corrections.
However, this does not make $\psi$-soft accurate at the next-to-soft (NS) level, since the original expression eq.~\eqref{eq:CataniTh1} was not.
Nevertheless, a class of NS terms can be predicted to all orders by performing a collinear improvement~\cite{Bonvini:2014joa}.
Indeed, the cusp anomalous dimension $A_{\text{cusp}}(\as(\mu^2))$ is the coefficient of the divergent part in the $z\rightarrow 1$ limit of the $P_{gg} (z,\as)$ splitting function.
Therefore, one can retain more terms in the soft expansion of $P_{gg}$ in powers of $1-z$. 
As shown in ref.~\cite{Bonvini:2014joa}, by retaining the LO gluon splitting function at order $(1-z)^{k-1}$ one accounts for the LL N$^k$S terms correctly to all orders.

Since these corrections correspond to additional powers of $1-z$, the net effect in Mellin space is to shift the value of $N$.
Here we use as our default prescription $\psi$-soft$_2$ (or $\psi$-soft AP2), obtained by expanding the $P_{gg}$ splitting function to second order in $1-z$, leading to the $N$-space expression
\begin{equation}
	\bm C_{\psi\text{-soft}_2} (N, \as) = g_0 (\as) \exp\left[ 2 \mathcal S(\as, \psi_0(N)) - 3 \mathcal S (\as, \psi_0(N+1)) + 2 \mathcal S(\as, \psi_0 (N+2)) \right].
\end{equation}
This prescription leads to more reliable results, in the sense that it offers a good approximation to the exact result when expanded at fixed order.
Alternatively, one can keep only the first order ($\psi$-soft$_1$ or $\psi$-soft AP1), thus obtaining an expression which differs from $\psi$-soft$_2$ from subdominant contributions,
\begin{equation}
	\bm C_{\psi\text{-soft}_1} (N, \as) = g_0 (\as) \exp \mathcal S(\as, \psi_0(N+1)).
\end{equation}
The comparison of the results obtained with the two prescriptions can be used as an estimate of the missing  $1/N$ terms.

Another source of uncertainty at the resummed level comes from subleading logarithmic terms.
It is possible to probe these subleading contributions by moving some or all constant terms from the exponent to the function in front.
Indeed, though the logarithmic accuracy does not change, different subleading terms are generated by interference with the constant contributions.
The default (most natural) choice for the position of the constant terms is determined by retaining in the exponent those terms which naturally arise there as in eq.~\eqref{eq:CataniTh1} and by collecting the remaining $N$-independent terms in the function $\bar g_0$. 
However, one can consider also variations in which all terms are put in $g_0$ --- as in ordinary $N$-soft --- or all the constant are exponentiated,
\begin{equation}
	\bm C_{\text{res}} (N, \as) =  \exp\left[\ln \bar g_0 (\as) +  \bar {\mathcal S} (\as, N) \right],
\end{equation}
where $\ln \bar g_0$ is meant to be expanded to the appropriate order. 
Since it is well known that the constant terms play a significant role in Higgs production~\cite{Ahrens:2008qu,Berger:2010xi,Stewart:2013faa}, to obtain a robust estimate of the perturbative uncertainty one can use these two options to assign an uncertainty to the default expression in which only the terms arising naturally at the exponent are kept exponentiated.

Before moving to discuss the results, let us recall that threshold resummation can also be performed in the SCET framework, as we discussed in sect.~\ref{sec:threshold}.
In the SCET formalism, the partonic coefficient function is written in a factorized form~\cite{Ahrens:2008nc}
\begin{equation}
	C(z, \as, \muF^2) = H (\muH^2) S(z, \muS^2) U(\muH^2, \muS^2, \muF^2),
\end{equation}
where the hard function $H$ and the soft function $S$ are evaluated at a hard and soft scales $\muH$ and $\muS$ respectively, such that their perturbative expansions are well behaved.
In the large-$\mt$ limit, the hard function is further factorized.
The evolution $U$ from the soft scale to the hard scale performs the resummation of the potentially large logarithms due to soft radiation.
Whereas the hard function has been known for a few years~\cite{Gehrmann:2010ue}, the soft function has only recently been computed~\cite{Bonvini:2014tea,Li:2014afw}.
These two ingredients allows for reaching N$^3$LL$^\prime$ accuracy also within a SCET approach (see appendix~\ref{app:SCET} for additional details).

Here we consider two variations of the resummation in the SCET formalism of ref.~\cite{Ahrens:2008nc} and we qualitatively compare the results with those obtained in direct QCD.
In particular, we consider both the variation of $1/N$ terms (corresponding to $(1-z)^0$ terms in direct space) and of subleading terms in the resummation.
The variation of the subdominant $1/N$  terms is obtained through the inclusion of a collinear improvement, thereby modifying the choice of soft logarithms adopted in ref.~\cite{Ahrens:2008nc}.
In ref.~\cite{Ahrens:2008nc} the soft logarithms correspond to plus distributions of the form $(1-z)^{-1} \ln^k \frac{1-z}{\sqrt{z}}$, where the presence of the $\sqrt{z}$ factor in the logarithms comes from kinematics.
With this choice, however, it turns out that the fixed-order expansion of the resummation systematically underestimates the full result and leads to a reduced impact of resummation at all orders~\cite{Ball:2013bra}.
By performing a collinear improvement (corresponding to the AP1 version of $\psi$-soft), which amounts to multiplying the soft-function $S$ by an overall factor $z$, the agreement with fixed order improves significantly~\cite{Bonvini:2014joa,Bonvini:2014tea}.
Finally, to probe the effects of subleading terms in the resummation, we consider the inclusion of the so-called $\pi^2$ resummation~\cite
{Ahrens:2008qu,Parisi:1979xd,Magnea:1990zb,Bakulev:2000uh,Eynck:2003fn,Stewart:2013faa}, adopted as default in ref.~\cite{Ahrens:2008nc}.

\subsection{The Higgs cross section at $N^3LO+N^3LL^\prime$ \amper its uncertainties}\label{sec:totalhiggsresult}

Having described various prescriptions for large-$x$ resummation, we now present our results at N$^3$LO+N$^3$LL$^\prime$ and we introduce a method to improve the MHOU estimate for the inclusive cross section.

The result at N$^3$LO+N$^3$LL$^\prime$ is obtained through an additive scheme.
Matching is achieved by adding together the fixed-order and the resummed expressions and subtracting the expansion of the resummation to avoid double counting.
Namely, the matched cross section at N$^j$LO+N$^k$LL$^\prime$ is written as
\begin{equation}
	\sigma_{\text{N$^k$LO+N$^j$LL$^\prime$}} = \sigma_{\text{N$^k$}LO} + \Delta_k \sigma_{\text{N$^j$LL$^\prime$}},
\end{equation}
having defined
\begin{equation}\label{eq:Deltasigma}
\Delta_k\sigma_{\text{N$^j$LL}} = \tau\, \sigma_0(\mH^2,\mt^2) \int_{c-i\infty}^{c+i\infty} \frac{dN}{2\pi i}\,\tau^{-N}\, \mathcal L_{gg}(N)\, \Delta_k \bm C_{\text{res,N$^j$LL}}(N,\as),
\end{equation}
where
\begin{equation}
	\Delta_k \bm C_{\text{res}}(N,\as) =\bm  C_{\text{res}}(N,\as) - \sum_{i=0}^k \as^i\, \bm C_{\text{res}}^{(i)}(N),
\end{equation}
$\bm C_{\text{res}}^{(i)}$ being the coefficients of the expansion in powers of $\as$ of $\bm C_{\text{res}}$.
By construction, $\Delta_k\sigma_{\text{N$^j$LL$^\prime$}}$ only contains higher-order corrections to $ \sigma_{\text{N$^k$}LO} $.

To compute the contribution $\Delta_k\sigma_{\text{N$^j$LL$^\prime$}}$ we use the code \texttt{TROLL} which has been introduced in sect.~\ref{sec:large-implementation}.
Since the code only computes the resummed contribution, the fixed order has to be provided using an independent code.
Here we use \texttt{ggHiggs4.0}~\cite{ggHiggs}, where the recent full N$^3$LO result of ref.~\cite{Mistlberger:2018etf} has been implemented.
We compute the results in the rEFT with $\sqrt{s}= 13$~TeV, using a top mass of $\mt = 172.5$ GeV in the pole scheme and a Higgs mass of $125$ GeV.
We use the NNLO PDF set \texttt{PDF4LHC15\_nnlo}~\cite
{Butterworth:2015oua,Carrazza:2015aoa,Ball:2014uwa,Harland-Lang:2014zoa,Dulat:2015mca} through the LHAPDF interface~\cite{Buckley:2014ana}, with $\as(m_Z^2) = 0.118$.

We consider four options for the central scale $\mu_0$:
\begin{equation}\label{eq:fourscales}
	\mu_0 = \mH/4, \qquad 	\mu_0 =\mH/2, \qquad 	\mu_0 = \mH, \qquad 	\mu_0 = 2\mH.
\end{equation}
For each scale, we then compute resummed results at N$^3$LO+N$^3$LL$^\prime$ using six different $\psi$-soft variants which differ by subdominant (AP1 vs. AP2) and subleading (default option for the constant terms, all constants in the exponent, or all constant in $g_0$) contributions.
In each case, we perform a canonical $7$-point scale variation varying the factorization and renormalization scales $\muF$ and $\muR$ about $\mu_0$ by a factor of two up and down, keeping the ratio $\muR/\muF$ between 1/2 and 2.
The central result is computed using $\psi$-soft$_2$ with the default option for the constants.
The final uncertainty is constructed as an envelope of scale variation and of the $\psi$-soft variants.
This corresponds to a total of $42$ cross section points (seven scales times six variants of resummation), for which we take the highest and lowest cross sections as the maximum and the minimum of the uncertainty band.
We will see shortly that this rather conservative procedure proves very powerful, overcoming the limitations of canonical scale variations, at least for the first orders in perturbation theory.
For reference, we also compute standard $N$-soft without collinear improvement (though we exclude it from the envelope).

\begin{figure}[t]
  \centering
  \includegraphics[width=0.4\textwidth,page=2]{figures/hadr_xsec_res_125_13_uncertainty.pdf} \qquad \qquad
  \includegraphics[width=0.4\textwidth,page=1]{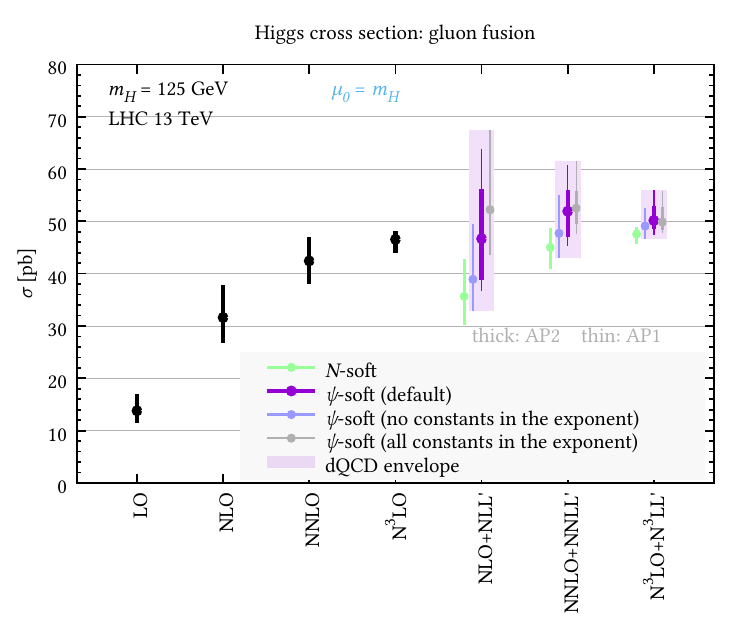}\\
  \includegraphics[width=0.4\textwidth,page=3]{figures/hadr_xsec_res_125_13_uncertainty.pdf}\qquad \qquad
  \includegraphics[width=0.4\textwidth,page=4]{figures/hadr_xsec_res_125_13_uncertainty.pdf}
  \caption{Higgs cross section at 13 TeV in the rescaled effective theory (rEFT), for four different choices of the central scale $\mu_0= \muF=\muR$:
    at the top we show $\mH/2$ and $\mH$, while at the bottom  $\mH/4$ and $2\mH$.
    The uncertainty on the fixed-order predictions and on $N$-soft comes solely from scale variation, as well as the thick uncertainty on the $\psi$-soft AP2 results. The thinner bands correspond to the 7-point scale variation
    envelope on the $\psi$-soft AP1 instead, whose central value is not shown.
    The light-purple rectangles are the envelope of all $\psi$-soft variants,
    corresponding to the 42-point uncertainty described in the text.}
  \label{fig:13TeV}
\end{figure}

We collect in tab.~\ref{tab:resultshiggs} the cross sections at fixed LO, NLO, NNLO and N$^3$LO accuracy and the resummed results at LO+LL$^\prime$, NLO+NLL$^\prime$, NNLO+NNLL$^\prime$ and N$^3$LO+N$^3$LL$^\prime$ accuracy for the four central scales eq.~\eqref{eq:fourscales}.
The error on the fixed-order computation is computed using the canonical 7-points variation, whereas we use the 42-point variations at the resummed level.
The same results are also shown in fig.~\ref{fig:13TeV}.

\begin{table}[t]
\footnotesize
  \centering
  \begin{tabular}{lllll}
    & $\mu_0=\mH/4$ & $\mu_0=\mH/2$ & $\mu_0=\mH$ & $\mu_0=2\mH$ \\
    \midrule
             LO & $ 18.6^{+5.8}_{-3.9} $ & $ 16.0^{+4.3}_{-3.1} $ & $ 13.8^{+3.2}_{-2.4} $ & $ 11.9^{+2.5}_{-1.9} $ \\[1ex]
            NLO & $ 44.2^{+12.0}_{-8.5} $ & $ 36.9^{+8.4}_{-6.2} $ & $ 31.6^{+6.3}_{-4.8} $ & $ 27.5^{+4.9}_{-3.9} $ \\[1ex]
           NNLO & $ 50.7^{+3.4}_{-4.6} $ & $ 46.5^{+4.2}_{-4.7} $ & $ 42.4^{+4.6}_{-4.4} $ & $ 38.6^{+4.4}_{-4.0} $ \\[1ex]
        N$^3$LO & $ 47.9^{+0.9}_{-2.9} $ & $ 48.2^{+0.1}_{-1.8} $ & $ 46.6^{+1.7}_{-2.6} $ & $ 44.3^{+2.5}_{-2.9} $ \\[1ex]
\midrule
          LO+LL$^\prime$ & $ 24.0^{+8.9}_{-6.8} $ & $ 20.1^{+6.2}_{-5.0} $ & $ 16.9^{+4.5}_{-3.7} $ & $ 14.3^{+3.3}_{-2.8} $ \\[1ex]
        NLO+NLL$^\prime$ & $ 46.9^{+15.1}_{-12.6} $ & $ 46.2^{+15.0}_{-13.2} $ & $ 46.7^{+20.8}_{-13.8} $ & $ 47.3^{+26.1}_{-15.8} $ \\[1ex]
      NNLO+NNLL$^\prime$ & $ 50.2^{+5.5}_{-5.3} $ & $ 50.1^{+3.0}_{-7.1} $ & $ 51.9^{+9.6}_{-8.9} $ & $ 54.9^{+17.6}_{-11.5} $ \\[1ex]
N$^3$LO+N$^3$LL$^\prime$ & $ 47.7^{+0.9}_{-2.9} $ & $ 48.7^{+1.5}_{-2.0} $ & $ 50.2^{+5.8}_{-3.5} $ & $ 52.9^{+13.1}_{-5.2} $ \\[1ex]
  \end{tabular}
  \caption{Fixed-order results and their scale uncertainty together with resummed results and their uncertainty for four choices of the central scale,
    for $\mH=125$~GeV at LHC with $\sqrt{s}=13$~TeV. All cross sections are in pb. 
    The N$^3$LO and the N$^3$LO+N$^3$LL$^\prime$ results slightly differ from those in~\cite{Bonvini:2016frm} as in ref.~\cite{Bonvini:2016frm} the N$^3$LO was computed as a threshold expansion~\cite{Anastasiou:2015ema,Anastasiou:2016cez}.    }
  \label{tab:resultshiggs}
\end{table}

Let us comment on the convergence of the fixed-order results, ignoring the LO result which contains too little information for being predictive.
We observe that for central scales $\mu_0=m_H$ and $\mu_0=2m_H$ the NNLO is a large correction, which is not covered by the NLO uncertainty band.
The N$^3$LO correction is smaller, thus indicating that the series is (at least asymptotically) converging, though at the largest scale $\mu_0=2m_H$ its central value is not covered by the NNLO uncertainty band.
The convergence pattern improves for smaller scales.
For $\mu_0=m_H/2$, the central NNLO is at the edge of the NLO uncertainty band and the central N$^3$LO is contained in the NNLO band.
However, the central N$^3$LO and its uncertainty band still do not overlap with the NLO band.
For $\mu_0=m_H/4$ the convergence pattern seems to improve further, though the N$^3$LO band becomes very asymmetric.
We also note that the N$^3$LO values computed at different central scales in table~\ref{tab:resultshiggs} are barely compatible with each other.
This analysis shows that the MHOU uncertainty using canonical 7-point scale variation is not reliable at fixed order.

Nonetheless, one can achieve a robust estimate of the missing higher order uncertainty using resummation.
On the one hand, resummed results have a better perturbative behaviour, thereby suggesting that convergence is improved when resummed contributions are included. 
On the other hand, the variation of subleading and subdominant contributions on top of canonical scale variation provides a more robust method for estimating the uncertainty from MHO.
For each choice of $\mu_0$, the uncertainty of the resummed result from NLO+NLL$^\prime$ onwards covers the central value and the band of the next matched result; in all cases, the N$^3$LO+N$^3$LL$^\prime$ band is fully contained in the NNLO+NNLL$^\prime$ band, which is in turn contained in the NLO+NLL$^\prime$ band.
A similar pattern is observed also if one considers a less conservative option, namely the $\psi$-soft$_2$ prescription with the default position of the constant terms.
With this choice, the errors look somewhat more natural, especially for large values of the central scale. 
We also observe a systematic reduction of the scale uncertainty band moving from one (logarithmic) order to the next.
Finally, we also observe that the resummed results at each order are compatible with each other for different choices of $\mu_0$, thus displaying a reduced sensitivity on the choice of the central scale.

We observe that in many respects the choice $\mu_0=\mH/2$ seems optimal, in agreement with previous analyses~\cite{Anastasiou:2016cez}.
For this scale, both the fixed order and especially the resummation converge nicely.
One could determine \textit{a priori} the optimal choice of scale by requiring that the partonic coefficient functions are free of large logarithms, such that possible logarithmic enhancements are minimized.
Factorization and renormalization scales typically appear together with threshold logarithms as $\ln (\mu^2 N^2/\mH^2)$ (in Mellin space, see eq.~\eqref{eq:CataniTh3}).
A saddle point analysis~\cite{Bonvini:2012an} shows that in Mellin space the cross section is dominated by a single value of $N=N_{\text{saddle}}$, defining a natural scale $\mu_0 \sim \mH/N_{\text{saddle}}$. 
Since for the process considered here $N_{\text{saddle}}\sim 2$, we find that the scale that minimizes the size of the logarithms is close to $\mu_0 = \mH/2$.
Similar conclusions were obtained from direct space arguments~\cite{Anastasiou:2002yz} and within a SCET framework~\cite{Ahrens:2008nc,Bonvini:2014qga}.

Here we have proposed a conservative way of estimating uncertainties, which proves successful at the previous known orders.  
For this reason, the uncertainty on the N$^3$LO+N$^3$LL$^\prime$ result at $\mu_0=\mH/2$ seems reasonably trustful.
To be even more conservative, one can symmetrize the error on the resummed result, leading to $48.7 \pm 2.0$ pb (rather then $ 48.7^{+1.5}_{-2.0} $) as the most reliable prediction of the inclusive cross section at 13 TeV in the rEFT setup.
Though the effect of resummation at N$^3$LO is small ($1.0\%$), we stress that it is \textit{not} covered by the fixed-order uncertainty.
Due to the proximity of a stationary point in the scale dependence, the N$^3$LO result is indeed very asymmetric and does not appear to be reliable.
If the error were at least symmetrized, the N$^3$LO result would be $48.2 \pm 1.8 $, compatible with the resummed result.

\begin{figure}[t]
  \centering
  \includegraphics[width=0.4\textwidth,page=2]{figures/hadr_xsec_res_125_13_uncertainty_SCET.pdf} \qquad \qquad
  \includegraphics[width=0.4\textwidth,page=1]{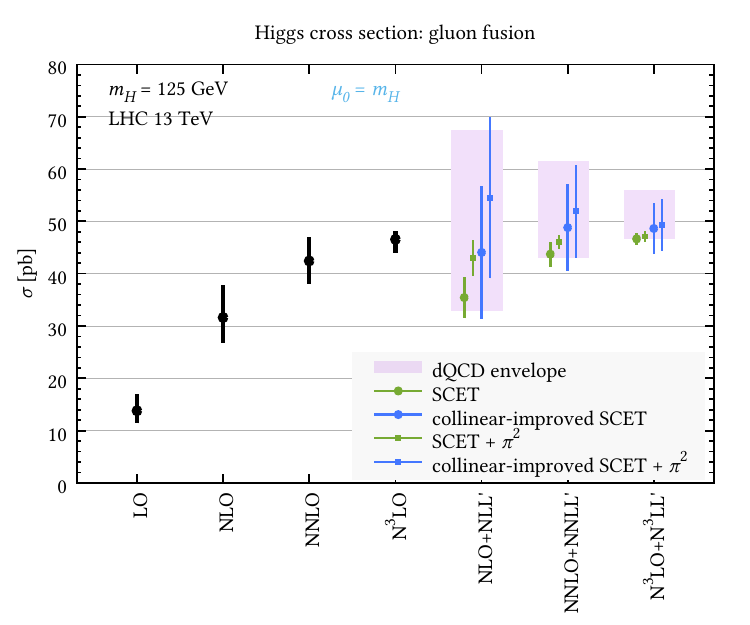}
  \caption{Same as fig.~\ref{fig:13TeV} now showing the SCET result for two different central scales. The dQCD envelope is also shown to facilitate the comparison.}
  \label{fig:13TeVSCET}
\end{figure}

To test the robustness of our result, we compare it with the predictions obtained using a SCET formalism in fig.~\ref{fig:13TeVSCET}.
As mentioned previously, we consider two different choices of soft-logarithms and we use $\pi^2$ resummation to probe subleading terms.
For each variant, we compute the uncertainties as in ref.~\cite{Ahrens:2008nc}. 
We vary independently $\muF$, $\muH$ and $\muS$ keeping the other scales fixed when one of them is varied.
Specifically, the scales $\muF$ and $\muH$ are varied by a factor of two up and down about $\mu_0$, which in this case we take to be either $\mH/2$ or $\mH$.
We refer the reader to ref.~\cite{Ahrens:2008nc} for a detailed explanation of the choice for the central value of $\muS$ and its range.
For each scale, the largest variation is symmetrized and the resulting uncertainty is computed by adding each individual uncertainty in quadrature.
To facilitate the comparison, we show in fig.~\ref{fig:13TeVSCET} the envelope of the $\psi$-soft variants. 
The N$^3$LO+N$^3$LL$^\prime$ result is shown here for the first time within a SCET formalism.

We observe that the original formulation of ref.~\cite{Ahrens:2008nc} (labelled `SCET' in the plot) leads to a small correction of the fixed-order result, which can be traced to the choice of the soft logarithms~\cite{Bonvini:2014qga,Bonvini:2014tea}.
Moreover, the uncertainty bands are comparable or smaller than their fixed-order counterparts, indicating that the MHOU are underestimated.
The impact of resummation is larger if subleading terms are included using a collinear improvement, indicating a better perturbative stability.
The convergence further improves if $\pi^2$ resummation is included, especially at $\mu_0=\mH/2$.
Remarkably, we observe a very good agreement between the spread of the SCET variants and the dQCD envelope, with only the N$^3$LO+N$^3$LL$^\prime$ central value without collinear improvement at $\mu_0=\mH$ being at the edge of the band. 
Also in this case, the choice $\mu_0=\mH/2$ seems optimal, leading to smaller errors and with an overall reduced impact of higher order terms.
This comparison provides a further confirmation of the reliability and robustness of the procedure detailed above to estimate MHOU.

In ref.~\cite{Bonvini:2016frm} the estimate of MHOU using scale variations and variation of subleading and power-suppressed terms in the all-order resummation was compared with other strategies, which do not rely on an arbitrary variation of the perturbative ingredients.
In particular, two complementary approaches were considered: the Cacciari-Houdeau method~\cite{Cacciari:2011ze,Forte:2013mda,Bagnaschi:2014wea}, which infers the uncertainty from the progression of the perturbative expansion within a Bayesian approach, and the use of algorithms to accelerate the convergence of the perturbative series and estimate their all-order sum.
The results were found to be fully compatible with the estimate from resummation and similar to that of the (symmetrized) fixed-order scale variation uncertainty.
The accelerated series exhibits good convergence properties at $\mH/2$ and the all-order estimate of the resummed and fixed-order series are both very close to the N$^3$LO+N$^3$LL$^\prime$ result.
These tests provide additional support that the all-order Higgs cross section in the rEFT lies within the uncertainties of the N$^3$LO+N$^3$LL$^\prime$ prediction.

\section{The Higgs transverse-momentum distribution}

Accurate predictions of transverse distributions are an essential ingredient in the LHC precision programme, as they allow one to discern whether experimental measurements differ from what the SM predicts.
In particular, the refined description of the Higgs transverse-momentum distribution plays an important role in constraining NP models which would modify its shape, for instance through modification of the Yukawa couplings of the Higgs to quarks~\cite{Bishara:2016jga,Soreq:2016rae}.
From a theoretical point of view, the description of this observable is particularly challenging, due to its sensitivity to different physics according to the kinematic region being probed.
This makes the transverse-momentum distribution an extremely interesting laboratory to study many non-trivial features of QCD, such as the effects related to the contributions of top and bottom quarks~\cite{Melnikov:2017pgf,Lindert:2017pky,Lindert:2018iug,Neumann:2018bsx,Caola:2018zye,Jones:2018hbb}.

In the large-$\mt$ limit, the transverse-momentum spectrum for Higgs production in gluon fusion is known at a very high level of accuracy.
At fixed-order, the state-of-the-art result is the NNLO calculations of refs.~\cite{Boughezal:2015dra,Boughezal:2015aha,Caola:2015wna,Chen:2016zka}.
As we discussed in sect.~\ref{sec:tmomres}, at small values of transverse momenta $\pth \ll \mH$ the differential distribution is enhanced by large logarithms $\ln \pth/\mH$, which must be resummed at all orders.
All the ingredients to compute the transverse-momentum distribution at N$^3$LL accuracy (with the only exception of the four-loop cusp anomalous dimension, which is known only numerically~\cite{Moch:2017uml,Moch:2018wjh}) are now known~\cite{Catani:2011kr,Catani:2012qa,Gehrmann:2014yya,Li:2016ctv,Vladimirov:2016dll}, which allows one to calculate results at NNLO+N$^3$LL accuracy~\cite{Bizon:2017rah,Chen:2018pzu,Bizon:2018foh} in the EFT.

In this section we will present matched results for the Higgs transverse momentum using the formalism introduced in sect.~\ref{sec:radish}.
We first present the relevant formul\ae\ for the resummation at N$^3$LL.
We then discuss how the matching to fixed-order is performed and we finally show matched predictions for $\pth$ both at the inclusive level and within the fiducial cuts in the $H\rightarrow \gamma\gamma$ channel.

\subsection{Transverse-momentum resummation at $N^3LL$}

In sect.~\ref{sec:radish} we introduced an approach which performs the resummation of large logarithms of the transverse momentum in direct space by exploiting the factorization properties of the QCD amplitudes.
The formalism allows for an efficient Monte Carlo implementation of the all-order calculation directly in momentum space, such that large logarithms are resummed by generating soft and/or collinear emissions in a fashion similar to that of an event generator.
In sect.~\ref{sec:radish} we derived eq.~\eqref{eq:master-NLL-kt-with-Lumi}, which contains an expression for the cumulative cross section differential over the Born phase space $d \Phi_B$ valid to NLL accuracy, and we discussed how to systematically include higher-order corrections.
Including terms up to N$^3$LL accuracy, one can recast the cumulative cross section as~\cite{Bizon:2017rah}
\begin{align}
\label{eq:master-kt-space}
&\frac{\rd\Sigma(v)}{\rd\Phi_B} =\int_0^\infty\frac{\rd k_{t,1}}{k_{t,1}}\frac{\rd
  \phi_1}{2\pi}\partial_{ L}\left(-e^{- R(k_{t,1})} {{\cal L}}_{\text{N$^3$LL}}(k_{t,1}) \right) \int \dZ\Theta\left(v-V(k_1,\dots, k_{n+1})\right)
                             \notag\\\notag\\
& + \int_0^\infty\frac{\rd k_{t,1}}{k_{t,1}}\frac{\rd
  \phi_1}{2\pi} e^{- R(k_{t,1})} \int \dZ\int_{0}^{1}\frac{\rd \zeta_{s}}{\zeta_{s}}\frac{\rd
  \phi_s}{2\pi}\Bigg\{\bigg({ R}' (k_{t,1}) {{\cal L}}_{\text{NNLL}}(k_{t,1}) - \partial_{ L} {{\cal L}}_{\text{NNLL}}(k_{t,1})\bigg)\notag\\
&\times\left({ R}'' (k_{t,1})\ln\frac{1}{\zeta_s} +\frac{1}{2} { R}'''
  (k_{t,1})\ln^2\frac{1}{\zeta_s} \right)\notag \\
  & - { R}' (k_{t,1})\left(\partial_{ L} {{\cal L}}_{\text{NNLL}}(k_{t,1}) - 2\frac{\beta_0}{\pi}\as^2(k^2_{t,1}) \hat{P}^{(0)}\otimes {{\cal L}}_{\text{NLL}}(k_{t,1}) \ln\frac{1}{\zeta_s}
\right)\notag\\
&+\frac{\as^2(k^2_{t,1}) }{\pi^2}\hat{P}^{(0)}\otimes \hat{P}^{(0)}\otimes {{\cal L}}_{\text{NLL}}(k_{t,1})\Bigg\}\notag\\
&\times \bigg\{\Theta\left(v-V(k_1,\dots,
  k_{n+1},k_s)\right) - \Theta\left(v-V(k_1,\dots,
  k_{n+1})\right)\bigg\}\notag\\\notag\\
& + \frac{1}{2}\int_0^\infty\frac{\rd k_{t,1}}{k_{t,1}}\frac{\rd
  \phi_1}{2\pi} e^{- R(k_{t,1})} \int \dZ\int_{0}^{1}\frac{\rd \zeta_{s1}}{\zeta_{s1}}\frac{\rd
  \phi_{s1}}{2\pi}\int_{0}^{1}\frac{\rd \zeta_{s2}}{\zeta_{s2}}\frac{\rd
  \phi_{s2}}{2\pi} { R}' (k_{t,1})\notag\\
&\times\Bigg\{ {{\cal L}}_{\text{NLL}}(k_{t,1}) \left({ R}'' (k_{t,1})\right)^2\ln\frac{1}{\zeta_{s1}} \ln\frac{1}{\zeta_{s2}} - \partial_{ L} {{\cal L}}_{\text{NLL}}(k_{t,1}) { R}'' (k_{t,1})\bigg(\ln\frac{1}{\zeta_{s1}}
  +\ln\frac{1}{\zeta_{s2}} \bigg)\notag\\
&+ \frac{\as^2(k^2_{t,1}) }{\pi^2}\hat{P}^{(0)}\otimes \hat{P}^{(0)}\otimes {{\cal L}}_{\text{NLL}}(k_{t,1})\Bigg\}\notag\\
&\times \bigg\{\Theta\left(v-V(k_1,\dots,
  k_{n+1},k_{s1},k_{s2})\right) - \Theta\left(v-V(k_1,\dots,
  k_{n+1},k_{s1})\right) -\notag\\ &\Theta\left(v-V(k_1,\dots,
  k_{n+1},k_{s2})\right) + \Theta\left(v-V(k_1,\dots,
  k_{n+1})\right)\bigg\} + {\cal O}\left(\as^n \ln^{2n -
                                    6}\frac{1}{v}\right),
\end{align}
where we defined $\zeta_{si} \equiv k_{t,si}/k_{t,1}$.
We further introduced the notation $\dZ$ to denote an ensemble that describes the emission of $n$ identical independent blocks, such that the average of a function $F(\{k_i\})$ over the measure $\rd {\cal Z}$ is 
\begin{equation}
\label{eq:dZ}
\begin{split}
\int \dZ  F(\{k_i\})=e^{-{ R}'(k_{t,1})\ln\frac{1}{\eps}}
   \sum_{n=0}^{\infty}\frac{1}{n!} \prod_{i=2}^{n+1}
    \int_{\eps}^{1} \frac{\rd\zeta_i}{\zeta_i}\int_0^{2\pi}
   \frac{\rd\phi_i}{2\pi} { R}'(k_{t,1})F(k_1,\dots,k_{n+1})\,.
\end{split}
\end{equation}
where $\zeta_{i} \equiv k_{t,i}/k_{t,1}$.
As in eq.~\eqref{eq:master-NLL-kt-with-Lumi}, the $\ln \frac{1}{\eps}$ divergence cancels exactly against that in the resolved real radiation, ensuring that the final result is $\eps$-independent.

\begin{mccorrection}
To highlight the different classes of terms which enter at a given logarithmic accuracy in eq.~\eqref{eq:master-kt-space} we have split the result into three terms. 
In particular, the first line includes all the N$^3$LL corrections to the hardest emission $k_{t,1}$.
As we mentioned in sect.~\ref{sec:radish}, at N$^3$LL one needs to account for the effect of up to two hard-collinear resolved partons.
These effects are collected in the second and in the third contributions of eq.~\eqref{eq:master-kt-space}.
In particular, the second contribution (lines two to six) encodes the corrections appearing when only a single emission $k_s$ of the ensemble is corrected, and the third contribution (lines seven to eleven) encodes the corrections due to two additional emissions $k_{s1}$ and $k_{s2}$.

Eq.~\eqref{eq:master-kt-space} depends on the luminosities ${\cal L}$, which are written in terms of the parton luminosities as well as the Born matrix element squared and the virtual amplitude.
At N$^3$LL, they read
\begin{align} 
\label{eq:luminosity-NLL}
{\cal L}_{\text{NLL}}(k_{t,1}) = \sum_{c, c'}\frac{\rd|\mathcal{M}_{B}|_{cc'}^2}{\rd\Phi_B} f_c\!\left(x_1, \muF^2 e^{-2 {L}}\right)f_{c'}\!\left(x_2, \muF^2 e^{-2{L}}\right),
\end{align}
\begin{align}
\label{eq:luminosity-NNLL}
&{\cal L}_{\text{NNLL}}(k_{t,1}) = \sum_{c, c'}\frac{\rd|\mathcal{M}_{B}|_{cc'}^2}{\rd\Phi_B} \sum_{i, j}\int_{x_1}^{1}\frac{\rd z_1}{z_1}\int_{x_2}^{1}\frac{\rd z_2}{z_2}f_i\!\left(\frac{x_1}{z_1}, \muF^2 e^{-2{L}}\right)f_{j}\!\left(\frac{x_2}{z_2},\muF^2 e^{-2{L}}\right)\notag\\&\times\Bigg\{\delta_{ci}\delta_{c'j}\delta(1-z_1)\delta(1-z_2)
\left(1+\frac{\as(\muR^2)}{2\pi} {H}^{(1)}(\muR,x_Q)\right) \notag\\
&+ \frac{\as(\muR^2)}{2\pi}\frac{1}{1-2\as(\muR^2)\beta_0
  {L}}\left({C}_{c i}^{(1)}(z_1,\muF,x_Q)\delta(1-z_2)\delta_{c'j}+
  \{z_1\leftrightarrow z_2; c,i \leftrightarrow c'j\}\right)\Bigg\},
\end{align}
\begin{align}
\label{eq:luminosity-N3LL}
&{\cal L}_{\text{N$^3$LL}}(k_{t,1})=\sum_{c,
  c'}\frac{\rd|\mathcal{M}_{B}|_{cc'}^2}{\rd\Phi_B} \sum_{i, j}\int_{x_1}^{1}\frac{\rd
  z_1}{z_1}\int_{x_2}^{1}\frac{\rd z_2}{z_2}f_i\!\left(\frac{x_1}{z_1},\muF^2 e^{-2{L}}\right)f_{j}\!\left(\frac{x_2}{z_2},\muF^2 e^{-2{L}}\right)\notag\\&\times\Bigg\{\delta_{ci}\delta_{c'j}\delta(1-z_1)\delta(1-z_2)
\left(1+\frac{\as(\muR^2)}{2\pi} {H}^{(1)}(\muR,x_Q) + \frac{\as^2(\muR^2)}{(2\pi)^2} {H}^{(2)}(\muR,x_Q)\right) \notag\\
&+ \frac{\as(\muR^2)}{2\pi}\frac{1}{1-2\as(\muR^2)\beta_0 {L}}\left(1- \as(\muR^2)\frac{\beta_1}{\beta_0}\frac{\ln\left(1-2\as(\muR^2)\beta_0 {L}\right)}{1-2\as(\muR^2)\beta_0 {L}}\right)\notag\\
&\times\left({C}_{c i}^{(1)}(z_1,\muF,x_Q)\delta(1-z_2)\delta_{c'j}+ \{z_1\leftrightarrow z_2; c,i \leftrightarrow c',j\}\right)\notag\\
& +
  \frac{\as^2(\muR^2)}{(2\pi)^2}\frac{1}{(1-2\as(\muR^2)\beta_0
  {L})^2}\Bigg({C}_{c i}^{(2)}(z_1,\muF,x_Q)\delta(1-z_2)\delta_{c'j} + \{z_1\leftrightarrow z_2; c,i \leftrightarrow c',j\}\Bigg) \notag\\&+  \frac{\as^2(\muR^2)}{(2\pi)^2}\frac{1}{(1-2\as(\muR^2)\beta_0 {L})^2}\Big({C}_{c i}^{(1)}(z_1,\muF,x_Q){C}_{c' j}^{(1)}(z_2,\muF,x_Q) + G_{c i}^{(1)}(z_1)G_{c' j}^{(1)}(z_2)\Big) \notag\\
& + \frac{\as^2(\muR^2)}{(2\pi)^2} {H}^{(1)}(\muR,x_Q)\frac{1}{1-2\as(\muR^2)\beta_0 {L}}\Big({C}_{c i}^{(1)}(z_1,\muF,x_Q)\delta(1-z_2)\delta_{c'j} + \{z_1\leftrightarrow z_2; c,i \leftrightarrow c',j\}\Big) \Bigg\}.
\end{align}
with
\begin{align}
x_1 &= \frac{\mH}{\sqrt{s}} \; e^{Y}, &
x_2 &= \frac{\mH}{\sqrt{s}} \; e^{-Y},
\end{align}
where $Y$ is the rapidity of the colour singlet in the centre-of-mass frame of the collision at Born level, $|{\cal M}_B|$ is the Born-level squared matrix element and $x_Q=Q/\mH$. 
Furthermore, the luminosities eq.~\eqref{eq:luminosity-NLL}-\eqref{eq:luminosity-N3LL} depends on the NLO and NNLO collinear coefficient function $ C_{ci}^{(n)} $ and $ G_{ci}^{(n)} $, and the hard virtual corrections $ H^{(n)}$ (see ref.~~\cite{Bizon:2017rah} for a precise definition).

Let us now briefly discuss the physical origin of the other terms appearing in eq.~\eqref{eq:master-kt-space}.
Thanks to rIRC safety, all inclusive blocks (see sect.~\ref{sec:radish}) with $k_{t,i} \ll k_{t,1}$ do not contribute to the value of the observable and are fully cancelled by the exponential factor in eq.~\eqref{eq:dZ}. 
This allowed us to expand all the ingredients in eq.~\eqref{eq:master-kt-space} around $k_{t,1}$ (which denotes the transverse momentum of the block with largest $k_t$) retaining all terms necessary to reach the desired logarithmic accuracy, as we did in sect.~\ref{sec:radish} when we derived the analogous expression valid at NLL accuracy.
Though this operation is not strictly necessary, the expansion around $k_{t,1}$ allows for an efficient numerical implementation, which has been exploited in ref.~\cite{Bizon:2018foh} to construct a resummed formula to simultaneously compute $p_t$ and the angular variable $\phi^*$~\cite{Banfi:2010cf}. 
At N$^3$LL accuracy, this amounts to retaining in the Taylor expansion of the Sudakov radiator ${R}$ the terms 
\begin{equation}
{ R}'= \rd {R}/\rd{L},\qquad { R}''= \rd
      { R}'/\rd{L},\qquad { R}'''= \rd { R}''/\rd{L},
\end{equation}
where ${R}$ at this order takes the form
\begin{align}
\label{eq:N3LLradiator}
{R}(k_{t,1}) &= - {L} g_1(\as \beta_0{L} ) -
  g_2(\as \beta_0{L} ) - \frac{\as}{\pi}
  g_3(\as \beta_0{L} ) - \frac{\as^2}{\pi^2}
  g_4(\as \beta_0{L} ),
\end{align}
with $\as = \as(\muR^2)$ (see appendix~\ref{app:sudakov-radiator}).
As we discussed in sect.~\ref{sec:radish}, at NNLL accuracy and beyond one has to take into account the exact rapidity bounds for a single emission in the generated soft-collinear ensemble, which give rise to subleading corrections which cannot be neglected.
In particular, at N$^3$LL one has
\begin{equation}
	R'(k_{t,i}) = R'(k_{t,1}) + R''(k_{t,1}) \ln \frac{1}{\zeta_i} + \frac{1}{2!} R'''(k_{t,1}) \ln^2 \frac{1}{\zeta_i} +\ldots\, ,
\end{equation}
where the dots indicate terms which enter at N$^4$LL.
We recognize the appearance of these corrections in eq.~\eqref{eq:master-kt-space}.

Along with the corrections encoded in the Sudakov radiator and its derivative, starting at NNLL one has to include the additional class of corrections included in what we called `exclusive DGLAP step' in sect.~\ref{sec:radish}.   
For this reason, eq.~\eqref{eq:master-kt-space} also contains the convolution of the regularized splitting functions $\hat P$~\cite{Moch:2004pa,Vogt:2004mw} with the parton luminosities (see appendix~\ref{app:sudakov-radiator}).
For the term $\hat P^{(0)} \otimes {\cal L}_{\text{NLL}}$ this convolution is defined as
\begin{align} 
\label{eq:Pluminosity-NLL}
\hat{P}^{(0)}\otimes{\cal L}_{\text{NLL}}(k_{t,1}) &\equiv \sum_{c,
  c'}\frac{\rd|\mathcal{M}_{B}|_{cc'}^2}{\rd\Phi_B} \bigg\{\left(\hat{P}^{(0)}\otimes
  f\right)_c\left(x_1,\muF^2 e^{-2{L}}\right)f_{c'}\!\left(x_2,\muF^2 e^{-2{L}}\right)
  \notag\\
&\hspace{3.5cm}+
  f_c \!\left(x_1,\muF^2 e^{-2{L}}\right) \left(\hat{P}^{(0)}\otimes f\right)_{c'}\left(x_2,\muF^2 e^{-2{L}}\right) \bigg\},
\end{align}
and analogously for the term $\hat{P}^{(0)}\otimes\hat{P}^{(0)}\otimes{\cal L}_{\text{NLL}}(k_{t,1})$.
Finally, the terms containing $\as(k^2_{t,1})$ in eq.~\eqref{eq:master-kt-space} are defined as
\begin{equation}
\label{eq:N3LLcoupling}
\as(k^2_{t,1})\equiv \frac{\as(\mu^2_R)}{1-2\as(\mu^2_R)\beta_0
  {L}}.
\end{equation}

In sect.~\ref{sec:radish}, our final expression at NLL eq.~\eqref{eq:master-NLL-kt-with-Lumi} contained resummed logarithms of the form $L = \ln (\mH/k_{t,1})$.
It is however customary to introduce a resummation scale $Q$, whose variation is used to probe the size of subleading logarithmic corrections which are not included in the resummed result.
As a consequence, the resummed expression contains resummed logarithms of the form $\ln (Q/k_{t,1})$. 
In the expression eq.~\eqref{eq:master-kt-space}, the resummed logarithms should be further replaced with \textit{modified} logarithms $\tilde L$, which are introduced to ensure that resummation does not affect the hard region of the spectrum when a matching to fixed order is performed.
To this end, the logarithms are supplemented by power-suppressed terms, which are negligible at small $k_{t,1}$ but ensure that resummation effects dwindle when $k_{t,1} \gg Q$.
One thus replaces the resummed logarithms of $\ln (Q/k_{t,1})$ with
\begin{equation}
\label{eq:modified-log}
\ln\frac{Q}{k_{t,1}} \to \tilde{L}=
\frac{1}{p}\ln\left(\left(\frac{Q}{k_{t,1}}\right)^{p} + 1\right),
\end{equation}
where $p$ is a positive parameter whose value is such that the resummed (differential) distribution vanishes faster that the fixed-order at large $\pth/\mH$, with slope $(1/(\pth/\mH))^{p+1}$.
To ensure that the single-emission event in the first line of eq.~\eqref{eq:master-kt-space} is a total derivative, one must  introduce the Jacobian factor
\begin{equation}
\label{eq:jakobian}
{\cal J}(k_{t,1}) = \left(\frac{Q}{k_{t,1}} \right)^p \left(1+\left(\frac{Q}{k_{t,1}} \right)^p\right)^{-1}
\end{equation}
in all integrals over $k_{t,1}$ in eq.~\eqref{eq:master-kt-space}.
This Jacobian does not modify the logarithmic structure and only affects the large-$\pth$ region.
Since this modification does not affect the measurement function in eq.~\eqref{eq:master-kt-space}, the final result depends on $p$ through power-suppressed terms.	
\end{mccorrection}

\subsection{Matching to fixed order}
\label{sec:matching}

To obtain a prediction valid at NNLO+N$^3$LL, we need to match the resummed result to the fixed-order result.
It is convenient to work at the level of the cumulative distribution $\Sigma$. 
The fixed-order cumulative distribution is defined as
\begin{equation}
	\Sigma^{\text{N$^3$LO}} (v) = \sigma_{\text{tot}}^{\text{N$^3$LO}}-\int_v^\infty dv' \frac{\rd \Sigma^{\text{NNLO}}(v') }{d v'},
\end{equation}
where $\sigma_{\text{tot}}^{\text{N$^3$LO}}$ is the total cross section for the process and $\rd \Sigma^{\text{NNLO}}/\rd v$ is the NNLO differential distribution, where $v=\pth/M$.
For inclusive Higgs production, the NNLO transverse-momentum distribution was computed in refs.~\cite{Boughezal:2015aha,Boughezal:2015dra,Caola:2015wna,Chen:2016zka}.
Currently, the N$^3$LO cross section within fiducial cuts is not yet known.
In the following, when considering differential distributions within fiducial cuts we approximate the N$^3$LO correction by rescaling the NNLO fiducial cross section by the N$^3$LO/NNLO $K$-factor computed using the inclusive cross section computed in refs.~\cite{Anastasiou:2015ema,Anastasiou:2016cez,Mistlberger:2018etf}. 
This approximation is a N$^4$LL effect, thus beyond the formal accuracy considered here.

Let us stress that though in this section we discuss the matching at the level of the cumulative distributions, in the next section we will show matched predictions at the level of differential distributions. 
Therefore, at the level of the spectrum, we drop one order in the fixed-order counting, so that the derivative of the $\Sigma^{\text{N$^3$LO}}$ cumulant will be referred to as NNLO, and similarly for the lower-order cases.

To construct the NNLO+N$^3$LL result, we consider two different matching prescriptions. 
The first scheme is a common additive scheme analogous to what we discussed above in the case of the inclusive cross section (and used for instance in ref.~\cite{Chen:2018pzu}):
\begin{equation}
\label{eq:additive}
\Sigma_{\text{add}}^{\text{matched}}(v) = \Sigma^{\text{N$^3$LL}}(v) + \Sigma^{\text{N$^3$LO}}(v) - \Sigma^{\text{expanded}}(v),
\end{equation}
where $\Sigma^{\text{expanded}}$ denotes the expansion of the resummation formula $\Sigma^{\text{N$^3$LL}}$ to N$^3$LO.
The second scheme belongs to the class of multiplicative schemes (see e.g.~\cite{Banfi:2012yh,Banfi:2012jm,Banfi:2015pju}) and is defined as
\begin{equation}
\label{eq:multiplicative0}
\Sigma_{\text{mult}}^{\text{matched}}(v) = \Sigma^{\text{N$^3$LL}}(v)
\left[\frac{\Sigma^{\text{N$^3$LO}}(v)}{\Sigma^{\text{expanded}}(v)}\right]_{\text{expanded to N$^3$LO}}.
\end{equation}
At the accuracy considered here, the two schemes are equivalent up to N$^4$LO and N$^4$LL terms.
However, in a multiplicative scheme higher-order corrections are damped by resummation at small values of $v$. 
Since this damping occurs in a region where numerical instabilities might spoil the fixed-order distributions, a multiplicative solution allows for stable matched distributions even in the presence of limited statistics in the fixed-order component.
An additional advantage of the multiplicative matching is that the N$^3$LO constant terms (formally entering at N$^4$LL) are automatically extracted from the fixed order in the matching procedure, whenever the N$^3$LO total cross section is known.
These terms, which contain the N$^3$LO collinear coefficient functions and the three-loop virtual corrections, would multiply the Sudakov radiator in the resummed formula eq.~\eqref{eq:master-kt-space} at N$^4$LL. 
As they are currently unknown analytically, in an additive matching they would disappear at the level of the differential distributions.
In a multiplicative scheme, however, they multiply the resummed cross section and allow one to extend the prediction to all terms of order $\as^n \ln^{2n -6} (1/v)$ in the expanded formula of $\Sigma$ --- thus extending the accuracy of the prediction to N$^3$LL$^\prime$, as we discussed in sect.~\ref{sec:logaccuracy}.
However, since the N$^3$LO cross section is currently unknown in the fiducial volumes, this tower of logarithms is correctly included only for the inclusive case, whereas for fiducial distributions the tower of $\as^n \ln^{2n -6} (1/v)$ terms in $\Sigma$ is not fully included.

The use of a multiplicative solution as defined in eq.~\eqref{eq:multiplicative0} has however a drawback, as in the limit $ L \rightarrow 0$ the cumulant $\Sigma^{\text{N$^3$LL}}$ tends to the integral of ${\mathcal  L}_{\text{N$^3$LL}} (\muF)$~\eqref{eq:luminosity-N3LL} over the Born phase space $\Phi_B$ evaluated at $ L = 0$. 
As a consequence, the fixed-order result at large $v$ receives spurious contributions of relative order $\as^4$
\begin{equation}
\Sigma_{\text{mult}}^{\text{matched}}(v) \sim \Sigma^{\text{N$^3$LO}}(v)\left(1+{\cal O}(\as^4)\right).
\end{equation}
Whilst this contribution is formally higher order, its effect can be sizeable for processes with large $K$-factors.
%
The problem can be addressed by defining the multiplicative matching scheme normalizing the resummed prefactor to its asymptotic value in the $ L \rightarrow 0$ limit as
\begin{equation}
\label{eq:asypt}
\Sigma^{\text{N$^3$LL}}_{\text{asym.}} = \int_{\text{with cuts}} \hspace{-1cm}\rd
\Phi_B \quad \left(\lim_{{L}\to 0}
{\cal
  L}_{\text{N$^3$LL}}\right).
\end{equation}
This way, the multiplicative matching scheme is defined as
\begin{equation}
\label{eq:multiplicative1}
\Sigma_{\text{mult}}^{\text{matched}}(v) = \frac{\Sigma^{\text{N$^3$LL}}(v)}{\Sigma^{\text{N$^3$LL}}_{\text{asym.}} } \left[\Sigma^{\text{N$^3$LL}}_{\text{asym.}} \frac{\Sigma^{\text{N$^3$LO}}(v)}{\Sigma^{\text{exp}}(v)}\right]_{\text{expanded to N$^3$LO}},
\end{equation}
where
\begin{equation}
\Sigma^{\text{N$^3$LL}}(v)\xrightarrow[v \gg Q/M]{} \Sigma^{\text{N$^3$LL}}_{\text{asym.}} ,
\end{equation}
and the whole term contained between square brackets is expanded at N$^3$LO.
This ensures that the matched result reproduces by construction the fixed-order one at large $v$ and that spurious higher-order contributions are absent in this region.
The results in the next section are computed using the multiplicative solution; a comparison between multiplicative and additive schemes was performed at NLO+N$^3$LL accuracy in ref.~\cite{Bizon:2018foh} and showed that the differences are rather moderate.
Some useful analytic formul\ae\ about the matching procedure are collected in appendix~\ref{app:matching}. 

\subsection{Matched predictions for inclusive \amper fiducial distributions}

We are now ready to present the predictions for the $\pth$ spectrum both for the inclusive $pp \rightarrow H$ production and in the $pp \rightarrow H \rightarrow \gamma\gamma$ channel within fiducial cuts.
We consider a centre-of-mass energy of 13 TeV, using the \texttt{PDF4LHC15\_nnlo\_mc} set~\cite
{Butterworth:2015oua,Carrazza:2015aoa,Ball:2014uwa,Harland-Lang:2014zoa,Dulat:2015mca}.
The results are computed in the large-$\mt$ limit, with $\mH=125$ GeV and $\mt=173.2$ GeV. 
The central factorization, renormalization and resummation scales are set to $\mH/2$.
The perturbative uncertainty is calculated by performing a canonical seven-scale variation of $\muR$ and $\muF$ by a factor of two in either direction, keeping $\muR/\muF$ between $1/2$ and $1$.
For the central scales, we also vary the resummation scale by a factor of two.
The scale uncertainty is then calculated as an envelope of all the above variations.
The resummation and matching described above have been implemented in the code \texttt{RadISH} (\texttt{Radiation off Initial State Hadrons}) and matched to the fixed-order results computed with the parton-level event generator \texttt{NNLOjet}~\cite{Chen:2014gva,Chen:2016zka}.

\paragraph{\itshape\mdseries Predictions for inclusive Higgs production.}

\begin{figure}[tp]
  \centering
  \includegraphics[width=0.495\columnwidth]{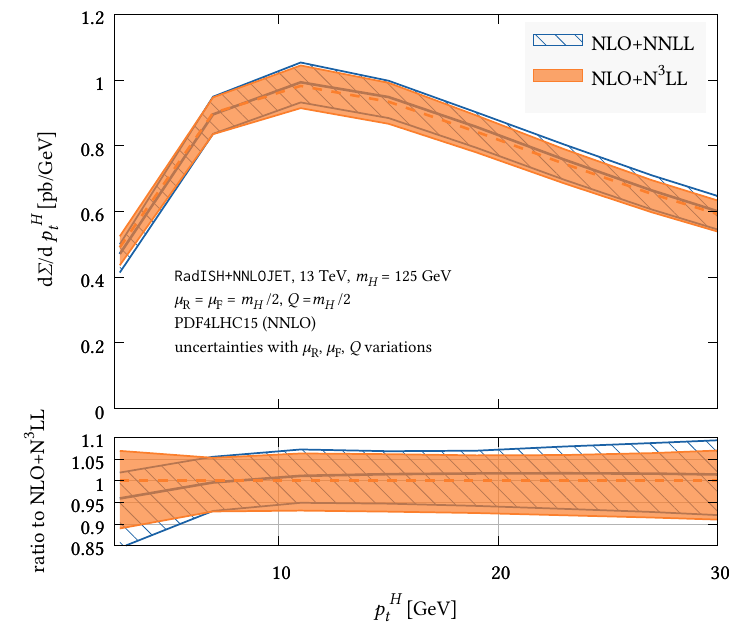} 
  \includegraphics[width=0.495\columnwidth]{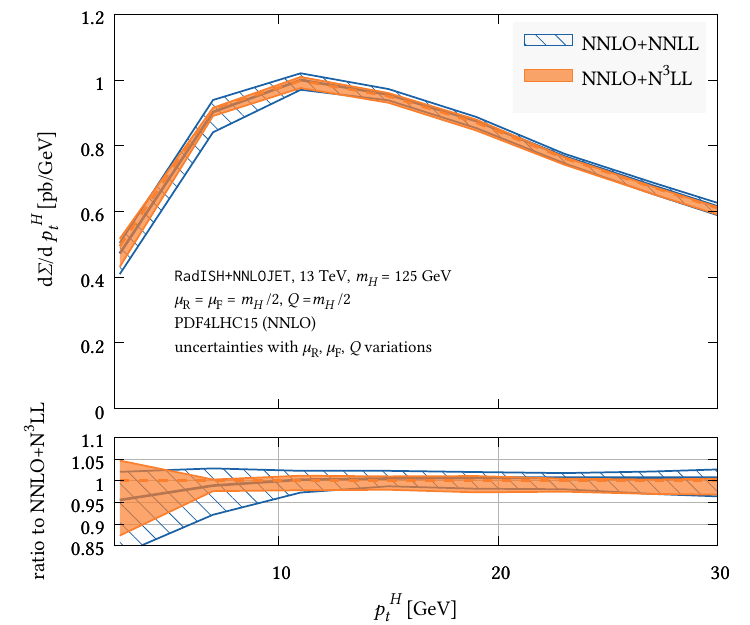} 
  \caption{Comparison between different combinations of fixed-order (NLO and NNLO) and resummation (NNLL and N$^3$LL) for the $\pth$ spectrum for Higgs boson production at $\sqrt{s} = 13$~TeV. Left: NLO; right: NNLO. The lower panel shows the ratio of the predictions to that obtained with N$^3$LL resummation.}
  \label{fig:n3ll_v_nnll}
\end{figure}

We start by discussing the matched result in the inclusive case. 
First, we discuss the size of the N$^3$LL correction with respect to the results at NNLL.
In fig.~\ref{fig:n3ll_v_nnll} we compare the differential distributions for the Higgs $\pth$ spectrum at (N)NLO+NNLL and (N)NLO+N$^3$LL in the left (right) plots.
In the lower panel, we show the ratio of both predictions to the (N)NLO+N$^3$LL band.
In the NLO case, we observe that the N$^3$LL corrections are tiny, with an effect which is at most $5\%$ only in the very small $\pth$ region and smaller at larger values of $\pth$. 
The central NLO+N$^3$LL result is entirely contained in the NLO+NNLL band.
The inclusion of the N$^3$LL correction reduces the perturbative uncertainties, though the effect is observed only for $\pth \lesssim 5$~GeV. 
We observe a similar trend in the NNLO case.
The effect of the N$^3$LL correction on the NNLO+NNLL central value is at the percent level over the whole $\pth$ range and reaches $5\%$ only at very small $\pth$.
In the NNLO case, the reduction of the perturbative uncertainty is more pronounced and starts already below $\sim 10$~GeV.
Let us observe that when NNLO is matched to NNLL the fixed order and the resummation differ by a divergent term $\sim 1/\pth$ at small $\pth$.
However, this divergence is absent if a multiplicative solution is chosen, which ensures that the matched prediction obeys the resummation scaling at small $\pth$.

\begin{figure}[tp]
  \centering
  \includegraphics[width=0.495\columnwidth]{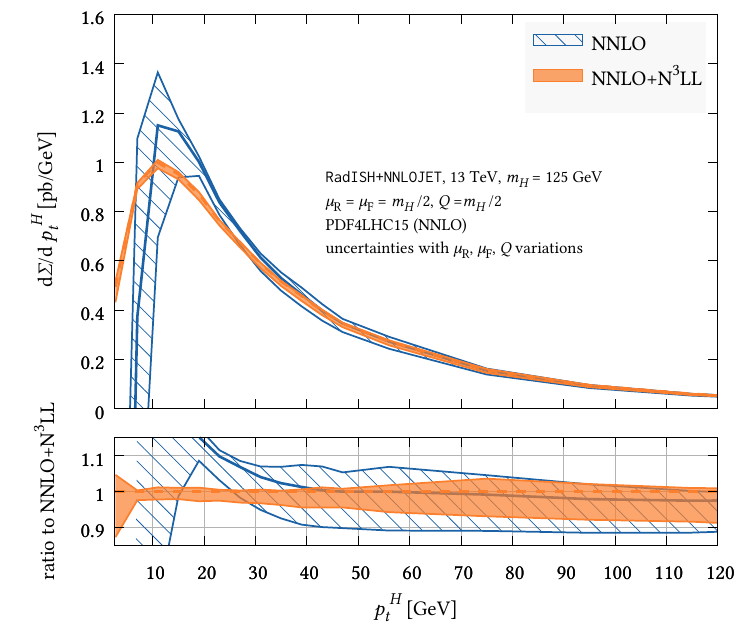} 
  \includegraphics[width=0.495\columnwidth]{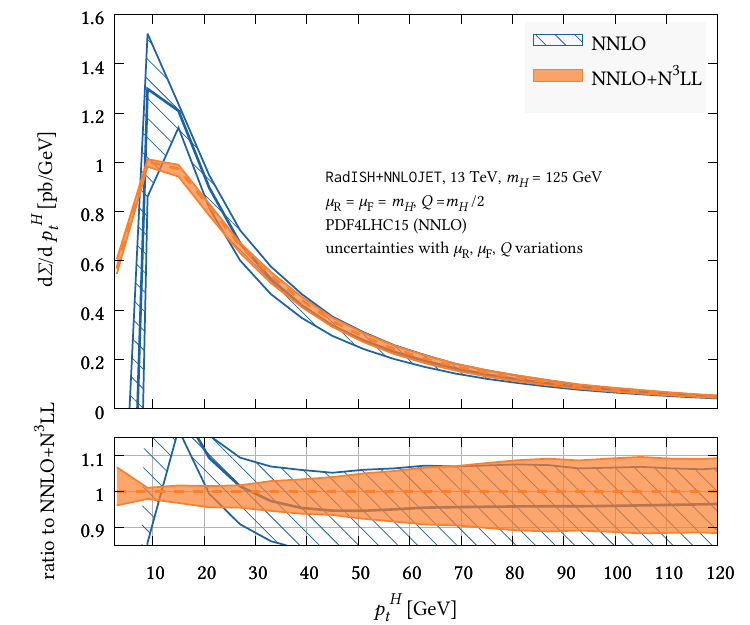} 
  \caption{Comparison of the  transverse-momentum distribution for Higgs boson production at NNLO and NNLO+N$^3$LL for a central scale choice of $\muR = \muF =\mH /2$ (left) and $\muR = \muF = \mH$ (right). In both cases, $Q=\mH/2$.}
  \label{fig:n3ll_nnlo_v_nnlo}
\end{figure}

In fig.~\ref{fig:n3ll_nnlo_v_nnlo} we now compare the matched prediction at NNLO+N$^3$LL to the fixed-order one for two different central scales $\mu_0$.
In the left plot, we show the two predictions for $\mu_0=\mH/2$, whereas in the right plot we show them for $\mu_0=\mH$.
In both cases, we observe that resummation effects are increasingly relevant for $\pth \lesssim 40$~GeV.
We observe that if the central scale is chosen as $\mH/2$ the uncertainty band is affected by large accidental cancellations in the scale variation terms, which lead to unnaturally small perturbative errors.
Indeed, as we observed in the case of the total cross section in sect.~\ref{sec:totalhiggsresult}, the scale-uncertainties band at $\mH/2$ are very asymmetric.
If the central scale is set to $\mH$, the uncertainty band is larger, thus allowing for a more robust estimate of the perturbative error.
The effect is more pronounced for $\pth \gtrsim 50$~GeV, where however the impact of resummation is reduced.
In the following, we will show results using  $\mH/2$ as a central scale, since the fixed-order runs with higher statistics have been obtained with this setup. 
However, a comparison of the theoretical predictions with experimental data would require a more thorough study of different central-scale choices.

\begin{figure}[tp]
  \centering
  \includegraphics[width=0.495\columnwidth]{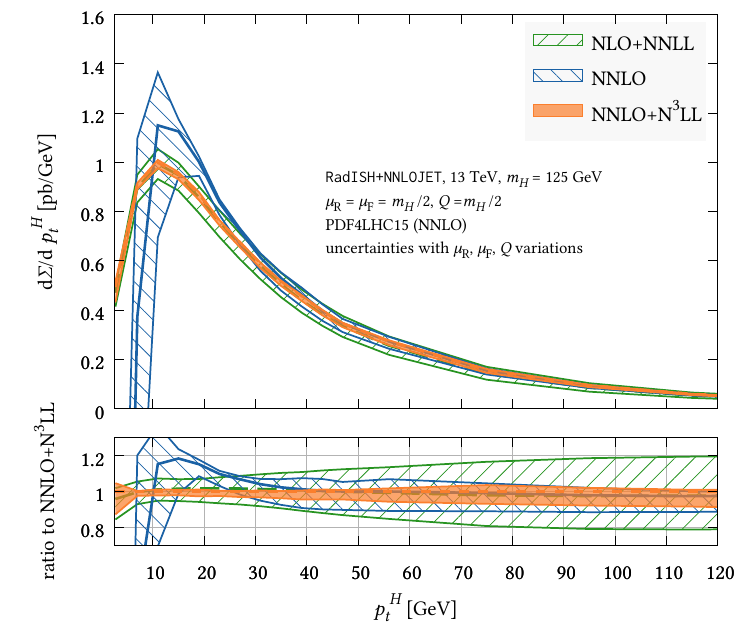} 
  \caption{Comparison of the  transverse-momentum distribution for Higgs boson production between NNLO+N$^3$LL, NLO+NNLL, and NNLO at central scale choice of $\muR = \muF =\mH /2$. The lower panel shows the ratio to the NNLO+N$^3$LL prediction.}
  \label{fig:n3ll_nnlo_v_nnlo_final}
\end{figure}

Finally, in fig.~\ref{fig:n3ll_nnlo_v_nnlo_final} we compare the prediction at NNLO+N$^3$LL with the NLO+NNLL and the NNLO distributions.
We observe a very good convergence of the perturbative uncertainties and a significant reduction of the scale uncertainty band in the overall kinematic range considered here. 

\paragraph{\itshape\mdseries Predictions for Higgs production within fiducial cuts.}

Theoretical predictions for inclusive Higgs production are highly valuable as they allow one to study the behaviour of the perturbative series and to understand many of its features.
Nevertheless, a comparison with data requires matched predictions in the fiducial volume where the experimental measurements are made.
The availability of matched predictions differential in the Born phase-space allows for a direct comparison, without relying on a Monte Carlo modeling of the experimental uncertainties.

Here we focus on the transverse-momentum distribution of the $\gamma \gamma $ system in the process $pp \rightarrow H \rightarrow \gamma \gamma$.
We generate events with an on-shell Higgs boson, followed by a decay into a photon pair using a narrow-width approximation, with a branching ratio of $2.35 \times 10^{-3}$.
The fiducial volume used here is defined by the cuts
\begingroup
\setlength{\jot}{8pt}
\begin{align}
\label{eq:Higgs_fiducial}
\min(\ptgo,\ptgt)> 31.25~\text{GeV},\qquad \max(\ptgo,\ptgt)> 43.75~\text{GeV},
\notag\\
0<|\eta^{\gamma_{1,2}}|<1.37 ~~{\rm or}~~ 1.52<|\eta^{\gamma_{1,2}}|<2.37,\qquad |Y_{\gamma\gamma}|<2.37
\,,
\end{align}
\endgroup
where $\ptg$ are the transverse momenta of the two photons,
$\etag$ are their pseudo-rapidities in the hadronic centre-of-mass
frame, and $Y_{\gamma\gamma}$ is the rapidity of the photon-pair.
The fiducial volume does not include the cuts for photon-isolation, as they would introduce additional non-global logarithmic corrections~\cite{Dasgupta:2001sh}, which would spoil the accuracy of the prediction. 
Nevertheless, since the photon-isolation criteria is not aggressive in this case, they could be included at fixed order.

\begin{figure}[tp]
  \centering
  \includegraphics[width=0.495\columnwidth]{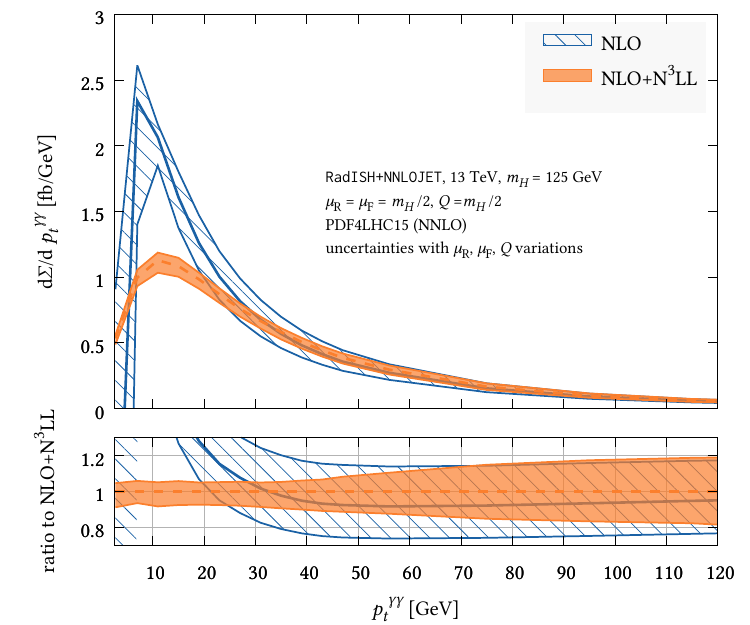} 
  \includegraphics[width=0.495\columnwidth]{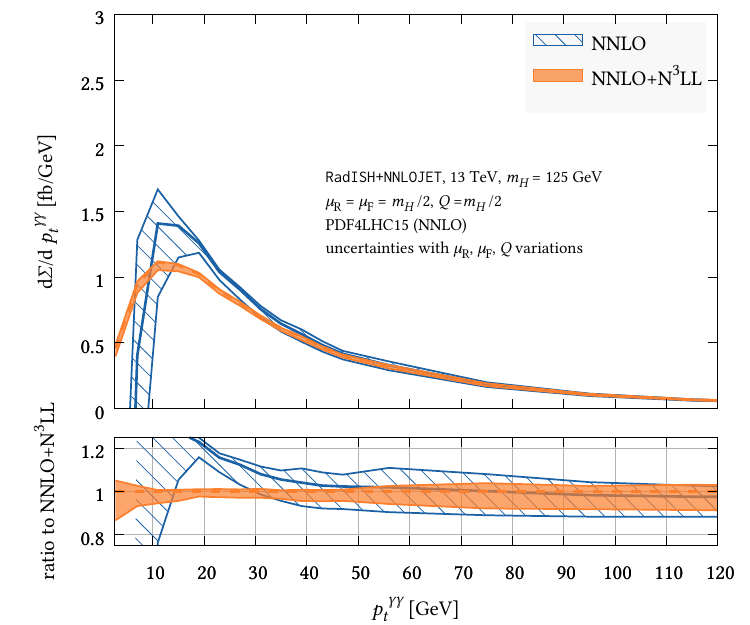} 
  \caption{Comparison of the transverse-momentum distribution for Higgs boson production at $\sqrt{s} = 13$~TeV in the fiducial volume defined by eq.~\eqref{eq:Higgs_fiducial} at NLO+N$^3$LL and NLO (left) and NNLO+N$^3$LL and NNLO (right). The lower panel shows the ratio to the NNLO+N$^3$LL prediction.}
  \label{fig:n3ll_nnlo_v_nnlo_gamgam}
\end{figure}

The comparison between fixed-order and matched predictions is shown in figure~\ref{fig:n3ll_nnlo_v_nnlo_gamgam} for different perturbative accuracies.
In the left (right) plot, we compare the (N)NLO and the (N)NLO+N$^3$LL distribution.
By comparing the left and the right plots, we observe a significant reduction of the scale uncertainty in the medium-high-$\ptgg$ region driven by the increased accuracy of the fixed-order computation.
The pattern observed in the right panel is very similar to what we observed in the inclusive case, and we note again a rather asymmetric scale uncertainty band, due to the large cancellations affecting the prediction using $\mu_0=\mH/2$ as a central scale.

These results are an important step in the LHC precision programme, where accurate predictions have become essential to fully exploit the experimental data. 
Additional effects must be included to improve on the predictions discussed here. 
At this level of accuracy, for Higgs production in gluon fusion the effect of heavy quarks, notably top and bottom, becomes relevant and must be taken into account.
Recent analyses have indeed shown that the effect of the top-bottom interference could modify the transverse-momentum spectrum with respect to the HEFT approximation~\cite{Lindert:2018iug,Caola:2018zye}.
The distortions at NLO+NNLL can be as large as $5\%$ (with an associated theory uncertainty of $\mathcal O(20\%)$) and are therefore comparable with the level of accuracy of the results presented in this section.
The inclusion of such effects is thus necessary for a consistent and accurate prediction of the transverse-momentum spectrum in the region $\pth \lesssim \mH$. 

\begin{savequote}[8cm]
An epilogue - Garp writes - is more than a body count. An epilogue, in the disguise of wrapping up the past, is really a way of warning about the future. 

\qauthor{--- John Irving, \textit{The world according to Garp}}

\end{savequote}

\chapter{\label{ch-summ}Summary \amper outlook} 



\lettrine[lines=2]{T}{he quest} for precision QCD at the LHC has multiple facets.
This thesis discusses two fundamental ingredients necessary to reduce theory errors to few percent: the computation of higher-order corrections to the perturbative series and the concurring improvement in the description of parton densities.
These two components are deeply intertwined, as the perturbative accuracy of the PDFs reflects that of the theory calculations used in their determination from experimental data. 
In particular, in this thesis we focused on all-order resummation techniques in perturbative QCD.
Thanks to resummation, the region of applicability of perturbative QCD extends also over regions where a fixed-order treatment is not sufficient.

The recent developments in threshold and high-energy resummation discussed in chapter~\ref{ch-res},  which lead to a series of public codes where resummed calculations have been implemented,  have been exploited in chapter~\ref{ch-respdfs} to obtain two determinations of parton densities in which fixed-order calculations are supplemented by threshold and high-energy resummation, respectively.
The advantage in such an improved description of PDFs is twofold.
Firstly, it is now possible to compute theoretical predictions which consistently include resummation effects both in the PDFs and in the coefficient functions.
This allows for a reliable description of kinematic regions where resummation is important, yet its inclusion in the PDFs may compensate for the effect of resummation in the sole coefficient function.
This has important implications for searches for new physics beyond the Standard Model at large invariant masses, since the consistent treatment of resummation effects generally reduces the overall enhancement.
Secondly, the quality of PDF fits improves if resummation effects are included, thereby providing an accurate description of the experimental data in regions where they are poorly described by a fixed-order treatment. 
Moreover, the inclusion of high-energy resummation provides convincing evidence for new physics effects \textit{within} the standard model, namely the onset of BFKL dynamics in the HERA structure function data.

The two analyses presented in chap.~\ref{ch-respdfs} share, however, a common limitation.
In both cases, it has indeed been necessary to use a reduced dataset, since resummed calculations are not yet available for a series of processes customarily included in PDF fits. 
Resummed calculations for the missing processes are thus required to obtain resummed PDFs competitive with state-of-the-art global PDF determinations.
Moreover, the accurate description of processes at high rapidity is likely to require the simultaneous resummation of threshold and high-energy logarithms.
Very recently, the first calculation where both classes of logarithms were resummed at the same time appeared in ref.~\cite{Bonvini:2018ixe}.  
Though much work remains to be done, the theoretical insight gained in the recent years is a promising indication that truly global PDF fits with high-energy and threshold logarithms resummed simultaneously might be viable in the foreseeable future. 

Resummed calculations are instrumental to reach an accurate description of many observables at the LHC. 
In this thesis, we have considered the case of Higgs boson production in gluon fusion and we have studied the effect of threshold resummation for the inclusive cross section, and of transverse-momentum resummation for the Higgs transverse-momentum spectrum.
In the case of the total cross section, the fixed-order perturbative series is almost pathological, with very large corrections at NLO and at NNLO. 
Moreover, scale variations are in this case a very poor estimator of the uncertainty due to missing higher orders. 
However, as discussed in chapter~\ref{ch-respheno}, threshold resummation proves a robust tool both to improve the convergence of the perturbative series and to obtain a robust estimate of the missing higher order uncertainty.
The N$^3$LO+N$^3$LL$^\prime$ prediction for the total cross section obtained in chapter~\ref{ch-respheno} is therefore both very precise and very accurate.

In chapter~\ref{ch-respheno} we also presented results at NNLO+N$^3$LL for the transverse-momentum distribution, using a novel approach formulated in direct space which we have introduced in chapter~\ref{ch-res}. 
The formalism can be extended to a wider class of observables and has been for instance exploited to resum the $\phi^*$ angle in Drell-Yan production with the same accuracy in ref.~\cite{Bizon:2018foh}.
This work opens up a variety of possible future directions.
Since it does not rely on a specific factorization theorem, the direct-space formalism can be extended to simultaneously resum observables which share the same Sudakov radiator.  
Moreover, it would be particularly interesting to consider the combined effect of high-energy, threshold, and transverse-momentum resummations on the Higgs $\pth$ spectrum. 
Such joint resummations have attracted quite some interest (see ref.~\cite{Muselli:2017xxx} for an in-depth overview of recent developments) and would allow for a very precise description of the transverse-momentum spectrum.
Finally, at this level of precision, it becomes necessary to start supplementing theoretical predictions with an estimate of the impact of heavy quark effects to improve on the results obtained using the Higgs effective field theory approximation.

With the experimental uncertainties reducing to a few percent, and in some cases even to a few permille, the challenge is now to reduce the theoretical predictions to a comparable or higher level of accuracy.
This would require a relentless effort to improve our understanding of QCD, by computing higher-order corrections for a larger number of processes and by refining our description of PDFs, parton showers, electroweak and non-perturbative effects.
The era of precision QCD has just begun...


\startappendices
\begin{savequote}[8cm]
\textlatin{With four parameters I can fit an elephant, and with five I can make him wiggle his trunk.}

  \qauthor{--- John von Neumann}
\end{savequote}

\chapter{\label{app:PDF}Review of PDF determination}

\lettrine[lines=2]{T}{he knowledge} of the functional form of parton distribution functions is essential for almost any theoretical prediction at hadron colliders.
Their determination from first principles, however, requires an understanding of QCD at non-perturbative scales $Q\sim \Lambda_{\rm QCD}$.
A possible approach to calculating PDFs is lattice QCD~\cite{Wilson:1974sk}.
In lattice QCD the QCD Lagrangian is discretized and the values of quark and gluon fields are considered in a finite-volume lattice.
A solution of QCD is then found by sampling the likelihood of different field configurations on the lattice.
Lattice QCD has been used successfully in several contexts, such as hadron spectroscopy, $\as(Q^2)$ and CKM determination.
However, the accuracy of PDFs based on lattice QCD is not yet competitive with that of PDFs extracted from a fit to experimental data\footnote{For a recent review on recent developments in PDF determinations in lattice QCD, see~\cite{Lin:2017snn}.}.
In these analyses, PDFs are parametrized at an initial scale and then evolved to the scale of the data using DGLAP evolution.
Theoretical predictions are then built by convoluting the PDFs with hard-scattering cross sections calculated using perturbative QCD and compared with data.
The best fit parameters are finally determined by minimizing an appropriate figure of merit.

Updated PDF sets are regularly being released by various collaborations.
These sets are determined using data from a variety of high-energy scattering processes in lepton-hadron and in hadron-hadron collisions. 
The analyses differ in several aspects, such as the input dataset used, the methodology, the computation of the data uncertainties, as well as theoretical details of the QCD analysis, like the computations of the cross sections and the treatment of heavy quarks. 
In this appendix, we first review the main ingredients for PDFs based on global analyses, i.e. the experimental input. 
Secondly, we briefly summarize the status of the theoretical calculations for the processes typically included in PDF analysis.
We then move to discuss the methodology employed in PDF fits and we summarize possible PDF parametrizations and minimization procedures.
We finally discuss various strategies for estimating PDF uncertainty.
We refer the interested reader to recent reviews on PDF determination~\cite{Forte:2010dt,DeRoeck:2011na,Perez:2012um,Forte:2013wc,Gao:2017yyd} for a more detailed description of the topics presented here.

\section{Experimental data}\label{sec:expdata}

One of the most important ingredients in the determination of parton densities is the dataset, whose constraining power depends on a combination of PDF sensitivity and experimental precision, as well as on the status of the theoretical calculations for the processes examined. 
To guarantee that leading-twist factorization can be used, kinematic cuts are often implemented in PDF analysis.

\begin{figure}[t]
\centering
   \includegraphics[]{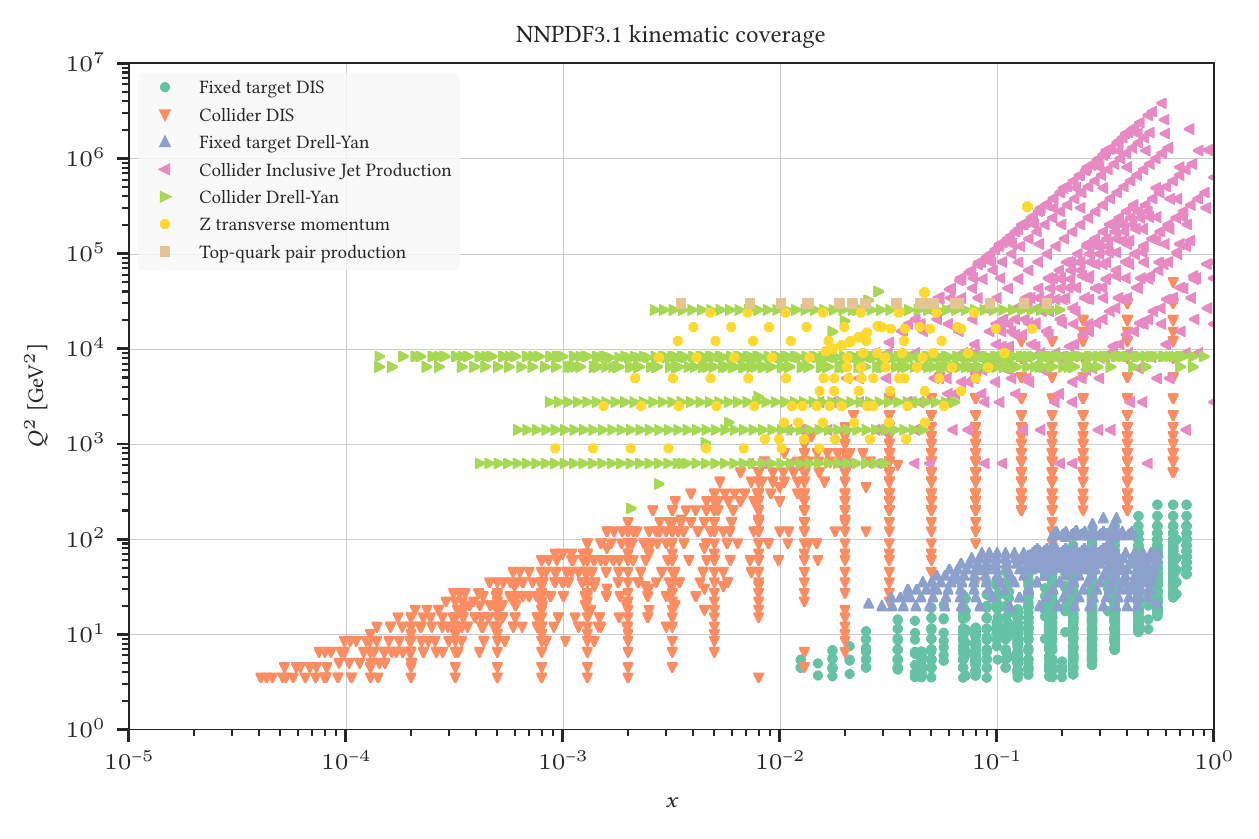}
  \caption{ Kinematic coverage of a state-of-the art global fit (in this case, ref.~\cite{Ball:2017nwa}) in the $(x,Q^2)$ plane.}
 \label{fig:kinplot31}
\end{figure}

Global PDF fits~\cite{Ball:2012cx,Jimenez-Delgado:2014zga,Harland-Lang:2014zoa,Dulat:2015mca,Accardi:2016qay,Alekhin:2017kpj,Ball:2017nwa} include data from a variety of measurements, from fixed-target DIS and DY processes, to jet and top-quark production at hadron colliders. 
A representative wealth of available data is displayed in fig.~\ref{fig:kinplot31}, where we show the approximate coverage in the $(x,Q^2)$ plane using LO kinematics of a state-of-the-art global analysis.
The high-$x$ and low-$Q^2$ region is covered by fixed target experiments, which were the backbone of the earliest PDF fits.
Collider DIS experiments populate the low-$x$ and low- and medium-$Q^2$ region, whereas the high-$Q^2$ region is dominated by recent data from collider experiments at the LHC and by older data from Tevatron. 
Here we briefly review some of the most relevant processes used in global PDF fits, discussing their PDF sensitivity and the available data, focusing in particular on the processes which are the mainstay of all PDF analyses: DIS and DY measurements. 

\subsection{Fixed-target \amper collider deep inelastic scattering}

Deep inelastic scattering experiments have been playing a key role in PDF determination since the first semi-quantitative determinations in the early '70s. 
In the LO approximation, the neutral current (NC) structure functions $F_2$ and $F_3$ (we recall that at LO $F_2= 2x F_1$) for the process $ep \rightarrow e X$ are given by (see e.g.~\cite{Patrignani:2016xqp})
\begin{align}
	\left[ F_2^\gamma, F_2^{\gamma Z}, F_2^Z\right] &=x \sum_q [e_q^2, 2 e_q g_V^q,(g_V^q)^2 + (g_A^q)^2] (q + \bar q) \\
		\left[ F_3^\gamma, F_3^{\gamma Z}, F_3^Z\right] &=x \sum_q [0, 2 e_q g_A^q,2 g_V^q g_A^q] (q - \bar q) 
\end{align}
where the sums over $q$ run over the active quarks of electric charge $e_q$, $g_A^q = \pm 1/2 $ and $g_V^q = \pm 1/2- 2 e_q \sin^2 \theta_W $, with $\pm$ corresponding to a $u$- or $d$-type quark, respectively.
The NC structure functions therefore provide only limited information on flavour separation, and in particular cannot separate quarks and antiquarks unless the scale $Q$ is sufficiently high such that the $Z$-mediated contribution becomes important.
However, NC DIS on deuterium targets provides a handle on the triplet contribution $T^3$ under the assumption of isospin symmetry which relates the quark and antiquark distributions of \mccorrect{the proton and the neutron, $u^{(p)}=d^{(n)}$, $u^{(n)}=d^{(p)}$, where $p$ is the proton and $n$ the neutron}. 

Additional constraint on the quark-antiquark separation is provided by the charged current (CC) processes $e^+ p \rightarrow \bar \nu X$ and $\nu p \rightarrow e^- X$.
In the LO approximation and neglecting CKM mixing, the structure functions are (above the charm threshold and below the top threshold)
\begin{align}
	&F_2^{W^-} = 2 x (u + \bar d + \bar s + c ), &F_3^{W^-} = 2 x (u - \bar d - \bar s + c ),
\end{align}
and the structure functions $F_2^{W^+},\, F_3^{W^+} $ for the processes $e^- p \rightarrow  \nu X$ and $\bar \nu \rightarrow e^+ X$ are obtained by the substitution $c\leftrightarrow s$, $u \leftrightarrow d$. 
As a consequence, neutrino data give access to the total valence component $V$.
Charm production in neutrino-induced DIS (known as dimuon production) is finally used to probe strangeness, since at LO and neglecting CKM mixing $F_2^{\nu p, c\bar c}=x F_3^{\nu p, c \bar c} = 2 x s$.

Since the gluon does not couple to electroweak final states, the gluon distribution is probed in DIS experiments directly through the small contribution which enters at  $\mathcal O(\as)$, or indirectly via scaling violations encoded in DGLAP evolution. 
In principle, $F_L$ measurements could be used to constrain the gluon PDF since $F_L$ is directly sensitive to the gluon PDF~\cite{Altarelli:1978tq}. 
The gluon PDF is also constrained by measurements of heavy-quark structure functions obtained in events where charm or bottom mesons are reconstructed in the final state, as the LO contribution (in absence of intrinsic components) is $g\gamma \rightarrow Q \bar Q$. 

There are many available data for DIS measurements, which are delivered either as cross sections or separated into structure functions.  
A prominent role is played by the combined NC, CC, charm and beauty structure function measurements at the HERA lepton-proton collider~\cite{Abramowicz:2015mha,H1:2018flt}. 
%

\subsection{Fixed-target \amper collider Drell-Yan}
The so-called Drell-Yan process --- i.e. the electroweak process in which a quark and antiquark pair annihilate to give a lepton pair --- plays a primary role in PDF analyses due to its constraining power on light flavour decomposition, including strangeness. 
At lowest order, and neglecting Cabibbo-suppressed contributions, the partonic subprocesses involved in inclusive EW boson production are
\begin{align}
	\qquad u \bar d, c \bar s \rightarrow W^+, \qquad d \bar u, s \bar c \rightarrow W^-, \qquad  q \bar q \rightarrow \gamma^*/Z,
\end{align}
which give access to the flavour composition of the proton since each flavour channel carries a different weight. 
Using LO kinematics, the rapidity and the invariant mass of the electroweak boson can be mapped onto the values of $x_1$, $x_2$ at which the PDFs are probed:
\begin{equation}\label{eq:DYkin}
	x_{1,2} = \frac{M}{\sqrt{s}}e^{\pm Y},
\end{equation}
where $M$ is the invariant mass of the virtual boson and $Y$ is its rapidity. 
Eq.~\eqref{eq:DYkin} therefore relates the PDF sensitivity of DY observables to kinematic variables which can be reconstructed in an experiment (for simplicity, in the following discussion we assume that in the CC case the experimentally inaccessible rapidity of the virtual boson coincides with the measured rapidity of the charged lepton in the final state).

Here we focus on the dominant $u$ and $d$ contributions in a $pp$ collider experiment.
Though the presence of a $\bar p$ beam provides additional information on the flavour structure of the proton, the overall constraints are analogous.
 In the CC case, information on the flavour separation can be obtained by considering the ratio of $W^+$ and $W^-$ production differential in rapidity
\begin{equation}
	R_{\pm} = \frac{d \sigma (W^+)/dY}{d \sigma (W^-)/dY},
\end{equation}
and the $W$ asymmetry
\begin{equation}
	A_{W} = \frac{d \sigma (W^+)/dY-d \sigma (W^-)/dY}{d \sigma (W^+)/dY+d \sigma (W^-)/dY}.
\end{equation}
In the central region, where $x_1=x_2=x_0=M/\sqrt{s}$ and one can approximate $\bar u = \bar d$,
\begin{equation}\label{eq:RandA}
	R_{\pm} \simeq \frac{u(x_0)}{d(x_0)}, \qquad A_{W} \simeq \frac{u_V(x_0)-d_V(x_0)}{u_V(x_0)+d_V(x_0)};
\end{equation}
therefore $R_{\pm}$ is sensitive to the ratio between $u$ and $d$, whilst $A_{W}$ is sensitive to the $u$ and $d$ valence difference.
Equivalent constraints can be derived in the forward region, see e.g.~\cite{Gao:2017yyd}.

NC DY provides similar information on the $u$ and $d$ content of the proton; there are additional constraints on PDFs at low and intermediate values of $x$ when one moves away from the $Z$ peak and explores the low-invariant mass region.
Here the virtual-photon exchange dominates
\begin{equation}
	\frac{d\sigma}{dY} \simeq \sum_q e_q^2 (q (x_1) \bar q(x_2) + q(x_2) \bar q(x_1)),
\end{equation}
and as a consequence the $u \bar u $ channel is enhanced with respect to the $d \bar d$. 
This region is also sensitive to the gluon PDF, especially if the cuts on final-state leptons tend to increase the importance of higher-order corrections.
Finally, NC DY production on fixed deuteron targets can be used to constrain the $\bar u/\bar d$ ratio.
In the valence region $x_1 \gtrsim 0.1$, where $q(x_1) \sim q_V(x_1)$, isospin symmetry implies that
\begin{equation}
\frac{\sigma^{pn}}{\sigma^{pp}}	\simeq \frac{\bar d (x_2)}{\bar u (x_2)}.
\end{equation}


Global PDF analyses use a variety of DY measurements, both on fixed target and at hadron colliders. 
Among the fixed-target experiments, an important role has been played by the E606 and E866 experiments at Fermilab~\cite{Towell:2001nh}.
There is a wealth of collider DY measurements, both at the $W$ and $Z$ peak, as well as off peak. 
A comprehensive list of recent datasets is presented in ref.~\cite{Gao:2017yyd}.


\subsection{Inclusive jet production}

Inclusive jet production measurements at hadron colliders have been used since the earliest measurements at the Tevatron to constrain the gluon PDF at large $x$.
Jet cross sections are experimentally reconstructed using a well-defined jet algorithm; to be compared with theoretical predictions, the jet algorithm used must be IRC safe.
The most common choice for data collected at the LHC is the anti-$k_T$ algorithm~\cite{Cacciari:2008gp}, though other algorithms are also used.

The PDF sensitivity of jet production varies according to the definition of the observable and the kinematics.
PDF fits typically include single-inclusive jet cross section data, which are double-differential in jet $p_T$ and rapidity $Y$. 
For each event, all jets are considered and included in the same distribution. 
The gluon-induced contribution in these measurements dominates at low $p_T$ but is also quite significant at high $p_T$. 
Since the quark content of the proton is already rather constrained by DIS and DY measurements, jet data provide a handle on the gluon PDF in the region of medium and large $x$. 
There are several double-differential single-inclusive datasets which can be included in PDF fits; again, we refer to ref.~\cite{Gao:2017yyd} for a comprehensive list.

\subsection{Transverse momentum of $Z$ boson}

Very precise measurements of the inclusive transverse momentum of the $Z$ boson have been recently released by the ATLAS, CMS and LHCb collaborations.
The dominant contribution in the region of moderate and large transverse momentum is gluon and quark scattering; as a consequence, the inclusion of $Z$ transverse momentum data can provide additional constraints to the gluon in the medium-$x$ region, which is only partly probed by collider DIS data and jets data. 

\subsection{Top quark production}

Another process which provides strong constraints on the gluon PDF, especially at large $x$, is top quark production.
At values of $\sqrt{s}$ probed at the LHC the gluon contribution to the top-pair cross section exceeds 80\%. 
Experimental measurements are presented at the total cross section level at Tevatron and both at the inclusive and differential level at the LHC.

\subsection{Other processes}

There are other processes which can provide additional constraints on several combinations of PDFs, which however we do not have time to discuss in detail.
The large-$x$ gluon can be probed also by prompt-photon production, whose dominant contribution is the QCD Compton scattering $qg \rightarrow q \gamma$.
$W^\pm+c$ production allows one to put constraint on the strangeness and on the strangeness asymmetry in the proton, whilst single-top production provides a handle on the $b$ quark content of the proton. 
Finally, the small-$x$ gluon can be further constrained by including in PDF fits charmed-meson production at hadron colliders, as the dominant channel for this process is $gg\rightarrow c\bar c$.
LHCb data~\cite{Aaij:2013mga,Aaij:2015bpa,Aaij:2015bpa} provide information on the gluon PDF down to very small values of $x\sim10^{-6}$, a region which is also important for neutrino astrophysics~\cite{CooperSarkar:2011pa,Gauld:2015yia,Gauld:2015kvh,Garzelli:2016xmx,Gauld:2016kpd}.

\section{Theoretical accuracy}

Since parton densities are based on a combination of experimental data and perturbative QCD calculations, the perturbative accuracy of a PDF fit depends on that of the theoretical predictions used in the analysis, that is the splitting functions eq.~\eqref{eq:AP} and the partonic cross sections which are convoluted with the PDFs. 

PDFs are typically determined using fixed-order perturbation theory.
For most of the processes we have reviewed in section~\ref{sec:expdata} the coefficient functions have been computed to NNLO accuracy and in some cases up to N$^3$LO.
In particular, DIS production is known up to N$^3$LO accuracy in the massless limit~\cite{Moch:2004xu,Vermaseren:2005qc} and to NNLO including mass effects~\cite{Laenen:1992zk,Laenen:1992xs}, and the coefficient functions are numerically implemented in a number of public codes such as \texttt{APFEL}~\cite{Bertone:2013vaa}, \texttt{QCDNUM}~\cite{Botje:2010ay} and \texttt{OpenQCDrad}~\cite{Alekhin:2010sv}. 
The NNLO calculation for differential $W$ and $Z$ rapidity distributions in DY production has been known since quite some time~\cite{Anastasiou:2003ds,Anastasiou:2003yy,Melnikov:2006di,Melnikov:2006kv}, and there exist several tools to produce theoretical predictions.
%
Single-inclusive jet and inclusive dijet production have been computed up to NNLO accuracy~\cite{Currie:2016bfm,Currie:2017eqf}, and Z $p_T$ distributions have been computed recently at NNLO by two independent groups~\cite{Boughezal:2015ded,Gehrmann-DeRidder:2016jns,Ridder:2016nkl,Boughezal:2016isb}.
The NNLO correction to the total cross section for top production has been known for a few years, whereas the full NNLO corrections for a variety of differential distribution have only recently become available~\cite{Czakon:2014xsa,Czakon:2015owf,Czakon:2016dgf}.
Let us note that, at this level of accuracy, electroweak (EW) corrections start to become relevant as $\alpha_{\rm EW}\sim \frac{1}{100}\sim \as^2$. 
EW corrections are not systematically included in current PDF analyses yet.
We refer the reader to ref.~\cite{Gao:2017yyd} for a thorough review of the status of the theoretical calculations for the various processes included in global analyses as well as for a discussion about the current status of EW corrections. 

By convoluting partonic cross sections computed at NNLO accuracy with DGLAP evolution computed using NNLO splitting functions PDFs can now be determined up to NNLO accuracy. 
NLO PDFs are widely used in conjunction with automatic NLO codes to produce NLO-accurate predictions for a variety of processes, whilst optimized LO PDFs are delivered to be used with Monte Carlo event generators~\cite{Sherstnev:2007nd}. 
For the majority of processes included in PDF analyses, fixed order theory accurately describes the data.
However, as discussed in chapter~\ref{ch-res}, in some cases logarithmic-enhanced terms appear at all orders and spoil the perturbative convergence of the series.
When this happens, one must supplement the fixed-order description with the resummation of the enhanced terms to all orders. 
PDF sets extracted using resummed calculations are discussed in chapter~\ref{ch-respdfs}.

\section{PDF parametrization}

The determination of a set of functions from a finite ensemble of experimental data points is a well-known mathematically ill-posed problem, as there is no unique solution. 
Therefore, to obtain a solution it is necessary to adopt an {\it ansatz}.
The PDFs are usually parametrized at the initial scale $Q_0 \sim 1-2$ GeV as
\begin{equation}\label{eq:PDFpar}
	x f_i(x,Q_0^2) = x^{\alpha_i} (1-x)^{\beta_i} \mathscr F_i (x, \{ \gamma_i \}),  
\end{equation}
where $\mathscr F_i$ is a smooth function and the parameters $\alpha_i, \,  \beta_i$ and $\{\gamma_i \}$ are determined by the fit. 
This choice is motivated by the theoretical, non-perturbative expectation that PDFs should behave as a power law at asymptotic values of $x$ (see e.g.~\cite{Ball:2016spl}).
In particular, the power behaviour at $x\rightarrow 0$ is predicted by Regge theory~\cite{Regge:1959mz}, whereas the behaviour at \mccorrect{$x\rightarrow 1$} is constrained by quark counting rules~\cite{Brodsky:1973kr,Devenish:2004pb,Roberts:1990ww}.

The parameterization eq.~\eqref{eq:PDFpar} is adopted for all the PDFs entering in the fit (or, typically, for the gluon and a convenient linear combination of the quark PDFs).   
In principle, the number of independent fitted PDFs should be equal to thirteen.
However, since heavy-quark PDFs are usually assumed to be generated by gluon splitting, it is customary to include seven independent PDFs in the fits (in recent fits by the NNPDF collaboration, the charm PDF is added alongside the other PDFs within the assumption that $c(x,Q_0^2)=\bar c(x,Q_0^2)$, thus the number of fitted PDFs is eight).

The form of $\mathscr F_i$ must be determined from the fit, with the only assumption that it tends to a constant in the limits $x\rightarrow 0$, $x \rightarrow 1$.
The specific choice for the interpolation function $\mathscr F_i$ is not uniquely defined and varies between different groups.
The simplest option, which has been largely used in the literature, is to assume a polynomial or an exponential polynomial of $x$ or $\sqrt{x}$.
However, the need for a more flexible and unbiased parameterization has emerged. 

In recent fits by the MMHT collaboration and the CT10 collaboration the function $\mathscr F_i$ is expanded on a basis of Chebyshev and Bernestein polynomials, respectively (see~\cite{Harland-Lang:2014zoa,Dulat:2015mca} for details).
This choice allows one to increase the number of free parameters whilst avoiding large cancellations amongst terms which plague fits with simpler functional forms, and at the same time it reduces the correlations between terms.

The NNPDF collaboration has adopted a somewhat different approach.
The functional dependence on $x$ is parametrized by neural networks, which allows for great flexibility and a reduced bias due to their redundancy~\cite{DelDebbio:2007ee,Hartland:2018}. 
Specifically, NNPDF uses multilayer feed-forward neural networks, whose architecture is fixed for all the PDFs entering in the fit. 
The number of free parameters is about one order of magnitude higher with respect to other PDF collaborations; in its latest fit, each PDF has 37 free parameters and the total number of parameters to be determined is thus 296. 

The rather different number of free parameters to be determined has important consequences on the minimization strategies used by each collaboration.
For instance, if the number of parameters becomes too large, the absolute minimum of the figure of merit used to determined the fit quality might not correspond to the best fit as one might start to fit the statistical noise rather than the underlying law.
We will discuss the choice of an appropriate figure of merit and various strategies for minimization in the next section.

\section{Fit quality \amper strategies for minimization}\label{sec:qualityandmin}

The figure of merit used to measure the fit quality is typically the log-likelihood function $\chi^2$
\begin{equation}\label{eq:chi2form1}
	\chi^2 (\{ \zeta_i\} ) = \sum_{i,j}^{N_{\rm dat }} (D_i - T_i(\{ \zeta_i\})) ({\rm cov}^{-1} )_{ij} (D_j - T_j(\{\zeta_i\})),
\end{equation}
where $D_i$ are the data points, $T_i$ are the theoretical predictions expressed in terms of the PDF parameters $\zeta_i = \{ \alpha_i, \beta_i, \{\gamma_i\}\}$, and the experimental covariance matrix ${\rm cov}$ and its inverse ${\rm cov}^{-1}$ are defined by 
\begin{align}
	({\rm cov})_{ij} &= (\sigma^{\rm uncorr}_{i})^2 \delta_{ij} + \sum_{\kappa=1}^{N_{\rm corr}} \sigma^{\rm corr}_{\kappa, i } \sigma^{\rm corr}_{\kappa, j },\\
	({\rm cov}^{-1})_{ij} &= \frac{\delta_{ij}}{(\sigma^{\rm uncorr}_{i})^2} - \sum_{\kappa,\lambda=1}^{N_{\rm corr}}  \frac{\sigma^{\rm corr}_{\kappa, i }}{(\sigma^{\rm uncorr}_{i})^2} A^{-1}_{\kappa \lambda}  \frac{\sigma^{\rm corr}_{\lambda, j }}{(\sigma^{\rm uncorr}_{j})^2}, & A_{\kappa \lambda} = \delta_{\kappa \lambda} + \sum_{i}^{N_{\rm dat }} \frac{\sigma^{\rm corr}_{\kappa, i } \sigma^{\rm corr}_{ \lambda,i } }{\sigma^{\rm uncorr}_{i}}.
\end{align}
Here $i=1, \ldots N_{\rm dat}  $ indicates the individual data points, affected by uncorrelated uncertainty $\sigma^{\rm uncorr}_{i}$ and $\kappa=1, \ldots {N_{\rm corr}}$ sources of correlated uncertainty $\sigma^{\rm corr}_{\kappa, i } $.
In the absence of correlated uncertainties the covariance matrix has non-zero entries only in the diagonal; however, a faithful figure of merit must include information on the correlations, since otherwise the uncertainty might be misestimated. 

The $\chi^2$ eq.~\eqref{eq:chi2form1} is sometimes written in terms of the nuisance parameters $r_\kappa$ as
\begin{equation}\label{eq:chi2form2}
	\chi^2 (\{\zeta_i\}, \{ r_\kappa \}) = \sum_{i=1}^{N_{\rm dat}}\left( \frac{\hat D_i - T_i (\{ \zeta_i\})}{(\sigma^{\rm uncorr}_{i})^2}\right)^2 + \sum_{\kappa=1}^{N_{\rm corr}} r_\kappa^2,
\end{equation}
where 
\begin{equation}\label{eq:sysshift}
	\hat D_i  = D_i -  \sum_{\kappa=1}^{N_{\rm corr}} r_\kappa \sigma^{\rm corr}_{\kappa, i } .
\end{equation}
Minimizing eq.~\eqref{eq:chi2form2} by assuming purely Gaussian errors with respect to the nuisance parameters one finds
\begin{align}
\tilde r_\kappa = \sum_{i=1}^{N_{\rm dat}}\frac{D_i - T_i}{(\sigma^{\rm uncorr}_{i})^2} \sum_\lambda^{N_{\rm corr}} A_{\kappa \lambda}^{-1} \sigma^{\rm corr}_{\lambda, i },
\end{align}
which can be substituted in eq.~\eqref{eq:chi2form2} to give back eq.~\eqref{eq:chi2form1}.  
Whilst formally equivalent to eq.~\eqref{eq:chi2form2} when $\{r_\kappa \}=\{\tilde r_\kappa \}$, the form~\eqref{eq:chi2form2} can be convenient to study the behaviour of the shifts at the minimum to check their gaussianity.
Moreover, figures which compare the data points shifted as in eq.~\eqref{eq:sysshift} to the theoretical predictions allow one to visualize the effect of the correlated uncertainties more effectively~\cite{Pumplin:2002vw}. 

Let us mention that there is a \mccorrect{subtlety} when multiplicative uncertainties (i.e. such that the size of the uncertainty is proportional to the measured value) are present in the definition of the $\chi^2$.
In this case, it has been shown that a $\chi^2$ defined as in \eqref{eq:chi2form1} leads to the so-called {\it d'Agostini bias}~\cite{DAgostini:1993arp}, as the data points subject to downward fluctuations have a smaller normalization uncertainty. 
As a consequence, if \eqref{eq:chi2form1} is used the fit tends to systematically undershoot the data.
Several solutions have been proposed in the literature.
For instance, the NNPDF collaboration uses the so-called $t_0$ definition of the covariance matrix~\cite{Ball:2009qv}: the covariance matrix is constructed by rescaling the multiplicative error by the theory predictions obtained from a previous iteration of the $\chi^2$ minimization.

Once a suitable figure of merit has been constructed, the next step is to find the global minimum in the parameter space spanned by the PDF parameters $\{ \zeta_i\}$.
The strategy to find the minimum varies among different groups.
If the number of free parameters is moderate, the methods used are typically gradient-based. 
Variants of these methods are employed in MMHT and CT analyses, where the Levenberg-Marquardt method~\cite{Levenberg,Marquardt} and the variable metric method provided by the \texttt{MINUIT} package~\cite{James:1975dr} are used, respectively. 

However, the computation of the gradient becomes computationally very challenging if the number of parameters is too large as in the case of NNPDF analyses, since it requires one to compute the inverse of the Hessian matrix many times. 
As a consequence, NNPDF follows a gradient-free minimization approach such as the Nodal Genetic Algorithm (NGA)~\cite{Montana:1989}.
The strategy of the algorithm is very simple: at each iteration of the minimization procedure, the parameters are varied gaussianly around the search centre and the best-fit parameters are selected for the next iteration.
Some care is needed to avoid over-learning, since the algorithm selects only the best candidates at each iteration and is therefore quite sensitive to noise. 
Recent studies based on the NNPDF framework~\cite{Bertone:2017tyb,Carrazza:2017udv} have explored another strategy --- the Covariance Matrix Adaption - Evolutionary Strategy (CMA-ES) algorithm --- which is commonly used when gradient descent methods are difficult to apply~\cite{Hansen2001ecj,Hansen16a} . 

\section{Error propagation: Hessian  \amper  Monte Carlo approaches}

Until the late '90s, PDF uncertainty could be assessed only by comparing predictions obtained using different PDF sets. 
When PDFs started to be a tool for precision physics, a reliable estimate of the uncertainty related to PDFs became mandatory. 
Therefore, state-of-the-art PDF sets are always delivered with an associated PDF uncertainty.
This currently includes the experimental uncertainty on the data used to extract the PDF (and possibly other methodological uncertainties), but does not include any theory error; for instance, since PDFs are extracted using fixed-order QCD, one should take into account the uncertainty due to missing higher orders.
The theory error has been always assumed to be negligible with respect to the experimental and methodological uncertainty.
Due to the wealth of very precise data and the constant methodological improvements, however, the theory error is now comparable to the PDF uncertainty in a wide range of $x$ and $Q^2$ and will eventually become dominant.
Whereas a thorough study of theory errors in PDF determination is still waited for, preliminary studies of theoretical uncertainties in PDF analyses have been presented in~\cite{Bendavid:2018nar}. 

In this section we will focus on the experimental uncertainty and we summarize two of the most common methods to evaluate the PDF error: the Hessian method and the Monte Carlo (MC) method.
Another method of error propagation, the Lagrange multiplier method, is reviewed in refs.~\cite{DeRoeck:2011na,Hartland:2014nha,Gao:2017yyd}. 

\subsection{The Hessian method}

The Hessian method is the most common method of uncertainty determination in PDF fitting. 
Qualitatively, the method is based on studying the perturbations of the $\chi^2$ around its minimum when the fitting parameters $\{\zeta_i \}$ are varied; the uncertainty on the physical observables is then determined geometrically by considering the values computed using the perturbed parameters. 

Let us discuss the method more quantitatively. 
Close to its minimum $\tilde \chi^2$, the variations of the $\chi^2$ can be  approximated as quadratic
\begin{equation}
	\Delta \chi^2 \equiv \chi^2 - \tilde \chi^2 = \sum_{i,j}^{N_{\rm par}} (\zeta_i - \tilde\zeta_i ) H_{ij} (\zeta_j - \tilde\zeta_j ),
	\end{equation}
where $\{ \tilde\zeta_i \} $ correspond to the best-fit parameters and we have defined the Hessian matrix
\begin{equation}\label{eq:hessianlinear}
	H_{ij} \equiv \frac{1}{2} \frac{\partial^2 \chi^2}{\partial \zeta_i \partial \zeta_j } \Bigg|_{\{  \zeta_i\} = \{ \tilde \zeta_i\}}.
\end{equation}
In principle, the error on a generic observable $\mathcal F$ which depends on the PDFs could be determined by simple linear error propagation as
\begin{equation}
	\sigma_{\mathcal F} = T \left( \sum_{i,j}^{N_{\rm par}} \frac{\partial \mathcal F}{\partial \zeta_i} (H)^{-1}_{ij}  \frac{\partial \mathcal F}{\partial \zeta_i} \right)^{1/2},
\end{equation}
where $T=\sqrt{\Delta \chi^2}$ is the so-called tolerance factor, which allows one to match the range of variation of the fit parameters to the confidence interval associated to the PDF uncertainties. 

However, eq.~\eqref{eq:hessianlinear} presents some limitations since the partial derivatives of the observables with respect to the fit parameters are generally unknown.
Moreover, numerically instabilities may arise due to very different variations of the $\chi^2$ in different directions in parameter space~\cite{DeRoeck:2011na}. 
One can overcome these limitation simply by diagonalizing the hessian matrix~\cite{Pumplin:2000vx,Pumplin:2001ct}.
After the diagonalization procedure, the error on an observable $\mathcal F$ is given by 
\begin{equation}
	\sigma_{\mathcal F}^2 = \frac{1}{2}  \sum_{i}^{N_{\rm par}} \left[ \mathcal F(S^+_i) - \mathcal F(S^-_i)  \right]^2,
\end{equation}
where $S^\pm$ are the PDF set constructed along the eigenvector directions, displaced by the desired $\Delta \chi^2$.  

The uncertainties determined from the Hessian procedure depend somewhat on the choice of the value of the tolerance $T$. 
For specialized PDF sets based on a restricted dataset one may use the textbook tolerance $\Delta \chi^2 = 1$, although the final uncertainty may be larger if methodological uncertainties are taken into account e.g. by considering variations of the parametrization as in the HERA PDF family~\cite{Abramowicz:2015mha}. 
However, in the context of global PDF fits the value of the tolerance is typically higher to take into account experimental inconsistencies or methodological uncertainties.
For this reason, CT and MMHT collaborations use a larger value for the tolerance, which in recent fits by the MMHT collaboration is determined dynamically (see~\cite{Martin:2009iq}).

\subsection{The Monte Carlo method} 

The Monte Carlo method is a complementary method for determination of the PDF uncertainty, which is based on a MC procedure in the space of the experimental data. 
The method is designed to construct a faithful representation of the uncertainties present in the initial data without any assumption on their nature.

The first step in the MC method is the construction of an ensemble of $N_{\rm rep}$ of artificial data replicas (dubbed {\it pseudodata} replicas) for every data point included in the fit, generated according to the probability distribution of the initial data. 
For a given experimental measurement of a generic observable $\mathcal F^{\rm exp}$, characterized by a total correlated uncertainty $\sigma^{\rm corr}$, $\kappa = 1, \ldots N_{\rm corr}$ uncorrelated uncertainties $\sigma^{\rm uncorr}_\kappa$ the artificial MC replicas $\mathcal F^{{\rm art},k}$ are constructed as~\cite{Forte:2002fg}
\begin{equation}
	\mathcal F^{{\rm art},\ell}  = \mathcal N^\ell \mathcal F^{{\rm art},\ell}  \left( 1 + \sum_{\kappa = 1}^{N_{\rm corr}} r^{(\ell)}_\kappa \sigma^{\rm corr}_\kappa + r^{(\ell)} \sigma^{ \rm uncorr}  \right) , \qquad \ell = 1, \ldots N_{\rm rep},
\end{equation}
where $r$ are random numbers (for instance, they can follow a gaussian distribution) and the normalization of the probability distribution is determined by the normalization pre-factor $\mathcal N^\ell$.
Several analyses have shown that a faithful representation of the data and their correlations requires $N_{\rm rep} \simeq 1000 $. 

Once a MC sample of the experimental data is available, one fit is performed for each pseudodata replica.
At the end of the procedure therefore there exist $N_{\rm rep}$ equally probable PDF sets which reliably describe the probability density of parton densities in PDF space based on the original experimental errors.
The central values and uncertainties of an observable $\mathcal F$ are finally obtained as the average and the variance over the replica ensemble
\begin{align}
	\langle \mathcal F \rangle &= \frac{1}{N_{\rm rep}} \sum_{\ell=1}^{N_{\rm rep}} \mathcal F^\ell,\\ 
	\sigma^2_{\mathcal F} &= \frac{1}{N_{\rm rep}-1} \sum_{\ell=1}^{N_{\rm rep}} (\mathcal F^\ell- \langle \mathcal F^{\ell} \rangle)^2,
\end{align}
where $\mathcal F^{\ell}$ denotes the value of the observable $\mathcal F$ evaluated with the PDF replica $\ell$. 
Since the MC method naturally propagates the experimental error without any assumption, there is no need to increase the tolerance.
In case of fully-consistent datasets, the MC method and the Hessian method with $\Delta \chi^2=1$ coincide~\cite{Dittmar:2009ii}.
Algorithms to construct a MC representation of a Hessian set~\cite{Watt:2012tq} and, conversely, a Hessian representation of a MC set~\cite{Carrazza:2015aoa} are available.

\begin{savequote}[8cm]
\textlatin{Someone told me that each equation I included in the book would halve the sales.}

  \qauthor{--- Stephen Hawking}
\end{savequote}

\chapter{\label{app:analitic}Analytical expressions}

\lettrine[lines=2]{I}{n this} section are collected various explicit formul\ae\ which complete the discussions throughout the main body of the thesis.

\section{Running coupling}

The convention for the RG equation of the strong coupling used below reads
\begin{align}
  \mu^2 \frac{\partial \as(\mu^2)}{\partial  \mu^2}
  =
  \beta(\as) &\equiv
  -\as^2\left( \beta_0  +\beta_1\as +\beta_2
  \as^2 +\beta_3 \as^3 + \dots\right)\\
  &= - \beta_0 \as^2 (1+ b_1 \as + b_2 \as^2 + b_3 \as^3 + \ldots ),
\end{align}
where $\beta_k=b_k \beta_0$ for $k \geq 1$. 
The coefficients of the $\beta$-function up to four loop are~\cite{Czakon:2004bu}
\begin{eqnarray}
  \beta_0 &=& \frac{11 C_A - 2 n_f}{12\pi}\,,\qquad 
  \beta_1 = \frac{17 C_A^2 - 5 C_A n_f - 3 C_F n_f}{24\pi^2}\,,\\
  \beta_2 &=& \frac{2857 C_A^3+ (54 C_F^2 -615C_F C_A -1415 C_A^2)n_f
       +(66 C_F +79 C_A) n_f^2}{3456\pi^3}\,,\\
\beta_3 &=& \frac{1}{(4\pi)^4}\Bigg\{C_A C_F n_f^2 \frac14\left(\frac{17152}{243} + \frac{448}9 \zeta_3\right) + 
C_A C_F^2 n_f \frac12\left(-\frac{4204}{27} + \frac{352}{9} \zeta_3\right)\nonumber\\
&&\hspace{10mm} + \frac{53}{243} C_A n_f^3 + C_A^2 C_F n_f\frac12 \left(\frac{7073}{243} - \frac{656}9 \zeta_3\right) + 
C_A^2 n_f^2 \frac14\left(\frac{7930}{81} + \frac{224}9 \zeta_3\right)\nonumber\\
&&\hspace{10mm} + \frac{154}{243} C_F n_f^3 + 
C_A^3 n_f \frac12\left(-\frac{39143}{81} + \frac{136}3 \zeta_3\right) + C_A^4 \left(\frac{150653}{486} - \frac{44}9 \zeta_3\right)\nonumber\\
&&\hspace{10mm} + C_F^2 n_f^2 \frac14\left(\frac{1352}{27} - \frac{704}9 \zeta_3 \right) + 23 C_F^3 n_f + n_f \frac{d_F^{abcd}d_A^{abcd}}{N_A} \left(\frac{512}9 - \frac{1664}3 \zeta_3\right)\nonumber\\
&&\hspace{10mm} + n_f^2\frac{d_F^{abcd}d_F^{abcd}}{N_A} \left(-\frac{704}9 + \frac{512}3 \zeta_3\right) + \frac{d_A^{abcd}d_A^{abcd}}{N_A} \left(-\frac{80}9 + \frac{704}3 \zeta_3\right)\Bigg\}\,,
\end{eqnarray}
with
\begin{eqnarray*}
\frac{d_F^{abcd}d_F^{abcd}}{N_A} = \frac{N_c^4 - 6 N_c^2 + 18}{96 N_c^2},\qquad
\frac{d_F^{abcd}d_A^{abcd}}{N_A} = \frac{N_c(N_c^2 + 6)}{48},\qquad
\frac{d_A^{abcd}d_A^{abcd}}{N_A} = \frac{N_c^2 (N_c^2 + 36)}{24},
\end{eqnarray*}
and $C_A = N_c$, $C_F = \frac{N_c^2-1}{2N_c}$, and $N_c = 3$.

\section{Threshold resummation formul\ae\ for the total cross section}

In this section we collect the analytical expressions for the quantities needed to achieve N$^3$LL$^\prime$ accuracy for the total cross section in the rEFT. 

\subsection{Resummed cross section in direct QCD}\label{app:dQCD}

The $N$-space resummed coefficient function has the form (see~\cite{Bonvini:2014joa} and reference therein)
\begin{align}
\bm C_{\rm res}\(N,\as\)&=\bar g_0\(\as\) \exp\bar\Sud(\as,N),\\
\bar\Sud(\as,N)
&= \int_0^1dz\, \frac{z^{N-1}-1}{1-z}
\( \int_{\mH^2}^{\mH^2\frac{(1-z)^2}{z}} \frac{d\mu^2}{\mu^2} 2A\(\as(\mu^2)\) + D\(\as([1-z]^2 \mH^2)\) \),\label{eq:ShatMellin}\\
\bar g_0(\as) &= 1+\sum_{k=1}^\infty \bar g_{0,k} \as^k,\\
A(\as)&=\sum_{k=1}^\infty A_k\as^k, \qquad
D(\as)=\sum_{k=1}^\infty D_k\as^k,\label{eq:A,D}
\end{align}
where $\bar g_0(\as)$ does not depend on $N$. 
In the full theory, all the top-mass dependence is in $\bar g_0$. 
In the rEFT assumption, its expression factorizes as
\beq
\bar g_0(\as) = W(\mH^2,\mt^2) \, \tilde{\bar{g}}_0(\as),
\eeq
where $\tilde{\bar g}_0(\as)$ does not depend on the top mass.

In standard $N$-soft resummation (see e.g. ref.~\cite{Catani:2014uta}) the Mellin transform in eq.~\eqref{eq:ShatMellin} is usually computed, to any finite logarithmic accuracy, in the large-$N$ limit, leading to the expression
\begin{align}
\label{eq:Cres2}
\bm C_{\text{$N-$soft}}\(N,\as\) &=g_0(\as) \exp\Sud(\as,\ln N),\\
\Sud(\as,\ln N) &= \left[\frac{1}{\bar \as} g_1\(\bar \alpha \ln \frac{1}{N} \)+ g_2\(\bar \alpha \ln \frac{1}{N}\)+ \bar \as g_3\(\bar \alpha \ln \frac{1}{N}\)+ \bar \as^2 g_4\(\bar \alpha \ln \frac{1}{N}\)+\dots\],\\
g_0(\as) &= 1+\sum_{k=1}^\infty g_{0,k} \as^k,\\
g_i(\lambda)&=(2 \beta_0 )^{2-i} \sum_{k=1}^\infty g_{i,k} \left(\frac{\lambda}{2\beta_0}\right)^k, \qquad g_{1,1}=0,
\end{align}
where
\beq
\bar \alpha \equiv 2 \beta_0 \as.
\eeq
Note that
\beq \label{eq:res-vs-Nsoft}
\bm C_{\text{$N-$soft}}\(N,\as\) =\bm C_\text{res}(N,\as) \[1+\mathcal O \(\frac{1}{N}\)\].
\eeq
The function $\bar g_0(\as)$ is related to $g_0(\as)$ by
\bea\label{eq:g0bardefapp}
\bar g_0(\as)=g_0(\as) 
\exp\left[-\sum_{n=1}^\infty \as^n \sum_{k=0}^nb_{n,k}
\frac{\Gamma^{(k+1)}(1)}{k+1}\right].
\eea
The coefficients $b_{n,k}$ for $n = 1, 2, 3$ can be found in appendix A.1 of ref.~\cite{Ball:2013bra}.

The functions $g_i$, $i=1,2,3,4$ can be found explicitly for many processes in ref.~\cite{Moch:2005ba}.
For Higgs production they read
\bea\label{eq:gi}
g_1(\lambda) =&\, \frac{2A_{\rm cusp}^{(1)}}{\beta_0} \left[ (1+\lambda)\ln(1+\lambda) -\lambda \right], \\
g_2(\lambda) =&\, \frac{A_{\rm cusp}^{(2)}}{\beta_0^2} \left[ \lambda - \ln(1+\lambda) \right]
   + \frac{A_{\rm cusp}^{(1)}}{\beta_0} \left[ \ln(1+\lambda) \left( \ln\frac{\mH^2}{\muR^2} -2\gamma_E \right) -\lambda \ln\frac{\muF^2}{\muR^2} \right] \nonumber\\
       &\,+ \frac{A_{\rm cusp}^{(1)}b_1}{\beta_0^2} \left[ \frac{1}{2}\ln^2(1+\lambda) +\ln(1+\lambda) -\lambda \right],\\
g_3(\lambda) =&\, \frac{1}{4\beta_0^3} \left( A_{\rm cusp}^{(3)} - A_{\rm cusp}^{(1)}  b_2 + A_{\rm cusp}^{(1)}  b_1^2 - A_{\rm cusp}^{(2)}  b_1 \right) \frac{\lambda^2}{1+\lambda} \nonumber\\
&\, + \frac{A_{\rm cusp}^{(1)} b_1^2}{2\beta_0^3} \; \frac{\ln(1+\lambda)}{1+\lambda} \left[ 1 + \frac{1}{2} \ln(1+\lambda)\right]
      + \frac{ A_{\rm cusp}^{(1)}  b_2 - A_{\rm cusp}^{(1)}  b_1^2 }{2\beta_0^3} \; \ln(1+\lambda) \nonumber\\
&\, +  \left( \frac{A_{\rm cusp}^{(1)}  b_1}{\beta_0^2}  \gamma_E 
          + \frac{A_{\rm cusp}^{(2)}  b_1 }{2\beta_0^3}
          \right) \left[ \frac{\lambda}{1+\lambda} 
- \frac{\ln(1+\lambda)}{1+\lambda} \right]
\nonumber\\
&\, - \left(
            \frac{A_{\rm cusp}^{(1)}  b_2 }{2\beta_0^3}
          + \frac{A_{\rm cusp}^{(1)}}{\beta_0}   (\gamma_E^2 + \zeta_2) 
          + \frac{A_{\rm cusp}^{(2)}}{\beta_0^2}  \gamma_E
          - \frac{D^{(2)}}{4\beta_0^2} 
         \right) 
           \frac{\lambda}{1+\lambda} 
\nonumber\\
&\,
       +   \left[
         \left(
            \frac{A_{\rm cusp}^{(1)}}{\beta_0}  \gamma_E
          + \frac{A_{\rm cusp}^{(2)} - A_{\rm cusp}^{(1)}  b_1 }{2\beta_0^2}
          \right) 
           \frac{\lambda}{1+\lambda} 
       +  \frac{A_{\rm cusp}^{(1)}  b_1}{2\beta_0^2} \;
          \frac{\ln(1+\lambda)}{1+\lambda}
          \right] \ln\frac{\mH^2}{\muR^2}
\nonumber\\
&\,
       -  \frac{A_{\rm cusp}^{(2)}}{2\beta_0^2} \, \lambda \, \ln\frac{\muF^2}{\muR^2}
       + \frac{A_{\rm cusp}^{(1)}}{4\beta_0} \left[
         \lambda\, \ln^2\frac{\muF^2}{\muR^2}
          - \frac{\lambda}{1+\lambda} \, \ln^2\frac{\mH^2}{\muR^2}
          \right],\\
g_4(\lambda) &= \frac{1}{{48 \beta_0^4
   (\lambda +1)^2}} \Bigg\{6 \beta_0 \ln \frac{\mH^2}{\muR^2} \Big[\lambda  \left(4 \gamma_E
   ^2 A_{\rm cusp}^{(1)} \beta_0^2 \lambda +4 A_{\rm cusp}^{(1)} \beta_0^2 \lambda  \zeta_2+8
   A_{\rm cusp}^{(1)} \beta_0^2 \zeta_2+8 \gamma_E ^2 A_{\rm cusp}^{(1)} \beta_0^2\right.\nonumber \\
&\quad\left.   +A_{\rm cusp}^{(1)}
   b_1^2 \lambda -A_{\rm cusp}^{(1)} b_2 \lambda -A_{\rm cusp}^{(2)} (\lambda +2)
   (b_1-4 \gamma_E  \beta_0)+A_{\rm cusp}^{(3)} (\lambda +2)-\beta_0 D^{(2)}
   \lambda -2 \beta_0 D^{(2)}\right)\nonumber \\
   &\quad +b_1 \ln (\lambda +1) (4 \gamma_E 
   A_{\rm cusp}^{(1)} \beta_0+2 A_{\rm cusp}^{(2)})-A_{\rm cusp}^{(1)} b_1^2 \ln ^2(\lambda
   +1)\Big]\nonumber \\
   &\quad-3 \beta_0^2 \ln ^2\frac{\mH^2}{\muR^2} (\lambda 
   (\lambda +2) (4 \gamma_E  A_{\rm cusp}^{(1)} \beta_0+2 A_{\rm cusp}^{(2)})+2 A_{\rm cusp}^{(1)} b_1
   \ln (\lambda +1))\nonumber \\
   &\quad +6 \beta_0^2 \lambda  (\lambda +1)^2 (A_{\rm cusp}^{(1)} b_1+2
   A_{\rm cusp}^{(2)}) \ln ^2\frac{\muF^2}{\muR^2}-16 \gamma_E ^3 A_{\rm cusp}^{(1)}
   \beta_0^3 \lambda ^2-32 \gamma_E ^3 A_{\rm cusp}^{(1)} \beta_0^3 \lambda\nonumber \\
   &\quad +2 A_{\rm cusp}^{(1)}
   \beta_0^3 \lambda  (\lambda +2) \ln ^3\frac{\mH^2}{\muR^2}-4
   A_{\rm cusp}^{(1)} \beta_0^3 \lambda  (\lambda +1)^2 \ln
   ^3\frac{\muF^2}{\muR^2}-48 \gamma_E  A_{\rm cusp}^{(1)} \beta_0^3
   \lambda ^2 \zeta_2\nonumber \\
   &\quad -96 \gamma_E  A_{\rm cusp}^{(1)} \beta_0^3 \lambda  \zeta_2-32
   A_{\rm cusp}^{(1)} \beta_0^3 \lambda ^2 \zeta_3-64 A_{\rm cusp}^{(1)} \beta_0^3 \lambda 
   \zeta_3-24 \gamma_E ^2 A_{\rm cusp}^{(1)} \beta_0^2 b_1 \ln (\lambda +1)\nonumber\\
   &\quad-24    A_{\rm cusp}^{(1)} \beta_0^2 b_1 \zeta_2 \ln (\lambda +1)-12 \gamma_E  A_{\rm cusp}^{(1)}
   \beta_0 b_1^2 \lambda ^2+12 \gamma_E  A_{\rm cusp}^{(1)} \beta_0 b_1^2
   \ln ^2(\lambda +1)\nonumber \\
   &\quad+12 \gamma_E  A_{\rm cusp}^{(1)} \beta_0 b_2 \lambda ^2-4
   A_{\rm cusp}^{(1)} b_1^3 \lambda ^3+6 A_{\rm cusp}^{(1)} b_1^3 \lambda ^2 \ln
   (\lambda +1)-2 A_{\rm cusp}^{(1)} b_1^3 \ln ^3(\lambda +1)\nonumber \\
   & \quad +8 A_{\rm cusp}^{(1)} b_1
   b_2 \lambda ^3+9 A_{\rm cusp}^{(1)} b_1 b_2 \lambda ^2-12 A_{\rm cusp}^{(1)}
   b_1 b_2 \lambda ^2 \ln (\lambda +1)+6 A_{\rm cusp}^{(1)} b_1
   b_2 \lambda\nonumber \\
   &\quad -6 A_{\rm cusp}^{(1)} b_1 b_2 \ln (\lambda +1)-12
   A_{\rm cusp}^{(1)} b_1 b_2 \lambda  \ln (\lambda +1)-4 A_{\rm cusp}^{(1)} b_3
   \lambda ^3-9 A_{\rm cusp}^{(1)} b_3 \lambda ^2\nonumber \\
   &\quad +6 A_{\rm cusp}^{(1)} b_3 \lambda ^2
   \ln (\lambda +1)-6 A_{\rm cusp}^{(1)} b_3 \lambda +6 A_{\rm cusp}^{(1)} b_3 \ln
   (\lambda +1)+12 A_{\rm cusp}^{(1)} b_3 \lambda  \ln (\lambda +1)\nonumber\\
   &\quad -24 \gamma_E ^2
   A_{\rm cusp}^{(2)} \beta_0^2 \lambda ^2-48 \gamma_E ^2 A_{\rm cusp}^{(2)} \beta_0^2 \lambda -24
   A_{\rm cusp}^{(2)} \beta_0^2 \lambda ^2 \zeta_2-48 A_{\rm cusp}^{(2)} \beta_0^2 \lambda 
   \zeta_2+12 \gamma_E  A_{\rm cusp}^{(2)} \beta_0 b_1 \lambda ^2\nonumber \\
   &\quad +24 \gamma_E 
   A_{\rm cusp}^{(2)} \beta_0 b_1 \lambda -24 \gamma_E  A_{\rm cusp}^{(2)} \beta_0
   b_1 \ln (\lambda +1)+4 A_{\rm cusp}^{(2)} b_1^2 \lambda ^3-3 A_{\rm cusp}^{(2)}
   b_1^2 \lambda ^2-6 A_{\rm cusp}^{(2)} b_1^2 \lambda \nonumber\\
   &\quad +6 A_{\rm cusp}^{(2)}
   b_1^2 \ln ^2(\lambda +1)+6 A_{\rm cusp}^{(2)} b_1^2 \ln (\lambda +1)-4
   A_{\rm cusp}^{(2)} b_2 \lambda ^3-12 \gamma_E  A_{\rm cusp}^{(3)} \beta_0 \lambda ^2\nonumber\\
   &\quad-24   \gamma_E  A_{\rm cusp}^{(3)} \beta_0 \lambda  -12 A_{\rm cusp}^{(3)} \beta_0 \lambda  (\lambda
   +1)^2 \ln \frac{\muF^2}{\muR^2}-4 A_{\rm cusp}^{(3)} b_1 \lambda
   ^3-3 A_{\rm cusp}^{(3)} b_1 \lambda ^2+6 A_{\rm cusp}^{(3)} b_1 \lambda\nonumber \\
   &\quad  -6 A_{\rm cusp}^{(3)}
   b_1 \ln (\lambda +1)+4 A_{\rm cusp}^{(4)} \lambda ^3+6 A_{\rm cusp}^{(4)} \lambda ^2+12
   \gamma_E  \beta_0^2 D^{(2)} \lambda ^2+24 \gamma_E  \beta_0^2 D^{(2)}\nonumber \\
   &\quad 
   \lambda -3 \beta_0 b_1 D^{(2)} \lambda ^2-6 \beta_0 b_1
   D^{(2)} \lambda +6 \beta_0 b_1 D^{(2)} \ln (\lambda +1)+3
   \beta_0 D^{(3)} \lambda ^2+6 \beta_0 D^{(3)} \lambda\Bigg\}.
\eea
The coefficients appearing in the previous functions are~\cite{Moch:2005ba,Moch:2005ky,Laenen:2005uz}
\bea
A_{\rm cusp}^{(1)} &= \frac{C_A}{\pi},\nonumber \\
A_{\rm cusp}^{(2)} &= \frac{C_A}{2\pi^2}
\left[ C_A \left(\frac{67}{18} - \frac{\pi^2}{6}\right) - \frac{10}{9}T_F\, n_f \right],\nonumber \\
A_{\rm cusp}^{(3)} &= \frac{C_A}{4\pi^3} \left[ C_A^2 
\left( \frac{245}{24} - \frac{67}{9}\,\zeta_2
+ \frac{11}{6}\,\zeta_3 + \frac{11}{5}\,\zeta_2^2 \right) 
+ \left( -  \frac{55}{24}  + 2\,\zeta_3 \right)C_F \,n_f  \right. 
\nonumber\\ & 
\left. \mbox{} \qquad
+ \left( - \frac{209}{108} + \frac{10}{9}\,\zeta_2 -\frac{7}{3}\zeta_3 \right)
C_A \,n_f - \frac{1}{27} \, n_f^2 \right],
\nonumber\\
D^{(2)} &= \frac{C_A}{16\pi^2}
\left[ C_A\left(-\frac{1616}{27}+\frac{88}{9}\pi^2+56\zeta_3\right) 
+ \left(\frac{224}{27}-\frac{16}{9}\pi^2\right) n_f \right],\\
D^{(3)} &= \frac{C_A}{64 \pi
   ^3}\left[ C_A^2 \left(-\frac{2992 \zeta_2^2}{15}-\frac{352
   \zeta_2 \zeta_3}{3}+1212.64 \zeta_2+\frac{40144 \zeta_3}{27}-384
   \zeta_5-\frac{594058}{729}\right)\right.\nonumber\\
   &\quad +C_A n_f \left(\frac{736
   \zeta_2^2}{15}-\frac{29392 \zeta_2}{81}-\frac{2480
   \zeta_3}{9}+\frac{125252}{729}\right)\nonumber \\
   &\left.\quad +C_F n_f \left(-\frac{64
   \zeta_2^2}{5}-32 \zeta_2-\frac{608
   \zeta_3}{9}+\frac{3422}{27}\right)+n_f^2 \left(\frac{640
   \zeta_2}{27}+\frac{320 \zeta_3}{27}-\frac{3712}{729}\right)\right].
\eea
The coefficient $A_{\rm cusp}^{(4)}$ is still unknown and has only recently been computed numerically~\cite{Moch:2017uml,Moch:2018wjh}.
In sect.~\ref{sec:totalhiggs} its value is estimated with the $[1,1]$ Pad\'e approximant~\cite{Moch:2005ba}
\begin{equation}
	A_{\rm cusp}^{(4)}=\frac {\(A_{\rm cusp}^{(3)}\)^2}{A_{\rm cusp}^{(2)}} \simeq 0.708187,  0.389178,  0.14013\quad  \text { for } \quad n_f=3,4,5.
\end{equation}
The coefficients $g_{0,i}$ with $i=1,2$ can be found with full scale dependence in ref.~\cite{Catani:2003zt}. 
Here we report also the $i=3$ coefficient for $\muR=\muF=\mH$, isolating the contributions from the Wilson coefficient squared $W$:
\begin{align}
g_{0,1} &=\frac{C_A \left(4 \zeta_2+2 \gamma_E ^2\right)}{\pi }+\frac{W_1}{\pi }\nonumber \\
&\simeq 6.91951 + 0.31831 W_1,\\
g_{0,2} &= \frac{C_A^2}{432 \pi ^2} \Big[432 \zeta_2^2+3216 \zeta_2+24 \gamma_E ^2 (126
   \zeta_2+67)\nonumber \\
   &\quad-924 \zeta_3-8 \gamma_E  (189 \zeta_3-202)+5130
   \zeta_4+864 \gamma_E ^4+528 \gamma_E ^3+2511\Big]\nonumber \\
   &\quad-\frac{C_A n_f}{54 \pi ^2}
   \Big[
   4 \gamma_E  (18 T_R \zeta_2-9 \zeta_2+7)+48 T_R
   \zeta_3+24 \gamma_E ^3 T_R+60 \zeta_2+9 \zeta_3+30 \gamma_E
   ^2+90\Big]\nonumber \\
   &\quad+\frac{C_F n_f (48 \zeta_3-67)}{48 \pi
   ^2}+\frac{W_2}{\pi ^2}+\frac{2 C_A W_1 \left(2 \zeta_2+\gamma_E ^2\right)}{\pi ^2}\nonumber\\
   &\simeq 33.1106 - 1.47186 n_f + 2.20255 W_1 + 0.101321 W_2,\\
g_{0,3} &= \frac{C_A^3}{1632960 \pi
   ^3} \Big\{-14 \gamma_E  \big[3485592 \zeta_2^2+360 \zeta_2 (3942
   \zeta_3-4049)\nonumber \\
   &\quad +5 (332856 \zeta_3-1710720 \zeta_4-139968
   \zeta_5-297029) \big]\nonumber \\
   &\quad -252 \gamma_E ^2 [316872 \zeta_2^2-168840
   \zeta_2+63360 \zeta_3-931500 \zeta_4-152845]\nonumber \\
   &\quad -9 [1874016
   \zeta_2^3-65520 \zeta_2^2+280 \zeta_2 (6831 \zeta_3-38718
   \zeta_4-23360)\nonumber \\
   &   \quad -35 \left(7776 \zeta_3^2-130828 \zeta_3+174858
   \zeta_4+31284 \zeta_5-216324 \zeta_6+215131\right)]\nonumber \\
   &\quad +20160 \gamma_E
   ^3 [297 \zeta_2-567 \zeta_3+1051]+15120 \gamma_E ^4 [648
   \zeta_2+925]+2177280 \gamma_E ^6+3991680 \gamma_E ^5\Big\}\nonumber \\
    &\quad-\frac{C_A^2 n_f}{233280 \pi ^3} \Big\{-45 [8 \zeta_2 (8 T_F
   (837 \zeta_3-202)-5994 \zeta_3-3963)-22912 T_F \zeta_3-25344
   T_F \zeta_4 \nonumber \\
   &\quad -110592 T_F \zeta_5+3456 \zeta_2^2+11752
   \zeta_3-3012 \zeta_4+62568 \zeta_5-98059]\nonumber \\
   &\quad+4 \gamma_E  [-1944
   (200 T_F+57) \zeta_2^2+180 (2148 T_F-505)\zeta_2\nonumber \\
   &\quad +540 (704
   T_F-465) \zeta_3+1555200 T_F \zeta_4+156565]\nonumber \\
   &\quad+360 \gamma_E ^2
   [8 T_F (198 \zeta_2-45 \zeta_3+202)+1728 \zeta_2+972
   \zeta_3+3403]\nonumber \\
   &\quad +2880 \gamma_E ^3 [T_F (324 \zeta_2+179)-108
   \zeta_2+139]+207360 \gamma_E ^5 T_F+17280 \gamma_E ^4 [11
   T_F+15]\Big\}\nonumber \\
   &\quad+C_A \Bigg\{\frac{n_f^2 }{46656 \pi ^3} [9
   (4608 T_F^2 \zeta_4+2304 (T_F-1) T_F \zeta_2^2+16
   (112 T_F-139) \zeta_2\nonumber \\
   &\quad+16 (160 T_F+129) \zeta_3-2736
   \zeta_4+7545)+ 64 \gamma_E  (27 \left(32 T_F^2-5\right)
   \zeta_3\nonumber \\
   &\quad+270 (2 T_F-1) \zeta_2+58)+288 \gamma_E ^2 \left(144
   T_F^2 \zeta_2-8 T_F (9 \zeta_2-7)-3\right)+6912 \gamma_E ^4
   T_F^2+11520 \gamma_E ^3 T_F]\nonumber \\
   &\quad-\frac{C_F n_f}{25920 \pi ^3}
   [-6 \gamma_E  \left(-2160 (2 T_F-1) \zeta_2+864 \zeta_2^2+4560
   \zeta_3-8555\right)\nonumber \\
   &\quad +5 (864 (4 T_F-39) \zeta_3-36 \zeta_2 (864
   \zeta_3-851)-792 \zeta_4-12960 \zeta_5+63991)\nonumber \\
   &\quad+8640 \gamma_E ^3
   T_F-4860 \gamma_E ^2 (16 \zeta_3-21)]\Bigg\}\nonumber \\
   &\quad-\frac{C_F
   n_f^2 (828 \zeta_2+3024 \zeta_3+72 \zeta_4-4481)}{2592 \pi
   ^3}+\frac{C_F^2
   n_f (111 \zeta_3-180 \zeta_5+38)}{36 \pi ^3}\nonumber \\
     &\quad +W_1 \left\{\frac{C_A^2}{432 \pi ^3} [432 \zeta_2^2+3216 \zeta_2+24
   \gamma_E ^2 (126 \zeta_2+67)\right.\nonumber \\
   &\quad-924 \zeta_3-8 \gamma_E  (189 \zeta_3-202)+5130
   \zeta_4+864 \gamma_E ^4+528 \gamma_E ^3+2511]\nonumber \\
   &\quad-\frac{C_A
   n_f}{54 \pi ^3} \left[4 \gamma_E  (9 (2 T_F-1) \zeta_2+7)+48 T_F
   \zeta_3+24 \gamma_E ^3 T_F+60 \zeta_2+9 \zeta_3+30 \gamma_E
   ^2+90\right]\nonumber \\
   &\quad \left.-\frac{C_F n_f}{48 \pi
   ^3} (67-48 \zeta_3)\right\}+\frac{2
   C_A W_2 \left(2 \zeta_2+\gamma_E ^2\right)}{\pi ^3}+\frac{W_3}{\pi ^3} \nonumber \\
   &\simeq 123.569 - 13.827 n_f + 0.123705 n_f^2\nonumber\\
   &\quad  + (10.5394 - 0.468507 n_f) W_1 +  0.701093 W_2 + 0.0322515 W_3.
\end{align}
The coefficients of the perturbative expansion of the Wilson coefficient squared $W$ in the pole-mass scheme are~\cite{Chetyrkin:1997un,Schroder:2005hy,Chetyrkin:2005ia} 
\begin{align}\label{eq:Wcoeff1}
	W_1 &= \frac{11}{2},\\
	\label{eq:Wcoeff2}
	W_2 &= \left(\frac{2}{3} \ln \frac{\mH^2}{\mt^2}-\frac{67}{48}\right) n_f+\frac{19}{8}
   \ln \frac{\mH^2}{\mt^2}+\frac{1933}{72}\nonumber\\
   &\simeq 26.8472 + 2.375 \ln \frac{\mH^2}{\mt^2}- 1.39583 n_f + 0.666667 \ln \frac{\mH^2}{\mt^2} n_f,\\
   W_3 &= \frac{1}{124416}\Bigg[-432 \ln^2 \frac{\mH^2}{\mt^2} \left(32 n_f^2-414 n_f-1881\right)\nonumber\\
  &\quad +144 \ln \frac{\mH^2}{\mt^2}\left(77 n_f^2+4496 n_f+20145\right)\nonumber \\
  &\quad -54920 n_f^2-6 n_f (332337
   \zeta_3-37850)+24244461 \zeta_3-9969834\Bigg]\nonumber\\
   &\simeq 154.107 + 23.316 \ln \frac{\mH^2}{\mt^2} + 
 6.53125 \ln^2 \frac{\mH^2}{\mt^2}\nonumber\\
 &\quad + n_f\left(-17.4401 + 5.2037 \ln \frac{\mH^2}{\mt^2} + 
    1.4375 \ln^2 \frac{\mH^2}{\mt^2}\right)  \nonumber\\
    &\quad + n_f^2\left(-0.441422 + 0.0891204 \ln \frac{\mH^2}{\mt^2} - 
    0.111111 \ln \frac{\mH^2}{\mt^2}^2\right) .
    \label{eq:Wcoeff3}
\end{align}
The $\bar g_{0,k}$ coefficients, $k=1,2,3$, with $\muR=\muF=\mH$ are (for compactness, here the colour factors are set to their numerical value)
\begin{align}
	\bar g_{0,1} &= \frac{W_1}{\pi }+\frac{6 \zeta_2}{\pi }\nonumber\\
	&\simeq 3.14159 + 0.31831 W_1,\\
	\bar g_{0,2} &=-\frac{n_f}{36 \pi ^2} (60 \zeta_2-30 \zeta_3+247)+\frac{-5184 \zeta_2^2+4824
   \zeta_2-5940 \zeta_3+15390 \zeta_4+7533}{144 \pi ^2}\nonumber\\
    &\quad +\frac{6 W_1
   \zeta_2}{\pi ^2}+\frac{W_2}{\pi ^2}\nonumber\\
   & \simeq 7.71025 - 0.871459 n_f + W_1 + 0.101321 W_2,\\
	\bar g_{0,3} &= -\frac{n_f^2}{15552 \pi ^3} (25776 \zeta_2+5616 \zeta_3+25200
   \zeta_4-103753)\nonumber \\
   &\quad -\frac{n_f}{108864 \pi ^3}
   [-435456 (5+12 \gamma_E ) \zeta_2^2+9072 \zeta_2 (117
   \zeta_3+188)\nonumber \\
   &\quad-6978888 \zeta_3+13063680 \gamma_E  \zeta_4+502740
   \zeta_4-1495368 \zeta_5+23704107]\nonumber \\
   &\quad-\frac{13996800}{108864 \pi ^3} (\zeta_2^3+141087744
   \gamma_E ^2 \zeta_2^2+86220288 \gamma_E  \zeta_2^2+43763328
   \zeta_2^2+9430344 \zeta_2 \zeta_3\nonumber\\
   &\quad-105815808 \zeta_2
   \zeta_4-36630468 \zeta_2-4408992 \zeta_3^2+106475796
   \zeta_3-352719360 \gamma_E ^2 \zeta_4\nonumber\\
   &\quad-215550720 \gamma_E  \zeta_4-79385670
   \zeta_4-17738028 \zeta_5+122655708 \zeta_6-121979277)\nonumber\\
   &\quad +\frac{W_1}{144 \pi^3} \left[-4 n_f (60
   \zeta_2-30 \zeta_3+247)-9(576 \zeta_2^2-536
   \zeta_2+660 \zeta_3-1710 \zeta_4-837)\right]\nonumber\\
   &\quad +\frac{6 W_2
   \zeta_2}{\pi ^3}+\frac{W_3}{\pi ^3}\nonumber\\
   &\simeq 8.6433 - 3.94659 n_f + 0.0566722 n_f^2 \nonumber\\
   &\quad + (2.45425 - 0.277394 n_f) W_1 + 
 0.31831 W_2 + 0.0322515 W_3.
\end{align}

\subsection{Resummed cross section in SCET}\label{app:SCET}

In the SCET formalism, the coefficient function in the $gg$ channel for inclusive Higgs production in gluon fusion can be written as~\cite{Ahrens:2008nc}
\begin{align}
&C_{gg}(z,\mH^2,\muF^2) = H(\muH^2) U(\muH^2,\muS^2,\muF^2)\tilde s_{\rm Higgs}\(\ln\frac{\mH^2}{\muS^2}+\partial_\eta,\muS^2\)
\frac{z^{-\eta}}{(1-z)^{1-2\eta}}  \frac{e^{2\gamma\eta}}{\Gamma(2\eta)},
\end{align}
where $\eta = 2 a_{\Gamma_{\rm cusp}}(\muS^2, \muF^2)$ and
\begin{align}
U(\muH^2, \muS^2, \muF^2) &= \frac{\as^2(\muS^2)}{\as^2(\muF^2)}
\left[ \frac{\beta(\as(\muS^2))/\as^2(\muS^2)}{\beta(\as(\muH^2))/\as^2(\muH^2)} \right]^2 \left|\left(\frac{-\mH^2 - i \epsilon}{\muH^2} \right)^{-2 a_{\Gamma_{\rm cusp}}(\muH^2,\muS^2)} \right| \nonumber \\
&\quad\times
 \left|\exp[4 S(\muH^2,\muS^2)-2 a_{\gamma^S} (\muH^2, \muS^2)+ 4 a_{\gamma^B}(\muS^2,\muF^2)]\right|,
\end{align}
having introduced the definitions
\begin{align}\label{eq:Sanda}
S(\nu^2, \mu^2) &= - \int_{\as(\nu^2)}^{\as(\mu^2)} d \alpha \frac{A_\textrm{cusp}(\alpha)}{\beta(\alpha)}
\int_{\as(\nu^2)}^\alpha \frac{d \alpha'}{\beta(\alpha')}, \nonumber\\
a_\gamma(\nu^2, \mu^2) &= - \int_{\as(\nu^2)}^{\as(\mu^2)} d \alpha \frac{\gamma(\alpha)}{\beta(\alpha)}.
\end{align}
Note that in ref.~\cite{Ahrens:2008nc} the perturbative expansions of the anomalous dimensions and $\beta$-function are written as
\begin{equation}
   \gamma(\alpha_s)
   = \sum_{n=0}^\infty\,\gamma_n
    \left( \frac{\alpha_s}{4\pi} \right)^{n+1} , \quad
       A_{\rm cusp}(\alpha_s)
   = \sum_{n=0}^\infty\,\tilde A_{\rm cusp}^{(n)}
    \left( \frac{\alpha_s}{4\pi} \right)^{n+1} , \quad
   \beta(\alpha_s) 
   = -\alpha_s \sum_{n=0}^\infty\,\tilde \beta_n 
    \left( \frac{\alpha_s}{4\pi} \right)^{n+1} .
\end{equation}
The coefficients of the cusp anomalous dimension and of the $\beta$-function within this convention can be found in Appendix~B of ref.~\cite{Becher:2007ty}. 
It is however straightforward to relate them to the coefficients defined above.
The explicit expression for the evolution functions $a_\gamma$ in eq.~\eqref{eq:Sanda} keeping terms through $O(\alpha_s^2)$ is
\begin{align}\label{asol}
   a_\gamma(\nu^2,\mu^2)
   &= \frac{\gamma_0}{2\tilde \beta_0}\,\left\{
    \ln\frac{\alpha_s(\mu^2)}{\alpha_s(\nu^2)}
    + \left( \frac{\gamma_1}{\gamma_0} - \frac{\tilde \beta_1}{\tilde \beta_0} 
    \right) \frac{\alpha_s(\mu^2) - \alpha_s(\nu^2)}{4\pi}\right. \nonumber\\ 
   &\quad\left.+ \left[ \frac{\gamma_2}{\gamma_0}
    - \frac{\tilde \beta_2}{\tilde \beta_0} - \frac{\tilde \beta_1}{\tilde \beta_0}
    \left( \frac{\gamma_1}{\gamma_0} - \frac{\tilde \beta_1}{\tilde \beta_0} 
    \right) \right]
    \frac{\alpha_s^2(\mu^2) - \alpha_s^2(\nu^2)}{32\pi^2} + \dots \right\}
    \,.
\end{align}
The expression for $S$ also involves the four-loop coefficients $\tilde A_{\rm cusp}^{(3)}$ and $\tilde \beta_3$ and reads
\begin{align}
   S(\nu^2,\mu^2) 
   &= \frac{\tilde A_{\rm cusp}^{(0)}}{4\tilde \beta_0^2}\,\Bigg\{
    \frac{4\pi}{\alpha_s(\nu^2)} \left( 1 - \frac{1}{r} - \ln r \right)
    + \left( \frac{\tilde A_{\rm cusp}^{(1)}}{\tilde A_{\rm cusp}^{(0)}} - \frac{\tilde \beta_1}{\tilde \beta_0}
    \right) (1-r+\ln r) + \frac{\tilde \beta_1}{2\tilde \beta_0} \ln^2 r \nonumber\\
   &\quad+ \frac{\alpha_s(\nu^2)}{4\pi} \Bigg[ 
    \left( \frac{\tilde \beta_1\tilde A_{\rm cusp}^{(1)}}{\tilde \beta_0\tilde A_{\rm cusp}^{(0)}} 
    - \frac{\tilde \beta_2}{\tilde \beta_0} \right) (1-r+r\ln r)
    + \left( \frac{\tilde \beta_1^2}{\tilde \beta_0^2} 
    - \frac{\tilde \beta_2}{\tilde \beta_0} \right) (1-r)\ln r \nonumber\\
   &\hspace{1.0cm}
    \mbox{}- \left( \frac{\tilde \beta_1^2}{\tilde \beta_0^2} 
    - \frac{\tilde \beta_2}{\tilde \beta_0}
    - \frac{\tilde \beta_1\tilde A_{\rm cusp}^{(1)}}{\tilde \beta_0\tilde A_{\rm cusp}^{(0)}} 
    + \frac{\tilde A_{\rm cusp}^{(2)}}{\tilde A_{\rm cusp}^{(0)}} \right) \frac{(1-r)^2}{2} \Bigg]
    \nonumber\\
   &\quad+ \left( \frac{\alpha_s(\nu^2)}{4\pi} \right)^2 \Bigg[
    \left( \frac{\tilde \beta_1\tilde \beta_2}{\tilde \beta_0^2} 
    - \frac{\tilde \beta_1^3}{2\tilde \beta_0^3}
    - \frac{\tilde \beta_3}{2\tilde \beta_0} + \frac{\tilde \beta_1}{\tilde \beta_0}
    \left( \frac{\tilde A_{\rm cusp}^{(2)}}{\tilde A_{\rm cusp}^{(0)}} - \frac{\tilde \beta_2}{\tilde \beta_0}
    + \frac{\tilde \beta_1^2}{\tilde \beta_0^2} 
    - \frac{\tilde \beta_1\tilde A_{\rm cusp}^{(1)}}{\tilde \beta_0\tilde A_{\rm cusp}^{(0)}} \right) 
    \frac{r^2}{2} \right) \ln r \nonumber\\
   &\hspace{1.0cm}
    \mbox{}+ \left( \frac{\tilde A_{\rm cusp}^{(3)}}{\tilde A_{\rm cusp}^{(0)}} 
    - \frac{\tilde \beta_3}{\tilde \beta_0}
    + \frac{2\tilde \beta_1\tilde \beta_2}{\tilde \beta_0^2} 
    + \frac{\tilde \beta_1^2}{\tilde \beta_0^2}
    \left( \frac{\tilde A_{\rm cusp}^{(1)}}{\tilde A_{\rm cusp}^{(0)}} - \frac{\tilde \beta_1}{\tilde \beta_0} \right)
    - \frac{\tilde \beta_2\tilde A_{\rm cusp}^{(1)}}{\tilde \beta_0\tilde A_{\rm cusp}^{(0)}}
    - \frac{\tilde \beta_1\tilde A_{\rm cusp}^{(2)}}{\tilde \beta_0\tilde A_{\rm cusp}^{(0)}} \right) 
    \frac{(1-r)^3}{3} \nonumber\\
   &\hspace{1.0cm}
    \mbox{}+ \left( \frac{3\tilde \beta_3}{4\tilde \beta_0} 
    - \frac{\tilde A_{\rm cusp}^{(3)}}{2\tilde A_{\rm cusp}^{(0)}} + \frac{\tilde \beta_1^3}{\tilde \beta_0^3}
    - \frac{3\tilde \beta_1^2\tilde A_{\rm cusp}^{(1)}}{4\tilde \beta_0^2\tilde A_{\rm cusp}^{(0)}}
    + \frac{\tilde \beta_2\tilde A_{\rm cusp}^{(1)}}{\tilde \beta_0\tilde A_{\rm cusp}^{(0)}}
    + \frac{\tilde \beta_1\tilde A_{\rm cusp}^{(2)}}{4\tilde \beta_0\tilde A_{\rm cusp}^{(0)}}
    - \frac{7\tilde \beta_1\tilde \beta_2}{4\tilde \beta_0^2} \right) (1-r)^2 \nonumber\\
   &\hspace{1.0cm}
    \mbox{}+ \left( \frac{\tilde \beta_1\tilde \beta_2}{\tilde \beta_0^2} 
    - \frac{\tilde \beta_3}{\tilde \beta_0}
    - \frac{\tilde \beta_1^2\tilde A_{\rm cusp}^{(1)}}{\tilde \beta_0^2\tilde A_{\rm cusp}^{(0)}}
    + \frac{\tilde \beta_1\tilde A_{\rm cusp}^{(2)}}{\tilde \beta_0\tilde A_{\rm cusp}^{(0)}} \right) \frac{1-r}{2}
    \Bigg] + \dots \Bigg\} \,,
\end{align}
where $r=\alpha_s(\mu^2)/\alpha_s(\nu^2)$. 
The first three expansion coefficients of the perturbative expansion of the anomalous dimension $\gamma^S$ are~\cite{Idilbi:2005ni,Idilbi:2005er}
\begin{align}
  \gamma^S_0 &= 0 \,, \\
  \gamma^S_1 
  &= C_A^2 \left( -\frac{160}{27} + \frac{11\pi^2}{9}
   + 4\zeta_3 \right) 
   + C_A T_F n_f \left( -\frac{208}{27} - \frac{4\pi^2}{9} \right) 
   - 8 C_F T_F n_f \,, \\
  \gamma^S_2 
  &= C_A^3 \left[ \frac{37045}{729} + \frac{6109\pi^2}{243}
   - \frac{319\pi^4}{135} 
   + \left( \frac{244}{3} - \frac{40\pi^2}{9} \right) \zeta_3 
   - 32\zeta_5 \right]\nonumber \\
  &\quad\mbox{}+ C_A^2 T_F n_f \left( -\frac{167800}{729}
   - \frac{2396\pi^2}{243} + \frac{164\pi^4}{135} 
   + \frac{1424}{27}\,\zeta_3 \right)\nonumber \\
  &\quad\mbox{}+ C_A C_F T_F n_f \left( \frac{1178}{27}
   - \frac{4\pi^2}{3} - \frac{16\pi^4}{45} 
   - \frac{608}{9}\,\zeta_3 \right) + 8 C_F^2 T_F n_f\nonumber \\  
  &\quad\mbox{}+ C_A T_F^2 n_f^2 \left( \frac{24520}{729}
   + \frac{80\pi^2}{81} - \frac{448}{27}\,\zeta_3 \right) 
   + \frac{176}{9} C_F T_F^2 n_f^2 \,.
\end{align}
Finally, the first three coefficients of the anomalous dimension $\gamma^B$ (which corresponds to one half of the coefficient $B$ of the $\delta(1-x)$ term in the splitting function $P_{gg}(x)$, cfr. eq.~\eqref{eq:splittinglimit}), read~\cite{Vogt:2004mw}
\begin{align}
  \gamma^B_0 
  &= \frac{11}{3}\,C_A - \frac{4}{3}\,T_F n_f \,, \\
  \gamma^B_1 
  &= 4 C_A^2 \left( \frac{8}{3} + 3\zeta_3 \right) 
   - \frac{16}{3}\,C_A T_F n_f - 4 C_F T_F n_f \,, \\
  \gamma^B_2 
  &= C_A^3 \left[ \frac{79}{2} + \frac{4\pi^2}{9} + \frac{11\pi^4}{54} 
   + \left( \frac{536}{3} - \frac{8\pi^2}{3} \right) \zeta_3 
   - 80\zeta_5 \right]\nonumber \\
  &\quad\mbox{}- C_A^2 T_F n_f \left( \frac{233}{9} + \frac{8\pi^2}{9}
   + \frac{2\pi^4}{27} + \frac{160}{3}\,\zeta_3 \right) \nonumber\\
  &\quad\mbox{}- \frac{241}{9}\,C_A C_F T_F n_f + 2 C_F^2 T_F n_f
   + \frac{58}{9}\,C_A T_F^2 n_f^2 + \frac{44}{9}\,C_F T_F^2 n_f^2 \,.
\end{align}

To reach N$^3$LL$^\prime$ one also needs the three loop hard and soft functions.
In the rEFT, the hard function further factorizes as 
\begin{equation}
	H(\muH^2) = \big[C_t(m_t^2)\big]^2\,     \left| C_S(-m_H^2-i\epsilon,\muH^2) \right|^2 .
\end{equation}
The coefficient $C_t$ admits the perturbative expansion
\begin{equation}
	C_t(m_t^2) = 1 + \as C^{(1)}_t(m_t^2) + \as^2 C^{(2)}_t(m_t^2) +\as^3  C^{(3)}_t(m_t^2) + \ldots
\end{equation}
and is related to the Wilson coefficient squared eqs.~\eqref{eq:Wcoeff1}-\eqref{eq:Wcoeff3} by
\begin{align}
	   C^{(1)}_t(m_t^2) &= \frac{W_1}{2}, \\
      C^{(1)}_t(m_t^2) & = \frac{W_2}{2} - \frac{W_1^2}{8}, \\
      C^{(3)}_t(m_t^2) &= \frac{W_3}{2}- \frac{W_1W_2}{4} + \frac{W_1^3}{16}. 
\end{align}
Explicitly,
\begin{align}
C^{(1)}_t(m_t^2) &=\frac{11}{4}\\
C^{(2)}_t(m_t^2)	&= \frac{2777}{288} + \frac{19}{16}  \ln\frac{\mH^2}{m_t^2} + n_f \left( -\frac{67}{96} + \frac{1}{3}\ln\frac{\mH^2}{m_t^2} \right)\\
C^{(3)}_t(m_t^2)   &=  n_f^2\left(-\frac{1}{18}\ln^2\frac{\mH^2}{m_t^2}+\frac{77  }{1728}\ln\frac{\mH^2}{m_t^2}-\frac{6865}{31104}\right)
  \nonumber \\
   &\quad+n_f \left(\frac{23 }{32} \ln^2\frac{\mH^2}{m_t^2}+\frac{91
    }{54}\ln\frac{\mH^2}{m_t^2}-\frac{110779
   \zeta_3}{13824}+\frac{58723}{20736}\right)\nonumber \\
   &\quad +\frac{209}{64} \ln^2\frac{\mH^2}{m_t^2}+\frac{2417 }{288} \ln\frac{\mH^2}{m_t^2}+\frac{897943
   \zeta_3}{9216}-\frac{2761331}{41472}.
\end{align}
The matching coefficient $C_S$ has been computed at three loop in ref.~\cite{Gehrmann:2010ue}. 
By writing its perturbative series in the form
\begin{equation}\label{CSexp}
   C_S(-m_H^2-i\epsilon,\mu^2) = 1 + \sum_{n=1}^\infty\,c_n(L)
   \left( \frac{\alpha_s(\mu^2)}{4\pi} \right)^n \! ,
\end{equation}
where $L=\ln[(-m_H^2-i\epsilon)/\mu^2]$, one has
\begin{align}\label{c1c2c3}
 c_1(L) &= C_A \bigg(- L^2 + \zeta_2  \bigg),  \\
c_2(L) &=
C_A^2 \bigg(
\frac{1}{2}L^4+\frac{11}{9}L^3+\bigg(-\frac{67}{9}+\zeta_2\bigg)L^2+\bigg(\frac{80}{27}-2\zeta_3-\frac{22}{3}\zeta_2\bigg)L\nonumber \\ &\hspace{2.5cm}+\frac{5105}{162}-\frac{143}{9}\zeta_3+\frac{67}{6}\zeta_2+\frac{1}{2}\zeta_2^2 \bigg)\nonumber \\ &\quad
+C_AN_F \bigg(
-\frac{2}{9}L^3+\frac{10}{9}L^2+\bigg(\frac{52}{27}+\frac{4}{3}\zeta_2\bigg)L-\frac{916}{81}-\frac{46}{9}\zeta_3-\frac{5}{3}\zeta_2 \bigg)\nonumber \\ &\quad
+C_FN_F \bigg(
2L -\frac{67}{6}+8\zeta_3 \bigg)  \\
c_3(L) &=
C_A^3 \bigg(
-\frac{1}{6}L^6-\frac{11}{9}L^5+\bigg(\frac{281}{54}-\frac{3}{2}\zeta_2\bigg)L^4+\bigg(\frac{11}{3}\zeta_2+\frac{1540}{81}+2\zeta_3\bigg)L^3\nonumber \\ &\hspace{2cm}+\bigg(\frac{143}{9}\zeta_3-\frac{6740}{81}+\frac{685}{18}\zeta_2-\frac{73}{10}\zeta_2^2\bigg)L^2\nonumber \\ &\hspace{2cm}+\bigg(\frac{2048}{27}\zeta_3+16\zeta_5+\frac{34}{3}\zeta_2\zeta_3-\frac{13420}{81}\zeta_2+\frac{176}{5}\zeta_2^2-\frac{373975}{1458}\bigg)L\nonumber \\ &\hspace{2cm}-\frac{1939}{270}\zeta_2^2+\frac{2222}{9}\zeta_5+\frac{105617}{729}\zeta_2-\frac{24389}{1890}\zeta_2^3-\frac{152716}{243}\zeta_3-\frac{605}{9}\zeta_2\zeta_3\nonumber \\ &\hspace{2cm}+\frac{29639273}{26244}-\frac{104}{9}\zeta_3^2 \bigg)\nonumber \\ &\quad
+C_A^2N_F \bigg(
\frac{2}{9}L^5-\frac{8}{27}L^4+\bigg(-\frac{734}{81}-\frac{2}{3}\zeta_2\bigg)L^3+\bigg(\frac{377}{27}+\frac{118}{9}\zeta_3-\frac{103}{9}\zeta_2\bigg)L^2\nonumber \\ &\hspace{2cm}+\bigg(\frac{133036}{729}+\frac{28}{9}\zeta_3+\frac{3820}{81}\zeta_2-\frac{48}{5}\zeta_2^2\bigg)L\nonumber \\ &\hspace{2cm}-\frac{3765007}{6561}+\frac{428}{9}\zeta_5-\frac{460}{81}\zeta_3-\frac{14189}{729}\zeta_2-\frac{82}{9}\zeta_2\zeta_3+\frac{73}{45}\zeta_2^2 \bigg)\nonumber \\ &
\quad+C_AN_F^2 \bigg(
-\frac{2}{27}L^4+\frac{40}{81}L^3+\bigg(\frac{116}{81}+\frac{8}{9}\zeta_2\bigg)L^2+\bigg(-\frac{14057}{729}-\frac{128}{27}\zeta_3-\frac{80}{27}\zeta_2\bigg)L\nonumber \\ &\hspace{2cm}+\frac{611401}{13122}+\frac{4576}{243}\zeta_3+\frac{4}{9}\zeta_2+\frac{4}{27}\zeta_2^2 \bigg)\nonumber \\ &
\quad+C_FN_F^2 \bigg(
\frac{4}{3}L^2+\bigg(-\frac{52}{3}+\frac{32}{3}\zeta_3\bigg)L+\frac{4481}{81}-\frac{112}{3}\zeta_3-\frac{20}{9}\zeta_2-\frac{16}{45}\zeta_2^2 \bigg)\nonumber \\ &\quad
+C_FC_AN_F \bigg(
-\frac{8}{3}L^3+\bigg(13-16\zeta_3\bigg)L^2+\bigg(\frac{3833}{54}-\frac{376}{9}\zeta_3+6\zeta_2+\frac{16}{5}\zeta_2^2\bigg)L\nonumber \\ &\hspace{2cm}-\frac{341219}{972}+\frac{608}{9}\zeta_5+\frac{14564}{81}\zeta_3-\frac{68}{9}\zeta_2+\frac{64}{3}\zeta_2\zeta_3-\frac{64}{45}\zeta_2^2 \bigg)\nonumber \\ &\quad
+C_F^2N_F \bigg(
-2L + \frac{304}{9}-160\zeta_5+\frac{296}{3}\zeta_3 \bigg).
\end{align}
Finally, writing the expansion of the soft function $\tilde s_{\rm Higgs}$ as
\begin{align}
\tilde s_{\rm Higgs}(L,\as) &= 1+ \frac{\as}\pi \tilde s_{\rm Higgs}^{(1)}(L)
+ \(\frac{\as}\pi\)^2 \tilde s_{\rm Higgs}^{(2)}(L)
+ \(\frac{\as}\pi\)^3 \tilde s_{\rm Higgs}^{(3)}(L) +\ldots
\end{align}
one has~\cite{Idilbi:2006dg,Ahrens:2008nc,Li:2014afw,Bonvini:2014tea}
 \begin{align}
 \tilde s_{\rm Higgs}^{(1)}(L) &= \frac{C_A}{2}\(L^2+\zeta_2\)\\
 \tilde s_{\rm Higgs}^{(2)}(L) &= 
 C_A^2
    \left[\frac{L^4}{8}-\frac{11 L^3}{72}+\frac{67 L^2}{72}+
    \left(\frac{7 \zeta_3}{4}-\frac{101}{54}\right)L -\frac{5
    \zeta_2^2}{8}+\frac{67 \zeta_2}{144}-\frac{11
    \zeta_3}{72}+\frac{607}{324}\right]
 \nonumber\\ &
 + C_A n_f
    \left(\frac{L^3}{36}-\frac{5 L^2}{36}+\frac{7 L}{27}-\frac{5
    \zeta_2}{72}+\frac{\zeta_3}{36}-\frac{41}{162}\right)\\
\tilde s_{\rm Higgs}^{(3)}(L) &=
C_A^3 \bigg[\frac{L^6}{48}-\frac{11 L^5}{144}+
   \left(\frac{925}{1728}-\frac{\zeta_2}{16}\right) L^4
   + \left(\frac{11
   \zeta_2}{144}+\frac{7 \zeta_3}{8}-\frac{1051}{648}\right) L^3\nonumber\\
   & \qquad\qquad+ \left(-\frac{13 \zeta_2^2}{80}-\frac{67 \zeta_2}{288}-\frac{209
   \zeta_3}{144}+\frac{20359}{5184}\right) L^2
\nonumber\\ & \qquad\qquad
   + \left(\frac{11
   \zeta_2^2}{40}+\zeta_2
   \left(\frac{193}{648}-\frac{\zeta_3}{24}\right)+\frac{1541
   \zeta_3}{216}-3 \zeta_5-\frac{297029}{46656}\right) L
\nonumber\\ & \qquad\qquad
   +\frac{11657
   \zeta_2^3}{15120}-\frac{4261 \zeta_2^2}{2160}+
   \frac{23333 \zeta_2}{46656}-\frac{11 \zeta_2\zeta_3}{24}\nonumber\\
   &\qquad\qquad+\frac{67
   \zeta_3^2}{36}-\frac{21763 \zeta_3}{3888}-\frac{121
   \zeta_5}{72}+\frac{5211949}{839808}\bigg]
\nonumber\\ &
+C_A^2 n_f
   \bigg[\frac{L^5}{72}-\frac{41 L^4}{432}+
   \left(\frac{457}{1296}-\frac{\zeta_2}{72}\right) L^3
   + \left(\frac{5
   \zeta_2}{144}+\frac{\zeta_3}{72}-\frac{793}{864}\right) L^2\nonumber\\
   &\qquad\qquad
   + \left(\frac{\zeta_2^2}{20}-\frac{19 \zeta_2}{648}-\frac{113
   \zeta_3}{216}+\frac{31313}{23328}\right) L
\nonumber\\ & \qquad\qquad
+\frac{389
   \zeta_2^2}{2160}-\frac{1633 \zeta_2}{23328}+\frac{19
   \zeta_3}{81}-\frac{\zeta_5}{12}-\frac{412765}{419904}\bigg]
\nonumber\\ &
+C_A C_F n_f \bigg[ \frac{L^3}{48}
  + \left(\frac{\zeta_3}{4}-\frac{55}{192}\right) L^2
  + \left(-\frac{\zeta_2^2}{10}-\frac{19\zeta_3}{36}+\frac{1711}{1728}\right) L\nonumber\\ &\qquad\qquad
  + \frac{19\zeta_2^2}{180} + \frac{\zeta_2\zeta_3}{12}-\frac{55 \zeta_2}{576}
  +\frac{355\zeta_3}{648}+\frac{7\zeta_5}{18}-\frac{42727}{31104}\bigg]
\nonumber\\ &
+C_A n_f^2
   \bigg[\frac{L^4}{432}-\frac{5 L^3}{324}+\frac{25 L^2}{648}+
   \left(-\frac{\zeta_3}{18}-\frac{29}{729}\right)L+\frac{13
   \zeta_2^2}{1080}-\frac{\zeta_2}{648}+\frac{55
   \zeta_3}{972}-\frac{4}{6561}\bigg].
\label{eq:stilde3}
\end{align}

\section{Transverse-momentum resummation formul\ae\ }

In this section are collected the expressions for the quantities needed for N$^3$LL resummation of the Higgs transverse-momentum distribution and for its matching to fixed order.
 
\subsection{Formul\ae\ for $N^3LL$ resummation}
\label{app:sudakov-radiator}

Including terms up to N$^3$LL, the cumulative cross section eq.~\eqref{eq:master-kt-space} written in terms of the modified logarithms defined in eq.~\eqref{eq:modified-log} is
\begin{align}
\label{eq:master-kt-space-modified}
&\frac{\rd\Sigma(v)}{\rd\Phi_B} =\int_0^\infty\frac{\rd k_{t,1}}{k_{t,1}}{\cal J}(k_{t,1})\frac{\rd
  \phi_1}{2\pi}\partial_{\tilde L}\left(-e^{-\tilde R(k_{t,1})} {\tilde{\cal L}}_{\rm
  N^3LL}(k_{t,1}) \right) \int \dZ\Theta\left(v-V(k_1,\dots, k_{n+1})\right)
                             \notag\\\notag\\
& + \int_0^\infty\frac{\rd k_{t,1}}{k_{t,1}}{\cal J}(k_{t,1})\frac{\rd
  \phi_1}{2\pi} e^{-\tilde R(k_{t,1})} \int \dZ\int_{0}^{1}\frac{\rd \zeta_{s}}{\zeta_{s}}\frac{\rd
  \phi_s}{2\pi}\Bigg\{\bigg({\tilde R}' (k_{t,1}) {\tilde{\cal L}}_{\rm
  NNLL}(k_{t,1}) - \partial_{\tilde L} {\tilde{\cal L}}_{\rm
  NNLL}(k_{t,1})\bigg)\notag\\
&\times\left({\tilde R}'' (k_{t,1})\ln\frac{1}{\zeta_s} +\frac{1}{2} {\tilde R}'''
  (k_{t,1})\ln^2\frac{1}{\zeta_s} \right)\notag \\
  & - {\tilde R}' (k_{t,1})\left(\partial_{\tilde L} {\tilde{\cal L}}_{\rm
  NNLL}(k_{t,1}) - 2\frac{\beta_0}{\pi}\as^2(k_{t,1}^2) \hat{P}^{(0)}\otimes {\tilde{\cal L}}_{\rm
  NLL}(k_{t,1}) \ln\frac{1}{\zeta_s}
\right)\notag\\
&+\frac{\as^2(k_{t,1}^2) }{\pi^2}\hat{P}^{(0)}\otimes \hat{P}^{(0)}\otimes {\tilde{\cal L}}_{\rm
  NLL}(k_{t,1})\Bigg\} \notag \\
  & \bigg\{\Theta\left(v-V(k_1,\dots,
  k_{n+1},k_s)\right) - \Theta\left(v-V(k_1,\dots,
  k_{n+1})\right)\bigg\}\notag\\\notag\\
& + \frac{1}{2}\int_0^\infty\frac{\rd k_{t,1}}{k_{t,1}}{\cal J}(k_{t,1})\frac{\rd
  \phi_1}{2\pi} e^{-\tilde R(k_{t,1})} \int \dZ\int_{0}^{1}\frac{\rd \zeta_{s1}}{\zeta_{s1}}\frac{\rd
  \phi_{s1}}{2\pi}\int_{0}^{1}\frac{\rd \zeta_{s2}}{\zeta_{s2}}\frac{\rd
  \phi_{s2}}{2\pi} {\tilde R}' (k_{t,1})\notag\\
&\times\Bigg\{ {\tilde{\cal L}}_{\rm
  NLL}(k_{t,1}) \left({\tilde R}'' (k_{t,1})\right)^2\ln\frac{1}{\zeta_{s1}} \ln\frac{1}{\zeta_{s2}} - \partial_{\tilde L} {\tilde{\cal L}}_{\rm
  NLL}(k_{t,1}) {\tilde R}'' (k_{t,1})\bigg(\ln\frac{1}{\zeta_{s1}}
  +\ln\frac{1}{\zeta_{s2}} \bigg)\notag\\
&+ \frac{\as^2(k_{t,1}^2) }{\pi^2}\hat{P}^{(0)}\otimes \hat{P}^{(0)}\otimes {\tilde{\cal L}}_{\rm
  NLL}(k_{t,1})\Bigg\}\notag\\
&\times \bigg\{\Theta\left(v-V(k_1,\dots,
  k_{n+1},k_{s1},k_{s2})\right) - \Theta\left(v-V(k_1,\dots,
  k_{n+1},k_{s1})\right) -\notag\\ &\Theta\left(v-V(k_1,\dots,
  k_{n+1},k_{s2})\right) + \Theta\left(v-V(k_1,\dots,
  k_{n+1})\right)\bigg\} + {\cal O}\left(\as^n \ln^{2n -
                                    6}\frac{1}{v}\right),
\end{align}
where the modified luminosities are defined as
\begin{align} 
\label{eq:mod-luminosity-NLL}
\tilde{\cal L}_{\rm NLL}(k_{t,1}) = \sum_{c, c'}\frac{\rd|\mathcal{M}_{B}|_{cc'}^2}{\rd\Phi_B} f_c\!\left(\mu_F e^{-\tilde{L}},x_1\right)f_{c'}\!\left(\mu_F e^{-\tilde{L}},x_2\right),
\end{align}
\begin{align}
\label{eq:mod-luminosity-NNLL}
&\tilde{\cal L}_{\rm NNLL}(k_{t,1}) = \sum_{c, c'}\frac{\rd|\mathcal{M}_{B}|_{cc'}^2}{\rd\Phi_B} \sum_{i, j}\int_{x_1}^{1}\frac{\rd z_1}{z_1}\int_{x_2}^{1}\frac{\rd z_2}{z_2}f_i\!\left(\mu_F e^{-\tilde{L}},\frac{x_1}{z_1}\right)f_{j}\!\left(\mu_F e^{-\tilde{L}},\frac{x_2}{z_2}\right)\notag\\&\times\Bigg\{\delta_{ci}\delta_{c'j}\delta(1-z_1)\delta(1-z_2)
\left(1+\frac{\as(\muR^2)}{2\pi} \tilde{H}^{(1)}(\muR,x_Q)\right) \notag\\
&+ \frac{\as(\muR^2)}{2\pi}\frac{1}{1-2\as(\muR^2)\beta_0
  \tilde{L}}\left(\tilde{C}_{c i}^{(1)}(z_1,\mu_F,x_Q)\delta(1-z_2)\delta_{c'j}+
  \{z_1\leftrightarrow z_2; c,i \leftrightarrow c'j\}\right)\Bigg\},
\end{align}
\begin{align}
\label{eq:mod-luminosity-N3LL}
&\tilde{\cal L}_{\rm N^3LL}(k_{t,1})=\sum_{c,
  c'}\frac{\rd|\mathcal{M}_{B}|_{cc'}^2}{\rd\Phi_B} \sum_{i, j}\int_{x_1}^{1}\frac{\rd
  z_1}{z_1}\int_{x_2}^{1}\frac{\rd z_2}{z_2}f_i\!\left(\mu_F e^{-\tilde{L}},\frac{x_1}{z_1}\right)f_{j}\!\left(\mu_F e^{-\tilde{L}},\frac{x_2}{z_2}\right)\notag\\&\times\Bigg\{\delta_{ci}\delta_{c'j}\delta(1-z_1)\delta(1-z_2)
\left(1+\frac{\as(\muR^2)}{2\pi} \tilde{H}^{(1)}(\muR,x_Q) + \frac{\as^2(\muR^2)}{(2\pi)^2} \tilde{H}^{(2)}(\muR,x_Q)\right) \notag\\
&+ \frac{\as(\muR^2)}{2\pi}\frac{1}{1-2\as(\muR^2)\beta_0 \tilde{L}}\left(1- \as(\muR^2)\frac{\beta_1}{\beta_0}\frac{\ln\left(1-2\as(\muR^2)\beta_0 \tilde{L}\right)}{1-2\as(\muR^2)\beta_0 \tilde{L}}\right)\notag\\
&\times\left(\tilde{C}_{c i}^{(1)}(z_1,\mu_F,x_Q)\delta(1-z_2)\delta_{c'j}+ \{z_1\leftrightarrow z_2; c,i \leftrightarrow c',j\}\right)\notag\\
& +
  \frac{\as^2(\muR^2)}{(2\pi)^2}\frac{1}{(1-2\as(\muR^2)\beta_0
  \tilde{L})^2}\Bigg(\tilde{C}_{c i}^{(2)}(z_1,\mu_F,x_Q)\delta(1-z_2)\delta_{c'j} + \{z_1\leftrightarrow z_2; c,i \leftrightarrow c',j\}\Bigg) \notag\\&+  \frac{\as^2(\muR^2)}{(2\pi)^2}\frac{1}{(1-2\as(\muR^2)\beta_0 \tilde{L})^2}\Big(\tilde{C}_{c i}^{(1)}(z_1,\mu_F,x_Q)\tilde{C}_{c' j}^{(1)}(z_2,\mu_F,x_Q) + G_{c i}^{(1)}(z_1)G_{c' j}^{(1)}(z_2)\Big) \notag\\
& + \frac{\as^2(\muR^2)}{(2\pi)^2} \tilde{H}^{(1)}(\muR,x_Q)\frac{1}{1-2\as(\muR^2)\beta_0 \tilde{L}}\Big(\tilde{C}_{c i}^{(1)}(z_1,\mu_F,x_Q)\delta(1-z_2)\delta_{c'j} + \{z_1\leftrightarrow z_2; c,i \leftrightarrow c',j\}\Big) \Bigg\}.
\end{align}

The modified radiator $\tilde{R}$ takes the form
\begin{align}
\label{eq:mod-radiator}
\tilde{R}(k_{t,1}) &= - \tilde{L} g_1(\as \beta_0\tilde{L} ) -
  g_2(\as \beta_0\tilde{L} ) - \frac{\as}{\pi}
  g_3(\as \beta_0\tilde{L} ) - \frac{\as^2}{\pi^2}
  g_4(\as \beta_0\tilde{L} ).
\end{align}
The functions $g_i(\lambda)$ entering in the N$^3$LL Sudakov radiator eq.~\eqref{eq:mod-radiator} and its derivative are defined as
\begin{align}
  g_{1}(\lambda) =& \frac{A^{(1)}}{\pi\beta_{0}}\frac{2 \lambda +\ln (1-2 \lambda )}{2  \lambda }, \\
  g_{2}(\lambda) =& \frac{1}{2\pi \beta_{0}}\ln (1-2 \lambda )
  \left(A^{(1)} \ln \frac{1}{x_Q^2}+B^{(1)}\right)
  -\frac{A^{(2)}}{4 \pi ^2 \beta_{0}^2}\frac{2 \lambda +(1-2
    \lambda ) \ln (1-2 \lambda )}{1-2
    \lambda} \notag\\
  &+A^{(1)} \bigg(-\frac{\beta_{1}}{4 \pi \beta_{0}^3}\frac{\ln
    (1-2 \lambda ) ((2 \lambda -1) \ln (1-2 \lambda )-2)-4
    \lambda}{1-2 \lambda}\notag\\
  &\hspace{10mm}-\frac{1}{2 \pi \beta_{0}}\frac{(2 \lambda(1
    -\ln (1-2 \lambda ))+\ln (1-2 \lambda ))}{1-2\lambda} \ln
  \frac{\muR^2}{x_Q^2 \mH^2}\bigg)\,,\\
  g_{3}(\lambda) =
  & \left(A^{(1)} \ln\frac{1}{x_Q^2}+B^{(1)}\right)
  \bigg(-\frac{\lambda }{1-2 \lambda} \ln
  \frac{\mu _{R}^2}{x_Q^2M^2}+\frac{\beta_{1}}{2 \beta_{0}^2}\frac{2 \lambda
    +\ln (1-2 \lambda )}{1-2 \lambda}\bigg)\notag\\
  &   -\frac{1}{2 \pi\beta_{0}}\frac{\lambda}{1-2\lambda}\left(A^{(2)}
  \ln\frac{1}{x_Q^2}+B^{(2)}\right)-\frac{A^{(3)}}{4 \pi ^2 \beta_{0}^2}\frac{\lambda ^2}{(1-2\lambda )^2} \notag\\
  &   +A^{(2)} \bigg(\frac{\beta_{1}}{4 \pi  \beta_{0}^3 }\frac{2 \lambda  (3
    \lambda -1)+(4 \lambda -1) \ln (1-2 \lambda )}{(1-2 \lambda
    )^2}-\frac{1}{\pi \beta_{0}}\frac{\lambda ^2 }{(1-2 \lambda )^2}\ln\frac{\muR^2}{x_Q^2 \mH^2}\bigg) \notag\\
  & +A^{(1)} \bigg(\frac{\lambda  \left(\beta_{0} \beta_{2} (1-3 \lambda
    )+\beta_{1}^2 \lambda \right)}{\beta_{0}^4 (1-2 \lambda)^2}
  +\frac{(1-2 \lambda) \ln (1-2 \lambda ) \left(\beta_{0} \beta_{2} 
    (1-2 \lambda )+2 \beta_{1}^2 \lambda \right)}{2\beta_{0}^4 (1-2 \lambda)^2} 
  \notag\\
  &\hspace{10mm}+\frac{\beta_{1}^2}{4 \beta_{0}^4}
  \frac{(1-4 \lambda ) \ln ^2(1-2 \lambda )}{(1-2 \lambda)^2}-\frac{\lambda ^2 }{(1-2 \lambda
    )^2} \ln ^2\frac{\muR^2}{x_Q^2 \mH^2}\notag\\
  &
  \hspace{10mm}   -\frac{\beta_{1}}{2 \beta_{0}^{2}}\frac{(2 \lambda  (1-2 \lambda)+(1-4 \lambda) \ln (1-2 \lambda ))
  }{(1-2\lambda )^2}\ln\frac{\muR^2}{x_Q^2 \mH^2}\bigg)\,,\\
  g_4(\lambda)  =& \frac{A^{(4)} (3-2 \lambda ) \lambda ^2}{24 \pi ^2 \beta_0^2 (2 \lambda -1)^3}\notag\\
  & + \frac{A^{(3)}}{48 \pi 
    \beta_0^3 (2 \lambda -1)^3}\Bigg\{3 \beta_1 (1-6 \lambda ) \ln (1-2 \lambda )+2 \lambda  \Bigg(\beta_1 (5 \lambda  (2 \lambda -3)+3)\notag\\
  &\hspace{10mm} +6 \beta_0^2 (3-2 \lambda ) \lambda  \ln
  \frac{\muR^2}{x_Q^2\mH^2}\Bigg)+12 \beta_0^2 (\lambda -1) \lambda
  (2 \lambda -1) \ln \frac{1}{x_Q^2}\Bigg\} \notag\\
  & + \frac{A^{(2)}}{24
    \beta_0^4 (2 \lambda -1)^3} \Bigg\{32 \beta_0 \beta_2 \lambda ^3-2 \beta_1^2 \lambda 
  (\lambda  (22 \lambda -9)+3)\notag\\
  &\hspace{10mm}+12 \beta_0^4 (3-2 \lambda ) \lambda ^2 \ln
  ^2\frac{\muR^2}{x_Q^2\mH^2}+6 \beta_0^2 \ln
  \frac{\muR^2}{x_Q^2\mH^2}\times\notag\\
  &\hspace{10mm}\left(\beta_1 (1-6 \lambda ) \ln (1-2
  \lambda )+2 (\lambda -1) \lambda  (2 \lambda -1) \left(\beta_1+2 \beta_0^2 \ln\frac{1}{x_Q^2}\right)\right)\notag\\
  &\hspace{10mm}+3 \beta_1 \Bigg(\beta_1 \ln (1-2
  \lambda ) (2 \lambda +(6 \lambda -1) \ln (1-2 \lambda )-1)\notag\\
  &\hspace{10mm}-2 \beta_0^2 (2 \lambda -1) (2
  (\lambda -1) \lambda -\ln (1-2 \lambda )) \ln \frac{1}{x_Q^2}\Bigg)\Bigg\}\notag\notag\\
  & + \frac{\pi  A^{(1)}}{12 \beta_0^5 (2 \lambda -1)^3} \Bigg\{\beta_1^3 (1-6 \lambda ) \ln ^3(1-2 \lambda )+3 \ln (1-2 \lambda )
  \Bigg(\beta_0^2 \beta_3 (2 \lambda -1)^3\notag\\
  &\hspace{10mm}+\beta_0 \beta_1
  \beta_2 \left(1-2 \lambda  \left(8 \lambda ^2-4 \lambda +3\right)\right)+4 \beta_1^3
  \lambda ^2 (2 \lambda +1)\notag\\
  &\hspace{10mm}+\beta_0^2 \beta_1 \ln \frac{\muR^2}{
    x_Q^2\mH^2} \left(\beta_0^2 (1-6 \lambda ) \ln \frac{\muR^2}{
    x_Q^2\mH^2}-4 \beta_1 \lambda \right)\Bigg)\notag\\
  &\hspace{10mm}+3 \beta_1^2 \ln ^2(1-2 \lambda
  ) \left(2 \beta_1 \lambda +\beta_0^2 (6 \lambda -1) \ln \frac{\muR^2}{x_Q^2\mH^2}\right)\notag\\
  &\hspace{10mm}+3 \beta_0^2 (2 \lambda -1) \ln
  \frac{1}{x_Q^2} \Bigg(-\beta_1^2 \ln ^2(1-2 \lambda ) +2 \beta_0^2
  \beta_1 \ln (1-2 \lambda ) \ln \frac{\muR^2}{
    x_Q^2\mH^2}\notag\\
  &\hspace{10mm}+4 \lambda 
  \left(\lambda  \left(\beta_1^2-\beta_0 \beta_2\right)+\beta_0^4
  (\lambda -1) \ln ^2\frac{\muR^2}{x_Q^2\mH^2}\right)\Bigg)\notag\\
  &\hspace{10mm}+2 \lambda  \Bigg(\beta_0^2 \beta_3 ((15-14 \lambda )
  \lambda -3)+\beta_0 \beta_1 \beta_2 (5 \lambda  (2 \lambda -3)+3)\notag\\
  &\hspace{10mm}+4
  \beta_1^3 \lambda ^2+2 \beta_0^6 (3-2 \lambda ) \lambda  \ln
  ^3\frac{\muR^2}{x_Q^2\mH^2}+3 \beta_0^4 \beta_1 \ln
  ^2\frac{\muR^2}{x_Q^2\mH^2}\notag\\
  &\hspace{10mm}+6 \beta_0^2 \lambda  (2 \lambda +1)
  \left(\beta_0 \beta_2-\beta_1^2\right) \ln \frac{\muR^2}{x_Q^2\mH^2}-8 \beta_0^6 \left(4 \lambda ^2-6 \lambda +3\right) \zeta_3\Bigg)\Bigg\}\notag\\
  & + \frac{B^{(3)} (\lambda -1) \lambda }{4 \pi  \beta_0 (1-2 \lambda )^2}+ \frac{B^{(2)} \left(\beta_1 \ln (1-2 \lambda )-2 (\lambda -1) \lambda  \left(\beta_1-2 \beta_0^2 \ln \frac{\muR^2}{x_Q^2\mH^2}\right)\right)}{4\beta_0^2
    (1-2\lambda )^2}\notag\\
  & + \frac{\pi  B^{(1)}}{4 \beta_0^3 (1-2 \lambda )^2} \Bigg\{4 \lambda  \left(\lambda 
  \left(\beta_1^2-\beta_0 \beta_2\right)+\beta_0^4 (\lambda -1) \ln
  ^2\frac{\muR^2}{x_Q^2\mH^2}\right)\notag\\
  &\hspace{10mm}-\beta_1^2 \ln ^2(1-2 \lambda )+2 \beta_0^2 \beta_1
  \ln (1-2 \lambda ) \ln \frac{\muR^2}{x_Q^2\mH^2}\Bigg\}.
\end{align}
where 
\begin{equation}
  \lambda = \as(\muR^2) \beta_0 \tilde L \,.
\end{equation}
The coefficients $A^{(i)}$ and $B^{(i)}$ which enter the formul\ae\ above are
\begin{align}
  A^{(1)} =& \,2 C_A,
  \notag\\
  \vspace{1.5mm}
  A^{(2)} =&
  \left( \frac{67}{9}-\frac{\pi ^2}{3} \right) C_A^2
  -\frac{10}{9} C_A n_f,
  \notag\\
  \vspace{1.5mm}
  A^{(3)} =&
   \left( -22 \zeta_3 - \frac{67 \pi^2}{27}+\frac{11 \pi^4}{90}+\frac{15503}{324} \right) C_A^3
  + \left( \frac{10 \pi^2}{27}-\frac{2051}{162} \right) C_A^2 n_f\notag\\
  &+ \left( 4 \zeta_3-\frac{55}{12} \right) C_A C_F n_f
  + \frac{50}{81} C_A n_f^2,
  \notag\\
  \vspace{1.5mm}
  A^{(4)} =&
     \left( \frac{121}{3} \zeta_3 \zeta_2-\frac{8789 \zeta_2}{162}-\frac{19093 \zeta_3}{54}-\frac{847 \zeta_4}{24}+132 \zeta_5+\frac{3761815}{11664} \right) C_A^4
   + \left( -\frac{4 \zeta_3}{9}-\frac{232}{729} \right) C_A n_f^3
  \notag\\&
   + \left( -\frac{22}{3} \zeta_3 \zeta_2+\frac{2731 \zeta_2}{162}+\frac{4955 \zeta_3}{54}+\frac{11 \zeta_4}{6}-24 \zeta_5-\frac{31186}{243} \right) C_A^3 n_f\notag\\
&   + \left( -\frac{38 \zeta_3}{9}-2 \zeta_4+\frac{215}{24} \right) C_A C_F n_f^2
   + \left( \frac{272 \zeta_3}{9}+11 \zeta_4-\frac{7351}{144} \right) C_A^2 C_F n_f\notag\\
&   + \left( -\frac{103 \zeta_2}{81}-\frac{47 \zeta_3}{27}+\frac{5
  \zeta_4}{6}+\frac{13819}{972} \right) C_A^2 n_f^2 + C_A \Delta {A}^{(4)},
  \notag\\
  \vspace{1.5mm}
  B^{(1)} =&
  -\frac{11}{3} C_A + \frac{2}{3}n_f,
  \notag\\
  \vspace{1.5mm}
  B^{(2)} =&
  \left( \frac{11 \zeta _2}{6}-6 \zeta _3-\frac{16}{3} \right) C_A^2
  + \left( \frac{4}{3}-\frac{\zeta _2}{3} \right) C_A n_f
  + C_A C_F,
  \notag\\
  \vspace{1.5mm}
  B^{(3)} =&
  \left( \frac{22 \zeta _3 \zeta _2}{3}-\frac{799 \zeta _2}{81}-\frac{5 \pi ^2 \zeta _3}{9}-\frac{2533 \zeta _3}{54}-\frac{77 \zeta _4}{12}+20 \zeta _5-\frac{319 \pi ^4}{1080}+\frac{6109 \pi
   ^2}{1944}+\frac{34219}{1944} \right) C_A^3
  \notag\\&
  + \left( \frac{103 \zeta _2}{81}+\frac{202 \zeta _3}{27}-\frac{5 \zeta _4}{6}+\frac{41 \pi ^4}{540}-\frac{599 \pi ^2}{972}-\frac{10637}{1944} \right) C_A^2 n_f\notag\\
&  + \left( -\frac{2 \zeta _3}{27}+\frac{5 \pi ^2}{162}+\frac{529}{1944} \right) C_A n_f^2
   + \left( 2 \zeta _4-\frac{\pi ^4}{45}-\frac{\pi ^2}{12}+\frac{241}{72} \right) C_A C_F n_f\notag\\
&  - \frac{1}{4} C_F^2 n_f
  - \frac{11}{36} C_A n_f^2+ C_A \Delta {B}^{(3)}.
\end{align}
The expressions for the coefficients $A^{(i)}$ and $B^{(i)}$ are
extracted from refs.~\cite{deFlorian:2001zd,Becher:2012yn,Li:2016ctv,Vladimirov:2016dll}.
The N$^3$LL anomalous dimension $A^{(4)}$ is at present incomplete since the analytical form of the four-loop cusp anomalous dimension is unknown (its numerical value was recently presented in ref.~\cite{Moch:2017uml,Moch:2018wjh}). 
In the above expressions, this contribution is set to zero. 

The expansion of hard-virtual coefficient function $H$ in powers of the strong coupling is
\begin{equation}
  H(M) =
  1
  +
  \sum_{n=1}^{2} \left( \frac{\as(\mH^2)}{2\pi} \right)^n \, H^{(n)}(\mH),
\end{equation}
with
\begin{align}
\label{eq:H-fun-G}
  H^{(1)}(\mH) =&  C_A\left(5+\frac{7}{6}\pi^2\right)-3 C_F,\notag\\
  H^{(2)}(\mH) =&   \frac{5359}{54} + \frac{137}{6}\ln\frac{m_H^2}{m_t^2} 
+ \frac{1679}{24}\pi^2 + \frac{37}{8}\pi^4- \frac{499}{6}\zeta_3
              + C_A \Delta {H}^{(2)} \,,\qquad n_f=5.
\end{align}
The factors $\tilde{H}$ that appear in the luminosity prefactors (eqs.~\eqref{eq:mod-luminosity-NLL},~\eqref{eq:mod-luminosity-NNLL},~\eqref{eq:mod-luminosity-N3LL}) are defined as
\begin{align}
\tilde{H}^{(1)}(\muR,&x_Q) = H^{(1)}(\muR) +
                             \left(-\frac{1}{2}A^{(1)}\ln x_Q^2 +
                             B^{(1)}\right) \ln x_Q^2,\notag\\
\tilde{H}^{(2)}(\muR,&x_Q) = H^{(2)}(\muR) +
                             \frac{(A^{(1)})^2}{8}\ln^4x_Q^2 - \left(\frac{A^{(1)}
                             B^{(1)}}{2}+\frac{A^{(1)}}{3}\pi\beta_0
                             \right)\ln^3 x_Q^2\notag\\
&+\left(\frac{-A^{(2)}+(B^{(1)})^2}{2} + \pi\beta_0
  \left(B^{(1)}+A^{(1)}\ln \frac{x_Q^2 \mH^2}{\muR^2}\right)\right)\ln^2 x_Q^2\notag\\
& - \left(-B^{(2)}+B^{(1)}2\pi\beta_0\ln \frac{x_Q^2
  \mH^2}{\muR^2}\right)\ln x_Q^2  + H^{(1)}(\muR)\ln x_Q^2\left( -\frac{1}{2}A^{(1)}\ln x_Q^2 +
                             B^{(1)} \right).
\end{align}
The new terms
\begin{equation}
\Delta{A}^{(4)} = -64 \pi^3 \beta_0^3 \zeta_3 ,\qquad \Delta
{B}^{(3)} = -32 \pi^2 \beta_0^2 \zeta_3, \qquad \Delta {H}^{(2)} =\frac{16}{3} \pi
  \beta_0 \zeta_3,
\end{equation}
are a feature due to performing the resummation in momentum space and do not appear in the anomalous dimensions in $b$ space (see ref.~\cite{Bizon:2017rah} for details). 

Finally, we also report the expansion of the collinear coefficient functions $C_{ab}$
\begin{align}
  C_{ab}(z) =& \delta(1-z)\delta_{ab} + \sum_{n=1}^{2} \left( \frac{\as(\mu^2)}{2\pi} \right)^n \,C_{ab}^{(n)}(z),
\end{align}
where $\mu$ is the same scale that enters parton densities. 
The first-order expansion has been known for a long time and reads 
\begin{equation}
\label{eq:coeff-fun}
C_{ab}^{(1)}(z)= - \hat P_{ab}^{(0),\epsilon}(z) - \delta_{ab}\delta(1-z)\frac{\pi^2}{12},
\end{equation}
where $\hat P_{ab}^{(0),\epsilon}(z)$ is the $\mathcal{O}(\epsilon)$
part of the leading-order regularized splitting functions $\hat P_{ab}^{(0)}(z)$
\begin{align}
  &\hat P^{(0)}_{qq}(z)=C_F\left[\frac{1+z^2}{(1-z)_+}+\frac32\delta(1-z)\right],  &\hat P^{(0),\epsilon}_{qq}(z) = -C_F (1-z), \nonumber\\
  &\hat P^{(0)}_{qg}(z)=\frac12\left[z^2+(1-z)^2\right],   &\hat P^{(0),\epsilon}_{qg}(z) = -z(1-z), \nonumber\\
  &\hat P^{(0)}_{gq}(z)=C_F\frac{1+(1-z)^2}{z},   &\hat P^{(0),\epsilon}_{gq}(z) = -C_F z,\nonumber\\
  &\hat P^{(0)}_{gg}(z)=2C_A\left[\frac z{(1-z)_+}+\frac{1-z}z+z(1-z)\right]+2\pi\beta_0\delta(1-z),  & \hat P^{(0),\epsilon}_{gg}(z) = 0.
\end{align}
The corresponding unregularised Altarelli-Parisi splitting functions in four dimensions are
\begin{eqnarray}
\label{eq:regAP}
\tilde  P^{(0)}_{qq}(z)&=&C_F\frac{1+z^2}{1-z},\nonumber\\
\tilde  P^{(0)}_{qg}(z)&=&\frac12\left[z^2+(1-z)^2\right],\nonumber\\
\tilde  P^{(0)}_{gq}(z)&=&C_F\frac{1+(1-z)^2}{z},\nonumber\\
\tilde  P^{(0)}_{gg}(z)&=&C_A\left[\frac z{1-z}+\frac{1-z}z+z(1-z)\right] \rightarrow C_A\left[2\frac{z}{1-z}+z(1-z)\right],
\end{eqnarray}
where in the last step we exploited the symmetry of the $\tilde  P^{(0)}_{gg}(z)$ splitting function in $z \rightarrow 1-z$. 
The second-order collinear coefficient functions $C_{ab}^{(2)}(z)$ and gluon collinear correlation coefficients $G$  (see eqs.~\eqref{eq:mod-luminosity-NLL}, \eqref{eq:mod-luminosity-NNLL}, \eqref{eq:mod-luminosity-N3LL}) for gluon-fusion processes were obtained in refs.~\cite{Catani:2011kr,Gehrmann:2014yya}.
Their expressions is extracted using the results of refs.~\cite{Catani:2011kr,Catani:2012qa}. 
For gluon-fusion processes, the $C^{(2)}_{gq}$ and $C^{(2)}_{gg}$ coefficients normalized as in eq.~\eqref{eq:coeff-fun} are extracted from eqs.~(30) and~(32) of ref.~\cite{Catani:2011kr}, respectively, where the hard coefficients of eqs.~\eqref{eq:H-fun-G} were used {\it without} the new term $\Delta { H}^{(2)}$ in the $H_g^{(2)}(M)$ coefficient\footnote{These must be replaced by $H^{(1)}\to H^{(1)}/2$   and $H^{(2)}\to H^{(2)}/4$ to match the convention of refs.~\cite{Catani:2011kr,Catani:2012qa}.}.
The coefficient $G^{(1)}$ is taken from eq.~(13) of ref.~\cite{Catani:2011kr}. 

The coefficients $\tilde{C}$ in eqs.~\eqref{eq:mod-luminosity-NLL},~\eqref{eq:mod-luminosity-NNLL},~\eqref{eq:mod-luminosity-N3LL} are defined as
\begin{align}
\tilde{C}_{ab}^{(1)}(z,&\muF,x_Q) = C_{ab}^{(1)}(z) +
                                    \hat{P}_{ab}^{(0)}(z)\ln\frac{x_Q^2 \mH^2}{\muF^2},\notag\\
\tilde{C}_{ab}^{(2)}(z,&\muF,x_Q) = C_{ab}^{(2)}(z) +
                                    \pi\beta_0 \hat{P}_{ab}^{(0)}(z)\left(
                                    \ln^2\frac{x_Q^2 \mH^2}{\muF^2} -
                                   2 \ln\frac{x_Q^2 \mH^2}{\muF^2}
                                    \ln\frac{x_Q^2
                                    \mH^2}{\muR^2}\right) +
                                    \hat{P}_{ab}^{(1)}(z)\ln\frac{x_Q^2
                                    \mH^2}{\muF^2} \notag\\
& + \frac{1}{2}(\hat{P}^{(0)}\otimes \hat{P}^{(0)})_{ab}(z) \ln^2\frac{x_Q^2
  \mH^2}{\muF^2} + (C^{(1)}\otimes \hat{P}^{(0)})_{ab}(z) \ln\frac{x_Q^2
  \mH^2}{\muF^2} - 2\pi\beta_0 C_{ab}^{(1)}(z) \ln\frac{x_Q^2
  \mH^2}{\muR^2}.
\end{align}

\subsection{Formul\ae\ for matching schemes}
\label{app:matching}

This section contains the necessary formul\ae\ to implement the matching schemes defined in eqs.~\eqref{eq:additive} and~\eqref{eq:multiplicative1}.
Let us first introduce a convenient notation for the perturbative expansion of the various ingredients, defining 
\begin{align}
\sigma_{\rm tot}^{\rm N^3LO}  = \sum_{i=0}^3\sigma^{(i)},\qquad
\Sigma^{\rm N^3LO}(v) = \sigma^{(0)} + \sum_{i=1}^3\Sigma^{(i)}(v),
\end{align}
where 
\begin{align}
\Sigma^{(i)}(v) = \sigma^{(i)} + \bar{\Sigma}^{(i)}(v),\qquad \bar{\Sigma}^{(i)}(v) \equiv - \int_{v}^{\infty} \rd v' \;\frac{\rd \Sigma^{(i)}(v')}{\rd v'}.
\end{align}
Moreover, we denote the perturbative expansion of the resummed cross section $\Sigma^{\rm   N^kLL}$ as
\begin{equation}
\Sigma^{\rm expanded}(v) = \sigma^{(0)} + \sum_{i=1}^3 \Sigma_{\rm N^kLL}^{(i)}(v).
\end{equation}
With this notation, the additive scheme of eq.~\eqref{eq:additive} becomes (for simplicity the explicit dependence on $v$ will be dropped in the following)
\begin{align}
          \Sigma_{\rm add}^{\rm matched} =& \Sigma^{\rm N^kLL}+ \left\{\sigma^{(1)}+\bar \Sigma^{(1)}- \Sigma_{\rm N^kLL}^{(1)}\right\}+  \left\{\sigma^{(2)}+\bar  \Sigma^{(2)}
                - \Sigma_{\rm N^kLL}^{(2)}\right\} +  \left\{\sigma^{(3)}+\bar  \Sigma^{(3)}
                - \Sigma_{\rm N^kLL}^{(3)}\right\},
\end{align}
where the three terms in curly brackets denote the NLO, NNLO and N$^3$LO contributions to the matching, respectively.

For the multiplicative scheme one needs to introduce the asymptotic expansion $\Sigma^{\rm N^kLL}_{\rm asym.} $, defined in eq.~\eqref{eq:asypt} (the definition for $k\neq 3$ is analogous with obvious replacements) in terms of the $\tilde{L}\to 0$ limit of the coefficients $\tilde{\cal L}_{\rm N^kLL}$ of eqs.~\eqref{eq:mod-luminosity-NLL},~\eqref{eq:mod-luminosity-NNLL},~\eqref{eq:mod-luminosity-N3LL}, which read
\begin{align}
\label{eq:lumi_asympt}
\tilde{\cal L}_{\rm NLL}^{\tilde{L}\to 0} &= \sum_{c,
  c'}\frac{\rd|\mathcal{M}_{B}|_{cc'}^2}{\rd\Phi_B}
  f_c\!\left(x_1,\muF^2\right)f_{c'}\!\left(x_2,\muF^2\right),\notag\\
\tilde{\cal L}_{\rm NNLL}^{\tilde{L}\to 0} &= \sum_{c,
                                             c'}\frac{\rd|\mathcal{M}_{B}|_{cc'}^2}{\rd\Phi_B}
                                             \sum_{i,
                                             j}\int_{x_1}^{1}\frac{\rd
                                             z_1}{z_1}\int_{x_2}^{1}\frac{\rd
                                             z_2}{z_2}f_i\!\left(\frac{x_1}{z_1},\muF^2\right)f_{j}\!\left(\frac{x_2}{z_2},\muF^2\right)\notag\\
&\times\Bigg\{\delta_{ci}\delta_{c'j}\delta(1-z_1)\delta(1-z_2)
\left(1+\frac{\as(\muR^2)}{2\pi} \tilde{H}^{(1)}(\muR,x_Q)\right) \notag\\
&+ \frac{\as(\muR^2)}{2\pi}\left(\tilde{C}_{c i}^{(1)}(z_1,\muF,x_Q)\delta(1-z_2)\delta_{c'j}+
  \{z_1\leftrightarrow z_2; c,i \leftrightarrow c'j\}\right)\Bigg\},\notag\\
\tilde{\cal L}_{\rm N^3LL}^{\tilde{L}\to 0} & =\sum_{c,
  c'}\frac{\rd|\mathcal{M}_{B}|_{cc'}^2}{\rd\Phi_B} \sum_{i, j}\int_{x_1}^{1}\frac{\rd
  z_1}{z_1}\int_{x_2}^{1}\frac{\rd z_2}{z_2}f_i\!\left(\frac{x_1}{z_1},\muF^2\right)f_{j}\!\left(\frac{x_2}{z_2},\muF^2\right)\notag\\&\times\Bigg\{\delta_{ci}\delta_{c'j}\delta(1-z_1)\delta(1-z_2)
\left(1+\frac{\as(\muR^2)}{2\pi} \tilde{H}^{(1)}(\muR,x_Q) + \frac{\as^2(\muR^2)}{(2\pi)^2} \tilde{H}^{(2)}(\muR,x_Q)\right) \notag\\
&+ \frac{\as(\muR^2)}{2\pi}\left(\tilde{C}_{c i}^{(1)}(z_1,\muF,x_Q)\delta(1-z_2)\delta_{c'j}+ \{z_1\leftrightarrow z_2; c,i \leftrightarrow c',j\}\right)\notag\\
& +
  \frac{\as^2(\muR^2)}{(2\pi)^2}\left(\tilde{C}_{c i}^{(2)}(z_1,\muF,x_Q)\delta(1-z_2)\delta_{c'j} + \{z_1\leftrightarrow z_2; c,i \leftrightarrow c',j\}\right) \notag\\&+  \frac{\as^2(\muR^2)}{(2\pi)^2}\Big(\tilde{C}_{c i}^{(1)}(z_1,\muF,x_Q)\tilde{C}_{c' j}^{(1)}(z_2,\muF,x_Q) + G_{c i}^{(1)}(z_1)G_{c' j}^{(1)}(z_2)\Big) \notag\\
& + \frac{\as^2(\muR^2)}{(2\pi)^2} \tilde{H}^{(1)}(\muR,x_Q)\Big(\tilde{C}_{c i}^{(1)}(z_1,\muF,x_Q)\delta(1-z_2)\delta_{c'j} + \{z_1\leftrightarrow z_2; c,i \leftrightarrow c',j\}\Big) \Bigg\}.
\end{align}
By denoting the perturbative expansion of $\Sigma^{\rm   N^kLL}_{\rm asym.}$ as
\begin{equation}
\Sigma^{\rm
  N^kLL}_{\rm asym.} = \sigma^{(0)} + \sum_{i=1}^{k-1}\Sigma_{\rm asym.}^{(i)},
\end{equation}
the matching formula~\eqref{eq:multiplicative1} reads
\begin{align}
&	\Sigma_{\rm mult}^{\rm matched}(v) =  \frac{\Sigma^{\rm N^kLL}}{\Sigma^{\rm
  N^kLL}_{\rm asym.}}\Bigg[\sigma^{(0)}+ \left\{\sigma^{(1)}+\bar \Sigma^{(1)} +\Sigma_{\rm asym.}^{(1)}- \Sigma_{\rm N^kLL}^{(1)} \right\}\nonumber \\ \nonumber \\
	&+ \left\{\sigma^{(2)} + \bar \Sigma^{(2)} +\Sigma_{\rm asym.}^{(2)}- \Sigma_{\rm N^kLL}^{(2)} +\frac{\Sigma_{\rm asym.}^{(1)}}{\sigma^{(0)}}\left(\sigma^{(1)}+\bar \Sigma^{(1)}\right) 
       + \frac{(\Sigma_{\rm N^kLL}^{(1)})^2}{\sigma^{(0)}} - \frac{\Sigma_{\rm N^kLL}^{(1)}}{\sigma^{(0)}}\left(\sigma^{(1)}+\bar \Sigma^{(1)}+\Sigma_{\rm asym.}^{(1)}\right) \right\}\nonumber \\ \nonumber \\
       &+\Bigg\{ \sigma^{(3)} + \bar \Sigma^{(3)} - \Sigma_{\rm N^kLL}^{(3)} 
  - \frac{(\Sigma_{\rm N^kLL}^{(1)})^3}{(\sigma^{(0)})^2} + \frac{(\Sigma_{\rm N^kLL}^{(1)})^2}{(\sigma^{(0)})^2}\left(\sigma^{(1)}+\bar \Sigma^{(1)} + \Sigma_{\rm asym.}^{(1)}\right) \nonumber \\
  &+\frac{1}{\sigma_0}\left((\sigma^{(1)}+\bar \Sigma^{(1)})(\Sigma_{\rm asym.}^{(2)}-\Sigma_{\rm N^kLL}^{(2)}) 
 +\Sigma_{\rm asym.}^{(1)}(\sigma^{(2)}+\bar \Sigma^{(2)} - \Sigma_{\rm N^kLL}^{(2)})\right) \nonumber\\
 &- \frac{1}{(\sigma^{(0)})^2}\Sigma_{\rm N^kLL}^{(1)}\left(\Sigma_{\rm asym.}^{(1)}(\sigma^{(1)}+\bar \Sigma^{(1)}) + \sigma^{(0)}(\sigma^{(2)}+\bar \Sigma^{(2)} + \Sigma_{\rm asym.}^{(2)}- 2\Sigma_{\rm N^kLL}^{(2)})\right) \Bigg\}\Bigg],
\end{align}
where, as above, the terms entering at NLO, NNLO, and N$^3$LO are grouped within curly brackets.


\begin{savequote}[8cm]
A man must read widely, a little of everything or whatever he can, but given the shortness of life and the verbosity of the world, not too much should be demanded of him. Let him begin with those titles no one should omit, commonly referred to as books for learning, as if not all books were for learning \dots
  \qauthor{--- Jos\'e Saramago, \textit{The Year of the Death of Ricardo Reis}}
\end{savequote}

\setlength{\baselineskip}{0pt} 


{\small
\providecommand{\href}[2]{#2}\begingroup\raggedright\endgroup

}

%

\end{document}